\catcode`!=11 %  ***** THIS MUST NEVER BE OMITTED
\def\PiC{P\kern-.12em\lower.5ex\hbox{I}\kern-.075emC}
\def\PiCTeX{\PiC\kern-.11em\TeX}

\def\!ifnextchar#1#2#3{%
  \let\!testchar=#1%
  \def\!first{#2}%
  \def\!second{#3}%
  \futurelet\!nextchar\!testnext}
\def\!testnext{%
  \ifx \!nextchar \!spacetoken 
    \let\!next=\!skipspacetestagain
  \else
    \ifx \!nextchar \!testchar
      \let\!next=\!first
    \else 
      \let\!next=\!second 
    \fi 
  \fi
  \!next}
\def\\{\!skipspacetestagain} 
  \expandafter\def\\ {\futurelet\!nextchar\!testnext} 
\def\\{\let\!spacetoken= } \\  %  ** set \spacetoken to a space token

\def\!tfor#1:=#2\do#3{%
  \edef\!fortemp{#2}%
  \ifx\!fortemp\!empty 
    \else
    \!tforloop#2\!nil\!nil\!!#1{#3}%
  \fi}
\def\!tforloop#1#2\!!#3#4{%
  \def#3{#1}%
  \ifx #3\!nnil
    \let\!nextwhile=\!fornoop
  \else
    #4\relax
    \let\!nextwhile=\!tforloop
  \fi 
  \!nextwhile#2\!!#3{#4}}

\def\!etfor#1:=#2\do#3{%
  \def\!!tfor{\!tfor#1:=}%
  \edef\!!!tfor{#2}%
  \expandafter\!!tfor\!!!tfor\do{#3}}

\def\!cfor#1:=#2\do#3{%
  \edef\!fortemp{#2}%
  \ifx\!fortemp\!empty 
  \else
    \!cforloop#2,\!nil,\!nil\!!#1{#3}%
  \fi}
\def\!cforloop#1,#2\!!#3#4{%
  \def#3{#1}%
  \ifx #3\!nnil
    \let\!nextwhile=\!fornoop 
  \else
    #4\relax
    \let\!nextwhile=\!cforloop
  \fi
  \!nextwhile#2\!!#3{#4}}

\def\!ecfor#1:=#2\do#3{%
  \def\!!cfor{\!cfor#1:=}%
  \edef\!!!cfor{#2}%
  \expandafter\!!cfor\!!!cfor\do{#3}}

\def\!empty{}
\def\!nnil{\!nil}
\def\!fornoop#1\!!#2#3{}

\def\!ifempty#1#2#3{%
  \edef\!emptyarg{#1}%
  \ifx\!emptyarg\!empty
    #2%
  \else
    #3%
  \fi}

\def\!getnext#1\from#2{%
  \expandafter\!gnext#2\!#1#2}%
\def\!gnext\\#1#2\!#3#4{%
  \def#3{#1}%
  \def#4{#2\\{#1}}%
  \ignorespaces}

\def\!getnextvalueof#1\from#2{%
  \expandafter\!gnextv#2\!#1#2}%
\def\!gnextv\\#1#2\!#3#4{%
  #3=#1%
  \def#4{#2\\{#1}}%
  \ignorespaces}

\def\!copylist#1\to#2{%
  \expandafter\!!copylist#1\!#2}
\def\!!copylist#1\!#2{%
  \def#2{#1}\ignorespaces}

\def\!wlet#1=#2{%
  \let#1=#2 
  \wlog{\string#1=\string#2}}

\def\!listaddon#1#2{%
  \expandafter\!!listaddon#2\!{#1}#2}
\def\!!listaddon#1\!#2#3{%
  \def#3{#1\\#2}}

\def\!rightappend#1\withCS#2\to#3{\expandafter\!!rightappend#3\!#2{#1}#3}
\def\!!rightappend#1\!#2#3#4{\def#4{#1#2{#3}}}

\def\!leftappend#1\withCS#2\to#3{\expandafter\!!leftappend#3\!#2{#1}#3}
\def\!!leftappend#1\!#2#3#4{\def#4{#2{#3}#1}}

\def\!lop#1\to#2{\expandafter\!!lop#1\!#1#2}
\def\!!lop\\#1#2\!#3#4{\def#4{#1}\def#3{#2}}

\def\!loop#1\repeat{\def\!body{#1}\!iterate}
\def\!iterate{\!body\let\!next=\!iterate\else\let\!next=\relax\fi\!next}

\def\!!loop#1\repeat{\def\!!body{#1}\!!iterate}
\def\!!iterate{\!!body\let\!!next=\!!iterate\else\let\!!next=\relax\fi\!!next}
\def\!removept#1#2{\edef#2{\expandafter\!!removePT\the#1}}
{\catcode`p=12 \catcode`t=12 \gdef\!!removePT#1pt{#1}}

\def\placevalueinpts of <#1> in #2 {%
  \!removept{#1}{#2}}

\def\!mlap#1{\hbox to 0pt{\hss#1\hss}}
\def\!vmlap#1{\vbox to 0pt{\vss#1\vss}}

\def\!not#1{%
  #1\relax
    \!switchfalse
  \else
    \!switchtrue
  \fi
  \if!switch
  \ignorespaces}

\let\!!!wlog=\wlog              % "\wlog" is defined in plain TeX
\def\wlog#1{}    

\newdimen\headingtoplotskip     %.A.................
\newdimen\linethickness         %.A..X....U........T
\newdimen\longticklength        %.A................T
\newdimen\plotsymbolspacing     %......D...L....Q...
\newdimen\shortticklength       %.A................T
\newdimen\stackleading          %.A..........P......
\newdimen\tickstovaluesleading  %.A................T
\newdimen\totalarclength        %......D...L....Q...
\newdimen\valuestolabelleading  %.A.................

\newbox\!boxA                   %.AW...............T
\newbox\!boxB                   %..W................
\newbox\!picbox                 %............P......
\newbox\!plotsymbol             %..........L..O.....
\newbox\!putobject              %............PO...S.
\newbox\!shadesymbol            %.................S.

\newcount\!countA               %.A....D..UL....Q.ST
\newcount\!countB               %......D..U.....Q.ST
\newcount\!countC               %...............Q..T
\newcount\!countD               %...................
\newcount\!countE               %.............O....T
\newcount\!countF               %.............O....T
\newcount\!countG               %..................T
\newcount\!fiftypt              %.........U.........
\newcount\!intervalno           %..........L....Q...
\newcount\!npoints              %..........L........
\newcount\!nsegments            %.........U.........
\newcount\!ntemp                %............P......
\newcount\!parity               %.................S.
\newcount\!scalefactor          %..................T
\newcount\!tfs                  %.......V...........
\newcount\!tickcase             %..................T

\newdimen\!Xleft                %............P......
\newdimen\!Xright               %............P......
\newdimen\!Xsave                %.A................T
\newdimen\!Ybot                 %............P......
\newdimen\!Ysave                %.A................T
\newdimen\!Ytop                 %............P......
\newdimen\!angle                %........E..........
\newdimen\!arclength            %..W......UL....Q...
\newdimen\!areabloc             %.A........L........
\newdimen\!arealloc             %.A........L........
\newdimen\!arearloc             %.A........L........
\newdimen\!areatloc             %.A........L........
\newdimen\!bshrinkage           %.................S.
\newdimen\!checkbot             %..........L........
\newdimen\!checkleft            %..........L........
\newdimen\!checkright           %..........L........
\newdimen\!checktop             %..........L........
\newdimen\!dimenA               %.AW.X.DVEUL..OYQRST
\newdimen\!dimenB               %....X.DVEU...O.QRS.
\newdimen\!dimenC               %..W.X.DVEU......RS.
\newdimen\!dimenD               %..W.X.DVEU....Y.RS.
\newdimen\!dimenE               %..W........G..YQ.S.
\newdimen\!dimenF               %...........G..YQ.S.
\newdimen\!dimenG               %...........G..YQ.S.
\newdimen\!dimenH               %...........G..Y..S.
\newdimen\!dimenI               %...BX.........Y....
\newdimen\!distacross           %..........L....Q...
\newdimen\!downlength           %..........L........
\newdimen\!dp                   %.A..X.......P....S.
\newdimen\!dshade               %.................S.
\newdimen\!dxpos                %..W......U..P....S.
\newdimen\!dxprime              %...............Q...
\newdimen\!dypos                %..WB.....U..P......
\newdimen\!dyprime              %...............Q...
\newdimen\!ht                   %.A..X.......P....S.
\newdimen\!leaderlength         %......D..U.........
\newdimen\!lshrinkage           %.................S.
\newdimen\!midarclength         %...............Q...
\newdimen\!offset               %.A................T
\newdimen\!plotheadingoffset    %.A.................
\newdimen\!plotsymbolxshift     %..........L..O.....
\newdimen\!plotsymbolyshift     %..........L..O.....
\newdimen\!plotxorigin          %..........L..O.....
\newdimen\!plotyorigin          %..........L..O.....
\newdimen\!rootten              %...........G.......
\newdimen\!rshrinkage           %.................S.
\newdimen\!shadesymbolxshift    %.................S.
\newdimen\!shadesymbolyshift    %.................S.
\newdimen\!tenAa                %...........G.......
\newdimen\!tenAc                %...........G.......
\newdimen\!tenAe                %...........G.......
\newdimen\!tshrinkage           %.................S.
\newdimen\!uplength             %..........L........
\newdimen\!wd                   %....X.......P....S.
\newdimen\!wmax                 %...............Q...
\newdimen\!wmin                 %...............Q...
\newdimen\!xB                   %...............Q...
\newdimen\!xC                   %...............Q...
\newdimen\!xE                   %..W.....E.L....Q.S.
\newdimen\!xM                   %..W.....E......Q.S.
\newdimen\!xS                   %..W.....E.L....Q.S.
\newdimen\!xaxislength          %.A................T
\newdimen\!xdiff                %..........L........
\newdimen\!xleft                %............P......
\newdimen\!xloc                 %..WB.....U.......S.
\newdimen\!xorigin              %.A........L.P....S.
\newdimen\!xpivot               %................R..
\newdimen\!xpos                 %..........L.P..Q.ST
\newdimen\!xprime               %...............Q...
\newdimen\!xright               %............P......
\newdimen\!xshade               %.................S.
\newdimen\!xshift               %..W.........PO...S.
\newdimen\!xtemp                %............P......
\newdimen\!xunit                %.AWBX...EUL.P..QRS.
\newdimen\!xxE                  %........E..........
\newdimen\!xxM                  %........E..........
\newdimen\!xxS                  %........E..........
\newdimen\!xxloc                %..WB....EU.........
\newdimen\!yB                   %...............Q...
\newdimen\!yC                   %...............Q...
\newdimen\!yE                   %..W.....E.L....Q...
\newdimen\!yM                   %..W.....E......Q...
\newdimen\!yS                   %..W.....E.L....Q...
\newdimen\!yaxislength          %.A................T
\newdimen\!ybot                 %............P......
\newdimen\!ydiff                %..........L........
\newdimen\!yloc                 %..WB.....U.......S.
\newdimen\!yorigin              %.A........L.P....S.
\newdimen\!ypivot               %................R..
\newdimen\!ypos                 %..........L.P..Q.ST
\newdimen\!yprime               %...............Q...
\newdimen\!yshade               %.................S.
\newdimen\!yshift               %..W.........PO...S.
\newdimen\!ytemp                %............P......
\newdimen\!ytop                 %............P......
\newdimen\!yunit                %.AWBX...EUL.P..QRS.
\newdimen\!yyE                  %........E..........
\newdimen\!yyM                  %........E..........
\newdimen\!yyS                  %........E..........
\newdimen\!yyloc                %..WB....EU.........
\newdimen\!zpt                  %.AWBX.DVEULGP.YQ.ST

\newif\if!axisvisible           %.A.................
\newif\if!gridlinestoo          %..................T
\newif\if!keepPO                %...................
\newif\if!placeaxislabel        %.A.................
\newif\if!switch                %H..................
\newif\if!xswitch               %.A................T

\newtoks\!axisLaBeL             %.A.................
\newtoks\!keywordtoks           %.A.................

\newwrite\!replotfile           %.............O.....

\newhelp\!keywordhelp{The keyword mentioned in the error message in unknown. 
Replace NEW KEYWORD in the indicated response by the keyword that 
should have been specified.}    %.A.................

\!wlet\!!origin=\!xM                   %.A................T
\!wlet\!!unit=\!uplength               %.A................T
\!wlet\!Lresiduallength=\!dimenG       %.........U.........
\!wlet\!Rresiduallength=\!dimenF       %.........U.........
\!wlet\!axisLength=\!distacross        %.A................T
\!wlet\!axisend=\!ydiff                %.A................T
\!wlet\!axisstart=\!xdiff              %.A................T
\!wlet\!axisxlevel=\!arclength         %.A................T
\!wlet\!axisylevel=\!downlength        %.A................T
\!wlet\!beta=\!dimenE                  %...............Q...
\!wlet\!gamma=\!dimenF                 %...............Q...
\!wlet\!shadexorigin=\!plotxorigin     %.................S.
\!wlet\!shadeyorigin=\!plotyorigin     %.................S.
\!wlet\!ticklength=\!xS                %..................T
\!wlet\!ticklocation=\!xE              %..................T
\!wlet\!ticklocationincr=\!yE          %..................T
\!wlet\!tickwidth=\!yS                 %..................T
\!wlet\!totalleaderlength=\!dimenE     %.........U.........
\!wlet\!xone=\!xprime                  %....X..............
\!wlet\!xtwo=\!dxprime                 %....X..............
\!wlet\!ySsave=\!yM                    %...................
\!wlet\!ybB=\!yB                       %.................S.
\!wlet\!ybC=\!yC                       %.................S.
\!wlet\!ybE=\!yE                       %.................S.
\!wlet\!ybM=\!yM                       %.................S.
\!wlet\!ybS=\!yS                       %.................S.
\!wlet\!ybpos=\!yyloc                  %.................S.
\!wlet\!yone=\!yprime                  %....X..............
\!wlet\!ytB=\!xB                       %.................S.
\!wlet\!ytC=\!xC                       %.................S.
\!wlet\!ytE=\!downlength               %.................S.
\!wlet\!ytM=\!arclength                %.................S.
\!wlet\!ytS=\!distacross               %.................S.
\!wlet\!ytpos=\!xxloc                  %.................S.
\!wlet\!ytwo=\!dyprime                 %....X..............

\!zpt=0pt                              % static
\!xunit=1pt
\!yunit=1pt
\!arearloc=\!xunit
\!areatloc=\!yunit
\!dshade=5pt
\!leaderlength=24in
\!tfs=256                              % static
\!wmax=5.3pt                           % static
\!wmin=2.7pt                           % static
\!xaxislength=\!xunit
\!xpivot=\!zpt
\!yaxislength=\!yunit 
\!ypivot=\!zpt
  \!dimenA=50pt \!fiftypt=\!dimenA     % static

\!rootten=3.162278pt                   % static
\!tenAa=8.690286pt                     % static  (A5)
\!tenAc=2.773839pt                     % static  (A3)
\!tenAe=2.543275pt                     % static  (A1)

\def\!cosrotationangle{1}      %................R..
\def\!sinrotationangle{0}      %................R..
\def\!xpivotcoord{0}           %................R..
\def\!xref{0}                  %............P......
\def\!xshadesave{0}            %.................S.
\def\!ypivotcoord{0}           %................R..
\def\!yref{0}                  %............P......
\def\!yshadesave{0}            %.................S.
\def\!zero{0}                  %..................T

\let\wlog=\!!!wlog
\def\normalgraphs{%
  \longticklength=.4\baselineskip
  \shortticklength=.25\baselineskip
  \tickstovaluesleading=.25\baselineskip
  \valuestolabelleading=.8\baselineskip
  \linethickness=.4pt
  \stackleading=.17\baselineskip
  \headingtoplotskip=1.5\baselineskip
  \visibleaxes
  \ticksout
  \nogridlines
  \unloggedticks}
\def\setplotarea x from #1 to #2, y from #3 to #4 {%
  \!arealloc=\!M{#1}\!xunit \advance \!arealloc -\!xorigin
  \!areabloc=\!M{#3}\!yunit \advance \!areabloc -\!yorigin
  \!arearloc=\!M{#2}\!xunit \advance \!arearloc -\!xorigin
  \!areatloc=\!M{#4}\!yunit \advance \!areatloc -\!yorigin
  \!initinboundscheck
  \!xaxislength=\!arearloc  \advance\!xaxislength -\!arealloc
  \!yaxislength=\!areatloc  \advance\!yaxislength -\!areabloc
  \!plotheadingoffset=\!zpt
  \!dimenput {{\setbox0=\hbox{}\wd0=\!xaxislength\ht0=\!yaxislength\box0}}
     [bl] (\!arealloc,\!areabloc)}
\def\visibleaxes{%
  \def\!axisvisibility{\!axisvisibletrue}}

\def\!fixkeyword#1{%
  \errhelp=\!keywordhelp
  \errmessage{Unrecognized keyword `#1': \the\!keywordtoks{NEW KEYWORD}'}}

\!keywordtoks={enter `i\fixkeyword}

\def\fixkeyword#1{%
  \!nextkeyword#1 }

\def\axis {%
  \def\!nextkeyword##1 {%
    \expandafter\ifx\csname !axis##1\endcsname \relax
      \def\!next{\!fixkeyword{##1}}%
    \else
      \def\!next{\csname !axis##1\endcsname}%
    \fi
    \!next}%
  \!offset=\!zpt
  \!axisvisibility
  \!placeaxislabelfalse
  \!nextkeyword}

\def\!axisbottom{%
  \!axisylevel=\!areabloc
  \def\!tickxsign{0}%
  \def\!tickysign{-}%
  \def\!axissetup{\!axisxsetup}%
  \def\!axislabeltbrl{t}%
  \!nextkeyword}

\def\!axistop{%
  \!axisylevel=\!areatloc
  \def\!tickxsign{0}%
  \def\!tickysign{+}%
  \def\!axissetup{\!axisxsetup}%
  \def\!axislabeltbrl{b}%
  \!nextkeyword}

\def\!axisleft{%
  \!axisxlevel=\!arealloc
  \def\!tickxsign{-}%
  \def\!tickysign{0}%
  \def\!axissetup{\!axisysetup}%
  \def\!axislabeltbrl{r}%
  \!nextkeyword}

\def\!axisright{%
  \!axisxlevel=\!arearloc
  \def\!tickxsign{+}%
  \def\!tickysign{0}%
  \def\!axissetup{\!axisysetup}%
  \def\!axislabeltbrl{l}%
  \!nextkeyword}

\def\!axisshiftedto#1=#2 {%
  \if 0\!tickxsign
    \!axisylevel=\!M{#2}\!yunit
    \advance\!axisylevel -\!yorigin
  \else
    \!axisxlevel=\!M{#2}\!xunit
    \advance\!axisxlevel -\!xorigin
  \fi
  \!nextkeyword}

\def\!axisvisible{%
  \!axisvisibletrue  
  \!nextkeyword}

\def\!axisinvisible{%
  \!axisvisiblefalse
  \!nextkeyword}

\def\!axislabel#1 {%
  \!axisLaBeL={#1}%
  \!placeaxislabeltrue
  \!nextkeyword}

\expandafter\def\csname !axis/\endcsname{%
  \!axissetup % This could done already by "ticks"; if so, now \relax
  \if!placeaxislabel
    \!placeaxislabel
  \fi
  \if +\!tickysign %                 ** (A "top" axis)
    \!dimenA=\!axisylevel
    \advance\!dimenA \!offset %      ** dimA = top of the axis structure
    \advance\!dimenA -\!areatloc %   ** dimA = excess over the plot area
    \ifdim \!dimenA>\!plotheadingoffset
      \!plotheadingoffset=\!dimenA % ** Greatest excess over the plot area
    \fi
  \fi}

\def\grid #1 #2 {%
  \!countA=#1\advance\!countA 1
  \axis bottom invisible ticks length <\!zpt> andacross quantity {\!countA} /
  \!countA=#2\advance\!countA 1
  \axis left   invisible ticks length <\!zpt> andacross quantity {\!countA} / }

\def\plotheading#1 {%
  \advance\!plotheadingoffset \headingtoplotskip
  \!dimenput {#1} [B] <.5\!xaxislength,\!plotheadingoffset>
    (\!arealloc,\!areatloc)}

\def\!axisxsetup{%
  \!axisxlevel=\!arealloc
  \!axisstart=\!arealloc
  \!axisend=\!arearloc
  \!axisLength=\!xaxislength
  \!!origin=\!xorigin
  \!!unit=\!xunit
  \!xswitchtrue
  \if!axisvisible 
    \!makeaxis
  \fi}

\def\!axisysetup{%
  \!axisylevel=\!areabloc
  \!axisstart=\!areabloc
  \!axisend=\!areatloc
  \!axisLength=\!yaxislength
  \!!origin=\!yorigin
  \!!unit=\!yunit
  \!xswitchfalse
  \if!axisvisible
    \!makeaxis
  \fi}

\def\!makeaxis{%
  \setbox\!boxA=\hbox{% (Make a pseudo-y[x] tick for an x[y]-axis)
    \beginpicture
      \!setdimenmode
      \setcoordinatesystem point at {\!zpt} {\!zpt}   
      \putrule from {\!zpt} {\!zpt} to
        {\!tickysign\!tickysign\!axisLength} 
        {\!tickxsign\!tickxsign\!axisLength}
    \endpicturesave <\!Xsave,\!Ysave>}%
    \wd\!boxA=\!zpt
    \!placetick\!axisstart}

\def\!placeaxislabel{%
  \advance\!offset \valuestolabelleading
  \if!xswitch
    \!dimenput {\the\!axisLaBeL} [\!axislabeltbrl]
      <.5\!axisLength,\!tickysign\!offset> (\!axisxlevel,\!axisylevel)
    \advance\!offset \!dp  % ** advance offset by the "tallness"
    \advance\!offset \!ht  % ** of the label
  \else
    \!dimenput {\the\!axisLaBeL} [\!axislabeltbrl]
      <\!tickxsign\!offset,.5\!axisLength> (\!axisxlevel,\!axisylevel)
  \fi
  \!axisLaBeL={}}

\def\arrow <#1> [#2,#3]{%
  \!ifnextchar<{\!arrow{#1}{#2}{#3}}{\!arrow{#1}{#2}{#3}<\!zpt,\!zpt> }}

\def\!arrow#1#2#3<#4,#5> from #6 #7 to #8 #9 {%
  \!xloc=\!M{#8}\!xunit   
  \!yloc=\!M{#9}\!yunit
  \!dxpos=\!xloc  \!dimenA=\!M{#6}\!xunit  \advance \!dxpos -\!dimenA
  \!dypos=\!yloc  \!dimenA=\!M{#7}\!yunit  \advance \!dypos -\!dimenA
  \let\!MAH=\!M%                         ** save current c/d mode
  \!setdimenmode%                        ** go into dimension mode
  \!xshift=#4\relax  \!yshift=#5\relax%  ** pick up shift
  \!reverserotateonly\!xshift\!yshift%   ** back rotate shift
  \advance\!xshift\!xloc  \advance\!yshift\!yloc
  \!xS=-\!dxpos  \advance\!xS\!xshift
  \!yS=-\!dypos  \advance\!yS\!yshift
  \!start (\!xS,\!yS)
  \!ljoin (\!xshift,\!yshift)
  \!Pythag\!dxpos\!dypos\!arclength
  \!divide\!dxpos\!arclength\!dxpos  
  \!dxpos=32\!dxpos  \!removept\!dxpos\!!cos
  \!divide\!dypos\!arclength\!dypos  
  \!dypos=32\!dypos  \!removept\!dypos\!!sin
  \!halfhead{#1}{#2}{#3}%                ** draw half of arrow head
  \!halfhead{#1}{-#2}{-#3}%              ** draw other half
  \let\!M=\!MAH%                         ** restore old c/d mode
  \ignorespaces}
  \def\!halfhead#1#2#3{%
    \!dimenC=-#1%                
    \divide \!dimenC 2 %                 ** half way back
    \!dimenD=#2\!dimenC%                 ** half the mid width
    \!rotate(\!dimenC,\!dimenD)by(\!!cos,\!!sin)to(\!xM,\!yM)
    \!dimenC=-#1%                        ** all the way back
    \!dimenD=#3\!dimenC
    \!dimenD=.5\!dimenD%                 ** half the full width
    \!rotate(\!dimenC,\!dimenD)by(\!!cos,\!!sin)to(\!xE,\!yE)
    \!start (\!xshift,\!yshift)
    \advance\!xM\!xshift  \advance\!yM\!yshift
    \advance\!xE\!xshift  \advance\!yE\!yshift
    \!qjoin (\!xM,\!yM) (\!xE,\!yE) 
    \ignorespaces}

\def\betweenarrows #1#2 from #3 #4 to #5 #6 {%
  \!xloc=\!M{#3}\!xunit  \!xxloc=\!M{#5}\!xunit%   
  \!yloc=\!M{#4}\!yunit  \!yyloc=\!M{#6}\!yunit%           
  \!dxpos=\!xxloc  \advance\!dxpos by -\!xloc
  \!dypos=\!yyloc  \advance\!dypos by -\!yloc
  \advance\!xloc .5\!dxpos
  \advance\!yloc .5\!dypos
  \let\!MBA=\!M%           ** save current coord\dimen mode
  \!setdimenmode%          ** express locations in dimens
  \ifdim\!dypos=\!zpt
    \ifdim\!dxpos<\!zpt \!dxpos=-\!dxpos \fi
    \put {\!lrarrows{\!dxpos}{#1}}#2{} at {\!xloc} {\!yloc}
  \else
    \ifdim\!dxpos=\!zpt
      \ifdim\!dypos<\!zpt \!dypos=-\!zpt \fi
      \put {\!udarrows{\!dypos}{#1}}#2{} at {\!xloc} {\!yloc}
    \fi
  \fi
  \let\!M=\!MBA%           ** restore previous c/d mode
  \ignorespaces}

\def\!lrarrows#1#2{% #1=width, #2=text
  {\setbox\!boxA=\hbox{$\mkern-2mu\mathord-\mkern-2mu$}%
   \setbox\!boxB=\hbox{$\leftarrow$}\!dimenE=\ht\!boxB
   \setbox\!boxB=\hbox{}\ht\!boxB=2\!dimenE
   \hbox to #1{$\mathord\leftarrow\mkern-6mu
     \cleaders\copy\!boxA\hfil
     \mkern-6mu\mathord-$%
     \kern.4em $\vcenter{\box\!boxB}$$\vcenter{\hbox{#2}}$\kern.4em
     $\mathord-\mkern-6mu
     \cleaders\copy\!boxA\hfil
     \mkern-6mu\mathord\rightarrow$}}}

\def\!udarrows#1#2{% #1=width, #2=text
  {\setbox\!boxB=\hbox{#2}%
   \setbox\!boxA=\hbox to \wd\!boxB{\hss$\vert$\hss}%
   \!dimenE=\ht\!boxA \advance\!dimenE \dp\!boxA \divide\!dimenE 2
   \vbox to #1{\offinterlineskip
      \vskip .05556\!dimenE
      \hbox to \wd\!boxB{\hss$\mkern.4mu\uparrow$\hss}\vskip-\!dimenE
      \cleaders\copy\!boxA\vfil
      \vskip-\!dimenE\copy\!boxA
      \vskip\!dimenE\copy\!boxB\vskip.4em
      \copy\!boxA\vskip-\!dimenE
      \cleaders\copy\!boxA\vfil
      \vskip-\!dimenE \hbox to \wd\!boxB{\hss$\mkern.4mu\downarrow$\hss}
      \vskip .05556\!dimenE}}}

\def\putbar#1breadth <#2> from #3 #4 to #5 #6 {%
  \!xloc=\!M{#3}\!xunit  \!xxloc=\!M{#5}\!xunit%   
  \!yloc=\!M{#4}\!yunit  \!yyloc=\!M{#6}\!yunit%           
  \!dypos=\!yyloc  \advance\!dypos by -\!yloc
  \!dimenI=#2  
  \ifdim \!dimenI=\!zpt %            ** If 0 breadth
    \putrule#1from {#3} {#4} to {#5} {#6} % ** Then draw line
  \else %                            ** Else, put in a rectangle
    \let\!MBar=\!M%                  ** save current c/d mode
    \!setdimenmode %                 ** go into dimension mode
    \divide\!dimenI 2
    \ifdim \!dypos=\!zpt             
      \advance \!yloc -\!dimenI %    ** Equal y coordinates
      \advance \!yyloc \!dimenI
    \else
      \advance \!xloc -\!dimenI %    ** Equal x coordinates
      \advance \!xxloc \!dimenI
    \fi
    \putrectangle#1corners at {\!xloc} {\!yloc} and {\!xxloc} {\!yyloc}
    \let\!M=\!MBar %                 ** restore c/d mode
  \fi
  \ignorespaces}

\def\setbars#1breadth <#2> baseline at #3 = #4 {%
  \edef\!barshift{#1}%
  \edef\!barbreadth{#2}%
  \edef\!barorientation{#3}%
  \edef\!barbaseline{#4}%
  \def\!bardobaselabel{\!bardoendlabel}%
  \def\!bardoendlabel{\!barfinish}%
  \let\!drawcurve=\!barcurve
  \!setbars}
\def\!setbars{%
  \futurelet\!nextchar\!!setbars}
\def\!!setbars{%
  \if b\!nextchar
    \def\!!!setbars{\!setbarsbget}%
  \else 
    \if e\!nextchar
      \def\!!!setbars{\!setbarseget}%
    \else
      \def\!!!setbars{\relax}%
    \fi
  \fi
  \!!!setbars}
\def\!setbarsbget baselabels (#1) {%
  \def\!barbaselabelorientation{#1}%
  \def\!bardobaselabel{\!!bardobaselabel}%
  \!setbars}
\def\!setbarseget endlabels (#1) {%
  \edef\!barendlabelorientation{#1}%
  \def\!bardoendlabel{\!!bardoendlabel}%
  \!setbars}

\def\!barcurve #1 #2 {%
  \if y\!barorientation
    \def\!basexarg{#1}%
    \def\!baseyarg{\!barbaseline}%
  \else
    \def\!basexarg{\!barbaseline}%
    \def\!baseyarg{#2}%
  \fi
  \expandafter\putbar\!barshift breadth <\!barbreadth> from {\!basexarg}
    {\!baseyarg} to {#1} {#2}
  \def\!endxarg{#1}%
  \def\!endyarg{#2}%
  \!bardobaselabel}

\def\!!bardobaselabel "#1" {%
  \put {#1}\!barbaselabelorientation{} at {\!basexarg} {\!baseyarg}
  \!bardoendlabel}

\def\!!bardoendlabel "#1" {%
  \put {#1}\!barendlabelorientation{} at {\!endxarg} {\!endyarg}
  \!barfinish}

\def\!barfinish{%
  \!ifnextchar/{\!finish}{\!barcurve}}

\def\putrectangle{%
  \!ifnextchar<{\!putrectangle}{\!putrectangle<\!zpt,\!zpt> }}
\def\!putrectangle<#1,#2> corners at #3 #4 and #5 #6 {%
  \!xone=\!M{#3}\!xunit  \!xtwo=\!M{#5}\!xunit%   
  \!yone=\!M{#4}\!yunit  \!ytwo=\!M{#6}\!yunit%           
  \ifdim \!xtwo<\!xone
    \!dimenI=\!xone  \!xone=\!xtwo  \!xtwo=\!dimenI
  \fi
  \ifdim \!ytwo<\!yone
    \!dimenI=\!yone  \!yone=\!ytwo  \!ytwo=\!dimenI
  \fi
  \!dimenI=#1\relax  \advance\!xone\!dimenI  \advance\!xtwo\!dimenI
  \!dimenI=#2\relax  \advance\!yone\!dimenI  \advance\!ytwo\!dimenI
  \let\!MRect=\!M%                  ** save current coord/dimen mode
  \!setdimenmode
  \!shaderectangle
  \!dimenI=.5\linethickness
  \advance \!xone  -\!dimenI%       ** adjust x-location to overlap corners
  \advance \!xtwo   \!dimenI%       ** ditto
  \putrule from {\!xone} {\!yone} to {\!xtwo} {\!yone} 
  \putrule from {\!xone} {\!ytwo} to {\!xtwo} {\!ytwo} 
  \advance \!xone   \!dimenI%       ** restore original x-values
  \advance \!xtwo  -\!dimenI% 
  \advance \!yone  -\!dimenI%       ** adjust y-location to overlap corners
  \advance \!ytwo   \!dimenI%       ** ditto
  \putrule from {\!xone} {\!yone} to {\!xone} {\!ytwo} 
  \putrule from {\!xtwo} {\!yone} to {\!xtwo} {\!ytwo} 
  \let\!M=\!MRect%                  ** restore coord/dimen mode
  \ignorespaces}

\def\shaderectanglesoff{%
  \def\!shaderectangle{}%
  \ignorespaces}

\shaderectanglesoff

\def\!!shaderectangle{%
  \!dimenA=\!xtwo  \advance \!dimenA -\!xone
  \!dimenB=\!ytwo  \advance \!dimenB -\!yone
  \ifdim \!dimenA<\!dimenB
    \!startvshade (\!xone,\!yone,\!ytwo)
    \!lshade      (\!xtwo,\!yone,\!ytwo)
  \else
    \!starthshade (\!yone,\!xone,\!xtwo)
    \!lshade      (\!ytwo,\!xone,\!xtwo)
  \fi
  \ignorespaces}
  
\def\frame{%
  \!ifnextchar<{\!frame}{\!frame<\!zpt> }}
\long\def\!frame<#1> #2{%
  \beginpicture
    \setcoordinatesystem units <1pt,1pt> point at 0 0 
    \put {#2} [Bl] at 0 0 
    \!dimenA=#1\relax
    \!dimenB=\!wd \advance \!dimenB \!dimenA
    \!dimenC=\!ht \advance \!dimenC \!dimenA
    \!dimenD=\!dp \advance \!dimenD \!dimenA
    \let\!MFr=\!M
    \!setdimenmode
    \putrectangle corners at {-\!dimenA} {-\!dimenD} and {\!dimenB} {\!dimenC}
    \!setcoordmode
    \let\!M=\!MFr
  \endpicture
  \ignorespaces}

\def\rectangle <#1> <#2> {%
  \setbox0=\hbox{}\wd0=#1\ht0=#2\frame {\box0}}

\def\plot{%
  \!ifnextchar"{\!plotfromfile}{\!drawcurve}}
\def\!plotfromfile"#1"{%
  \expandafter\!drawcurve \input #1 /}

\def\setquadratic{%
  \let\!drawcurve=\!qcurve
  \let\!!Shade=\!!qShade
  \let\!!!Shade=\!!!qShade}

\def\setlinear{%
  \let\!drawcurve=\!lcurve
  \let\!!Shade=\!!lShade
  \let\!!!Shade=\!!!lShade}

\def\sethistograms{%
  \let\!drawcurve=\!hcurve}

\def\!qcurve #1 #2 {%
  \!start (#1,#2)
  \!Qjoin}
\def\!Qjoin#1 #2 #3 #4 {%
  \!qjoin (#1,#2) (#3,#4)             % \!qjoin  is defined in QUADRATIC
  \!ifnextchar/{\!finish}{\!Qjoin}}

\def\!lcurve #1 #2 {%
  \!start (#1,#2)
  \!Ljoin}
\def\!Ljoin#1 #2 {%
  \!ljoin (#1,#2)                    % \!ljoin  is defined in LINEAR
  \!ifnextchar/{\!finish}{\!Ljoin}}

\def\!finish/{\ignorespaces}

\def\!hcurve #1 #2 {%
  \edef\!hxS{#1}%
  \edef\!hyS{#2}%
  \!hjoin}
\def\!hjoin#1 #2 {%
  \putrectangle corners at {\!hxS} {\!hyS} and {#1} {#2}
  \edef\!hxS{#1}%
  \!ifnextchar/{\!finish}{\!hjoin}}

\def\vshade #1 #2 #3 {%
  \!startvshade (#1,#2,#3)
  \!Shadewhat}

\def\hshade #1 #2 #3 {%
  \!starthshade (#1,#2,#3)
  \!Shadewhat}

\def\!Shadewhat{%
  \futurelet\!nextchar\!Shade}
\def\!Shade{%
  \if <\!nextchar
    \def\!nextShade{\!!Shade}%
  \else
    \if /\!nextchar
      \def\!nextShade{\!finish}%
    \else
      \def\!nextShade{\!!!Shade}%
    \fi
  \fi
  \!nextShade}
\def\!!lShade<#1> #2 #3 #4 {%
  \!lshade <#1> (#2,#3,#4)                 % \!lshade is defined in SHADING
  \!Shadewhat}
\def\!!!lShade#1 #2 #3 {%
  \!lshade (#1,#2,#3)
  \!Shadewhat} 
\def\!!qShade<#1> #2 #3 #4 #5 #6 #7 {%
  \!qshade <#1> (#2,#3,#4) (#5,#6,#7)      % \!qshade is defined in SHADING
  \!Shadewhat}
\def\!!!qShade#1 #2 #3 #4 #5 #6 {%
  \!qshade (#1,#2,#3) (#4,#5,#6)
  \!Shadewhat} 

\setlinear

\def\setdashpattern <#1>{%
  \def\!Flist{}\def\!Blist{}\def\!UDlist{}%
  \!countA=0
  \!ecfor\!item:=#1\do{%
    \!dimenA=\!item\relax
    \expandafter\!rightappend\the\!dimenA\withCS{\\}\to\!UDlist%
    \advance\!countA  1
    \ifodd\!countA
      \expandafter\!rightappend\the\!dimenA\withCS{\!Rule}\to\!Flist%
      \expandafter\!leftappend\the\!dimenA\withCS{\!Rule}\to\!Blist%
    \else 
      \expandafter\!rightappend\the\!dimenA\withCS{\!Skip}\to\!Flist%
      \expandafter\!leftappend\the\!dimenA\withCS{\!Skip}\to\!Blist%
    \fi}%
  \!leaderlength=\!zpt
  \def\!Rule##1{\advance\!leaderlength  ##1}%
  \def\!Skip##1{\advance\!leaderlength  ##1}%
  \!Flist%
  \ifdim\!leaderlength>\!zpt 
  \else
    \def\!Flist{\!Skip{24in}}\def\!Blist{\!Skip{24in}}\ignorespaces
    \def\!UDlist{\\{\!zpt}\\{24in}}\ignorespaces
    \!leaderlength=24in
  \fi
  \!dashingon}

\def\!dashingon{%
  \def\!advancedashing{\!!advancedashing}%
  \def\!drawlinearsegment{\!lineardashed}%
  \def\!puthline{\!putdashedhline}%
  \def\!putvline{\!putdashedvline}%
  \ignorespaces}% 
\def\!dashingoff{%
  \def\!advancedashing{\relax}%
  \def\!drawlinearsegment{\!linearsolid}%
  \def\!puthline{\!putsolidhline}%
  \def\!putvline{\!putsolidvline}%
  \ignorespaces}

\def\setdots{%
  \!ifnextchar<{\!setdots}{\!setdots<5pt>}}
\def\!setdots<#1>{%
  \!dimenB=#1\advance\!dimenB -\plotsymbolspacing
  \ifdim\!dimenB<\!zpt
    \!dimenB=\!zpt
  \fi
\setdashpattern <\plotsymbolspacing,\!dimenB>}

\def\setdotsnear <#1> for <#2>{%
  \!dimenB=#2\relax  \advance\!dimenB -.05pt  
  \!dimenC=#1\relax  \!countA=\!dimenC 
  \!dimenD=\!dimenB  \advance\!dimenD .5\!dimenC  \!countB=\!dimenD
  \divide \!countB  \!countA
  \ifnum 1>\!countB 
    \!countB=1
  \fi
  \divide\!dimenB  \!countB
  \setdots <\!dimenB>}

\def\setdashes{%
  \!ifnextchar<{\!setdashes}{\!setdashes<5pt>}}
\def\!setdashes<#1>{\setdashpattern <#1,#1>}

\def\setdashesnear <#1> for <#2>{%
  \!dimenB=#2\relax  
  \!dimenC=#1\relax  \!countA=\!dimenC 
  \!dimenD=\!dimenB  \advance\!dimenD .5\!dimenC  \!countB=\!dimenD
  \divide \!countB  \!countA
  \ifodd \!countB 
  \else 
    \advance \!countB  1
  \fi
  \divide\!dimenB  \!countB
  \setdashes <\!dimenB>}

\def\setsolid{%
  \def\!Flist{\!Rule{24in}}\def\!Blist{\!Rule{24in}}%  
  \def\!UDlist{\\{24in}\\{\!zpt}}%
  \!dashingoff}  
\setsolid

\def\!divide#1#2#3{%
  \!dimenB=#1%                      **  dimB  holds current remainder (r)
  \!dimenC=#2%                      **  dimC  holds divisor (d)
  \!dimenD=\!dimenB%                **  dimD  holds quotient q=r/d for this 
  \divide \!dimenD \!dimenC%        **    step, in units of scaled pts
  \!dimenA=\!dimenD%                **  dimA  eventually holds answer (a)
  \multiply\!dimenD \!dimenC%       **  r <-- r - dq
  \advance\!dimenB -\!dimenD%       **  First step complete. Have integer part
  \!dimenD=\!dimenC%                **  Temporarily use dimD to hold |d|
    \ifdim\!dimenD<\!zpt \!dimenD=-\!dimenD 
  \fi
  \ifdim\!dimenD<64pt%              **  Branch on the magnitude of |d|
    \!divstep[\!tfs]\!divstep[\!tfs]%
  \else 
    \!!divide
  \fi
  #3=\!dimenA\ignorespaces}

\def\!!divide{%
  \ifdim\!dimenD<256pt
    \!divstep[64]\!divstep[32]\!divstep[32]%
  \else 
    \!divstep[8]\!divstep[8]\!divstep[8]\!divstep[8]\!divstep[8]%
    \!dimenA=2\!dimenA
  \fi}

\def\!divstep[#1]{%                 **  #1 = "B"
  \!dimenB=#1\!dimenB%              **  r <-- B*r
  \!dimenD=\!dimenB%                **  dimD  holds quotient q=r/d for this 
    \divide \!dimenD by \!dimenC%   **    step, in units of scaled pts
  \!dimenA=#1\!dimenA%              **  a <-- B*a + q
    \advance\!dimenA by \!dimenD%
  \multiply\!dimenD by \!dimenC%    **  r <-- r - dq
    \advance\!dimenB by -\!dimenD}

\def\Divide <#1> by <#2> forming <#3> {%
  \!divide{#1}{#2}{#3}}

\def\circulararc{%
  \ellipticalarc axes ratio 1:1 }

\def\ellipticalarc axes ratio #1:#2 #3 degrees from #4 #5 center at #6 #7 {%
  \!angle=#3pt\relax%                    ** get angle
  \ifdim\!angle>\!zpt 
    \def\!sign{}%                        ** counterclockwise
  \else 
    \def\!sign{-}\!angle=-\!angle%       ** clockwise
  \fi
  \!xxloc=\!M{#6}\!xunit%                ** convert CENTER to dimension
  \!yyloc=\!M{#7}\!yunit     
  \!xxS=\!M{#4}\!xunit%                  ** get STARTing point on rim of ellipse
  \!yyS=\!M{#5}\!yunit
  \advance\!xxS -\!xxloc%                ** make center of ellipse (0,0)
  \advance\!yyS -\!yyloc
  \!divide\!xxS{#1pt}\!xxS %             ** scale point on ellipse to point on 
  \!divide\!yyS{#2pt}\!yyS %                 corresponding circle
  \let\!MC=\!M%                          ** save current c/d mode
  \!setdimenmode%                        ** go into dimension mode
  \!xS=#1\!xxS  \advance\!xS\!xxloc
  \!yS=#2\!yyS  \advance\!yS\!yyloc
  \!start (\!xS,\!yS)%
  \!loop\ifdim\!angle>14.9999pt%         ** draw in major portion of ellipse 
    \!rotate(\!xxS,\!yyS)by(\!cos,\!sign\!sin)to(\!xxM,\!yyM) 
    \!rotate(\!xxM,\!yyM)by(\!cos,\!sign\!sin)to(\!xxE,\!yyE)
    \!xM=#1\!xxM  \advance\!xM\!xxloc  \!yM=#2\!yyM  \advance\!yM\!yyloc
    \!xE=#1\!xxE  \advance\!xE\!xxloc  \!yE=#2\!yyE  \advance\!yE\!yyloc
    \!qjoin (\!xM,\!yM) (\!xE,\!yE)
    \!xxS=\!xxE  \!yyS=\!yyE 
    \advance \!angle -15pt
  \repeat
  \ifdim\!angle>\!zpt%                   ** complete remaining arc, if any
    \!angle=100.53096\!angle%            ** convert angle to radians, divide
    \divide \!angle 360 %                **   by 2, and multiply by 32
    \!sinandcos\!angle\!!sin\!!cos%      ** get 32*sin & 32*cos
    \!rotate(\!xxS,\!yyS)by(\!!cos,\!sign\!!sin)to(\!xxM,\!yyM) 
    \!rotate(\!xxM,\!yyM)by(\!!cos,\!sign\!!sin)to(\!xxE,\!yyE)
    \!xM=#1\!xxM  \advance\!xM\!xxloc  \!yM=#2\!yyM  \advance\!yM\!yyloc
    \!xE=#1\!xxE  \advance\!xE\!xxloc  \!yE=#2\!yyE  \advance\!yE\!yyloc
    \!qjoin (\!xM,\!yM) (\!xE,\!yE)
  \fi
  \let\!M=\!MC%                          ** restore c/d mode
  \ignorespaces}%                        **   if appropriate

\def\!rotate(#1,#2)by(#3,#4)to(#5,#6){% 
  \!dimenA=#3#1\advance \!dimenA -#4#2%   ** Rcos(x+t)=Rcosx*cost - Rsinx*sint
  \!dimenB=#3#2\advance \!dimenB  #4#1%   ** Rsin(x+t)=Rsinx*cost + Rcosx*sint
  \divide \!dimenA 32  \divide \!dimenB 32 
  #5=\!dimenA  #6=\!dimenB
  \ignorespaces}
\def\!sin{4.17684}%                       ** 32*sin(pi/24) (pi/24=7.5deg)
\def\!cos{31.72624}%                      ** 32*cos(pi/24)

\def\!sinandcos#1#2#3{%
 \!dimenD=#1%                **  angle is expressed in radians/32: 1pt = 1/32rad
 \!dimenA=\!dimenD%          **  dimA will eventually contain 32sin(angle)in pts
 \!dimenB=32pt%              **  dimB will eventually contain 32cos(angle)in pts
 \!removept\!dimenD\!value%  **  get value of 32*angle, without "pt"
 \!dimenC=\!dimenD%          **  holds 32*angle**i/i! in pts
 \!dimenC=\!value\!dimenC \divide\!dimenC by 64 %   ** now 32*angle**2/2
 \advance\!dimenB by -\!dimenC%                     ** 32-32*angle**2/2
 \!dimenC=\!value\!dimenC \divide\!dimenC by 96 %   ** now 32*angle**3/3!
 \advance\!dimenA by -\!dimenC%                     ** now 32*(angle-angle**3/6)
 \!dimenC=\!value\!dimenC \divide\!dimenC by 128 %  ** now 32*angle**4/4!
 \advance\!dimenB by \!dimenC%
 \!removept\!dimenA#2%                              ** set 32*sin(angle)
 \!removept\!dimenB#3%                              ** set 32*cos(angle)
 \ignorespaces}

\def\putrule#1from #2 #3 to #4 #5 {%
  \!xloc=\!M{#2}\!xunit  \!xxloc=\!M{#4}\!xunit%   
  \!yloc=\!M{#3}\!yunit  \!yyloc=\!M{#5}\!yunit%           
  \!dxpos=\!xxloc  \advance\!dxpos by -\!xloc
  \!dypos=\!yyloc  \advance\!dypos by -\!yloc
  \ifdim\!dypos=\!zpt
    \def\!!Line{\!puthline{#1}}\ignorespaces
  \else
    \ifdim\!dxpos=\!zpt
      \def\!!Line{\!putvline{#1}}\ignorespaces
    \else 
       \def\!!Line{}
    \fi
  \fi
  \let\!ML=\!M%           ** save current coord\dimen mode
  \!setdimenmode%         ** express locations in dimens
  \!!Line%
  \let\!M=\!ML%           ** restore previous c/d mode
  \ignorespaces}

\def\!putsolidhline#1{%
  \ifdim\!dxpos>\!zpt 
    \put{\!hline\!dxpos}#1[l] at {\!xloc} {\!yloc}
  \else 
    \put{\!hline{-\!dxpos}}#1[l] at {\!xxloc} {\!yyloc}
  \fi
  \ignorespaces}

\def\!putsolidvline#1{%
  \ifdim\!dypos>\!zpt 
    \put{\!vline\!dypos}#1[b] at {\!xloc} {\!yloc}
  \else 
    \put{\!vline{-\!dypos}}#1[b] at {\!xxloc} {\!yyloc}
  \fi
  \ignorespaces}

\def\!hline#1{\hbox to #1{\leaders \hrule height\linethickness\hfill}}
\def\!vline#1{\vbox to #1{\leaders \vrule width\linethickness\vfill}}

\def\!putdashedhline#1{%
  \ifdim\!dxpos>\!zpt 
    \!DLsetup\!Flist\!dxpos
    \put{\hbox to \!totalleaderlength{\!hleaders}\!hpartialpattern\!Rtrunc}
      #1[l] at {\!xloc} {\!yloc} 
  \else 
    \!DLsetup\!Blist{-\!dxpos}
    \put{\!hpartialpattern\!Ltrunc\hbox to \!totalleaderlength{\!hleaders}}
      #1[r] at {\!xloc} {\!yloc} 
  \fi
  \ignorespaces}

\def\!putdashedvline#1{%
  \!dypos=-\!dypos%            ** vertical leaders go from top to bottom
  \ifdim\!dypos>\!zpt 
    \!DLsetup\!Flist\!dypos 
    \put{\vbox{\vbox to \!totalleaderlength{\!vleaders}
      \!vpartialpattern\!Rtrunc}}#1[t] at {\!xloc} {\!yloc} 
  \else 
    \!DLsetup\!Blist{-\!dypos}
    \put{\vbox{\!vpartialpattern\!Ltrunc
      \vbox to \!totalleaderlength{\!vleaders}}}#1[b] at {\!xloc} {\!yloc} 
  \fi
  \ignorespaces}

\def\!DLsetup#1#2{%            ** Dashed-Line set up
  \let\!RSlist=#1%             ** set !Rule-Skip list
  \!countB=#2%                 ** convert rule length to integer (number of sps)
  \!countA=\!leaderlength%     ** ditto, leaderlength
  \divide\!countB by \!countA% ** number of complete leader units
  \!totalleaderlength=\!countB\!leaderlength
  \!Rresiduallength=#2%
  \advance \!Rresiduallength by -\!totalleaderlength%  \** excess length
  \!Lresiduallength=\!leaderlength
  \advance \!Lresiduallength by -\!Rresiduallength
  \ignorespaces}

\def\!hleaders{%
  \def\!Rule##1{\vrule height\linethickness width##1}%
  \def\!Skip##1{\hskip##1}%
  \leaders\hbox{\!RSlist}\hfill}

\def\!hpartialpattern#1{%
  \!dimenA=\!zpt \!dimenB=\!zpt 
  \def\!Rule##1{#1{##1}\vrule height\linethickness width\!dimenD}%
  \def\!Skip##1{#1{##1}\hskip\!dimenD}%
  \!RSlist}

\def\!vleaders{%
  \def\!Rule##1{\hrule width\linethickness height##1}%
  \def\!Skip##1{\vskip##1}%
  \leaders\vbox{\!RSlist}\vfill}

\def\!vpartialpattern#1{%
  \!dimenA=\!zpt \!dimenB=\!zpt 
  \def\!Rule##1{#1{##1}\hrule width\linethickness height\!dimenD}%
  \def\!Skip##1{#1{##1}\vskip\!dimenD}%
  \!RSlist}

\def\!Rtrunc#1{\!trunc{#1}>\!Rresiduallength}
\def\!Ltrunc#1{\!trunc{#1}<\!Lresiduallength}

\def\!trunc#1#2#3{%          
  \!dimenA=\!dimenB         
  \advance\!dimenB by #1%
  \!dimenD=\!dimenB  \ifdim\!dimenD#2#3\!dimenD=#3\fi
  \!dimenC=\!dimenA  \ifdim\!dimenC#2#3\!dimenC=#3\fi
  \advance \!dimenD by -\!dimenC}

\def\!start (#1,#2){%
  \!plotxorigin=\!xorigin  \advance \!plotxorigin by \!plotsymbolxshift
  \!plotyorigin=\!yorigin  \advance \!plotyorigin by \!plotsymbolyshift
  \!xS=\!M{#1}\!xunit \!yS=\!M{#2}\!yunit
  \!rotateaboutpivot\!xS\!yS
  \!copylist\!UDlist\to\!!UDlist% **\!UDlist has the form \\{dimen1}\\{dimen2}..
  \!getnextvalueof\!downlength\from\!!UDlist
  \!distacross=\!zpt%             ** 1st point goes at start of curve
  \!intervalno=0 %                ** initialize interval counter
  \global\totalarclength=\!zpt%   ** initialize distance traveled along curve
  \ignorespaces}

\def\!ljoin (#1,#2){%
  \advance\!intervalno by 1
  \!xE=\!M{#1}\!xunit \!yE=\!M{#2}\!yunit
  \!rotateaboutpivot\!xE\!yE
  \!xdiff=\!xE \advance \!xdiff by -\!xS%**  xdiff = xE - xS
  \!ydiff=\!yE \advance \!ydiff by -\!yS%**  ydiff = yE - yS
  \!Pythag\!xdiff\!ydiff\!arclength%     **  arclength = sqrt(xdiff**2+ydiff**2) 
  \global\advance \totalarclength by \!arclength%
  \!drawlinearsegment%   ** set by dashpat to \!linearsolid or \!lineardashed
  \!xS=\!xE \!yS=\!yE%   ** shift ending points to starting points
  \ignorespaces}

\def\!linearsolid{%
  \!npoints=\!arclength
  \!countA=\plotsymbolspacing
  \divide\!npoints by \!countA%      ** now #pts =. arclength/plotsymbolspacing
  \ifnum \!npoints<1 
    \!npoints=1 
  \fi
  \divide\!xdiff by \!npoints
  \divide\!ydiff by \!npoints
  \!xpos=\!xS \!ypos=\!yS
  \loop\ifnum\!npoints>-1
    \!plotifinbounds
    \advance \!xpos by \!xdiff
    \advance \!ypos by \!ydiff
    \advance \!npoints by -1
  \repeat
  \ignorespaces}

\def\!lineardashed{%
  \ifdim\!distacross>\!arclength
    \advance \!distacross by -\!arclength  %nothing to plot in this interval
  \else
    \loop\ifdim\!distacross<\!arclength
      \!divide\!distacross\!arclength\!dimenA%  ** dimA = across/arclength
      \!removept\!dimenA\!t%  ** \!t holds value in dimA, without the "pt"
      \!xpos=\!t\!xdiff \advance \!xpos by \!xS
      \!ypos=\!t\!ydiff \advance \!ypos by \!yS
      \!plotifinbounds
      \advance\!distacross by \plotsymbolspacing
      \!advancedashing
    \repeat  
    \advance \!distacross by -\!arclength%    ** prepare for next interval 
  \fi
  \ignorespaces}

\def\!!advancedashing{%
  \advance\!downlength by -\plotsymbolspacing
  \ifdim \!downlength>\!zpt
  \else
    \advance\!distacross by \!downlength
    \!getnextvalueof\!uplength\from\!!UDlist
    \advance\!distacross by \!uplength
    \!getnextvalueof\!downlength\from\!!UDlist
  \fi}

\def\inboundscheckoff{%
  \def\!plotifinbounds{\!plot(\!xpos,\!ypos)}%
  \def\!initinboundscheck{\relax}\ignorespaces}
 
\inboundscheckoff

\def\!!plotifinbounds{%
  \ifdim \!xpos<\!checkleft
  \else
    \ifdim \!xpos>\!checkright
    \else
      \ifdim \!ypos<\!checkbot
      \else
         \ifdim \!ypos>\!checktop
         \else
           \!plot(\!xpos,\!ypos)
         \fi 
      \fi
    \fi
  \fi}

\def\!!initinboundscheck{%
  \!checkleft=\!arealloc     \advance\!checkleft by \!xorigin
  \!checkright=\!arearloc    \advance\!checkright by \!xorigin
  \!checkbot=\!areabloc      \advance\!checkbot by \!yorigin
  \!checktop=\!areatloc      \advance\!checktop by \!yorigin}

\def\!logten#1#2{%
  \expandafter\!!logten#1\!nil
  \!removept\!dimenF#2%
  \ignorespaces}

\def\!!logten#1#2\!nil{%
  \if -#1%
    \!dimenF=\!zpt
    \def\!next{\ignorespaces}%
  \else
    \if +#1%
      \def\!next{\!!logten#2\!nil}%
    \else
      \if .#1%
        \def\!next{\!!logten0.#2\!nil}%
      \else
        \def\!next{\!!!logten#1#2..\!nil}%
      \fi
    \fi
  \fi
  \!next}

\def\!!!logten#1#2.#3.#4\!nil{%
  \!dimenF=1pt %                 ** DimF holds log10 original argument
  \if 0#1%                      
    \!!logshift#3pt %            ** Argument < 1
  \else %                        ** Argument >= 1
    \!logshift#2/%               ** Shift decimal pt as many places
    \!dimenE=#1.#2#3pt %         **   as there are figures in #2
  \fi %                          ** Now dimE holds revised X want log10 of
  \ifdim \!dimenE<\!rootten%          ** Transform X to XX between sqrt(10) 
    \multiply \!dimenE 10 %           **   and 10*sqrt(10)
    \advance  \!dimenF -1pt
  \fi
  \!dimenG=\!dimenE%                  ** dimG <- (XX + 10)
    \advance\!dimenG 10pt
  \advance\!dimenE -10pt %            ** dimE <- (XX - 10)
  \multiply\!dimenE 10 %              ** dimE = 10*(XX-10)
  \!divide\!dimenE\!dimenG\!dimenE%   ** Now dimE=10t==10*(XX-10)/(XX+10)
  \!removept\!dimenE\!t%              ** !t=10t, with "pt" removed
  \!dimenG=\!t\!dimenE%               ** dimG=100t**2
  \!removept\!dimenG\!tt%             ** !tt=100t**2, with "pt" removed
  \!dimenH=\!tt\!tenAe%               ** dimH=10*a5*(10t)**2 /100
    \divide\!dimenH 100
  \advance\!dimenH \!tenAc%           ** ditto + 10*a3
  \!dimenH=\!tt\!dimenH%              ** ditto * (10t)**2 /100
    \divide\!dimenH 100   
  \advance\!dimenH \!tenAa%           ** ditto + 10*a1
  \!dimenH=\!t\!dimenH%               ** ditto * 10t / 100
    \divide\!dimenH 100 %             ** Now dimH = log10(XX) - 1
  \advance\!dimenF \!dimenH}%         ** dimF = log10(X)

\def\!logshift#1{%
  \if #1/%
    \def\!next{\ignorespaces}%
  \else
    \advance\!dimenF 1pt 
    \def\!next{\!logshift}%
  \fi 
  \!next}

 \def\!!logshift#1{%
   \advance\!dimenF -1pt
   \if 0#1%
     \def\!next{\!!logshift}%
   \else
     \if p#1%
       \!dimenF=1pt
       \def\!next{\!dimenE=1p}%
     \else
       \def\!next{\!dimenE=#1.}%
     \fi
   \fi
   \!next}

\def\beginpicture{%
  \setbox\!picbox=\hbox\bgroup%
  \!xleft=\maxdimen  
  \!xright=-\maxdimen
  \!ybot=\maxdimen
  \!ytop=-\maxdimen}

\def\endpicture{%
  \ifdim\!xleft=\maxdimen%  ** check if nothing was put in picbox
    \!xleft=\!zpt \!xright=\!zpt \!ybot=\!zpt \!ytop=\!zpt 
  \fi
  \global\!Xleft=\!xleft \global\!Xright=\!xright
  \global\!Ybot=\!ybot \global\!Ytop=\!ytop
  \egroup%
  \ht\!picbox=\!Ytop  \dp\!picbox=-\!Ybot
  \ifdim\!Ybot>\!zpt
  \else 
    \ifdim\!Ytop<\!zpt
      \!Ybot=\!Ytop
    \else
      \!Ybot=\!zpt
    \fi
  \fi
  \hbox{\kern-\!Xleft\lower\!Ybot\box\!picbox\kern\!Xright}}

\def\endpicturesave <#1,#2>{%
  \endpicture \global #1=\!Xleft \global #2=\!Ybot \ignorespaces}

\def\setcoordinatesystem{%
  \!ifnextchar{u}{\!getlengths }
    {\!getlengths units <\!xunit,\!yunit>}}
\def\!getlengths units <#1,#2>{%
  \!xunit=#1\relax
  \!yunit=#2\relax
  \!ifcoordmode 
    \let\!SCnext=\!SCccheckforRP
  \else
    \let\!SCnext=\!SCdcheckforRP
  \fi
  \!SCnext}
\def\!SCccheckforRP{%
  \!ifnextchar{p}{\!cgetreference }
    {\!cgetreference point at {\!xref} {\!yref} }}
\def\!cgetreference point at #1 #2 {%
  \edef\!xref{#1}\edef\!yref{#2}%
  \!xorigin=\!xref\!xunit  \!yorigin=\!yref\!yunit  
  \!initinboundscheck % ** See linear.tex
  \ignorespaces}
\def\!SCdcheckforRP{%
  \!ifnextchar{p}{\!dgetreference}%
    {\ignorespaces}}
\def\!dgetreference point at #1 #2 {%
  \!xorigin=#1\relax  \!yorigin=#2\relax
  \ignorespaces}

\long\def\put#1#2 at #3 #4 {%
  \!setputobject{#1}{#2}%
  \!xpos=\!M{#3}\!xunit  \!ypos=\!M{#4}\!yunit  
  \!rotateaboutpivot\!xpos\!ypos%
  \advance\!xpos -\!xorigin  \advance\!xpos -\!xshift
  \advance\!ypos -\!yorigin  \advance\!ypos -\!yshift
  \kern\!xpos\raise\!ypos\box\!putobject\kern-\!xpos%
  \!doaccounting\ignorespaces}

\long\def\multiput #1#2 at {%
  \!setputobject{#1}{#2}%
  \!ifnextchar"{\!putfromfile}{\!multiput}}
\def\!putfromfile"#1"{%
  \expandafter\!multiput \input #1 /}
\def\!multiput{%
  \futurelet\!nextchar\!!multiput}
\def\!!multiput{%
  \if *\!nextchar
    \def\!nextput{\!alsoby}%
  \else
    \if /\!nextchar
      \def\!nextput{\!finishmultiput}%
    \else
      \def\!nextput{\!alsoat}%
    \fi
  \fi
  \!nextput}
\def\!finishmultiput/{%
  \setbox\!putobject=\hbox{}%
  \ignorespaces}

\def\!alsoat#1 #2 {%
  \!xpos=\!M{#1}\!xunit  \!ypos=\!M{#2}\!yunit  
  \!rotateaboutpivot\!xpos\!ypos%
  \advance\!xpos -\!xorigin  \advance\!xpos -\!xshift
  \advance\!ypos -\!yorigin  \advance\!ypos -\!yshift
  \kern\!xpos\raise\!ypos\copy\!putobject\kern-\!xpos%
  \!doaccounting
  \!multiput}

\def\!alsoby*#1 #2 #3 {%
  \!dxpos=\!M{#2}\!xunit \!dypos=\!M{#3}\!yunit 
  \!rotateonly\!dxpos\!dypos
  \!ntemp=#1%
  \!!loop\ifnum\!ntemp>0
    \advance\!xpos by \!dxpos  \advance\!ypos by \!dypos
    \kern\!xpos\raise\!ypos\copy\!putobject\kern-\!xpos%
    \advance\!ntemp by -1
  \repeat
  \!doaccounting 
  \!multiput}

\def\accountingon{\def\!doaccounting{\!!doaccounting}\ignorespaces}

\accountingon
\def\!!doaccounting{%
  \!xtemp=\!xpos  
  \!ytemp=\!ypos
  \ifdim\!xtemp<\!xleft 
     \!xleft=\!xtemp 
  \fi
  \advance\!xtemp by  \!wd 
  \ifdim\!xright<\!xtemp 
    \!xright=\!xtemp
  \fi
  \advance\!ytemp by -\!dp
  \ifdim\!ytemp<\!ybot  
    \!ybot=\!ytemp
  \fi
  \advance\!ytemp by  \!dp
  \advance\!ytemp by  \!ht 
  \ifdim\!ytemp>\!ytop  
    \!ytop=\!ytemp  
  \fi}

\long\def\!setputobject#1#2{%
  \setbox\!putobject=\hbox{#1}%
  \!ht=\ht\!putobject  \!dp=\dp\!putobject  \!wd=\wd\!putobject
  \wd\!putobject=\!zpt
  \!xshift=.5\!wd   \!yshift=.5\!ht   \advance\!yshift by -.5\!dp
  \edef\!putorientation{#2}%
  \expandafter\!SPOreadA\!putorientation[]\!nil%
  \expandafter\!SPOreadB\!putorientation<\!zpt,\!zpt>\!nil\ignorespaces}

\def\!SPOreadA#1[#2]#3\!nil{\!etfor\!orientation:=#2\do\!SPOreviseshift}

\def\!SPOreadB#1<#2,#3>#4\!nil{\advance\!xshift by -#2\advance\!yshift by -#3}

\def\!SPOreviseshift{%
  \if l\!orientation 
    \!xshift=\!zpt
  \else 
    \if r\!orientation 
      \!xshift=\!wd
    \else 
      \if b\!orientation
        \!yshift=-\!dp
      \else 
        \if B\!orientation 
          \!yshift=\!zpt
        \else 
          \if t\!orientation 
            \!yshift=\!ht
          \fi 
        \fi
      \fi
    \fi
  \fi}

\long\def\!dimenput#1#2(#3,#4){%
  \!setputobject{#1}{#2}%
  \!xpos=#3\advance\!xpos by -\!xshift
  \!ypos=#4\advance\!ypos by -\!yshift
  \kern\!xpos\raise\!ypos\box\!putobject\kern-\!xpos%
  \!doaccounting\ignorespaces}

\def\!setdimenmode{%
  \let\!M=\!M!!\ignorespaces}
\def\!setcoordmode{%
  \let\!M=\!M!\ignorespaces}
\def\!ifcoordmode{%
  \ifx \!M \!M!}
\def\!ifdimenmode{%
  \ifx \!M \!M!!}
\def\!M!#1#2{#1#2} 
\def\!M!!#1#2{#1}
\!setcoordmode
\let\setdimensionmode=\!setdimenmode
\let\setcoordinatemode=\!setcoordmode

\def\!stack[#1]{%
  \let\!lglue=\hfill \let\!rglue=\hfill
  \expandafter\let\csname !#1glue\endcsname=\relax
  \!ifnextchar<{\!!stack}{\!!stack<\stackleading>}}
\def\!!stack<#1>#2{%
  \vbox{\def\!valueslist{}\!ecfor\!value:=#2\do{%
    \expandafter\!rightappend\!value\withCS{\\}\to\!valueslist}%
    \!lop\!valueslist\to\!value
    \let\\=\cr\lineskiplimit=\maxdimen\lineskip=#1%
    \baselineskip=-1000pt\halign{\!lglue##\!rglue\cr \!value\!valueslist\cr}}%
  \ignorespaces}

\def\!lines[#1]#2{%
  \let\!lglue=\hfill \let\!rglue=\hfill
  \expandafter\let\csname !#1glue\endcsname=\relax
  \vbox{\halign{\!lglue##\!rglue\cr #2\crcr}}%
  \ignorespaces}

\def\!Lines[#1]#2{%
  \let\!lglue=\hfill \let\!rglue=\hfill
  \expandafter\let\csname !#1glue\endcsname=\relax
  \vtop{\halign{\!lglue##\!rglue\cr #2\crcr}}%
  \ignorespaces}

\def\setplotsymbol(#1#2){%
  \!setputobject{#1}{#2}
  \setbox\!plotsymbol=\box\!putobject%
  \!plotsymbolxshift=\!xshift 
  \!plotsymbolyshift=\!yshift 
  \ignorespaces}

\def\!!plot(#1,#2){%
  \!dimenA=-\!plotxorigin \advance \!dimenA by #1%    ** over
  \!dimenB=-\!plotyorigin \advance \!dimenB by #2%    ** up
  \kern\!dimenA\raise\!dimenB\copy\!plotsymbol\kern-\!dimenA%
  \ignorespaces}

\def\!!!plot(#1,#2){%
  \!dimenA=-\!plotxorigin \advance \!dimenA by #1%    ** over
  \!dimenB=-\!plotyorigin \advance \!dimenB by #2%    ** up
  \kern\!dimenA\raise\!dimenB\copy\!plotsymbol\kern-\!dimenA%
  \!countE=\!dimenA
  \!countF=\!dimenB
  \immediate\write\!replotfile{\the\!countE,\the\!countF.}%
  \ignorespaces}

\def\savelinesandcurves on "#1" {%
  \immediate\closeout\!replotfile
  \immediate\openout\!replotfile=#1%
  \let\!plot=\!!!plot}

\def\dontsavelinesandcurves {%
  \let\!plot=\!!plot}
\dontsavelinesandcurves

{\catcode`\%=11\xdef\!Commentsignal{%}}
\def\writesavefile#1 {%
  \immediate\write\!replotfile{\!Commentsignal #1}%
  \ignorespaces}

\def\replot"#1" {%
  \expandafter\!replot\input #1 /}
\def\!replot#1,#2. {%
  \!dimenA=#1sp
  \kern\!dimenA\raise#2sp\copy\!plotsymbol\kern-\!dimenA
  \futurelet\!nextchar\!!replot}
\def\!!replot{%
  \if /\!nextchar 
    \def\!next{\!finish}%
  \else
    \def\!next{\!replot}%
  \fi
  \!next}
\def\!Pythag#1#2#3{%
  \!dimenE=#1\relax                                     
  \ifdim\!dimenE<\!zpt 
    \!dimenE=-\!dimenE 
  \fi%                                            ** dimE = |x|
  \!dimenF=#2\relax
  \ifdim\!dimenF<\!zpt 
    \!dimenF=-\!dimenF 
  \fi%                                            ** dimF = |y|
  \advance \!dimenF by \!dimenE%                  ** dimF = s = |x|+|y|
  \ifdim\!dimenF=\!zpt 
    \!dimenG=\!zpt%                               ** dimG = z = sqrt(x**2+y**2)
  \else 
    \!divide{8\!dimenE}\!dimenF\!dimenE%          ** now dimE = 8t = (8|x|)/s
    \advance\!dimenE by -4pt%                     ** 8tau = (8t-4)*2
      \!dimenE=2\!dimenE%                         **   (tau = 2*t - 1)
    \!removept\!dimenE\!!t%                       ** 8tau, without "pt"
    \!dimenE=\!!t\!dimenE%                        ** (8tau)**2, in pts
    \advance\!dimenE by 64pt%                     ** u = [64 + (8tau)**2]/2
    \divide \!dimenE by 2%                        **   [u = (8f)**2]
    \!dimenH=7pt%                                 ** initial guess g at sqrt(u)
    \!!Pythag\!!Pythag\!!Pythag%                  ** 3 iterations give sqrt(u)
    \!removept\!dimenH\!!t%                       ** 8f=sqrt(u), without "pt"
    \!dimenG=\!!t\!dimenF%                        ** z = (8f)*s/8
    \divide\!dimenG by 8
  \fi
  #3=\!dimenG
  \ignorespaces}

\def\!!Pythag{%                                   ** Newton-Raphson for sqrt
  \!divide\!dimenE\!dimenH\!dimenI%               ** v = u/g
  \advance\!dimenH by \!dimenI%                   ** g <-- (g + u/g)/2
    \divide\!dimenH by 2}

\def\placehypotenuse for <#1> and <#2> in <#3> {%
  \!Pythag{#1}{#2}{#3}}

\def\!qjoin (#1,#2) (#3,#4){%
  \advance\!intervalno by 1
  \!ifcoordmode
    \edef\!xmidpt{#1}\edef\!ymidpt{#2}%
  \else
    \!dimenA=#1\relax \edef\!xmidpt{\the\!dimenA}%
    \!dimenA=#2\relax \edef\!ymidpt{\the\!dimenA}%
  \fi
  \!xM=\!M{#1}\!xunit  \!yM=\!M{#2}\!yunit   \!rotateaboutpivot\!xM\!yM
  \!xE=\!M{#3}\!xunit  \!yE=\!M{#4}\!yunit   \!rotateaboutpivot\!xE\!yE
  \!dimenA=\!xM  \advance \!dimenA by -\!xS%   ** dimA = I = xM - xS
  \!dimenB=\!xE  \advance \!dimenB by -\!xM%   ** dimB = II = xE-xM
  \!xB=3\!dimenA \advance \!xB by -\!dimenB%   ** b=3I-II
  \!xC=2\!dimenB \advance \!xC by -2\!dimenA%  ** c=2(II-I)
  \!dimenA=\!yM  \advance \!dimenA by -\!yS%   
  \!dimenB=\!yE  \advance \!dimenB by -\!yM%  
  \!yB=3\!dimenA \advance \!yB by -\!dimenB%  
  \!yC=2\!dimenB \advance \!yC by -2\!dimenA% 
  \!xprime=\!xB  \!yprime=\!yB%          ** x'(t) = b + 2ct
  \!dxprime=.5\!xC  \!dyprime=.5\!yC%    ** dt=1/4 ==> dx'(t) = c/2
  \!getf \!midarclength=\!dimenA
  \!getf \advance \!midarclength by 4\!dimenA
  \!getf \advance \!midarclength by \!dimenA
  \divide \!midarclength by 12
  \!arclength=\!dimenA
  \!getf \advance \!arclength by 4\!dimenA
  \!getf \advance \!arclength by \!dimenA
  \divide \!arclength by 12%             ** Now have arc length over [1/2,1]
  \advance \!arclength by \!midarclength
  \global\advance \totalarclength by \!arclength
  \ifdim\!distacross>\!arclength 
    \advance \!distacross by -\!arclength%   ** nothing 
  \else
    \!initinverseinterp%  ** initialize for inverse interpolation on arc length
    \loop\ifdim\!distacross<\!arclength%     ** loop over points on arc 
      \!inverseinterp%    ** find  t  such that arc length[0,t] = distacross,
      \!xpos=\!t\!xC \advance\!xpos by \!xB
        \!xpos=\!t\!xpos \advance \!xpos by \!xS
      \!ypos=\!t\!yC \advance\!ypos by \!yB
        \!ypos=\!t\!ypos \advance \!ypos by \!yS
      \!plotifinbounds%                       ** plot point if in bounds
      \advance\!distacross \plotsymbolspacing%** advance arc length for next pt
      \!advancedashing%                       ** see "linear"
    \repeat  
    \advance \!distacross by -\!arclength%    ** prepare for next interval 
  \fi
  \!xS=\!xE%              ** shift ending points to starting points
  \!yS=\!yE
  \ignorespaces}

\def\!getf{\!Pythag\!xprime\!yprime\!dimenA%
  \advance\!xprime by \!dxprime
  \advance\!yprime by \!dyprime}

\def\!initinverseinterp{%
  \ifdim\!arclength>\!zpt
    \!divide{8\!midarclength}\!arclength\!dimenE% ** dimE=8w=8r/s, where  r 
    \ifdim\!dimenE<\!wmin \!setinverselinear
    \else 
      \ifdim\!dimenE>\!wmax \!setinverselinear
      \else%                                    ** w  in range: initialize
        \def\!inverseinterp{\!inversequad}\ignorespaces
         \!removept\!dimenE\!Ew%           **  8w, without "pt"
         \!dimenF=-\!Ew\!dimenE%           **  -(8w)**2
         \advance\!dimenF by 32pt%         **  32 - (8w)**2
         \!dimenG=8pt 
         \advance\!dimenG by -\!dimenE%    **  8 - 8w
         \!dimenG=\!Ew\!dimenG%            **  (8w)*(8-8w)
         \!divide\!dimenF\!dimenG\!beta%   **  beta = (32-(8w)**2)/(8w(8-8w))
         \!gamma=1pt
         \advance \!gamma by -\!beta%      **  gamma = 1-beta
      \fi%       ** end of the \ifdim\!dimenE>\!wmax
    \fi%         ** end of the \ifdim\!dimenE<\!wmin
  \fi%           ** end of the \ifdim\!arclength>\!zpt
  \ignorespaces}

\def\!inversequad{%
  \!divide\!distacross\!arclength\!dimenG%   ** dimG = v = distacross/arclength
  \!removept\!dimenG\!v%                     ** v, without "pt"
  \!dimenG=\!v\!gamma%                       ** gamma*v
  \advance\!dimenG by \!beta%                ** beta + gamma*v
  \!dimenG=\!v\!dimenG%                      ** t = v*(beta + gamma*v)
  \!removept\!dimenG\!t}%                    ** t, without "pt"

\def\!setinverselinear{%
  \def\!inverseinterp{\!inverselinear}%
  \divide\!dimenE by 8 \!removept\!dimenE\!t
  \!countC=\!intervalno \multiply \!countC 2
  \!countB=\!countC     \advance \!countB -1
  \!countA=\!countB     \advance \!countA -1
  \wlog{\the\!countB th point (\!xmidpt,\!ymidpt) being plotted 
    doesn't lie in the}%
  \wlog{ middle third of the arc between the \the\!countA th 
    and \the\!countC th points:}%
  \wlog{ [arc length \the\!countA\space to \the\!countB]/[arc length 
    \the \!countA\space to \the\!countC]=\!t.}%
  \ignorespaces}

\def\!inverselinear{% 
  \!divide\!distacross\!arclength\!dimenG
  \!removept\!dimenG\!t}

\def\startrotation{%
  \let\!rotateaboutpivot=\!!rotateaboutpivot
  \let\!rotateonly=\!!rotateonly
  \!ifnextchar{b}{\!getsincos }%
    {\!getsincos by {\!cosrotationangle} {\!sinrotationangle} }}
\def\!getsincos by #1 #2 {%
  \edef\!cosrotationangle{#1}%
  \edef\!sinrotationangle{#2}%
  \!ifcoordmode 
    \let\!ROnext=\!ccheckforpivot
  \else
    \let\!ROnext=\!dcheckforpivot
  \fi
  \!ROnext}
\def\!ccheckforpivot{%
  \!ifnextchar{a}{\!cgetpivot}%
    {\!cgetpivot about {\!xpivotcoord} {\!ypivotcoord} }}
\def\!cgetpivot about #1 #2 {%
  \edef\!xpivotcoord{#1}%
  \edef\!ypivotcoord{#2}%
  \!xpivot=#1\!xunit  \!ypivot=#2\!yunit
  \ignorespaces}
\def\!dcheckforpivot{%
  \!ifnextchar{a}{\!dgetpivot}{\ignorespaces}}
\def\!dgetpivot about #1 #2 {%
  \!xpivot=#1\relax  \!ypivot=#2\relax
  \ignorespaces}

\def\stoprotation{%
  \let\!rotateaboutpivot=\!!!rotateaboutpivot
  \let\!rotateonly=\!!!rotateonly
  \ignorespaces}

\def\!!rotateaboutpivot#1#2{%
  \!dimenA=#1\relax  \advance\!dimenA -\!xpivot
  \!dimenB=#2\relax  \advance\!dimenB -\!ypivot
  \!dimenC=\!cosrotationangle\!dimenA
    \advance \!dimenC -\!sinrotationangle\!dimenB
  \!dimenD=\!cosrotationangle\!dimenB
    \advance \!dimenD  \!sinrotationangle\!dimenA
  \advance\!dimenC \!xpivot  \advance\!dimenD \!ypivot
  #1=\!dimenC  #2=\!dimenD
  \ignorespaces}

\def\!!rotateonly#1#2{%
  \!dimenA=#1\relax  \!dimenB=#2\relax 
  \!dimenC=\!cosrotationangle\!dimenA
    \advance \!dimenC -\!rotsign\!sinrotationangle\!dimenB
  \!dimenD=\!cosrotationangle\!dimenB
    \advance \!dimenD  \!rotsign\!sinrotationangle\!dimenA
  #1=\!dimenC  #2=\!dimenD
  \ignorespaces}
\def\!rotsign{}
\def\!!!rotateaboutpivot#1#2{\relax}
\def\!!!rotateonly#1#2{\relax}
\stoprotation

\def\!reverserotateonly#1#2{%
  \def\!rotsign{-}%
  \!rotateonly{#1}{#2}%
  \def\!rotsign{}%
  \ignorespaces}

\def\!getspan span <#1>{%
  \!dshade=#1\relax
  \!ifcoordmode 
    \let\!GRnext=\!GRccheckforAP
  \else
    \let\!GRnext=\!GRdcheckforAP
  \fi
  \!GRnext}
\def\!GRccheckforAP{%
  \!ifnextchar{p}{\!cgetanchor }
    {\!cgetanchor point at {\!xshadesave} {\!yshadesave} }}
\def\!cgetanchor point at #1 #2 {%
  \edef\!xshadesave{#1}\edef\!yshadesave{#2}%
  \!xshade=\!xshadesave\!xunit  \!yshade=\!yshadesave\!yunit
  \ignorespaces}
\def\!GRdcheckforAP{%
  \!ifnextchar{p}{\!dgetanchor}%
    {\ignorespaces}}
\def\!dgetanchor point at #1 #2 {%
  \!xshade=#1\relax  \!yshade=#2\relax
  \ignorespaces}

\def\setshadesymbol{%
  \!ifnextchar<{\!setshadesymbol}{\!setshadesymbol<,,,> }}

\def\!setshadesymbol <#1,#2,#3,#4> (#5#6){%
  \!setputobject{#5}{#6}%                        
  \setbox\!shadesymbol=\box\!putobject%
  \!shadesymbolxshift=\!xshift \!shadesymbolyshift=\!yshift
  \!dimenA=\!xshift \advance\!dimenA \!smidge% ** default LS = xshift - smidge
  \!override\!dimenA{#1}\!lshrinkage%         
  \!dimenA=\!wd \advance \!dimenA -\!xshift%   ** default RS = width - xshift
    \advance\!dimenA \!smidge%                                  - smidge
    \!override\!dimenA{#2}\!rshrinkage
  \!dimenA=\!dp \advance \!dimenA \!yshift%    ** default BS = depth + yshift
    \advance\!dimenA \!smidge%                                  - smidge
    \!override\!dimenA{#3}\!bshrinkage
  \!dimenA=\!ht \advance \!dimenA -\!yshift%   ** default TS = height - yshift
    \advance\!dimenA \!smidge%                                  - smidge
    \!override\!dimenA{#4}\!tshrinkage
  \ignorespaces}
\def\!smidge{-.2pt}%

\def\!override#1#2#3{%
  \edef\!!override{#2}% 
  \ifx \!!override\empty
    #3=#1\relax
  \else
    \if z\!!override
      #3=\!zpt
    \else
      \ifx \!!override\!blankz
        #3=\!zpt
      \else
        #3=#2\relax
      \fi
    \fi
  \fi
  \ignorespaces}
\def\!blankz{ z}

\def\!startvshade#1(#2,#3,#4){%
  \let\!!xunit=\!xunit%
  \let\!!yunit=\!yunit%
  \let\!!xshade=\!xshade%
  \let\!!yshade=\!yshade%
  \def\!getshrinkages{\!vgetshrinkages}%
  \let\!setshadelocation=\!vsetshadelocation%
  \!xS=\!M{#2}\!!xunit
  \!ybS=\!M{#3}\!!yunit
  \!ytS=\!M{#4}\!!yunit
  \!shadexorigin=\!xorigin  \advance \!shadexorigin \!shadesymbolxshift
  \!shadeyorigin=\!yorigin  \advance \!shadeyorigin \!shadesymbolyshift
  \ignorespaces}

\def\!starthshade#1(#2,#3,#4){%
  \let\!!xunit=\!yunit%
  \let\!!yunit=\!xunit%
  \let\!!xshade=\!yshade%
  \let\!!yshade=\!xshade%
  \def\!getshrinkages{\!hgetshrinkages}%
  \let\!setshadelocation=\!hsetshadelocation%
  \!xS=\!M{#2}\!!xunit
  \!ybS=\!M{#3}\!!yunit
  \!ytS=\!M{#4}\!!yunit
  \!shadexorigin=\!xorigin  \advance \!shadexorigin \!shadesymbolxshift
  \!shadeyorigin=\!yorigin  \advance \!shadeyorigin \!shadesymbolyshift
  \ignorespaces}

\def\!lattice#1#2#3#4#5{%
  \!dimenA=#1%                        ** dimA = ANCHOR
  \!dimenB=#2%                        ** dimB = SPAN  (assumed > 0pt)
  \!countB=\!dimenB%                  ** ctB  = SPAN, as a count
  \!dimenC=#3%                        ** dimC = LOCATION
  \advance\!dimenC -\!dimenA%         ** now dimC = LOCATION-ANCHOR
  \!countA=\!dimenC%                  ** ctA = above, as a count
  \divide\!countA \!countB%           ** now ctA = desired index, if dimC <= 0
  \ifdim\!dimenC>\!zpt
    \!dimenD=\!countA\!dimenB%        ** (tentative k)*span
    \ifdim\!dimenD<\!dimenC%          ** if this is false, ctA = desired index
      \advance\!countA 1 %            ** if true, have to add 1
    \fi
  \fi
  \!dimenC=\!countA\!dimenB%          ** lattice location = anchor + ctA*span
    \advance\!dimenC \!dimenA
  #4=\!countA%                        ** the desired index
  #5=\!dimenC%                        ** corresponding lattice location
  \ignorespaces}

\def\!qshade#1(#2,#3,#4)#5(#6,#7,#8){%
  \!xM=\!M{#2}\!!xunit
  \!ybM=\!M{#3}\!!yunit
  \!ytM=\!M{#4}\!!yunit
  \!xE=\!M{#6}\!!xunit
  \!ybE=\!M{#7}\!!yunit
  \!ytE=\!M{#8}\!!yunit
  \!getcoeffs\!xS\!ybS\!xM\!ybM\!xE\!ybE\!ybB\!ybC%**Get coefficients B & C for
  \!getcoeffs\!xS\!ytS\!xM\!ytM\!xE\!ytE\!ytB\!ytC%**y=y0 + B(x-X0) + C(x-X0)**2
  \def\!getylimits{\!qgetylimits}%
  \!shade{#1}\ignorespaces}

\def\!lshade#1(#2,#3,#4){%
  \!xE=\!M{#2}\!!xunit
  \!ybE=\!M{#3}\!!yunit
  \!ytE=\!M{#4}\!!yunit
  \!dimenE=\!xE  \advance \!dimenE -\!xS%   ** xE-xS
  \!dimenC=\!ytE \advance \!dimenC -\!ytS%  ** ytE-ytS
  \!divide\!dimenC\!dimenE\!ytB%            ** ytB = (ytE-ytS)/(xE-xS)
  \!dimenC=\!ybE \advance \!dimenC -\!ybS%  ** ybE-ybS
  \!divide\!dimenC\!dimenE\!ybB%            ** ybB = (ybE-ybS)/(xE-xS)
  \def\!getylimits{\!lgetylimits}%
  \!shade{#1}\ignorespaces}

\def\!getcoeffs#1#2#3#4#5#6#7#8{% 
  \!dimenC=#4\advance \!dimenC -#2%            ** dimC=Y1-Y0
  \!dimenE=#3\advance \!dimenE -#1%            ** dimE=X1-X0
  \!divide\!dimenC\!dimenE\!dimenF%            ** dimF=S1
  \!dimenC=#6\advance \!dimenC -#4%            ** dimC=Y2-Y1
  \!dimenH=#5\advance \!dimenH -#3%            ** dimH=X2-X1
  \!divide\!dimenC\!dimenH\!dimenG%            ** dimG=S2
  \advance\!dimenG -\!dimenF%                  ** dimG=S2-S1
  \advance \!dimenH \!dimenE%                  ** dimH=X2-X0
  \!divide\!dimenG\!dimenH#8%                  ** C=(S2-S1)/(X2-X0)
  \!removept#8\!t%                             ** C, without "pt"
  #7=-\!t\!dimenE%                             ** -C*(X1-X0)
  \advance #7\!dimenF%                         ** B=S1-C*(X1-X0)
  \ignorespaces}

\def\!shade#1{%
  \!getshrinkages#1<,,,>\!nil% %       ** now effective LS=dimE, RS=dimF,
  \advance \!dimenE \!xS%              ** now dimE=xS+LS
  \!lattice\!!xshade\!dshade\!dimenE%  ** set parity=index of left-mst x-lattice
    \!parity\!xpos%                    **   point >= xS+LS, xpos=its location
  \!dimenF=-\!dimenF%                  ** set dimF=xE-RS
    \advance\!dimenF \!xE
  \!loop\!not{\ifdim\!xpos>\!dimenF}%  ** loop over x-lattice points <= xE-RS
    \!shadecolumn%                 
    \advance\!xpos \!dshade%           ** move over to next column
    \advance\!parity 1%                ** increase index of x-point
  \repeat
  \!xS=\!xE%                           ** shift ending values to starting values
  \!ybS=\!ybE
  \!ytS=\!ytE
  \ignorespaces}

\def\!vgetshrinkages#1<#2,#3,#4,#5>#6\!nil{%
  \!override\!lshrinkage{#2}\!dimenE
  \!override\!rshrinkage{#3}\!dimenF
  \!override\!bshrinkage{#4}\!dimenG
  \!override\!tshrinkage{#5}\!dimenH
  \ignorespaces}
\def\!hgetshrinkages#1<#2,#3,#4,#5>#6\!nil{%
  \!override\!lshrinkage{#2}\!dimenG
  \!override\!rshrinkage{#3}\!dimenH
  \!override\!bshrinkage{#4}\!dimenE
  \!override\!tshrinkage{#5}\!dimenF
  \ignorespaces}

\def\!shadecolumn{%
  \!dxpos=\!xpos
  \advance\!dxpos -\!xS%            ** dx = x - xS
  \!removept\!dxpos\!dx%            ** ditto, without "pt"
  \!getylimits%                     ** get top and bottom y-values
  \advance\!ytpos -\!dimenH%        ** less TS
  \advance\!ybpos \!dimenG%         ** plus BS
  \!yloc=\!!yshade%                 ** get anchor point for this column
  \ifodd\!parity 
     \advance\!yloc \!dshade
  \fi
  \!lattice\!yloc{2\!dshade}\!ybpos%
    \!countA\!ypos%                 ** ypos=smallest y point for this column
  \!dimenA=-\!shadexorigin \advance \!dimenA \!xpos%      ** over
  \loop\!not{\ifdim\!ypos>\!ytpos}% ** loop over ypos <= yt(t)
    \!setshadelocation%             ** vmode: xloc=xpos, yloc=ypos 
    \!rotateaboutpivot\!xloc\!yloc%
    \!dimenA=-\!shadexorigin \advance \!dimenA \!xloc%    ** over
    \!dimenB=-\!shadeyorigin \advance \!dimenB \!yloc%    ** up
    \kern\!dimenA \raise\!dimenB\copy\!shadesymbol \kern-\!dimenA
    \advance\!ypos 2\!dshade
  \repeat
  \ignorespaces}

\def\!qgetylimits{%
  \!dimenA=\!dx\!ytC              
  \advance\!dimenA \!ytB%         ** yt(t)=ytS + dx*(Bt + dx*Ct)
  \!ytpos=\!dx\!dimenA
  \advance\!ytpos \!ytS
  \!dimenA=\!dx\!ybC              
  \advance\!dimenA \!ybB%         ** yb(t)=ybS + dx*(Bb + dx*Cb)
  \!ybpos=\!dx\!dimenA
  \advance\!ybpos \!ybS}

\def\!lgetylimits{%
  \!ytpos=\!dx\!ytB%              ** yt(t)=ytS + dx*Bt
  \advance\!ytpos \!ytS
  \!ybpos=\!dx\!ybB%              ** yb(t)=ybS + dx*Bb
  \advance\!ybpos \!ybS}

\def\!vsetshadelocation{%         ** vmode: xloc=xpos, yloc=ypos 
  \!xloc=\!xpos
  \!yloc=\!ypos}
\def\!hsetshadelocation{%         ** hmode: xloc=ypos, yloc=xpos 
  \!xloc=\!ypos
  \!yloc=\!xpos}

\def\!axisticks {%
  \def\!nextkeyword##1 {%
    \expandafter\ifx\csname !ticks##1\endcsname \relax
      \def\!next{\!fixkeyword{##1}}%
    \else
      \def\!next{\csname !ticks##1\endcsname}%
    \fi
    \!next}%
  \!axissetup
    \def\!axissetup{\relax}%
  \edef\!ticksinoutsign{\!ticksinoutSign}%
  \!ticklength=\longticklength
  \!tickwidth=\linethickness
  \!gridlinestatus
  \!setticktransform
  \!maketick
  \!tickcase=0
  \def\!LTlist{}%
  \!nextkeyword}

\def\ticksout{%
  \def\!ticksinoutSign{+}}

\ticksout

\def\nogridlines{%
  \def\!gridlinestatus{\!gridlinestoofalse}}
\nogridlines

\def\loggedticks{%
  \def\!setticktransform{\let\!ticktransform=\!logten}}
\def\unloggedticks{%
  \def\!setticktransform{\let\!ticktransform=\!donothing}}
\def\!donothing#1#2{\def#2{#1}}
\unloggedticks

\expandafter\def\csname !ticks/\endcsname{%
  \!not {\ifx \!LTlist\empty}
    \!placetickvalues
  \fi
  \def\!tickvalueslist{}%
  \def\!LTlist{}%
  \expandafter\csname !axis/\endcsname}

\def\!maketick{%
  \setbox\!boxA=\hbox{%
    \beginpicture
      \!setdimenmode
      \setcoordinatesystem point at {\!zpt} {\!zpt}   
      \linethickness=\!tickwidth
      \ifdim\!ticklength>\!zpt
        \putrule from {\!zpt} {\!zpt} to
          {\!ticksinoutsign\!tickxsign\!ticklength}
          {\!ticksinoutsign\!tickysign\!ticklength}
      \fi
      \if!gridlinestoo
        \putrule from {\!zpt} {\!zpt} to
          {-\!tickxsign\!xaxislength} {-\!tickysign\!yaxislength}
      \fi
    \endpicturesave <\!Xsave,\!Ysave>}%
    \wd\!boxA=\!zpt}
  
\def\!ticksin{%
  \def\!ticksinoutsign{-}%
  \!maketick
  \!nextkeyword}

\def\!ticksout{%
  \def\!ticksinoutsign{+}%
  \!maketick
  \!nextkeyword}

\def\!tickslength<#1> {%
  \!ticklength=#1\relax
  \!maketick
  \!nextkeyword}

\def\!tickslong{%
  \!tickslength<\longticklength> }

\def\!ticksshort{%
  \!tickslength<\shortticklength> }

\def\!tickswidth<#1> {%
  \!tickwidth=#1\relax
  \!maketick
  \!nextkeyword}

\def\!ticksandacross{%
  \!gridlinestootrue
  \!maketick
  \!nextkeyword}

\def\!ticksbutnotacross{%
  \!gridlinestoofalse
  \!maketick
  \!nextkeyword}

\def\!tickslogged{%
  \let\!ticktransform=\!logten
  \!nextkeyword}

\def\!ticksunlogged{%
  \let\!ticktransform=\!donothing
  \!nextkeyword}

\def\!ticksunlabeled{%
  \!tickcase=0
  \!nextkeyword}

\def\!ticksnumbered{%
  \!tickcase=1
  \!nextkeyword}

\def\!tickswithvalues#1/ {%
  \edef\!tickvalueslist{#1! /}%
  \!tickcase=2
  \!nextkeyword}

\def\!ticksquantity#1 {%
  \ifnum #1>1
    \!updatetickoffset
    \!countA=#1\relax
    \advance \!countA -1
    \!ticklocationincr=\!axisLength
      \divide \!ticklocationincr \!countA
    \!ticklocation=\!axisstart
    \loop \!not{\ifdim \!ticklocation>\!axisend}
      \!placetick\!ticklocation
      \ifcase\!tickcase
          \relax %  Case 0: no labels
        \or
          \relax %  Case 1: numbered -- not available here
        \or
          \expandafter\!gettickvaluefrom\!tickvalueslist
          \edef\!tickfield{{\the\!ticklocation}{\!value}}%
          \expandafter\!listaddon\expandafter{\!tickfield}\!LTlist%
      \fi
      \advance \!ticklocation \!ticklocationincr
    \repeat
  \fi
  \!nextkeyword}

\def\!ticksat#1 {%
  \!updatetickoffset
  \edef\!Loc{#1}%
  \if /\!Loc
    \def\next{\!nextkeyword}%
  \else
    \!ticksincommon
    \def\next{\!ticksat}%
  \fi
  \next}    
      
\def\!ticksfrom#1 to #2 by #3 {%
  \!updatetickoffset
  \edef\!arg{#3}%
  \expandafter\!separate\!arg\!nil
  \!scalefactor=1
  \expandafter\!countfigures\!arg/
  \edef\!arg{#1}%
  \!scaleup\!arg by\!scalefactor to\!countE
  \edef\!arg{#2}%
  \!scaleup\!arg by\!scalefactor to\!countF
  \edef\!arg{#3}%
  \!scaleup\!arg by\!scalefactor to\!countG
  \loop \!not{\ifnum\!countE>\!countF}
    \ifnum\!scalefactor=1
      \edef\!Loc{\the\!countE}%
    \else
      \!scaledown\!countE by\!scalefactor to\!Loc
    \fi
    \!ticksincommon
    \advance \!countE \!countG
  \repeat
  \!nextkeyword}

\def\!updatetickoffset{%
  \!dimenA=\!ticksinoutsign\!ticklength
  \ifdim \!dimenA>\!offset
    \!offset=\!dimenA
  \fi}

\def\!placetick#1{%
  \if!xswitch
    \!xpos=#1\relax
    \!ypos=\!axisylevel
  \else
    \!xpos=\!axisxlevel
    \!ypos=#1\relax
  \fi
  \advance\!xpos \!Xsave
  \advance\!ypos \!Ysave
  \kern\!xpos\raise\!ypos\copy\!boxA\kern-\!xpos
  \ignorespaces}

\def\!gettickvaluefrom#1 #2 /{%
  \edef\!value{#1}%
  \edef\!tickvalueslist{#2 /}%
  \ifx \!tickvalueslist\!endtickvaluelist
    \!tickcase=0
  \fi}
\def\!endtickvaluelist{! /}

\def\!ticksincommon{%
  \!ticktransform\!Loc\!t
  \!ticklocation=\!t\!!unit
  \advance\!ticklocation -\!!origin
  \!placetick\!ticklocation
  \ifcase\!tickcase
    \relax % Case 0: no labels
  \or %      Case 1: numbered
    \ifdim\!ticklocation<-\!!origin
      \edef\!Loc{$\!Loc$}%
    \fi
    \edef\!tickfield{{\the\!ticklocation}{\!Loc}}%
    \expandafter\!listaddon\expandafter{\!tickfield}\!LTlist%
  \or %      Case 2: labeled
    \expandafter\!gettickvaluefrom\!tickvalueslist
    \edef\!tickfield{{\the\!ticklocation}{\!value}}%
    \expandafter\!listaddon\expandafter{\!tickfield}\!LTlist%
  \fi}

\def\!separate#1\!nil{%
  \!ifnextchar{-}{\!!separate}{\!!!separate}#1\!nil}
\def\!!separate-#1\!nil{%
  \def\!sign{-}%
  \!!!!separate#1..\!nil}
\def\!!!separate#1\!nil{%
  \def\!sign{+}%
  \!!!!separate#1..\!nil}
\def\!!!!separate#1.#2.#3\!nil{%
  \def\!arg{#1}%
  \ifx\!arg\!empty
    \!countA=0
  \else
    \!countA=\!arg
  \fi
  \def\!arg{#2}%
  \ifx\!arg\!empty
    \!countB=0
  \else
    \!countB=\!arg
  \fi}

\def\!countfigures#1{%
  \if #1/%
    \def\!next{\ignorespaces}%
  \else
    \multiply\!scalefactor 10
    \def\!next{\!countfigures}%
  \fi
  \!next}

\def\!scaleup#1by#2to#3{%
  \expandafter\!separate#1\!nil
  \multiply\!countA #2\relax
  \advance\!countA \!countB
  \if -\!sign
    \!countA=-\!countA
  \fi
  #3=\!countA
  \ignorespaces}

\def\!scaledown#1by#2to#3{%
  \!countA=#1\relax%                          ** get original #
  \ifnum \!countA<0 %                         ** take abs value,
    \def\!sign{-}%                            **   remember sign
    \!countA=-\!countA
  \else
    \def\!sign{}%
  \fi
  \!countB=\!countA%                          ** copy |#|
  \divide\!countB #2\relax%                   ** integer part (|#|/sf)
  \!countC=\!countB%                          ** get sf * (|#|/sf)
    \multiply\!countC #2\relax
  \advance \!countA -\!countC%                ** ctA is now remainder
  \edef#3{\!sign\the\!countB.}%               ** +- integerpart.
  \!countC=\!countA %                         ** Tack on proper number
  \ifnum\!countC=0 %                          **   of zeros after .
    \!countC=1
  \fi
  \multiply\!countC 10
  \!loop \ifnum #2>\!countC
    \edef#3{#3\!zero}%
    \multiply\!countC 10
  \repeat
  \edef#3{#3\the\!countA}%                    ** Add on rest of remainder
  \ignorespaces}

\def\!placetickvalues{%
  \advance\!offset \tickstovaluesleading
  \if!xswitch
    \setbox\!boxA=\hbox{%
      \def\\##1##2{%
        \!dimenput {##2} [B] (##1,\!axisylevel)}%
      \beginpicture 
        \!LTlist
      \endpicturesave <\!Xsave,\!Ysave>}%
    \!dimenA=\!axisylevel
      \advance\!dimenA -\!Ysave
      \advance\!dimenA \!tickysign\!offset
      \if -\!tickysign
        \advance\!dimenA -\ht\!boxA
      \else
        \advance\!dimenA  \dp\!boxA
      \fi
    \advance\!offset \ht\!boxA 
      \advance\!offset \dp\!boxA
    \!dimenput {\box\!boxA} [Bl] <\!Xsave,\!Ysave> (\!zpt,\!dimenA)
  \else
    \setbox\!boxA=\hbox{%
      \def\\##1##2{%
        \!dimenput {##2} [r] (\!axisxlevel,##1)}%
      \beginpicture 
        \!LTlist
      \endpicturesave <\!Xsave,\!Ysave>}%
    \!dimenA=\!axisxlevel
      \advance\!dimenA -\!Xsave
      \advance\!dimenA \!tickxsign\!offset
      \if -\!tickxsign
        \advance\!dimenA -\wd\!boxA
      \fi
    \advance\!offset \wd\!boxA
    \!dimenput {\box\!boxA} [Bl] <\!Xsave,\!Ysave> (\!dimenA,\!zpt)
  \fi}

\normalgraphs
\catcode`!=12 %  *****  THIS MUST NEVER BE OMITTED

\documentstyle[mysprocl,epsf]{article}

\def\mysection#1{\setcounter{footnote}{0} \section{#1}}

\def\mum{{\mu_{\mbox{\tiny M}}}}
\def\M2{\mbox{\tiny M$^2$RPA}}
\def\vec#1{{\bf#1}}
\def\hatn#1{{\bf{\hat{#1}}}}
\def\vf{v_{\mbox{\tiny{F}}}^*}
\def\kf{k_{\mbox{\tiny{F}}}}
\def\pf{p_{\mbox{\tiny{F}}}}

\def\vfns{v_{\mbox{\tiny{F}}}}
\def\nusub#1{\nu_{{}_{#1}}}

\def\vec#1{{\bf #1}}

\def\be{\begin{equation}}

\bibliographystyle{unsrt}    
\begin{document}
\title{The Chern-Simons Fermi Liquid Description of Fractional Quantum
  Hall States}

\author{Steven H. Simon} \address{Bell Laboratories, Lucent Technologies \\
 Murray Hill, NJ 07974}
\maketitle\abstracts{The composite fermion picture has had a
  remarkable number of recent successes both in the description of the
  fractional quantized Hall states and in the description of the even
  denominator Fermi-liquid like states.  In this chapter, we give an
  introductory account of the  Chern-Simons fermion theory, focusing
  on the description of the even denominator states as unusual
  Fermi liquids.}

\tableofcontents

\newpage

\mysection{Introduction}

Since its discovery, the field of quantum Hall
physics~\cite{Prange,Book2,DasSarmabook,Chakraborty} has had more than
its share of surprising discoveries.  Many, if not most, of the
important concepts in the field seem so exotic that it is almost
certain that they never would have been taken seriously were they not
so strongly supported by experiments and exact diagonalizations.  The
composite fermion is one of these unbelievable concepts.

The field of composite fermion physics began in 1989 with a paper by
Jainendra Jain~\cite{Jain0,Jain} who pointed out that there is a
mapping between the wavefunctions of integer quantized Hall states and
approximate --- but extremely good --- wavefunctions for fractional
quantized Hall states.  This wavefunction mapping can be thought of as
binding an even number of vortices (zeros) of the wavefunction to each
electron, turning it into a ``composite'' fermion.  

Although the idea of the composite fermion initially met with a fair
amount of resistance from the theoretical community, slowly its
validity was accepted and the idea was further developed by several
groups. Using machinery developed earlier in the bosonic picture of
the fractional quantum Hall
effect~\cite{GirvinMacDonald,NickReadOld,ZhangReview} and in work on
anyon superconductivity~\cite{Anyons,AnyonSupercon,AnyonicCS}, a field
theoretical formalism was developed that roughly corresponded to
Jain's composite fermion wavefunction approach~\cite{Lopez,HLR,Zhang}.
In this ``Chern-Simons'' approach, the electron is exactly modeled as
as fermion bound to an even number of fictitious flux quanta which in
some sense represent the even number of vortices of Jain's composite
fermion.  The great advantage of this Chern-Simons approach is that it
allows for simple yet systematic (or at least semi-systematic)
calculations of quantities such as conductivities that are measurable
experimentally.

The Chern-Simons fermionic field theory was first used by Lopez and
Fradkin~\cite{Lopez} to describe the Jain series of incompressible
fractional quantized Hall states.  Shortly thereafter, it was used by
Halperin, Lee, and Read (known as HLR)~\cite{HLR} and Kalmeyer and
Zhang~\cite{Zhang} to describe the even denominator states.  The most
surprising result of this approach is that the even denominator states
are compressible and are Fermi-liquid-like.

In this chapter, I will review recent work on the Chern-Simons fermion
description of fractional Hall states at a level that should be
accessible to readers who are not experts in the field.  I hope that
this chapter will serve, at least partially, as an introduction to the
other chapters in this book.  Since the field of composite fermion
physics is quite large, no single article could hope to discuss all of
the work that has been done in the field~\footnote{This entire {\it
    book} does not even come close to discussing all of the work done
  in this field.}.  I have thus chosen to discuss some subset of
topics that fit together to form a coherent story, and hopefully I
will touch on many of the more important issues that have been raised
in the last few years.  However, many other important works will certainly
be neglected and I will apologize in advance for these omissions.

The outline of this chapter is as follows.  Section \ref{sec:review} is
an introduction to the Chern-Simons theory.  We begin with a review of
integer quantized Hall effect and move on to the Chern-Simons theory and
the Chern-Simons mean field approximation.  Within this approximation,
we will discuss both the incompressible fractional quantized Hall
states as well as the compressible Fermi-liquid-like even denominator
states.  The mean field, of course, is extremely crude and in
particular does not correctly predict response function such as the
Hall conductivity.  To fix this problem, in section \ref{sec:RPA} we
will discuss the more sophisticated Chern-Simons RPA approximation in
great depth.  Although the RPA repairs some of the problems of mean
field theory, it cannot be made to give the correct energy scale for
low energy excitations while maintaining Galilean invariance.
Furthermore, as we will see in section \ref{sec:Pert}, attempts to
systematically calculate corrections to RPA are plagued with infrared
divergences of quantities such as the effective mass.

The problems with the RPA description encourage us to turn to a
phenomenological Landau-Fermi liquid theory approach in section
\ref{sec:FLT}.  We begin by giving a detailed review of the Landau
description and describe how we expect the Chern-Simons fermi liquid
to fit into this picture.  We then develop the MRPA~\cite{Simonhalp}, a
phenomenological approximation motivated by Fermi liquid theory that
repairs the RPA's problems with energy scales.  We realize that even
this improved approximation does not properly represent so-called
magnetization effects, requiring us to propose yet another
approximation~\cite{SimonStern,SimonNFL}, the $\mbox{M}^2\mbox{RPA}$
and show how this approximation can also fit into the Landau picture.

In section \ref{sec:Pert} we attempt a more systematic perturbation
expansion and wrestle with the pathologies of this Chern-Simons
theory.  In particular, we are concerned with to what extent the
Landau fermi liquid picture we developed above is consistent with the
results of perturbative calculations. 

In section \ref{sec:wavefunction} we discuss the wavefunction approach
to composite fermions, which leads us to a somewhat different picture
of {\it neutral} dipole fermions at even denominator filling fractions
(compared to the Chern-Simons fermions which are charged), and we
relate this neutral dipole picture to the Chern-Simons picture.
Finally, in section \ref{sec:exp} we will critically discuss some of
the experimental results, and in section \ref{sec:theend} we will
briefly mention some other directions and summarize what we have
learned.

\mysection{Introduction to Chern-Simons Fermions}
\label{sec:review}

In this section background material will be given in detail.  Readers
who desire a more thorough review of previous works in quantum Hall
physics are referred to References
\onlinecite{Prange,Book2,DasSarmabook,Chakraborty}.

\subsection{Quantum Hall Effect Basics}
\label{sub:Basics}

This section is written for the reader who needs to be reminded of a
few of the essentials of quantum Hall physics.  The experienced reader
is encouraged to skip to section \ref{sub:CS} referring back to this
primer only when necessary.

We begin by considering a two dimensional electron gas (2DEG)
consisting of $N$ interacting electrons of band mass $m_{\rm b}$ in a
magnetic field $B = \nabla \times \vec A$ perpendicular to the plane
of the system (We will call the normal to the plane the $\hatn z$
direction).  We will always neglect the spin degree of freedom of the
electrons, assuming that the magnetic field is sufficiently high such
that the electrons are spin-polarized.  This assumption is reasonable
for many quantum Hall experiments, and it will make our discussion
much simpler.

The Hamiltonian for such a spin-polarized (or spinless) system of
electrons is written as
\begin{equation}
  \label{eq:Hamiltonian}
  H_{\rm e} = \sum_{j} \frac{\left[\vec p_j + \frac{e}{c} \vec A(\vec
    r_j)\right]^2}{2m_{\rm b}} 
+ \sum_{i<j} v(\vec r_i - \vec r_j).
\end{equation}
where $v$ is the two body interaction potential, $c$ is the speed of
light and $-e$ is the charge of the electron~\footnote{This
  inconvenient convention is the source of endless sign problems.  We
  have Ben Franklin to blame for this headache.}.  Here, $m_{\rm b}$
is the bare electron band mass~\footnote{It is convenient that the
  subscript ${\rm b}$ can stand for `bare' or `band'}.  We will often
specialize to the physical case of Coulombic interaction
\begin{equation}
v(\vec r) = v(|\vec r|) = \frac{e^2}{\epsilon
|\vec r|}
\end{equation}
with $\epsilon$ the background dielectric constant.  However, it will
also be useful at times to consider other forms of electron-electron
interaction.  It is amusing that so much interesting quantum Hall
physics will come from such a simple looking Hamiltonian. 

\subsubsection{Integer Quantized Hall Effect : Single Electron Physics}
\label{sub:IQHE}

If we ignore the interactions between electrons (setting $v=0$) the
Hamiltonian breaks up into the sum of single particle Hamiltonians
which can be easily diagonalized (See Ref.~\onlinecite{Prange} for
details).  We can solve for the eigenfunctions $\varphi_{kn}(\vec r)$
whose eigenenergies are given by $E_{nk} = E_n = \hbar \omega_{\rm c} (n +
\frac{1}{2})$ where
\begin{equation}
  \omega_{\rm c} = \frac{eB}{m_{\rm b} c}
\end{equation}
is the cyclotron frequency~\footnote{The discrete energy levels can be
  understood semiclassically as being the Bohr quantization of the
  electron making cyclotron orbits in a magnetic field.  The index $k$
  corresponds to the degeneracy of the many places we can put the
  center of the cyclotron orbit.}.  For each value of $n$, the index
$k$ can take $B/\phi_0$ different values per unit area of the system
where~\footnote{The reader is warned that $\phi_0 = 2 \pi$ is used as
  often as not in the literature, and factors of $\hbar$, $c$ and $e$
  tend to appear and disappear almost randomly.}
\begin{equation}
  \phi_0 = \frac{2 \pi \hbar c}{e} = 2 \pi   \,\,\,\,\,\  \mbox{(in
  units with} \,\,\,\, \hbar = e = c = 1) 
\end{equation}
is the quantum mechanical unit of flux.  Thus the spectrum breaks up
into highly degenerate ``Landau bands'' whose degeneracy is given by
the value of the magnetic field and the area of the system.  We can
then define a natural magnetic length scale
\begin{equation}
 l_B = \sqrt{\frac{\phi_0}{2 \pi B}}, 
\end{equation}
such that the filling fraction
\begin{equation}
  \label{eq:nudef}
  \nu = \frac{\phi_0 n_{\rm e}}{B} = 2 \pi n_{\rm e} l_B^2
\end{equation}
with $n_{\rm e}$ the electron density  gives the number of Landau levels
completely filled.  Note that when an integer number of Landau bands
are completely filled there is a discontinuity in the chemical
potential ({\em i.e.\/}, when $\nu$ is an integer, adding the $(n+1)^{st}$
electron costs $\hbar \omega_{\rm c}$ more energy than adding the $n^{th}$
electron).  This discontinuity in the chemical potential, or
``thermodynamic incompressibility'' is the trademark of a quantized Hall
state. Another way to describe this incompressibility is to note that
when there are an integer number of Landau levels filled there is an
energy gap of
\begin{equation}
\label{eq:IQHEgap}
 E_g = \hbar \omega_{\rm c} = \hbar e B/(m_{\rm b} c)
\end{equation}
between the ground state and the lowest excited states since any
excitation of this system must involve promoting an electron to a
higher Landau level.  

By using Galilean invariance, one can easily show that a perfectly
clean system at finite filling fraction $\nu$ has zero diagonal
(longitudinal) DC resistivity $\rho_{xx}$ and a DC Hall resistivity
given by $ \rho_{xy} = \frac{1}{\nu}\frac{h}{e^2}$ with $h$ Planck's
constant and $e$ the electron charge~\footnote{To show this, consider
  applying an electric field $\vec E$ to the system.  In a reference
  frame moving at a velocity $\vec v$, the electric field is $\vec E -
  \frac{1}{c} \vec v \times \vec B$.  Choosing $\vec v$ appropriately
  (such that the Lorentz force $\vec F= e \vec E + \frac{e}{c} \vec v
  \times \vec B$ vanishes) then in this new frame, we simply have a
  system of electrons in magnetic field $\vec B$ but in zero electric
  field so there is no net current in this frame.  The current in the
  original frame is then just the boost velocity times the charge
  density of the system.  Thus, we find that $\vec j = \hatn z \times
  \vec E (n_{\rm e} e c /B)$ or $\rho_{xy} = n_{\rm e} e c/B = h/(e^2 \nu)$ and
  $\rho_{xx} =0$.  Clearly, any amount of disorder will ruin this
  argument.}.  A general theorem~\cite{Prange} then states that when a
perfectly clean system forms an incompressible state at some filling
fraction $\nu_0$ (an integer for example), then a system with small
but nonzero disorder will display a quantized Hall state for some
range of filling fractions around $\nu_0$, with zero diagonal
resistivity (despite the disorder) and quantized Hall resistivity.
The DC resistivity matrix of this quantized Hall state for a range of
fillings around filling fraction $\nu_0$ is thus given
by~\footnote{Note that the inverse of this matrix (the conductivity
  matrix) also has zero diagonal components.  Thus, we have the
  interesting case of having zero longitudinal resistivity as well as
  zero longitudinal conductivity.  This, of course, is just the
  statement that the current runs precisely perpendicular to the
  voltage.}
\begin{equation}
  \rho = \frac{h}{e^2} \left[ \begin{array}{cc} 0 & 1/\nu_0 \\ -1/\nu_0 &
  0
\end{array} \right]_.
\end{equation}
where the resistivity matrix is defined to relate the local current
density to the electric field via
\begin{equation}
  \vec j  = \rho \vec E.
\end{equation}

It has been shown that gauge invariance guarantees the {\it precise}
quantization of the resistance~\footnote{For most systems there is an
  important distinction between the {\it resistivity} (in this case a
  matrix) which relates local currents to local electric fields and
  the various possible {\it resistances} of the system which relate
  the voltage measured between two contacts to the currents passing
  through two leads.  However, for quantized Hall states, due to the
  zero longitudinal resistivity, it can easily be shown~\cite{Prange}
  that any such resistance measurement (ratio of a current to a
  voltage) must either be zero (for any longitudinal measurement) or
  must be equal to the quantized Hall value of the resistivity
  $\frac{1}{\nu} h/e^2$ (for any Hall measurement).}  of the quantized
Hall state~\cite{LaughlinInt}.  Indeed, in Von Klitzing's now famous
paper first demonstrating the integer quantized Hall
effect~\cite{Klitzing}, it was proposed that this effect could be used
for precision measurements of the resistance quantum $h/e^2$.  Such
measurements have now been performed to a precision of a part in
$10^9$ and are now used as an international
metrological standard~\cite{Metrology}.  This incredible precision is
roughly analogous to measuring the circumference of the earth to
within a single centimeter~\footnote{Some atomic physics experiments
  have achieved precisions of a part in $10^{14}$ or even better.
  However, when one recalls that the quantum Hall system is full of
  all sorts of impurities and other garbage, even atomic physicists
  are impressed.}.  Fundamentally, this precise quantization is based
on the incompressibility, or rigidity, of the state.  When the clean
system is incompressible, small changes in filling fraction result in
defects in the state (quasiparticles), rather than a global
destruction of the state.  So long as these defects in the state
become localized due to disorder they do not ruin the macroscopic
integrity of the quantum Hall state, such that for a range of filling
fractions around the quantized value, the resistance of the state
remains unchanged.  This situation is quite reminiscent of the
situation in a type II superconductor~\cite{ZhangReview} where applying
a magnetic field creates a vortex; and so long as the vortex remains
pinned, the system remains  superconductive.

\subsubsection{Fractional Quantized Hall Effect : Interactions}

In the single electron picture at filling fractions that are not
integers, there is an enormous degeneracy of states associated with
the Landau level degeneracy.  Consider for example, the filling
fraction $\nu=\frac{1}{3}$.  Here, we have a macroscopic number $N$ of
electrons and $3N$ states in the lowest Landau level.  Thus, there are
$(3N)!/(N! (2N)!)$ ways to distribute the electrons which is an immensely
huge number.  Neglecting interactions, all of these ways have
precisely the same energy ($\frac{N}{2} \hbar \omega_{\rm c}$).  Of course,
when one includes interactions, some of these ways of distributing the
electrons will be found to be more favorable than other ways.
However, owing to the enormous degeneracy of states, one might naively
guess that the state at filling fraction $\nu=\frac{1}{3}$ would be
quite compressible because of the high density of states all with very
similar energies.  It thus came as quite a surprise when it was first
discovered~\cite{firstFQHE} that incompressible ``fractional''
quantized Hall states can form at many filling fractions (such as
$\nu=\frac{1}{3}$) that are not integers.  The existence of these fractional
Hall states tells us that somehow, interactions are causing
incompressible states to be formed at these non-integer filling
fractions.  In order to understand fractional Hall states, we will
have to find a way to treat inter-electron interactions as well as the
kinetic energy of the Landau levels.  This, however, is a difficult
challenge.

To begin to address this challenge we should carefully consider the
two terms in the Hamiltonian (Eq. \ref{eq:Hamiltonian}).  Since the
kinetic term of the Hamiltonian is easily diagonalized, one might
consider treating the interaction term in some sort of perturbation
theory.  However, the rules of degenerate perturbation theory require
us to diagonalize each degenerate subspace first before proceeding.
In our case, the degenerate subspace of a partially filled Landau
level is immensely huge and this first step is almost as impossible as
solving the entire problem, thus rendering such a perturbative
approach hopeless.  The interaction term in the Hamiltonian, on the
other hand, if treated alone, results in a Wigner crystal which is
quite a different state from the fractional Hall state.  It is only
the interplay between the kinetic and potential terms that yields the
fractional quantized Hall effect.  Thus, it seems that the traditional
systematic perturbative approaches for understanding the fractional
quantum Hall effect are out of the question.

Indeed, much of our understanding of fractional Hall effect is based
on understanding the properties of trial wavefunctions, and not on any
systematic perturbation approach~\cite{Laughlin,Jain}.  Only recently
the Chern-Simons field theories~\cite{ZhangReview,Lopez,HLR} have
provided a starting point for a semi-systematic perturbative approach
to understanding these states. As we will see below, these approaches
have their share of complications (hence the prefix ``semi'' before
the word ``systematic'').  Nonetheless, the Chern-Simons field
theories have led to a much deeper understanding of many aspects of
the fractional quantum Hall effect.  Most of the remainder of this
paper will be devoted to elucidating aspects of the so-called
``Chern-Simons-Fermionic'' description of the fractional
quantum Hall regime.

\subsection{Chern-Simons Transformation}
\label{sub:CS}

The Chern-Simons approach is employed by making a transformation on
the phase of the many-electron wavefunction.  Writing the electron
wavefunction as $\Psi_{\rm e}(\vec r_1,\vec r_2,\ldots \vec r_N)$ with $\vec
r_j$ the position of the $j^{th}$ electron, we define a new
transformed wavefunction
\begin{equation}
 \label{eq:noanal1}
  \Phi(\vec r_1,\vec r_2,\ldots,\vec r_N) = 
 \left[ \prod_{i<j}  e^{- i \tilde \phi \theta(\vec r_i - \vec r_j)} \right] 
 \Psi_{\rm e}(\vec r_1,\vec r_2,\ldots,\vec r_N)
\end{equation}
where
\begin{equation}
  \tilde \phi = 2m
\end{equation}
is an even integer, and the `function' $\theta(\vec r_i - \vec r_j)$
is the angle formed by the vector $\vec r_i - \vec r_j$ with the $\hatn
x$ axis.  Note that the `function' $\theta$ is defined only modulo
$2 \pi$, but becomes well defined once exponentiated (one should
consider it to be multiple valued rather than to have branch cuts).

It is then easy to see that~\cite{Lopez,HLR} if $\Psi_{\rm e}$ is a solution of
the Schroedinger equation $H_{\rm e} \Psi_{\rm e} = E \Psi_{\rm e}$, then $\Phi$ is a
solution to the Schroedinger equation $H \Phi = E \Phi$ with
\begin{equation}
  \label{eq:transformedH}
  H = \sum_{j} \frac{\left[\vec p_j + \frac{e}{c} \vec A(\vec r_j) -
    \frac{e}{c} \vec a(\vec r_j)\right]^2}{2m_{\rm b}} + \sum_{i<j} v(\vec
  r_i - \vec r_j)
\end{equation} 
where $\vec a$ is the ``Chern-Simons'' vector potential 
\begin{equation}
\label{eq:csvec1}
\vec a(\vec r_i) =  i 
\nabla_i \left[ \prod_{j} 
 e^{- i \tilde \phi   \theta(\vec r_i - \vec r_j)} \right] =
\frac{\tilde \phi \phi_0}{2 \pi} \sum_{j=1}^{N}
\frac{\hatn{z} \times (\vec r_i - \vec r_j)}{|\vec r_i - \vec r_j|^2},
\end{equation}
and $\tilde \phi = 2m$.  

Since $\vec a$ is a gradient, $\nabla \times \vec a(\vec r) = 0$ for
all $\vec r$ not equal to the position of one of the electrons.  Thus
we might think of this transformation as being just a gauge
transformation.  
However, due to the nonsinglevaluedness of the function $\theta$, the
gauge transformation is singular and  we
have a singularity, $\nabla \times \vec a(\vec r) = \tilde \phi \phi_0
\delta(\vec r - \vec r_j)$ at the position of each electron.  The
Chern-Simons magnetic field $b(\vec r)$ associated with the vector
potential $\vec a$ is then given by
\begin{equation}
\label{eq:last1}
b(\vec r) = \nabla \times \vec a(\vec r) =   \phi_0 \tilde \phi n(\vec r) 
 = 2 \pi \tilde \phi n(\vec r)
\end{equation} 
where $n(\vec r) = \sum_{j} \delta(\vec r - \vec r_j)$ is the local
particle density.  Note that since the magnitude of the wavefunction
is not changed by this Chern-Simons transformation, the electron
density and the transformed fermion density are equal, as can easily
be seen by writing
\begin{eqnarray} \nonumber
  n_{\rm e}(\vec r_1) &=&  \int \! d\vec r_2 \int \! d\vec r_3 \cdots \int \! d\vec r_N 
\left|     \Phi(\vec r_1,\vec r_2,\ldots,\vec r_N) \right|^2 \\
&=& \int \! d\vec r_2 \int \! d\vec r_3 \cdots  \int \! d\vec r_N 
\left|     \Psi(\vec r_1,\vec r_2,\ldots,\vec r_N) \right|^2 =
n_{\rm f}(\vec r_1).
\end{eqnarray}

It is also easy to see that if $\Psi$ obeys fermionic statistics in
the sense that the wavefunction is antisymmetric under exchange of any
two particles, then the transformed wavefunction $\Phi$ is similarly
antisymmetric and hence represents a fermionic wavefunction.  Thus we
should think of the Hamiltonian $H$ as being the Hamiltonian for $N$
interacting transformed fermions.  For $\tilde \phi$ not equal to an
even integer, the resulting wavefunction $\Phi$ would not be
antisymmetric. In this case, the resulting Hamiltonian represents
particles of non-fermionic statistics --- bosonic for $\tilde \phi$ an
odd integer and anyonic~\cite{AnyonicCS} for non-integer values.
Indeed this type of Chern-Simons transformation had been used
previously to develop a bosonic description of Quantum Hall
states~\cite{GirvinMacDonald,NickReadOld,ZhangReview}, and in the description
of anyon superconductors~\cite{Anyons,AnyonSupercon,AnyonicCS}.

In summary, the Chern-Simons transformation can be described as the
exact modeling of an electron as a fermion attached to $\tilde \phi =
2m$ flux quanta.  We call these fermions ``singularly gauge
transformed'',``Chern-Simons'', or ``composite''
fermions.~\footnote{The loose use of nomenclature here is somewhat
  unfortunate.  Jain~\cite{Jain0} originally used the term `composite
  fermion' to describe a related --- but nonetheless distinctly
  different --- wavefunction transformation.  For a brief period of
  time, some members of the community made an effort to distinguish
  between Jain's composite fermions and these
  Chern-Simons-singularly-gauge-transformed fermions.  However, this
  more specific nomenclature was apparently too clumsy and now the
  simpler terminology `composite fermion' is used as often as not to
  describe the transformed fermions.  The relation between Jain's
  approach and the Chern-Simons approach will be elaborated in section
  \ref{sec:wavefunction} below.} We emphasize that this fictitious
Chern-Simons magnetic field (which in truth is just a mathematical
convenience) certainly does not exist outside of the two dimensional
electron system.

At this point, one might note that we have transformed a relatively
simple looking system (electrons in a strong magnetic field
interacting via the Coulomb interaction) into a much more complicated
looking system (transformed fermions in a strong magnetic field
interacting via the Coulomb interaction {\it and} via a Chern-Simons
gauge field).  However, as we will see below, this transformation is
actually very useful since the simple looking electron system is
actually totally intractable (due to the huge degeneracy of states),
whereas the more complicated looking transformed fermion system is
actually something we will be able to work with.

\subsection{Mean Field Theory}
\label{sec:MFT}

The simplest approach to analyzing this transformed Chern-Simons
fermion system is to make the mean field approximation in which
density is assumed uniform and the Chern-Simons flux quanta attached
to the fermions are smeared out into a uniform magnetic field of
magnitude
\begin{equation}
  \label{eq:mfield1}
  \langle b \rangle = n_{\rm e} \tilde \phi \phi_0 = 2 \pi n_{\rm e} \tilde \phi
\end{equation}
with $n_{\rm e}$ the average density, and $\tilde \phi = 2m$ the even
number of flux quanta attached to each fermion.  Choosing the
Chern-Simons flux to be in the opposite direction as the applied
magnetic field, this field $\langle b \rangle$ cancels off part of the
external magnetic field $B$ leaving a (mean) residual field seen by
the transformed fermions
\begin{equation}
  \label{eq:DeltaB1}
  \Delta B = B - \langle b \rangle = B - n_{\rm e} \tilde \phi \phi_0 
\end{equation}

Making this approximation of uniform density or ``mean field'', the
Hamiltonian (Eq.  \ref{eq:transformedH}) is approximated as
\begin{equation} \label{eq:Hmf}
  H_{\mbox{\scriptsize{mean-field}}} = \sum_{j} \frac{\left[\vec p_j + \frac{e}{c} 
\vec{\Delta A}(\vec r_j)
    \right]^2}{2m_{\rm b}} 
\end{equation} 
where $\vec {\Delta A}$ is the vector potential associated with the
mean magnetic field $\Delta B$ ({\em i.e.\/}, $\nabla \times \vec {\Delta A} =
\Delta B$), which simply describes free fermions in a uniform magnetic
field.  Note that the Coulomb interaction also disappears at the mean
field level where the density is assumed completely uniform throughout
the system since it only contributes an (albeit infinite)
constant~\footnote{We have similarly assumed that no equilibrium currents
  are flowing in the system at mean field level.  As we will see below
  in section \ref{sub:RPA}, such currents would induce a Chern-Simons
  electric field that would then have to be included too.  Such
  equilibrium currents are important when one thinks about systems
  with nonuniform density~\cite{Edges,SimonStern}.}.

\subsubsection{Even Denominator Fractions}

At some special value of the filling fraction, when $B = \langle b
\rangle$, the applied magnetic field precisely cancels the
Chern-Simons flux at the mean field level ({\em i.e.\/}, $\Delta B = 0$).
This exact cancellation occurs when $B = \langle b \rangle = \tilde
\phi \phi_0 n_{\rm e}$ or equivalently at the filling fraction
\begin{equation}
\nu = \frac{n_{\rm e} \phi_0}{B} = \frac{1}{\tilde \phi} 
   = \frac{1}{2m}_.
\end{equation}
Thus for even denominator filling fractions, at the mean field level
we describe the ground state of the system as fermions in zero
magnetic field, which is just a filled Fermi sea with Fermi
momentum~\footnote{This value of the Fermi momentum differs by a factor
  of $\sqrt{2}$ from the Fermi momentum of the electron gas in zero
  magnetic field since in zero field there two spin states whereas
  here we have assumed fully polarized spins.}
\begin{equation}
\kf = \sqrt{4 \pi n_{\rm e}} = \frac{1}{l_B \sqrt{m}}_.
\end{equation}
The existence of this Fermi-liquid like state at even denominator
filling fractions was predicted by Kalmeyer and Zhang~\cite{Zhang} and
by Halperin, Lee, and Read~\cite{HLR}.  It should be emphasized that it
is an extremely surprising result that one can add a huge magnetic
field to a system and end up with an effective system that behaves in
some ways as if it were in zero magnetic field.

If we are at a magnetic field such that we are close to (but not
exactly at) such an even denominator filling fraction, then the
applied magnetic field and the Chern-Simons flux do not quite exactly
cancel and $\Delta B$ is nonzero.  Thus, we have a Fermi sea in a
small magnetic field in which case the elementary quasiparticle
excitations above the Fermi sea travel in large cyclotron orbits of
radius~\footnote{For general circular motion $R=v^2 m/F$. In a magnetic
  field, the Lorentz force is $F = e B v/c$.  Using $v=\hbar \kf/m$,
  we find that all factors of the mass $m$ cancel from this expression.}
\begin{equation}
\label{eq:Rc}
  R_{\rm c}^* = \frac{\hbar c \kf}{e \Delta B}
\end{equation}
This new length scale has been clearly observed
experimentally~\cite{WillettReview} giving strong support to the
composite fermion picture (See sections \ref{sub:saw} and \ref{sub:Rc}
below for discussion of the relevant experiments).

Although the mean field description gives one a starting point for
understanding the physics of the putative composite fermion Fermi
liquid, it is clear that mean field theory is quite crude.  For
example, at the mean field level, since the system is described as
being in zero effective magnetic field, one would predict that the
Hall conductivity of the system is zero --- which is clearly absurd
for a system in extremely high magnetic field.  Thus we must think
about treating our Chern-Simons Hamiltonian (Eq.
\ref{eq:transformedH}) in an approximation beyond mean field.  It
should be noted, however, that this mean field description of the even
denominator $\nu=\frac{1}{2m}$ states is a non-degenerate starting
point for attempting a controlled perturbation theory --- unlike the
original highly degenerate partially filled Landau level.  In section
\ref{sec:RPA} below, we will begin the discussion of approximations
that go beyond this mean field description.

\subsubsection{Jain Series of Fractional Hall States}
\label{sub:jain}

For completeness, we also consider the case when the filling fraction
is further away from $\nu = \frac{1}{2m}$.  Here, after canceling the
mean Chern-Simons field $\langle b \rangle$ with the external field
$B$, there is some residual field $\Delta B = B - \langle b \rangle$
left over which, in general, can itself be large.  The mean field
system is described as noninteracting fermions in the uniform nonzero
field $\Delta B$.  Fortunately, such a noninteracting system is
completely soluble (see section \ref{sub:IQHE} above).  The effective
filling fraction $p$ for these gauge transformed fermions is given by
(compare with Eq. \ref{eq:nudef})
\begin{equation}
\label{eq:onepoint}
p = \frac{n_{\rm e} \phi_0}{\Delta B.}
\end{equation}
(Note that $p$ can be negative corresponding to a negative $\Delta B$
or a situation where $\langle b \rangle > B$).  When $p$ is a small
integer, at the mean field level, this is just a system of $|p|$
filled Landau levels of fermions, and one should observe the integer
quantized Hall effect of transformed fermions.  Using Eq.
\ref{eq:DeltaB1} as well as the definition of the filling fraction
(Eq. \ref{eq:nudef}), this condition (Eq. \ref{eq:onepoint} with $p$
an integer) yields precisely the Jain series~\cite{Jain0,Jain} of fractional
quantized Hall states
\begin{equation}
  \label{eq:jainseries}
  \nu = \frac{p}{2mp+1.}
\end{equation}
Thus, the fractional quantized Hall effect at these filling fractions
is identified with an integer quantized Hall effect of gauge
transformed fermions~\cite{Jain,Lopez,HLR}.  The most striking early
success of the composite fermion theory was simply that this Jain
series (Eq. \ref{eq:jainseries}) correctly predicts precisely those
fractional quantized Hall states in the lowest Landau level that are
seen experimentally in roughly the correct order of stability (the
most stable states having small $|p|$ and small $|m|$).~\footnote{In
  very high quality samples at low temperature, additional fractional
  Hall states are seen that do not fit this formula.  Many of the
  strongest of these states (such as $\nu = 4/5$) can be simply
  described as the particle-hole conjugate of simple state (in this
  case $4/5 = 1 - 1/5$).  Others observed states are thought to be
  spin unpolarized states.  Other states may require a hierarchical
  construction even within the composite fermion
  framework~\cite{Jain}.}

In mean field theory the excitation gaps for these fractional
quantized Hall states are naturally given by the corresponding
effective cyclotron frequency of the composite fermions (analogous to
Eq. \ref{eq:IQHEgap})
\begin{equation}
  \label{eq:gap}
  E_{\rm gap} = \hbar \Delta \omega_{\rm c}^* = \frac{\hbar e \Delta
    B}{m_{\rm gap}^*(\nu) c}
\end{equation}
where $m_{\rm gap}^*(\nu)$ is an effective mass of the transformed
fermion.  There has been increasing experimental
evidence~\cite{WillettReview} that the fractional Hall gaps do indeed
increase (at least roughly) linearly with $\Delta B$ (See section
\ref{subsub:expgaps} below for discussion of the relevant experiments).

At the mean field level (See Eq. \ref{eq:Hmf}) we have $m_{\rm gap}^* =
m_{\rm b}$, the bare mass of the electron.  However, being that the
fractional Hall gap must be set by the inter-electron interaction
strength, we realize that the mean field result is not accurate and we
should expect the value of $m^*_{\rm gap}$ to be highly renormalized.  In
section \ref{sub:effmass} below, we will more thoroughly discuss the
energy scales in the problem and the expected value of the effective
mass.  Unfortunately, we will find that this discrepancy is not just a
problem in mean field --- it will persist even in the more
sophisticated RPA calculation. Furthermore, in section
\ref{sec:infrared} below we will find that systematic attempts to
actually calculate the effective mass are plagued with infrared
divergences.

Another problem with the mean field theory is that the quantized Hall
resistivity is not correctly obtained.  Since the $\nu =
\frac{p}{2mp+1}$ state is just an integer $\nu = p$ state of
transformed fermions, we have at the mean field level that the Hall
resistivity is given by $\rho_{xy} = h/(p e^2)$ rather than the
correct quantized value of $\rho_{xy} = h/(\nu e^2) = (2mp+1) h/(p
e^2)$ which we know we should obtain by Galilean invariance (see
section \ref{sub:IQHE} above).  Similarly, in mean field theory the
finite frequency response will be incorrectly predicted as well (as we
will see in section \ref{subsub:sumrules} below).  Fortunately, some
of these problems with mean field response functions are corrected in
more sophisticated theories such as RPA that we will consider next.

Despite these shortcoming, however, the Chern-Simons mean field
description is quite appealing due to its impressive simplicity.
Furthermore, as mentioned above, the mean field theory is a
nondegenerate state around which to attempt a controlled (or
semi-controlled) perturbation theory or to calculate systematic (or
semi-systematic) corrections.

\mysection{RPA}
\label{sec:RPA}

One of the great appeals of the Chern-Simons Fermion theory is that it
allows for the analytic calculation of physically relevant quantities.
Indeed, an argument can be made that a theory is only useful if one
can use it to make predictions for physical quantities.  In section
\ref{sec:review} above we used mean field theory to make an number of
nontrivial predictions (scaling of gaps, effective cyclotron radius,
existence of certain types of states at certain filling fractions,
etc.).  However as we saw above, mean field --- being the simplest
possible approximation --- also gets a number of important physical
results wrong.

At the mean field level, one considers the density (and current) of
the system to be everywhere constant.  When the density is everywhere
constant, the Coulomb interaction has no effect on the response of the
system (contributing just an infinite constant to the energy of the
system).  Similarly at mean field level, the effect of the bound
Chern-Simons flux is just to impose a constant effective magnetic
field on the system.  When we go beyond mean field level, we must
treat fluctuations in densities and currents and find that the Coulomb
and Chern-Simons terms have much more nontrivial effects.  For example
individual fermions should certainly interact via the Coulomb
interaction such that when there is a local density fluctuation in one
place, the nearby fermions feel the resulting Coulomb potential and
respond to it. Fortunately, however, so long as we confine our
attention to calculations of resistivities and conductivities (and not
the full electromagnetic response), we will find that this Coulomb
term will not be too important.  Similarly, and more importantly, the
bound Chern-Simons flux also causes an interaction --- as a fermion
moves, it carries its flux quanta with it and the other fermions feel
the resulting change in the Chern-Simons vector potential.  Proper
treatment of this Chern-Simons interaction is required to obtain the
proper Hall conductivity for the composite fermion system.

The simplest approximation beyond mean field is the RPA or Random
Phase Approximation~\footnote{The RPA is also sometimes given the more
  descriptive name of ``Time Dependent Hartree'' approximation.  The
  name `Random Phase' is almost a historical accident.}.  The
RPA for Chern-Simons theories was first developed in the context of
anyon superconductivity~\cite{Anyons,AnyonSupercon,AnyonicCS}.  A
similar approximation was then used by Lopez and Fradkin~\cite{Lopez}
(See also the Chapter by Lopez and Fradkin in this book) to study the
Jain series of fractional Hall states.  Most recently, the RPA was
exploited by Kalmeyer and Zhang~\cite{Zhang} and extensively by
Halperin, Lee, and Read~\cite{HLR} for the study the even denominator
states~\footnote{It should be noted that the form of the RPA used in
  Refs.~\onlinecite{Anyons,AnyonSupercon,AnyonicCS} appear somewhat
  different from the form used by HLR~\cite{HLR} (which is what we
  follow here).  In appendix \ref{app:RPAs} we will show that these
  prescriptions are equivalent.}.

\subsection{Chern-Simons RPA Basics}
\label{sub:RPA}

Below, in sections \ref{subsub:response}-\ref{sub:CSagain}, we will give
a detailed description of the RPA, along with more formal definitions
of various useful response functions.  Here, we will give a much
rougher description that, although less formal, will be sufficient for
many of our needs. 

In the Chern-Simons theory, an excess density $\delta n$ carries an
excess Chern-Simons flux of $\tilde \phi = 2m$ flux quanta per
particle resulting in an excess Chern-Simons magnetic field 
\begin{equation} 
\label{eq:bcs}
\delta b
= \phi_0 \tilde \phi \,\,  \delta n = 2 \pi \tilde \phi \,\,  \delta n.
\end{equation} 
Similarly, a composite fermions'
current $\vec j$ carries a current of flux tubes, thus inducing a
Chern-Simons electric field given by
\begin{equation}
  \label{eq:ecs}
  \vec e = \phi_0 \tilde \phi \frac{e}{c}  (\hatn z \times \vec j) = 2
  \pi \tilde \phi (\hatn z \times \vec j)
\end{equation}
To see where this electric field comes from, we consider a system with
a current $\vec j$ running in it locally.  In a frame moving with
velocity $\vec v = \vec j/n_{\rm e}$, the particles are stationary and there
is only a Chern-Simons magnetic field.  When we boost back to the
laboratory frame, some of the magnetic field is transformed into this
Chern-Simons electric field~\footnote{My appreciation to Ady Stern for
  pointing out this clear argument.  See also section
  \ref{sec:infrared} below for another derivation of this interaction
  through the Lagrangian approach.  See also Ref.~\onlinecite{Lopez}.}.  We
might also have guessed this result using Faraday's law by considering
a closed loop and moving a current of flux tubes across the loop ---
the EMF being proportional to and perpendicular to the current
crossing across the loop.  It should be noted that this Chern-Simons
field --- like the Chern-Simons flux --- is fictitious in the sense
that it is not measurable by a voltmeter --- only the other
transformed fermions in the 2DEG will see this field.

We now declare that the Chern-Simons fermions have some conductivity
matrix $\sigma_{\rm CF}$ (the so-called ``composite fermion
conductivity''), but that they should respond, not only to the
physical electric field $\vec E$ (that measured by a voltmeter), but
also to the self-consistently induced Chern-Simons electric field
$\vec e$.  Thus, we should have
\begin{equation}
  \label{eq:sigmacf}
 \vec j = \sigma_{\rm CF} \left( \vec E + \vec e(\vec j) \right).
\end{equation}
We now rewrite Eq. \ref{eq:ecs} as
\begin{equation}
  \vec e = - \rho_{\rm CS} \vec j
\end{equation}
with
\begin{equation}
  \label{eq:rhocsdef2}
  \rho_{\rm CS} = \frac{2 \pi \hbar \tilde \phi}{e^2} \left[
  \begin{array}{cc} 0 & 1 \\ -1 &
    0
\end{array} \right]
\end{equation}
such that we can convert Eq. \ref{eq:sigmacf} to an expression for the
resistivity matrix (defined by $\vec E = \rho \vec j$) given by 
\begin{equation}
  \label{eq:splitrho}
  \rho = \rho_{\rm CF} + \rho_{\rm CS}
\end{equation}
where $\rho_{\rm CF} = [\sigma_{\rm CF}]^{-1}$ is the ``composite fermion
resistivity'' matrix.  

One quantity of particular interest is the longitudinal conductivity
$\sigma_{xx}$ of the electrons.  This is obtained by inverting the
matrix $\rho$ to yield, $\sigma_{xx} = \rho^{\rm
  CF}_{yy}/\mbox{Det}[\rho^{\rm CF} + \rho^{\rm CS}]$.  For a typical
experimental system, near the even denominator state,
$\nu=\frac{1}{\tilde \phi}$, the $\rho^{\rm CS}$ dominates the
denominator giving
\begin{equation}
\label{eq:CSrhoyy}
  \sigma_{xx} \approx \rho^{\rm CF}_{yy} \left( \frac{e^2}{\tilde \phi h}
  \right)^2_.  
\end{equation}

Thus far, all of these manipulations might be thought of as a
complicated way to {\it define} the composite fermion conductivity
$\sigma_{\rm CF}$.  However, in terms of this quantity, the Chern-Simons
RPA can be given a very simple definition.  Here, the RPA
approximation is simply the statement that the composite fermion
conductivity should be approximated as the mean field conductivity
({\em i.e.\/}, that of a system of noninteracting fermions of mass $m_{\rm b}$ in
magnetic field $\Delta B$ --- see Eq. \ref{eq:Hmf}).  Thus, at RPA
level, we think of the composite fermions as being free fermions that
respond to the {\it total} electric field --- including both the
physical electric field and the self-consistently induced Chern-Simons
field.

As an example, we consider the DC response of a system of electrons at
filling fraction $\nu=\frac{p}{2mp+1}$.  We make the composite fermion
transformation and map this problem to a system of fermions at filling
fraction $p$ which displays the integer quantized Hall effect.  As
mentioned above in section \ref{sub:jain}, the mean field value of
this resistivity ( $\rho_{xy} = \frac{1}{p} h/e^2$) is not the correct
value for the fractionally quantized $\nu=\frac{p}{2mp+1}$ state.
However, using the RPA approximation and replacing $\rho_{\rm CF}$ by
$\rho_{\mbox{\scriptsize{mean-field}}}$ now yields (using $\frac{1}{\nu} = \frac{1}{p} + 2m$),
\begin{eqnarray}
  \rho &=& \rho_{\mbox{\scriptsize{mean-field}}} + \rho_{\rm CS}  \\
  &=& \frac{h}{e^2} \left[
  \begin{array}{cc} 0 & 1/p \\ -1/p &
    0
\end{array} \right] + 
 \frac{h}{e^2} \left[
  \begin{array}{cc} 0 & 2m \\ -2m &
    0
\end{array} \right] 
= \frac{h}{e^2} \left[
  \begin{array}{cc} 0 & 1/\nu \\ -1/\nu &
    0
\end{array} \right] \nonumber
\end{eqnarray}
which is the correct quantized Hall resistivity for the state.  
Thus, at the RPA level (but not at mean field level) we recover the
fundamental thermodynamical quantity of the quantized Hall resistance.

On a physical level, the way the Hall resistance occurs is the
following.  We run a current through the sample which carries the
Chern-Simons flux quanta through the system creating a Chern-Simons
electric field (by the Faraday effect).  The other fermions see this
electric field, and in order for them to continue to move straight
there must be no net electric field, so there must
be a physical electric field built up to cancel this Chern-Simons
field.

We note that using this approach, we could similarly calculate any
finite frequency or finite wavevector response by simply inserting the
finite frequency mean field (noninteracting) resistivity into Eq.
\ref{eq:splitrho}.  Finite frequency and wavevector conductivities can
be calculated explicitly for noninteracting systems and are discussed
further below.  A convenient approximate calculation
of these conductivities can be made using the Boltzmann approach,
which will be discussed in section \ref{sec:FLT}.

As another example, at finite $q$, and low frequency, $\rho_{yy}$ for
a noninteracting system in zero magnetic is given roughly by (this
result is from section \ref{sec:FLT} below where we obtain the result
Eq.  \ref{eq:rhoyyex} which yields the following when expanded for
small $\Omega$ and $F_1 = 0$)
\begin{eqnarray}
  \label{eq:rhoform}
  \rho_{yy}(q)  &=& \frac{q}{\kf} \frac{h}{e^2}  \,\,\,\,\,\,
  \mbox{for} \,\,\,\,\,\, q \gg 2/l
  \\
   &=&  \frac{2}{\kf l} \frac{h}{e^2} \,\,\,\,\,\, \mbox{for} \,\,\,\,\,\, q < 2/l
\end{eqnarray}
where $l = \vf \tau$ is some disorder scattering length.  Using these
expressions as $\rho^{\rm CF}$, we would then predict that at
$\nu=\frac{1}{2m}$, the low frequency conductivity is given by (using
Eq. \ref{eq:CSrhoyy})
\begin{eqnarray}
\label{eq:sigres}
    \sigma_{xx}(q) &\approx& \frac{e^2}{h} \frac{q}{\tilde \phi^2 \kf}
     \,\,\,\,\,\,
  \mbox{for} \,\,\,\,\,\, q \gg 2/l \\ \label{eq:sigres2}
   &\approx&  \frac{e^2}{h}\frac{2}{\kf l \tilde \phi^2}  \,\,\,\,\,\, \mbox{for} \,\,\,\,\,\, q < 2/l
\end{eqnarray}
This enhanced finite wavevector conductivity linear in $q$ has indeed
been observed experimentally~\cite{WillettReview} (See section \ref{sub:saw}
below) for large $q$.  

The astute reader might notice that in our above discussion of the
RPA, we did not ever mention the effects of the Coulomb interaction.
Fortunately, so long as we confine our attention to calculations of
resistivities and conductivities we will find that this Coulomb term
will not be too important.  However, in calculations of the full
electromagnetic response function (which we will define next) the
Coulomb interaction will be important.

\subsection{Electromagnetic Response $K_{\mu\nu}$}
\label{subsub:response}

Thus far, we have concentrated on calculating the resistivity, or
conductivity tensors.  There are, however, many other important
response functions that we might consider calculating --- the most
important of which is known as the electromagnetic response matrix
$K$.  Below, we will relate this object to a number of more familiar
quantities, such as the conductivity of the system.  

To define $K$, a weak vector potential
$A_\mu^{\mbox{\scriptsize{ext}}}$ is externally applied to the system
inducing a current $j_\mu$.  Here we define $A_0$ to be the scalar
potential, and $j_0$ to be the induced density modulation $n - n_{\rm e}$
(with $n_{\rm e} = \langle n \rangle$).  In linear response theory, we can
define the response kernel $K_{\mu \nu}(\vec r,t; \vec r',t')$ such
that
\begin{equation}
  \label{eq:kernal}
 j_\mu(\vec r, t) = e \int_{-\infty}^t dt' \int d\vec r'
  \, \, K_{\mu \nu}(\vec r, t; \vec r', t')
A_\nu^{\mbox{\scriptsize{ext}}}
(\vec r',t')
\end{equation}
where the sum over the repeated index ($\nu$) is implied.  Assuming
translational invariance (in time as well as space), we can write
$K_{\mu \nu}(\vec r-\vec r', t-t')$, indicating that Eq.
\ref{eq:kernal} is just a convolution which simplifies in Fourier
space to
\begin{equation} 
  \label{eq:Kdef}
j_{\mu}(\vec q, \omega) =  e K_{\mu \nu}(\vec q, \omega)
A_{\nu}^{\mbox{\scriptsize{ext}}}(\vec q, \omega)
\end{equation}
(Again the repeated index is summed). Another way to think about this
response matrix is to imagine applying an external potential
$A_\nu^{\mbox{\scriptsize{ext}}}$ to the system at wavevector $\vec q$
and frequency $\omega$.  Then in linear response, a current $j_\mu(\vec
q,\omega)$ is induced given by Eq.  \ref{eq:Kdef}.  We will always
choose the convention that the $\vec q$ is parallel to the $\hatn x$
axis such that perturbations are proportional to $e^{i \omega t - i q
  x}$.

Most generally, we will attempt to calculate the full electromagnetic
response matrix $K_{\mu\nu}$ where $\mu$ and $\nu$ take the values
$0,x,y$.  However, we can simplify our life by using current
conservation $\nabla \cdot j + \frac{d}{dt} n =0 $ or $q j_x = \omega
j_0$ and choosing the gauge $A_x = 0$.  With $j_x$ and $A_x$ thus
determined, we can then treat $K_{\mu\nu}$ as a $2\times 2$ matrix
with indices taking the values 0 or 1 denoting the time (0) or
transverse space ($y$) components respectively.  In this notation the
current vector $j_\mu$ is $(j_0,j_y)$, and the vector potential
$A_\mu$ is $(A_0,A_y)$.  Note that from here on, we will routinely
drop the explicit matrix subscripts $\mu$ and $\nu$ as well as the
explicit $q$ and $\omega$ dependences.

Although all of the elements of the response matrix $K_{\mu \nu}$ are
genuine physical quantities, the so-called density density response
$K_{00}$ (also called $\chi$) is perhaps most important for quantum
Hall physics because it can often be measured experimentally (See
section \ref{sub:saw} below).  Theoretically, $K_{00}$ is also an
important quantity, being related to the more familiar dynamical
structure factor $S(q,\omega)$ via
\begin{equation}
 -\frac{1}{\pi}\mbox{Im} K_{00}(\vec q,\omega + i 0^{{}^+}) = 
  S(\vec q,\omega) = \sum_{n} \left|
    \left\langle j | \hat n(\vec q) | 0 \right\rangle \right|^2 
  \delta(E_j - E_0 - \omega)
\end{equation}
where $|j \rangle$ is the $j^{th}$ many body eigenstate whose energy
is $E_j$, $|0 \rangle$ is the ground state (with energy $E_0$), and
$\hat n(\vec q)$ is the density operator at wavevector $\vec q$.
Thus, $-\frac{1}{\pi}\mbox{Im} K_{00} = S$ has poles at the
frequencies of each excitation mode of the system, and the residue (or
weight) of this pole indicates how easy it is to create this excited
state from the ground state by applying a finite wavevector
perturbation.  (Note that $K_{\mu\nu}$ satisfies Kramers-Kr\"onig
relations so the imaginary part completely determines the real part).

We will also define $K^0$ to be the response function for
noninteracting fermions (The ${}^0$ means noninteracting).  In
appendix \ref{app:non} we show the full result for $K^0$ for fermions
in zero field.  (See also Refs.~\onlinecite{Anyons} and
\onlinecite{Simonhalp} for explicit expressions for the response in
finite magnetic field.)  The actual calculation of $K^0$ is described
in more detail in section \ref{sub:diag} below.  It is often
convenient to use a Boltzmann approach to approximate $K^0$. The
advantage of this approach (besides its simplicity) is that it is
quite easy to approximate the effects of disorder or Fermi liquid
interactions at a phenomenological level.  This will be discussed in
more depth in section \ref{sec:FLT} below.  Using the Boltzmann
approach, one directly obtains the noninteracting conductivity
$\sigma^0$ which can be converted to the response using $K^0 = T
\sigma^0 T$ with the conversion matrices $T$ given by Eq. \ref{eq:TT}
below (we will discuss this more in a moment).  Such a Boltzmann
calculation is performed for the case of fermions in zero field in
appendix \ref{app:non}.

\subsection{Hartree Part of the Coulomb Interaction
  : Separating $\Pi^v$}
\label{subsub:separatingcoulomb}

We now turn to consider the effect of the Coulomb interaction. 
The Coulomb term in the Hamiltonian (compare Eq.
\ref{eq:Hamiltonian}) is given by
\begin{equation}
        H_{\rm int} = \sum_{i < j} v(\vec r_i - \vec r_j) = 
\frac{1}{2} \int d\vec r_1 \! \! \int d\vec r_2   \,\, 
n(\vec r_1) n(\vec r_2) v(\vec r_1 - \vec r_2).
\end{equation}
with $n(\vec r) = \sum_i \delta(\vec r - \vec r_i)$ the density
operator.  The Hartree approximation is then written as~\footnote{The
  reader may notice that a factor of $\frac{1}{2} $ has disappeared.
  This is a result of taking $\langle n(\vec r_1) \rangle n(\vec r_2)
  + n(\vec r_1) \langle n(\vec r_2) \rangle$ as Wick's theorem would
  suggest.}
\begin{eqnarray}
  H_{\rm int} &=&   - \int d\vec r \, \, n(\vec r) \,  e
 A^{v-\mbox{\scriptsize{ind}}}_0  (\vec r)
 \\
  - e   A^{v-\mbox{\scriptsize{ind}}}_0(\vec r)   
&=& \int d\vec r' \, \, \langle n(\vec r')
  \rangle \, \, v(\vec r - \vec r') \label{eq:self1}
\end{eqnarray}
Physically, the Hartree approximation is just the simple statement
that any (expectation value of the) electron density in the system
will induce a Coulomb potential which the other electrons then respond
to.  (Exchange interaction, or ``Fock terms'' are neglected in the
Hartree approximation). 

In Fourier space, Eq. \ref{eq:self1} simplifies to
\begin{equation}
  \label{eq:phiind}
 - e  A^{v-\mbox{\scriptsize{ind}}}_0(\vec q, \omega)   
 = v(\vec
  q) \, \, \langle n(\vec q,\omega) \rangle 
\end{equation}
where $v(q)$ is the Fourier transformed interaction, given by $2 \pi
e^2/(\epsilon q)$ for the usual Coulomb case.  Suppressing wavevector
and frequency dependences, we now write this in the $2 \times 2$
matrix notation (for later convenience),
\begin{equation}
  -e A^{v-\mbox{\scriptsize{ind}}} =  V \, j \\ \label{eq:ACJC}
  \label{eq:AUJC}
\end{equation}
with
\begin{equation}
\label{eq:Vdef}
V=\left[ \begin{array}{cc} v(q) & 0 \\ 0 & 0 \end{array} \right]_.
\end{equation}

We now define the total-physical vector potential that includes both
the external part and that induced by the Coulomb
interaction~\footnote{The use of the long phrase ``total physical'' may
  seem a bit cumbersome here, but it will help keep us from getting
  confused later.}
\begin{equation}
\label{eq:phitotal}
 A^{\mbox{\scriptsize{total-physical}}} = A^{\mbox{\scriptsize{ext}}} +
  A^{v-\mbox{\scriptsize{ind}}}  
\end{equation}
and define a response matrix $\Pi^v$ (sometimes called the
Polarization) that gives the proportionality between the current and
the total-physical vector potential via
\begin{equation}
    \label{eq:PiV}
  j = e \Pi^v  A^{\mbox{\scriptsize{total-physical}}}.
\end{equation}
The superscript ${}^v$ is included here to distinguish this
polarization from one defined below when we consider the Chern-Simons
problem, and also to remind us that we are looking at a response to
the total field that includes the self-consistent Coulomb ($v$)
interaction.  In other words, $\Pi^v$ gives the density and
current response to the {\it total} physical field including both the external
field and the field internally induced by the Coulomb interaction.
Indeed, when one attaches leads to the sample and measures a voltage,
one is actually measuring this total field.  Since the conductivity
is defined as the response to the total field (being that that is what
is measured with a voltmeter), $\Pi^v$ must be closely
related to conductivity.  This connection will be made more clear in
the next section.

Using Eqs. \ref{eq:Kdef}, \ref{eq:AUJC}, \ref{eq:phitotal}, and
\ref{eq:PiV} we can easily derive the matrix equation (dependences on
$q$ an $\omega$ are implied)
\begin{equation}
  \label{eq:K2sep}
  K^{-1} = [\Pi^v]^{-1} + V
\end{equation}
which can also be written as~\footnote{The $-$ signs here are a
  function of our conventions for the charge of the electron and our
  use of $e$ in Eq. \ref{eq:Kdef}.}
\begin{equation}
\label{eq:expand}
  K = \Pi^v -  \Pi^v \,  V \, \Pi^v + \Pi^v \,
  V \, \Pi^v \, V \, \Pi^v-  \ldots
\end{equation}
In terms of diagrams, we say that $K$ is a chain sum of $\Pi^v$
connected by $V$ propagators, or equivalently, that
$\Pi^v$ is the $V$-irreducible part of $K$.  Sometimes it is said that
$\Pi^v$ is the response $K$ with the Hartree part of the interaction
``separated out''.  All of these are equivalent ways of saying that
$K$ is the response to the externally applied field, whereas $\Pi^v$
is the response to the total field including external field and that
induced internally by $v(\vec q)$.

Note that with some amount of algebra one can derive from the matrix
equation Eq.  \ref{eq:K2sep} the simpler statement 
\begin{equation}
  \label{eq:Ksep}
  K_{00}^{-1}  = [\Pi^v_{00}]^{-1} + v
\end{equation} 
which is what we would have naturally found if we had only kept track
of the density $j_0$ and the scalar potential $A_0$ and not considered
the full matrix response $K_{\mu \nu}$.  It is worth noting that the
relation derived here between $K$ and $\Pi^v$ is a very general
relation that is always true for any system (by definition).

\subsection{Relation of Response to Resistivity and Conductivity}
\label{subsub:cond}

As mentioned above, the response function $\Pi^v$ is the response to
the total physical field including both the external field and the
field internally induced by the Coulomb interaction.  Since this total
physical field is exactly the field that is measured experimentally by
a voltmeter, we suspect that $\Pi^v$ can be related to the
conductivity of the system.  To make this connection apparent, we
write the electric field in terms of the vector potential as $ \vec E
= \nabla A_0 - \frac{d}{dt} \vec A$ or in Fourier space as (recall
that we have chosen Coulomb gauge $\vec A = A_1 \hatn y$) $ E_x = -i q
A_0 $ and $ E_y = -i \omega A_1$.  We will also use the current
conservation equation $q j_x = \omega j_0$ to convert between $j_x$
and $j_0$. It is now convenient to define a conversion matrix
\begin{equation}
  \label{eq:TT}
  T= e \left[ \begin{array}{cc} \frac{i\sqrt{i\omega}}{q} & 0 \\ 0 &
  \frac{1}{\sqrt{i\omega}}
\end{array} \right]
\end{equation}
This matrix will be used to apply the correct factors of $q$ and
$\omega$ for converting between vector potential $A=(A_0,A_y)$ and
electric field $\vec E=(E_x,E_y)$ and for converting between the
vector $j = (j_0,j_y)$ and the vector $\vec j = (j_x,j_y)$.  With this
definition, we have 
\begin{eqnarray}
\label{eq:eT1}
  \vec E  &=& \left( -i  \sqrt{-i \omega }
   \right) T^{-1} A 
  \\
  \vec j &=& \left(-i \sqrt{-i \omega } 
\right) T \, j 
\end{eqnarray}
This conversion matrix will be used to switch freely between the two
notations.

Using this conversion, we can rewrite Eq. \ref{eq:PiV} as
\begin{equation}
  \vec j = \sigma \vec E^{\mbox{\scriptsize{total-physical}}}
\end{equation}
where the matrix $\sigma$ is given by~\footnote{In
  reference~\onlinecite{HLR} a slightly different definition of the
  conductivity was used so as to obtain a result whose semiclassical
  approximation is well defined in the low frequency limit.}
\begin{equation}
  \label{eq:sigmaPiv}
    \sigma = \rho^{-1} = T \Pi^v T.
\end{equation}
Since the matrix $\sigma$ relates the current to the total physical
electric field (that measured by a voltmeter) this is what we call the
conductivity matrix.  It will be convenient at times to note that the
longitudinal ($00$ or $xx$) part of this matrix equation implies
\begin{equation}
  \label{eq:sigPiV}
  \sigma_{xx} = \frac{\omega}{i \vec q^2}\Pi^v_{00} \,\,\,\,\, {}_.
\end{equation}

\subsection{General Theme and Simple Example of RPA}

So far, we have not made any progress in calculating the response $K$;
all we have done is to separate $K$ into pieces.  Actually calculating
$K$ or $\Pi^v$ is impossible for most nontrivial systems without
making some sort of approximation.  One of the simplest nontrivial
approximations we can use to calculate a response is the RPA.  The
philosophy behind this approximation is to treat the interaction (such
as $v$) by separating out the Hartree piece (such as in Eq.
\ref{eq:K2sep}).  Once this interaction piece is ``taken care of'',
one approximates that which is left behind ($\Pi^v$) as the response
of {\it noninteracting} fermions.  Another way of saying this is to say
that at RPA level the electrons respond just as if they were
noninteracting electrons --- only they respond to both the external
field and the self-consistent induced field also (reminiscent of our
Chern-Simons RPA from section \ref{sub:RPA} above). 

The general rule for performing RPA is then the following.  First we should
write our Hamiltonian as
\begin{equation}
  H = H_0 + H_{\rm int},
\end{equation}
where $H_0$ is the Hamiltonian for noninteracting fermions.
We then separate out $H_{\rm int}$ at Hartree level to define a
polarization $\tilde \Pi$
\begin{equation}
\label{eq:theme2}
  K^{-1} = \tilde \Pi^{-1}  + [{\rm Interaction}].
\end{equation}
For example, above, we have separated out the Coulomb term $v(q)$ and
defined the polarization $\Pi^v$ in Eq. \ref{eq:K2sep}.  However, when
we have a system with more complicated interactions, we would want to
separate out these pieces too.  Once we have made this separation,
approximating $\tilde \Pi$ as the response of noninteracting (mean
field system) fermions (which we call $K^0$) yields the RPA response.

As a demonstration of this approach, we will show how to calculate $K$
for a system of electrons at $B=0$ (leaving the world of quantum Hall
for the time being).  Here, $v(q)$ is the only interaction.  As
described above, in the RPA, the polarization $\Pi^v$ is approximated
as the response of noninteracting electrons, $K^0$ (discussed above in
section \ref{subsub:response}).  Using Eq.  \ref{eq:Ksep}, we then
obtain the RPA expression
\begin{equation}
\label{eq:eRPA}
  K_{00} = \frac{1}{[K^0_{00}]^{-1} + v(q)}
\end{equation}
This approach is a time honored method for understanding the $B=0$
electron gas.

\subsection{Chern-Simon RPA (Again)}
\label{sub:CSagain}
\label{subsub:rhoK}

We now return to the world of high magnetic fields and Chern-Simons
fermions.  We will re-derive the Chern-Simons RPA discussed above in
section \ref{sub:RPA} in terms of the response function $K$.  The theme
of separating out the Hartree part of the interaction will pervade the
rest of this discussion --- only now we will worry about the
Chern-Simons interaction as well as the Coulomb interaction.

In the Chern-Simons approach, when we make the mean field
approximation, we fix the density and currents in the system to
be everywhere constant such that the Chern-Simons magnetic field is a
fixed constant and the Coulomb interaction is a constant.  When we go
beyond mean field, we look at the effect of the density or currents
fluctuations coupling to the interactions.  Analogous to the Hartree
separation of the Coulomb interaction that we discussed above, we will
define new response functions in this section that separate out the
Chern-Simons interaction.

As we discussed above (See Eqs \ref{eq:bcs} and \ref{eq:ecs}), in the
Chern-Simons theory an excess density $j_0$ carries an excess
Chern-Simons flux of $\tilde \phi = 2m$ flux quanta per particle
resulting in an excess Chern-Simons magnetic field $ \delta b = \tilde
\phi \phi_0 j_0$.  Similarly a composite fermions' current $\vec
j$ carries a current of flux tubes, inducing a Chern-Simons electric
field $ \vec e = -2 \pi \tilde \phi \frac{e}{c} \hatn z \times \vec
j(\vec q)$.

It is convenient now to write these results in terms of the vector
potential, (Again, we have the convention that the wavevector $\vec q
\| \hatn x$; we also will typically drop the matrix indices and the
explicit $q$ and $\omega$ dependences)
\begin{eqnarray}
\label{eq:ACS}
  e A^{CS-\mbox{\scriptsize{ind}}} &=& - C \, j
\end{eqnarray}
where the superscript ${}^{\rm CS}$ stands for Chern-Simons and the
interaction matrix is given by~\footnote{Note that HLR~\cite{HLR} call
  this $C^{-1}$.}  (See Eqns. \ref{eq:ecs} and \ref{eq:eT1})
\begin{equation}
\label{eq:Cdef}
C= T \rho_{\rm CS} T =  
\frac{2 \pi \tilde \phi \hbar}{e} \left[ \begin{array}{cc} 0 &
\frac{i}{q}
  \\ \frac{-i}{q} & 0 \end{array} \right]_. 
\end{equation}
 
Recalling that a vector potential $  A^{\mbox{\scriptsize{ind}}}_{v} 
 = V j$ is induced by the Coulomb interaction,  we have 
\begin{equation}
  \label{eq:Atotal} 
  A^{\mbox{\scriptsize{total}}} = A^{\mbox{\scriptsize{ext}}} +
  A^{v-\mbox{\scriptsize{ind}}}  +  A^{\mbox{\scriptsize{CS}}-\mbox{\scriptsize{ind}}}.
\end{equation}
We emphasize again that the induced Chern-Simons vector potential is
not a physical field measured by a voltmeter since it is only seen by
other {\it transformed} fermions within the two dimensional electron
gas.  However, the transformed fermions do respond to this
Chern-Simons piece.

We now define a response function $\Pi$ that gives the current and
density response to this total field ~\footnote{Our matrix $\Pi$ is
  written as $\tilde K$ in references~\onlinecite{HLR} and~\onlinecite{Simonhalp}.  
  However, our notation for $\Pi$ agrees with
  that used in 
  references~\onlinecite{SimonStern},~\onlinecite{SimonNFL},~\onlinecite{Ady}, 
  and~\onlinecite{Kim}.  The
  other common notation is to define $\Pi = T \sigma T$ which is what
  we call $\Pi^v$}.
\begin{equation} 
  \label{eq:Pidef}
  j = e \Pi A^{\mbox{\scriptsize{total}}}. 
\end{equation}
We can now combine Eqns. \ref{eq:Pidef}, \ref{eq:Kdef},
\ref{eq:AUJC}, and \ref{eq:ACS} to obtain the relation
\begin{equation}
  \label{eq:KPi}
  K^{-1} = \Pi^{-1} + V + C
\end{equation}
Thus the definition of $\Pi$ separates out the Hartree part of both
the Coulomb and the Chern-Simons interactions.  In terms of a
diagrammatic expansion, $\Pi$ is defined as the contribution from all
Feynman diagrams for $K$ that are irreducible with respect to both the
Coulomb interaction $V$ and the Chern-Simons interaction $C$.  Here
$\Pi$ should be though of as the composite fermion response, analogous
to $\sigma_{\rm CF}$ discussed above in section \ref{sub:RPA}.  Indeed,
using Eqs. \ref{eq:KPi}, \ref{eq:Cdef}, \ref{eq:sigmaPiv},
\ref{eq:K2sep}, and \ref{eq:splitrho} we have
\begin{equation}
\label{eq:Pirho}
 \rho_{\rm CF} =  [ T \Pi T ]^{-1}
\end{equation}

We note that Eq. \ref{eq:KPi} has the general form of separation as
defined in Eq.  \ref{eq:theme2}. Thus, in terms of the response
function $K$, the RPA is analogously obtained by setting $\Pi$ to be
$K^0$ the response for the noninteracting mean field system, such that
we have the RPA prescription
\begin{equation}
  \label{eq:BigRPA}
  K^{-1} = [K^0]^{-1} + V + C
\end{equation}
Results of RPA calculations for the Jain fractions
$\nu=\frac{p}{2mp+1}$ such that $\Delta B \ne 0$ are discussed at
length in Refs.~\onlinecite{Lopez} and~\onlinecite{Simonhalp}.

\subsection{Results of Chern-Simons RPA at $\nu=\frac{1}{2m}$}

At this point, we will turn our focus to the even denominator states
$\nu=\frac{1}{2m}$.  The mean field system is then in zero magnetic
field (such that $K^0$ is diagonal).  We can now plug $K^0$ into Eq.
\ref{eq:BigRPA} to obtain
\begin{equation}
  \label{eq:K00half}
  K_{00} = \frac{1}{\left(\frac{2 \pi \tilde \phi}{q}\right)^2
    K^0_{11} 
+ [K^0_{00}]^{-1} +  v(q)}
\end{equation}
We note that this looks exactly like the RPA response $K_{00}$ for a
normal Fermi gas in zero magnetic field (Eq. \ref{eq:eRPA}) except for
the presence of the $K_{11}^0$ term in the denominator.  The full
forms of $K^0$ for fermions in zero magnetic field are given in
appendix \ref{app:non}.  

We will now examine the results of the Chern-Simons RPA at low
frequency $\omega \ll q \vfns$.  Plugging in the low frequency form of $K^0$
(given by Eqns \ref{eq:non1} and \ref{eq:non2}) into Eq.
\ref{eq:K00half} yields (here we set the mass of the fermion to be
$m_{\rm b}$ as it should be in the RPA approximation),
\begin{equation} \label{eq:lowenergyRPA}
  K_{00}(q,\omega)  = \frac{1}{\frac{2 \pi}{m_{\rm b}} [ 1 + \frac{\tilde
  \phi^2}{12} ] + v(q) + i \left(\frac{2 \pi \tilde \phi}{q} \right)^2
  \left( \frac{2 \hbar n_{\rm e}}{m_{\rm b}} \right) \frac{\omega}{q \vfns}}
\end{equation}
At zero frequency, this expression looks very similar to the RPA
response of fermions in zero magnetic field~\footnote{At zero field,
  using Eq. \ref{eq:non1} in Eq.  \ref{eq:eRPA}, we obtain
$$
  K_{00}(q,\omega)  = \frac{1}{\frac{2 \pi}{m_{\rm b}} [ 1 + \frac{i \omega}{q
  \vfns}] + v(q) }
$$
which is quite similar to the HLR-RPA form above at $\omega = 0$
(the only difference being that the term $\tilde \phi^2/12$ is
missing).  We note that the imaginary frequency pole here is at
$\omega \sim i v(q) q$ which yields very different low energy physics
from that of the Chern-Simons Fermi liquid.}. However at finite
frequency this response has a pole at imaginary frequency
\begin{equation}
   \omega \sim i q^3 v(q)
\end{equation}
which does not occur for the usual Fermi gas at zero field.  This
so-called ``overdamped mode'' is diffusive ({\em i.e.\/}, $\sim q^2$) for
Coulomb interactions and sub-diffusive ($\sim q^3$) for short range
interactions, and represents a slow (or extremely slow in the case of
short range interactions) relaxation of density fluctuations.  One
might expect this slow relaxation in large magnetic field since an
electron being pushed away from a region of high density by the
Coulomb interaction ends up moving perpendicular to the force, thus
not relaxing the density at all.  It is also natural that the density
relaxation should be slower in the case of short range interaction
where the interaction cannot move density at greater distances.  The
existence of this overdamped mode is what causes the infra-red divergences
in section \ref{sec:infrared} below.

 \begin{figure}[htbp]
  \begin{center}
    \leavevmode
    \epsfxsize=3.5in
     \epsfbox{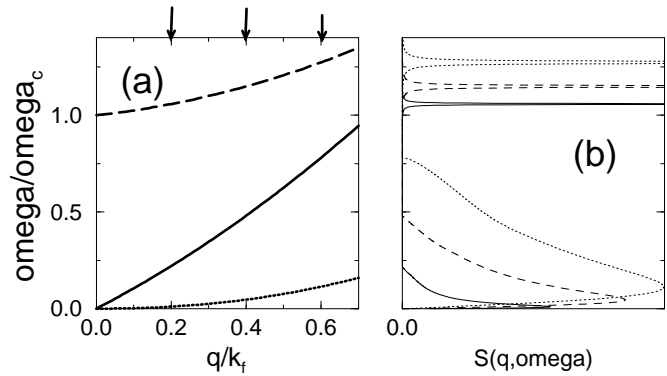}
  \caption{\protect{\small Excitation Spectrum of $\nu=\frac{1}{2}$ in RPA}}
  \label{fig:RPA}
  \begin{minipage}[t]{4.5in}  
    {\small In (a) the solid line is the edge of the low energy
      continuum of quasiparticles, the dotted line is the location of
      the peak in the weight (the maximum of the structure factor) of
      the low energy continuum ($\omega \sim q^3 v(q)$), and the
      dashed line is the cyclotron mode.  In (b) the structure factor
      $S(q,\omega)$ is shown explicitly for three different
      wavevectors $q/\kf = .2$ (solid), $.4.$ (dashed), and $.6$
      (dotted).  The amplitude of $S$ is the horizontal direction and
      the frequency axis is the same scale as for (a).  Note that the
      peak is at very low frequency.  A small broadening is used so the
      sharp cyclotron mode can be seen.  Note that the integrated
      weight of the structure factor is proportional to $q^2$ in
      accordance with the $f$-sum rule.  Here we have used $E_{\rm c}/(\hbar
      \omega_{\rm c}) = 5$ which is large if we use the bare mass, but
      reasonable if we use a renormalized mass  in $\omega_{\rm c}$.}
  \end{minipage} 
   \end{center}
 \end{figure}

For higher frequencies, we look back to the noninteracting case for
insight.  For noninteracting fermions, at finite wavevector $q$, there
is a continuum of low energy excitations that one can make by exciting
a single fermion out of the Fermi sea by wavevector $q$.  This can
create excitations up to a maximum energy $\omega_{cutoff} =
\frac{1}{2m} [(q+\kf)^2 -\kf^2 ] \approx q \vfns$.   Similarly for the
composite fermion Fermi liquid, we have the same cutoff to the low
energy band (this is obvious, since in Eq. \ref{eq:BigRPA} the
noninteracting $K^0$ must have an imaginary part in order for $K$ to
have an imaginary part).   However, unlike the noninteracting case,
there is also a cyclotron mode at high energy, and the weight of the
low energy continuum is bunched down at very low energy $\omega \sim
q^3 v(q)$. 

In Fig. \ref{fig:RPA} we show the excitation spectrum of the
$\nu=\frac{1}{2}$ state calculated in RPA.   In (a) the edge of the
continuum is shown solid, the cyclotron mode (See Eq. \ref{eq:cyc}) is
shown dashed, and the maximum of the weight in the continuum is
shown dotted.   In (b) the structure factor is shown explicitly as a
function of frequency for several wavevectors.   Note that the RPA is
only expected to be valid at small $q$, but we have displayed it here
to larger $q$ for clarity.

\subsection{Sum Rules}
\label{subsub:sumrules}

It is instructive to stop at this point and examine some of the
properties we expect our response functions to have --- particularly
at high frequency.  To begin with, we recall that Kohn's
theorem~\cite{Kohn} (a result of Galilean invariance) requires that the
$q \rightarrow 0$ behavior of our system be determined by the band
mass $m_{\rm b}$ rather than any renormalized mass.  One can imagine all of
the electrons in the system oscillating in unison so that
electron-electron interactions have no effect.  More formally, one
finds that the motion of the center of mass degree of freedom of the
system decouples from any interactions between the particles so that
in the long wavelength limit, interactions do not effect the response
of the system.

Similarly the $f$-sum rule~\cite{Pines,Nozieres,LandauII} simply says
that the behavior of our system in the $\omega \rightarrow \infty$
limit is also determined by the band mass $m_{\rm b}$.  This is easily
imagined since at high frequency one can think of the electrons
oscillating very quickly with very small magnitude so that these
oscillations do not appreciably change the positions of the electrons
or couple to the electron-electron interaction.  We note that the
$f$-sum rule is stronger than Kohn's theorem in the sense that it hold
also in a disordered system.

We can summarize both of these rules by stating that in the long
wavelength or high frequency limit, the resistivity of any system
should look like
\begin{equation}   \label{eq:fsum1}
   \rho \sim \frac{m_{\rm b}}{e^2 n_{\rm e}} 
   \left[ \begin{array}{cc} -i\omega & \omega_{\rm c} \\
                         -\omega_{\rm c} &  -i\omega \end{array} \right]
+ {\cal O}(q^2/\omega).
\end{equation}
If the resistivity of our system indeed has this high frequency, long
wavelength limit, then Kohn's theorem and the $f$-sum rules are
satisfied~\footnote{Often the $f$-sum rule is stated in terms of the
  conductivity $\sigma$ or the electromagnetic response $K$. By using
  a Kramers-Kr\"onig relation, this high frequency condition can be
  written as an integral over frequency such as
\begin{equation}
  \int_{-\infty}^{\infty} d \omega \,  \mbox{Re} 
  \sigma_{xx}(q,\omega) = \pi n_{\rm e}/m_{\rm b}. 
\end{equation}
It is tempting to think that the name ``$f$-sum'' comes from this sum
over frequencies, but in truth the name is (once again) more of a
historical accident}.  

We now check that the RPA approximation satisfies this condition.  The
mean field resistivity of the system is just the resistivity of a
system of free fermions of mass $m_{\rm b}$ in a  field $\Delta B$, and
should thus (by Kohn's theorem and the $f$-sum rule for free fermions)
have the form 
\begin{equation} 
  \label{eq:meansum}
   \rho_{\mbox{\scriptsize{mean-field}}} \sim \frac{m_{\rm b}}{e^2 n_{\rm e}} 
   \left[ \begin{array}{cc} -i\omega & \Delta \omega_{\rm c} \\
                         -\Delta \omega_{\rm c} &  -i\omega \end{array} \right]
+ {\cal O}(q^2/\omega).
\end{equation}
where $\Delta \omega_{\rm c} = e\Delta B/(m_{\rm b} c)$ is the cyclotron frequency
associated with the effective magnetic field and the band mass.
Defining the mean field resistivity to be the composite fermion
resistivity $\rho_{\rm CF}$, using Eq. \ref{eq:splitrho} to calculate the
electron resistivity $\rho$, and using the fact that (Eq.
\ref{eq:DeltaB1}) $\Delta B = B - \phi_0 \tilde \phi n_{\rm e}$, we
can easily show that the RPA resistivity also satisfies Kohn's theorem
and the $f$-sum rule.  Indeed, what we find is that the RPA
resistivity satisfies these sum rules if and only if the composite
fermion resistivity $\rho_{\rm CF}$ satisfies these sum rules with
respect to the effective magnetic field ({\em i.e.\/}, if the composite
fermion resistivity has same large frequency small wavevector form as
the mean field result of Eq. \ref{eq:meansum}).

It is interesting to note that by using the long wavelength, high
frequency form of the resistivity fixed by the $f$-sum rule (Eq.
\ref{eq:fsum1}),   we can  calculate the response function $K_{00}$
by using Eqs.  \ref{eq:fsum1}, \ref{eq:sigmaPiv}, and \ref{eq:Ksep}.
Once we have this form we can find the excitation spectrum by looking
for poles of the response matrix $K_{00}$, which are then given by
\begin{equation}
  \label{eq:cyc}
  \omega^2 = \omega_{\rm c}^2 + \frac{q^2 v(q) n_{\rm e}}{m_{\rm b}}
\end{equation}
which is a completely general result~\footnote{Other corrections occur
  at order $q^2$.}.  Note that in zero magnetic field $\omega_{\rm c}
\rightarrow 0$, this Kohn mode goes continuously into the plasma mode.

\subsection{Energy Scales, Effective Mass, and a Problem with RPA}
\label{sub:effmass}

As mentioned above in section \ref{sub:jain}, at mean field level, the
excitation gap for the fractional Hall state is given by $\Delta
\omega_{\rm c} = \frac{\hbar e \Delta B}{m_{\rm b} c}$ which is not correct.  At
RPA level, this remains a problem.  To explain in more detail why this
is a such a serious problem, we should carefully examine the energy
scales of the problem.

The natural energy scale for the interaction strength~\footnote{Note
  that $l_B$ is considered to be the natural length scale of the
  problem.  One might also consider defining $1/\sqrt{n_{\rm e}}$ to be the
  natural length scale.  This differs from $l_B$ only by a factor of
  $(2 \pi/\nu)^{1/2}$.} is $v(l_B)$ with $v$ the interelectron
interaction potential and $l_B$ the magnetic length. For the case of
the Coulomb interaction, this energy scale is thus (the subscript
  ${}_{\rm c}$ is for Coulomb)
\begin{equation}
  \label{eq:einteraction}
 E_{c} = e^2/(\epsilon l_B) 
\end{equation}
which is proportional to $\sqrt{B}$.  On the other hand, the cyclotron
energy (the spacing between Landau levels) is $\hbar \omega_{\rm c} \sim
B/m_{\rm b}$.  Thus, a natural simplification to make is to assume the limit
of large magnetic field (or sometimes $m_{\rm b} \rightarrow 0$) such that
the cyclotron energy is much greater than the interaction
energy~\footnote{In practice, for typical GaAs band mass, the cyclotron
  energy becomes greater than the interaction energy at roughly 8
  Tesla.  However, it is also thought that the main effect of the
  higher Landau levels is to screen the Coulomb interaction (by
  allowing virtual inter-Landau-level transitions) so that even when
  $\hbar \omega_{\rm c}$ is not much greater than the interaction energy, we
  can still consider only the physics of the lowest Landau level.}.
In this case, we can assume that the inter-Landau-level excitations
are energetically forbidden so that we need only consider states
within a single Landau level (within the lowest Landau level for the
case of $\nu < 1$).  Once the problem is reduced to a single Landau
level, the only energy scale remaining in the problem is the Coulomb
energy~\footnote{It should be noted, however, that such projection to
  the lowest Landau level is not without risk.  See section
  \ref{sec:Mag} below as well as Ref.~\onlinecite{Sondhi} for some
  problems that occur when one tries to work in only a single Landau
  level.}.  It is worth noting that all of the exact diagonalizations
calculations, as well as all trial wavefunctions, are studied in this
limit~\cite{Jain,Fano,HaldaneChapter,Laughlin}.  In this limit, the
excitation gap for the fractional quantum Hall state must clearly be
on order of the interaction energy $(e^2/\epsilon l_B)$.  This is
quite natural being that the correlation effect, and the fractional
Hall state itself, is entirely due to the interelectron interactions.

As discussed above, however, in the Chern-Simons picture at the mean
field level, the fractional quantum Hall gap is given by $\Delta
\omega_{\rm c} = \hbar e \Delta B/(m_{\rm b} c)$ which is completely independent
of the interaction strength, and is thus clearly incorrect.  At RPA
level, the low energy excitations remains on the scale of $\Delta
\omega_{\rm c}$.  It should be immediately clear that we can not obtain a
gap on the scale of the interaction strength at RPA level since we
were able to define the essence of the RPA (in section \ref{sub:RPA})
without even mentioning the Coulomb interaction!  It is also clear
that this problem is serious since for typical experimental
parameters~\cite{WillettReview} $\Delta \omega_{\rm c}$ can be 4 to 15 times
the measured energy gap! In order to fix this problem we will need an
approximation beyond RPA.

One naive approach we might immediately try is to just assume that the
composite fermion mass becomes renormalized from the bare band mass
$m_{\rm b}$ to some renormalized mass $m^*$.  This is not such a strange
thing to assume since we know that this is a highly interacting system
and interactions often renormalize quantities such as the mass of
particles.  It thus seems reasonable to assume the mass is somehow
renormalized and to treat $m^*$ as an experimentally measured
phenomenological parameter without worrying too much about exactly how
it gets renormalized.  Indeed, such an assumption has been made in
calculations by many different groups.  The problem with such an
approach, however, is that the resulting calculations violate the
$f$-sum rule and Kohn's theorem.  To see this, we imagine replacing
$m_{\rm b}$ with $m^*$ everywhere it occurs.  We then calculate the
composite fermion resistivity $\rho_{\rm CF}^{\rm naive}$ and find that it is
now given by (compare to Eq. \ref{eq:meansum})
\begin{equation} 
   \rho_{\rm CF}^{\rm naive} \sim \frac{m_{\rm b}}{e^2 n_{\rm e}} 
   \left[ \begin{array}{cc} -i\omega & \Delta \omega_{\rm c}^* \\
                         -\Delta \omega_{\rm c}^* &  -i\omega \end{array} \right]
+ {\cal O}(q^2/\omega).
\end{equation}
where now we have the effective cyclotron frequency
\begin{equation}
  \Delta \omega_{\rm c}^* = \frac{e \Delta B}{m^* c}
\end{equation}
which now differs from the mean field result.  As discussed above, the
electron system will satisfy Kohn's theorem and the $f$-sum rule, if
and only if the composite fermion resistivity $\rho_{\rm CF}$ has the same
limiting form as the mean field result Eq. \ref{eq:meansum}.  Since
this is not the case, we see that this naive approach must violate
these sum rules.  In order to satisfy these sum rules and obtain the
correct energy scales for excitations, we will have to consider a more
complicated, and somewhat phenomenological, approximation which we
will discuss in the next section.

% give some semiclassical results.  derive sigma ~ q etc. ... perhaps
% put this above in the RPA section.  Also, the relaxation in a dirty
% system. Also the 2 expressions for current with m* and m_{\rm b}

\mysection{Landau Fermi Liquid Theory and MRPA}
\label{sec:FLT}

In section \ref{sec:RPA}, we found that the RPA gave many
results correctly, but it still did not give us the correct energy
scale for low energy excitations.  Furthermore, as we saw in section
\ref{sub:effmass} above, straightforward renormalization of the
fermion mass $m_{\rm b}$ to some renormalized value $m^*$ resulted in
violation of Kohn's theorem and the $f$-sum rule.

Fundamentally, the reason we are having trouble with RPA is that the
Chern-Simons system is highly interacting.  In particular, the
Chern-Simons interaction (the fact that one fermion sees, and responds
to, the flux attached to all of the other fermions) is quite strong.
In perturbative approaches such as RPA~\footnote{RPA can be thought of
  as the lowest order perturbative correction to mean field made
  self-consistent}, we attempt to identify some small
parameter (much less than one) in which to expand.  Here, there is no
small parameter.  We will see below in section \ref{sec:infrared} that
the closest thing we have to a small parameter is $\tilde \phi$, the
number of flux quanta attached to each fermion, which is at least 2
and is clearly not small. 

Fortunately, many highly interacting fermion systems have been
successfully studied in the past.  Although the Chern--Simons
interaction between composite fermions is different from the
interactions encountered in well studied fermionic systems such as
Helium--3 or electrons in metals, it is reasonable to expect that some
version of Fermi liquid theory~\cite{Pines,Nozieres} will be
applicable.  In this section, we will discuss the phenomenological
Fermi liquid theory and how it can cure some of the illnesses of the
Chern-Simons RPA.

The knowledgeable reader might object that framing the Chern-Simons
fermion theory in the language of traditional simple Fermi liquid
theory completely misses the most interesting pieces of physics.  In
particular, one might object that infra-red divergences actually make
the system non-Fermi liquid (or ``marginally'' Fermi liquid) in
several ways, and therefore should not be described within the Fermi
liquid framework.  Admittedly, these infra-red divergences do lead to
a very non-traditional system and we will discuss these issues in
depth in section \ref{sec:infrared} below.  However, for now, we will
be concerned with seeing how well we can describe the system within
this traditional framework.  For the reader who is worried about our
blithe neglect of these important issues, we point out that we can
justify this neglect and sweep these infra-red issue under the rug by
considering the case of long range interactions.  If we assume that
the inter-electron interaction $v(r)$ is longer range than Coulomb
(for example, a power law form $v(r) \sim r^{-\eta}$ with $\eta < 1$)
then the long wavelength density fluctuations become suppressed and
the system no longer has these infra-red problems (We will discuss
this issue more in section \ref{sec:infrared} below).  Although this
hypothetical long range interaction clearly does not correspond to any
real electron system~\footnote{See however,
  Ref.~\onlinecite{SternVortex} in which it is shown that a Josephson
  junction array is analogous to a quantum Hall system with long range
  interactions.} it is nonetheless useful to think about this simple
case.  Once we have fully analyzed this more simple case, we will
return to the physical situation and worry about these additional
complications in sections \ref{sub:div} and \ref{sub:dFLT} below.

The essence of Landau's Fermi liquid theory is that the long
wavelength low frequency response of a system of fermions in a Fermi
liquid phase can be deduced from an effective low energy description
of weakly interacting quasiparticle excitations near a Fermi surface.
As with free fermions, it is assumed that the ground state is a filled
Fermi sea, and quasiparticles excited from below the Fermi level to
above constitute the elementary excitations.  These quasiparticles are
allowed to have a renormalized effective mass $m^*$, and some local
interactions described by a function $f$ that we will define more
carefully below.  In general, the properties of these quasiparticles
can be quite different from the properties of the bare particles in
the system.  For example, in the Fermi liquid phase of Helium-3, the
effective mass of the quasiparticle is over 3 times as large as that
of the bare Helium-3 fermion.  A similar thing occurs for the
Chern-Simons fermion system.  We we will find that, within a Fermi
liquid theory description, we can have a quasiparticle mass that is
very different from the bare mass (being set by the interaction
energy) and we can still satisfy the necessary sum rules (Kohn's
theorem and the $f$-sum rule).  Furthermore, we will see in section
\ref{sec:Mag} that an additional complication associated with what we
call magnetization currents can also be solved within this Fermi
liquid format by allowing the Fermi liquid quasiparticle to become
magnetized.  Perhaps most importantly, this Fermi liquid description
will allow us to calculate response functions that can then be
compared to experiment.

Section \ref{sub:conventional} is an extended review of basic Fermi
liquid theory.  Those who are quite familiar with these topics are
encouraged to skip over this section, reading only that which is
necessary to become familiar with the notation.  In
section \ref{sub:silin} we will discuss the Landau-Silin extension of Fermi
liquid theory for Fermi liquids with long range interactions.  This
will, of course, be directly applied to the case at hand of our
composite fermions interacting via a long ranged Coulomb interaction
as well as a Chern-Simons interaction.   Finally, in section
\ref{sub:MRPA} we discuss the MRPA approximation for calculation of
response functions for the Chern-Simons Fermi liquid.   The main
physics of MRPA is that by including a Landau $F_1$ coefficient we can
renormalize the mass scale of the RPA calculation while preserving
Kohn's theorem and the $f$-sum rule. 

\subsection{Conventional Fermi Liquid Theory}
\label{sub:conventional}

In the Landau Fermi liquid theory, the state of the system is described by
the function $\delta n(\vec p,\vec r)$, characterizing the difference
of the density of fermions at the phase space point $\vec p, \vec r$
between an excited state and the ground state~\footnote{Needless to
  say, such a phase space density is only defined semiclassically
  since $\vec p$ and $\vec r$ don't commute as operators.}.  The
function $\delta n(\vec p,\vec r)$ fully describes the density and
current density anywhere in the system. Thus, to calculate the
response of the system to a force, one need only find the behavior of
this function $\delta n(\vec p,\vec r)$.  Note that in the case of a
charged Fermi liquid, we should consider $\vec p$ to be the
canonical~\cite{Nozieres,Pines} momentum $\hbar \vec k - \frac{e}{c}
\vec A$ with $\vec A$ the vector potential.

In this section we will review the basics of Landau's theory for
systems with short range interactions and we will worry about the
complications involved in adding the long range Coulomb and
Chern-Simons interactions below in section \ref{sub:silin}.

\subsubsection{Fermi Liquid Basics}

Fermi liquid theory states that a weak long wavelength low frequency
driving force affects only the occupation of states near the Fermi
momentum. The energy cost associated with a deviation $\delta n(\vec
p,\vec r)$ of the phase space distribution function from its ground
state value is given by a phenomenological energy density functional
\begin{eqnarray}
  \label{eq:energyfunc} \nonumber
  {\cal E}\left[\delta n(\vec p,\vec r)\right] &=& \int \frac{d\vec p}{(2 \pi
    \hbar)^2} \epsilon_0(\vec p) \delta n(\vec p,\vec r) \\ &+& \frac{1}{2} \int
  \frac{d\vec p}{(2 \pi \hbar)^2} \int \frac{d\vec p'}{(2 \pi
    \hbar)^2} f(\vec p,\vec p') \delta n(\vec p,\vec r) \delta n(\vec
  p',\vec r).
\end{eqnarray}
Here $\epsilon_0 = \vec p^2/2 m^*$.  Thus, this energy functional
depends on only two parameters --- the quasiparticle effective mass,
$m^*$ which determines the energy of exciting a single quasiparticle
above the Fermi surface, and the Landau interaction function $f(\vec
p,\vec p')$ that gives the added quasiparticle-quasiparticle
interaction energy associated with having more than one quasiparticle
excited.  Note that this interaction functional can only describe
short range interactions (those local in $\vec r$).  The two
parameters $m^*$ and $f$ in the energy functional (Eq.
\ref{eq:energyfunc}) are formally defined in terms of one and two
particle Green's functions~\cite{Nozieres}. These definitions are not
essential for this part of our discussion, though we will return to
this issue in section \ref{sec:infrared} below.

The quasiparticle energy $\tilde \epsilon$ is obtained by
differentiating Eq.  \ref{eq:energyfunc} with respect to $\delta
n(\vec p,\vec r)$, yielding,
\begin{equation}
  \tilde \epsilon(\vec p,\vec r) = \epsilon_0(\vec p,\vec r) + 
 \int\frac{d\vec p'}{(2
      \pi\hbar)^2} f(\vec p, \vec p\,') \delta n(\vec p\,',\vec r).
\end{equation}
The quasiparticle velocity is then naturally given as $\vec u(\vec p)
= \nabla_{\vec p} \tilde \epsilon$, such that we can write the
particle density and current in terms of the phase space density
\begin{eqnarray}
                  \label{eq:j0j1}
                  n(\vec r) &=&  
                \int \frac{d\vec p}{(2 \pi \hbar)^2} \delta n(\vec
                  p,\vec r)  \\ \label{eq:j1j1}
                  \vec j(\vec r) &=& 
                \int \frac{d\vec p}{(2 \pi \hbar)^2} \delta [\vec u(\vec p)
                  n(\vec p,\vec r)].  
                \end{eqnarray}

To calculate a response, we should use a Boltzmann transport equation
derived from the energy functional.  Most generally, a Boltzmann
equation is difficult to solve.  However, we will be able to make use
of the fact that in Fermi liquid theory, the relevant~\footnote{Here we
  use the word ``relevant'' in the renormalization group sense as well
  as in the conventional sense of the word.  A good review of how
  renormalization group can be applied to Fermi systems to obtain
  Fermi liquid theory is given in Ref.~\onlinecite{ShankarRG}.}
low energy excitations are smooth fluctuations of the shape of the
Fermi surface.  To this end, we consider fluctuations of the shape of
the Fermi surface that we describe as~\footnote{Note that definition of
  $\nu$ used here agrees with that in 
  Refs.~\onlinecite{Ady},~\onlinecite{Nozieres}, and~\onlinecite{Pines} , 
  but differs from the
  function $f$ used in references~\onlinecite{Lee} 
  and~\onlinecite{Simonhalp} by a factor of $\vf = \pf/m^*$.}
 \begin{equation}
   \label{eq:nudef0}
  n(\vec p, \vec r) = n_0[ p - \pf - \nu(\theta, \vec r) ]
\end{equation}
where $\theta$ is the direction of $\vec p$ on the Fermi surface,
$\pf$ is the Fermi momentum, and $n_0$ is the equilibrium phase space
density (taken to be a step function at zero temperature, and an
appropriate Fermi occupation at finite temperature).  Thus, the
function $\nu(\theta, \vec r)$ describes the shape of the Fermi
surface~\footnote{Sometimes it is useful to assume that $\nu$ is small
  and expand to yield
$$
  \delta n(\vec p, \vec r) = \nu(\theta, \vec r) \delta(|\vec p| - \pf).
$$
However, this must be used with caution since we may sometimes want
the second order contributions.}.  We will use the convention that
$\theta = 0$ points in the $\hatn x$ direction.

Since we are only concerned with small excitations of the Fermi
surface, we assume that the interaction function $f(\vec p, \vec p')$
is not a function of the magnitude of the vectors $\vec p$ and $\vec
p'$ (since all magnitudes are approximately $\pf$) and is only a
function of their directions~\footnote{In the Chern-Simons case (with
  Coulomb or short range interactions), it is not so clear that
  $f(\vec p,\vec p')$ should depend only on the angle since the
  behavior of quasiparticles is singular as they approach the Fermi
  surface.  This complication is discussed further in sections
  \ref{sec:infrared} and \ref{sub:dFLT} below, although our
  understanding of this is far from complete.  One hopes that in a
  fully renormalized theory (nonperturbatively) these singularities do
  not prevent us from writing a Boltzmann transport equation.  We note
  that Kim et al~\cite{Kim} recently showed that a form of quantum
  Boltzmann equation can be derived independent of these
  singularities.}.  Furthermore, by rotational invariance, the
function $f$ should only depend on only the angle between $\vec p$ and
$\vec p'$ so that in two dimensions we can write $f(\theta - \theta')$
where $\theta$ is the angle of $\vec p$ and $\theta'$ is the angle of
$\vec p'$.

\subsubsection{Fourier Space and Restrictions on Fermi Liquid Coefficients}
\label{subsub:flcs}

It is often more convenient to work with Fourier transformed
quantities.  We now define
\begin{eqnarray}
  \label{eq:fldef}
  f_l &=& \frac{1}{2 \pi} \int_0^{2 \pi} \! \! d \theta \, f(\theta) e^{i l
    \theta} \\
  \nusub{l}(\vec r) &=& \frac{1}{2 \pi} \! \! \int_0^{2 \pi} d \theta
    \, 
  \nu(\theta,\vec r) e^{i l \theta}.  \label{eq:nuldef}
\end{eqnarray}
Note that due to the symmetry of the interaction function $f(\theta) =
f(2\pi-\theta)$ we expect that $f_l = f_{-l}$.  Note that a very
common notation is to define the dimensionless quantity~\footnote{In
  references~\onlinecite{Simonhalp},~\onlinecite{Platz1}, 
  and~\onlinecite{Lee}, $F_l$ is called $A_l$.}
\begin{equation}
   F_l = \frac{m^* f_l}{2 \pi \hbar^2}
\end{equation}

Clearly, the zeroth Fourier mode corresponds to a uniform compression
of the Fermi surface, which corresponds to the density change
(See Eq.  \ref{eq:j0j1})
\begin{equation}
  \label{eq:nnu0}
  \delta n(\vec r) = \frac{\pf \nusub{0}(\vec r) 
}{2 \pi \hbar^2} +
  {\cal O}(\nusub{0}^2). 
\end{equation}
The first Fourier mode, on the other hand, corresponds to a uniform
boost of the Fermi surface resulting in the current
density~\footnote{To see this result more clearly, consider for example
  boosting the entire Fermi sea by a small vector $\vec q$,
  corresponding to a current density $\vec j =  n_{\rm e} \vec q/m_{\rm b}$.  A
  point $\vec p$ (at angle $\theta$) on the Fermi surface, then gets
  boosted to a point $\vec p+ \vec q$ whose magnitude is $\pf + \vec p
  \cdot \vec q + {\cal O}(q^2)$ thus giving $\nu(\theta) = \cos(\theta
  - \theta_q)$.  Converting to Fourier space then yields $\nusub{1} +
  \nusub{-1} = q_x$ and $\nusub{1} - \nusub{-1} = i q_y$.  One could
  have alternately inserted $\nu(\theta)$ into Eq. \ref{eq:j1j1}.
  However, in order to get the same result, we would then need
  Eq. \ref{eq:f1effmass} below.}
\begin{eqnarray}
 \label{eq:currentfourx0} j_x(\vec r) &=& \frac{n_{\rm e}}{m_{\rm b}}
  (\nusub{1}(\vec r) + \nusub{-1}(\vec r)) + {\cal O}(\nu^2) \\
    \label{eq:currentfoury0} j_y(\vec r) &=& \frac{n_{\rm e}}{i m_{\rm b}} (\nusub{1}(\vec r) - \nusub{-1}(\vec r)) + {\cal O}(\nu^2).
\end{eqnarray}
Note that here we use the bare mass of the fermion since we
are boosting the entire Fermi sea.  Higher Fourier modes correspond to
more complex deformations of the Fermi surface that carry no net current
and no net charge. 

In terms of these Fourier transformed quantities, the energy-density
functional (Eq. \ref{eq:energyfunc}) can be written simply as
\begin{equation}
  \label{eq:membrane}
  {\cal E}[\nusub{l}(\vec r)]  = \epsilon_0(\pf) \delta n(\vec r) + 
\sum_l  \frac{n_{\rm e}}{m^*} 
\left( 1 + F_l \right)
|\nusub{l}(\vec r) |^2.
\end{equation}
This expression assigns to each mode of deformation of the Fermi
surface a specific energy density which is made up of a noninteracting
part (the ``1'' in the parenthesis) and an interaction contribution
(the $F_l$ piece).  It is worth noting that the $l^{th}$ Fourier
deformation mode couples only to the interaction coefficient $F_l$.
Thus, $F_0$ (or $f_0$) describes the interaction energy associated
with the compression mode $l=0$ of the Fermi surface, whereas $F_1$
(or $f_1$) describes the interaction energy associated with the boost
mode ($l=1$).  As such, we expect that $f_0$ should be related to the
compressibility of the system, whereas $f_1$ should be related to the
bare mass of the system (which should tell us the energy of boosting
the entire mass of the system).  Indeed, these expectations are true
as we will now see.

By noting that $\mu = d {\cal E}/d (\delta n(\vec r))$, explicitly
differentiating Eq. \ref{eq:membrane}, and making use of Eq.
\ref{eq:nnu0} to relate $\nusub{0}$ to $\delta n(\vec r)$, we obtain
the compressibility equation~\footnote{In the case of the Chern-Simons
  Fermi liquid, the compressibility derivative $\frac{d\mu}{dn}$ is
  taken at fixed $\Delta B$. This point is somewhat subtle, and is
  discussed in detail in sections \ref{sec:sep} and \ref{sub:dFLT}
  below.}
\begin{equation}
  \frac{d \mu}{d n} = \frac{2 \pi \hbar^2}{m^*} + f_0.
  \label{eq:f0comp}
\end{equation}

Similarly, boosting the entire system a momentum $\vec q$, by Galilean
invariance, should cost an energy density of $n_{\rm e} q^2/(2 m_{\rm b})$.  For
convenience we choose $\vec q$ to be in the $\hatn x$ direction such
that have $\nusub{1} = \nusub{-1} = \frac{1}{2} q$.  We can now
calculate the energy density of this boost in a different way by using
Eq.  \ref{eq:membrane} to obtain $\frac{2 n_{\rm e}}{m^*} (1 + F_1)
(\frac{q}{2})^2$.  Equating these two equivalent expression for the
energy then yields
\begin{equation}
  \label{eq:F1part2}
  \frac{m^*}{m_{\rm b}} = 1 + F_1
\end{equation}
or equivalently
\begin{equation}
  \frac{1}{m_{\rm b}} = \frac{1}{m^*} + \frac{f_1}{2 \pi \hbar^2}.
  \label{eq:f1effmass}
\end{equation}

We note in passing that one can derive an additional sum rule
restriction on the value of the Fermi liquid coefficients based on the
Pauli exclusion principle~\cite{Platzman,SimonNFL}.  These restrictions
will not be particularly important for our present discussion.

This relation (Eq. \ref{eq:f1effmass}) between the $f_1$ coefficient,
the effective mass and the bare mass is an extremely important result.
This tells us that within Fermi liquid theory, whenever the effective
mass is renormalized from the bare mass, then in order to have a
consistent theory one must also include an interaction of the $f_1$
type.  In other words, if we think of the effective mass of the
quasiparticle becoming different from the bare mass in some sort of
renormalized theory~\cite{ShankarRG}, then the renormalization
procedure should also generate an $f_1$ type interaction.  In section
\ref{sub:effmass} above, we had trouble renormalizing the composite
fermion mass in a consistent way.  The reason for this trouble was
that we neglected to consider the possibility of an $f_1$ interaction.
By including such an interaction term in sections \ref{sub:silin} and
\ref{sub:MRPA} below, we will be able to fix this problem.

\subsubsection{Boltzmann Transport}

We now consider the derivation of a transport equation.  The
quasiparticle energy $\tilde \epsilon$ can be used as an effective one
particle Hamiltonian to derive a Landau-Boltzmann
transport~\footnote{The Boltzmann equation is nothing more than
  conservation of particle number in phase space.  The current in the
  real space direction is $\vec j_r = n \vec u = n \frac{d\vec r}{dt}
  = n \nabla_{\vec p} \tilde \epsilon$.  Analogously the current in
  the momentum space direction is $\vec j_p = n \frac{d \vec p}{d t} =
  n(\vec F - \nabla_{\vec r} \tilde \epsilon)$.  Thus we obtain the
  Boltzmann equation by conserving particle number $\nabla_{\vec p}
  \cdot \vec j_p + \nabla_{\vec r} \cdot \vec j_r + \frac{\partial
    n}{\partial t} = 0$. } equation~\cite{Pines,Nozieres}
\begin{equation}
  \label{eq:trans2}
    \frac{\partial n}{\partial t} + \nabla_{\vec r} n \cdot
    \nabla_{\vec p} \tilde \epsilon - \nabla_{\vec p} n \cdot
    \nabla_{\vec r} \tilde \epsilon + \vec F \cdot
    \nabla_{\vec p} n= \left( \frac{\partial n}{\partial t}
  \right)_{\rm coll} 
\end{equation}
where $\vec F(\vec r)$ is an applied force and the right hand
side is a collision term. 

We now consider applying a uniform magnetic field $B$ to the system as
well as a weak perturbing electric field $\vec E$ at wavevector $q$
and frequency $\omega$.  We will maintain the convention that $\vec q
\| \hatn x$ such that all perturbations are proportional to
%%%%%%%%%%%%*** CHECK SIGNS HERE
$e^{iqx-i\omega t }$.  
%%%%%%%%%%%%%%%%%%%%%%%%
With some effort, the Boltzmann equation can be
then be linearized and written in terms of the small fluctuation of
the Fermi surface as~\cite{Lee,Pines}
\begin{equation}
\label{eq:kin1}
-i\omega \nu(\theta) +
\left(iq\vf \cos(\theta) -
 \omega_{\rm c}^* \frac{\partial}{\partial \theta} \right)
[\nu(\theta) + \delta \epsilon_1(\theta)] = -e \vec{E} \cdot
\hatn{n}(\theta)  + \mbox{$I$}(\theta)
\end{equation}
where 
\begin{equation}
  \label{eq:eps12}
  \delta \epsilon_1(\theta) = \frac{m^*}{(2 \pi \hbar)^2} \int
  d\theta' f(\theta - \theta') \nu(\theta'),
\end{equation}
where $\hatn n(\theta) = (\cos \theta, \sin \theta)$, with $\omega_{\rm c}^*
= \frac{e B}{m^*c}$, the mass renormalized cyclotron frequency, and
$I(\theta)$ a scattering term whose form will be discussed in the next
section.  In the case of the Chern-Simons Fermi liquid, we should use
the effective cyclotron frequency $ \Delta \omega_{\rm c}^* = \frac{e \Delta
  B}{m^*c}$ seen by the transformed fermions.

\subsubsection{Scattering}
\label{subsub:scat}

As a model of scattering, one might hope to use the 
phenomenological relaxation form 
\begin{equation} 
\label{eq:badscat}
 I(\theta) = -\tau^{-1} \nu(\theta).
\end{equation}
Although this is very simple, and clearly corresponds to an
analytic continuation of $\omega \rightarrow \omega + \frac{i}{\tau}$
in Eq. \ref{eq:kin1}, we will see that this simple form is not
allowed.

If we apply the operator $\int_0^{2 \pi} d\theta$ to both sides of the
kinetic equation (Eq. \ref{eq:kin1}), with some algebra we obtain
\begin{equation}
  \frac{4 \pi^2 \hbar^{2}}{\pf} \left( i \omega \delta n - i q j_x
  \right)
  =  \int_0^{2 \pi} d
  \theta I(\theta)
\end{equation}
Thus, in order to assure local current conservation $\omega \delta n =
q j_x$, we must have a ``current conserving'' scattering
term~\footnote{Scattering due to quasi--particle collisions would be
  subject to conservation of momentum density too.  This would require
  also $\int_0^{2 \pi} d\theta \cos(\theta) I(\theta) = 0$.} that
satisfies $\int_0^{2 \pi} I(\theta)d\theta =0$.  Thus, the simplest
model of scattering that one may consider is given by~\cite{Platz1}
\begin{equation}
  \label{eq:sctheta}
  I(\theta) = - \frac{1}{\tau} \left( \nu(\theta)
 - \frac{1}{2\pi}\int_0^{2 \pi}
  d\theta \nu(\theta)\right) 
\end{equation}
The use of such a density preserving scattering term has been
emphasized in Ref.~\onlinecite{WolfleScat}.  However, we will see
below that this current conservation correction to the scattering term
typically has a pretty small effect, and sometimes has exactly no
effect.

\subsubsection{Results of Conventional Fermi Liquid Theory}
\label{sub:simple}

In this section, we quote results regarding the calculation of
conductivities from the Boltzmann equation.  The general method of
calculation is to impose some electric field $\vec E(q,\omega)$ in Eq.
\ref{eq:kin1} and solve for the resulting fluctuation of the Fermi
surface $\nu(q,\omega)$.  Once this is obtained, the current is given
by Eqs. \ref{eq:nuldef}, \ref{eq:currentfourx0}, and
\ref{eq:currentfoury0}, which enables us to extract the conductivity.
In appendix \ref{app:non} we have shown such a
calculation for the case of a noninteracting Fermi system ({\em i.e.\/}, all
$f_l = 0$) in zero magnetic field.  Reasonably simple results can also
be obtained for free fermions in a magnetic
field~\cite{Harrison,Simonhalp,HLR}.

The inclusion of nonzero Fermi liquid coefficients in general makes
the solution of the Boltzmann equation more difficult.  However, the
effects of $f_0$ and $f_1$, as well as the effects of the scattering
model discussed above (Eq. \ref{eq:sctheta}) can be treated explicitly
without re-solving the full Boltzmann equation~\cite{Simonun}.   These
``tricks'' for solving more complicated Boltzmann equations will be
described here and the proofs are left for the adventurous readers. 

For treating $f_0$, we relate the resistivity matrix of an arbitrary
system to the resistivity matrix of a system that is the same except
that the zero${}^{th}$ Landau coefficient $f_0$ has been artificially
set to zero.  We call the resistivity of this artificial system
$\rho^{f_0=0}$, and we have the relation
\begin{equation}
  \label{eq:tf0t}
  \rho = \rho^{f_0 =0} + \frac{i f_0 q^2}{\omega e^2} \left(
  \begin{array}{cc} 1 & 0 \\ 0 & 0 \end{array} \right)_.
\end{equation}
In terms of a polarization matrix (just by applying factors of $T$),
we might write the analogous
\begin{equation}
\label{eq:Pi-1}
  \Pi^{-1} =  \Pi^{-1}_{f_0 = 0} +  \left(
  \begin{array}{cc} f_0 & 0 \\ 0 & 0 \end{array} \right)_.
\end{equation}  
The form of this result should be clear.  This is just an RPA-like
treatment of the $f_0$ interaction which is a
density-density interaction (like the Coulomb interaction) and
therefore enters only in one component of the inverse response
(compare Eqs. \ref{eq:K2sep} and \ref{eq:Vdef}).  

Similarly, for $f_1$, we can relate the resistivity of an arbitrary
system to the resistivity of an artificial system for which the first
Landau coefficient $f_1$ has been set to zero and the particle mass is
set equal to $m^*$.  We call the resistivity of the artificial system
$\rho_{f_1=0}^*$ and derive the result~\footnote{In the appendix of 
Ref.~\onlinecite{Simonhalp} this is derived in an extremely complicated
  way.  A much cleaner, and more general, RPA-like
  derivation~\cite{Simonun} of this result can also be performed.}
\begin{equation}
  \label{eq:tf1t}
  \rho = \rho^*_{f_1=0}  + \left[ i \omega \frac{f_1 m^*}{ 2 \pi \hbar^2}
  \frac{m_{\rm b}}{e^2 n_{\rm e}} \right] I 
\end{equation}
with $I$ the identity matrix.  In terms of a polarization again we
have
\begin{equation}
  \label{eq:Pi-2}
  \Pi^{-1} = [\Pi^*_{f1=0}]^{-1} +  \frac{m^* - m_{\rm b}}{n_{\rm e} e^2} \left( \begin{array}{cc} \frac{\omega^2}{q^2} & 0 \\ 0 & -1
  \end{array} \right)
\end{equation}
The ${}^*$ here reminds us that this polarization should be calculated
using the effective mass (but with $f_1 = 0$).  The form of this
equation is similarly a RPA-like treatment of the current-current
interaction $f_1$.

We can also consider the effect of a scattering term.  We recall that
the simplest scattering model (Eq. \ref{eq:badscat}) does not conserve
current.  However, the solution of such a nonphysical model is
obtained very simply by analytic continuation ($\omega \rightarrow
\omega + \frac{i}{\tau}$) of the resistivity of a system without
scattering (this should be obvious from Eq. \ref{eq:kin1}).  The
solution to the current conserving model (Eq.  \ref{eq:sctheta}) can
also be obtained in a similar manner.  The additional term in the
current conserving model has the effect of an $f_0$ coefficient
yielding the result
\begin{equation}
\label{eq:correction}
  \rho(\omega) = \rho^{f_0 =0,\tau = \infty}(\omega + \frac{i}{\tau})
  + \frac{i \kappa q^2}{(\omega + \frac{i}{\tau}) e^2} \left(
  \begin{array}{cc} 1 & 0 \\ 0 & 0 \end{array} \right).
\end{equation}
with 
\begin{equation}
  \kappa = (1 + \frac{i}{\omega \tau}) f_0 + \frac{2 \pi
    \hbar^2}{m^*} \frac{i}{\omega\tau}.
\end{equation}
where $\rho^{f_0 =0,\tau = \infty}(\omega)$ is the resistance of a
system identical to our actual system, but without scattering
($\tau=\infty$) and with the zeroth Fermi liquid coefficient set to
zero.  Once this resistance is calculated for real $\omega$, it must
be analytically continued to $\omega + \frac{i}{\tau}$.  We note that
only the $\rho_{xx}$ term of the resistivity matrix distinguishes
between the current conserving and the non-current conserving models of
scattering.  

Thus, we have here several rules for relating resistivities of certain
systems to resistivities of simpler systems.  As a simple example of
using some of these results, we consider a system of particles in zero
magnetic field with a Landau interaction function that has nonzero
$f_0$ and $f_1$ coefficients (but with all other coefficients zero)
and with the current conserving scattering term (Eq.
\ref{eq:sctheta}).  We will focus on obtaining $\rho_{yy}$ since this
is what will be most relevant for the Chern-Simons problem (See Eq.
\ref{eq:CSrhoyy}).  From Eqs. \ref{eq:tf0t} and \ref{eq:correction} it
is clear that $f_0$ and the current conservation correction has no
effect on $\rho_{yy}$.  Thus, we can obtain $\rho_{yy}$ for this model
by simply analytically continuing the results of Eq. \ref{eq:tf1t} via
$\omega \rightarrow \omega+\frac{i}{\tau}$.  Using Eq.
\ref{eq:appres2} for the Boltzmann conductivity of free fermions with
Eq. \ref{eq:tf1t} then yields the result (recall $F_1 = \frac{m^*
  f_1}{2 \pi \hbar^2}$)
\begin{equation}
  \label{eq:rhoyyex}
  \rho_{yy} =  i \frac{2 \pi \hbar}{e^2} \frac{q}{\kf} \left[
  -\frac{1}{\Omega} \left( 1  - \left[1 - \frac{1}{\Omega^2} \right]^{1/2}
 \right)^{-1} + 2 \Omega \left(\frac{F_1}{1 + F_1} \right)
  \right].
\end{equation}
with $\Omega = (\omega + \frac{i}{\tau})/(q \vf)$.  It is worth noting
that the form of the resistivity is given by $\rho_{yy} = q
G_{yy}(\Omega)$ with $G_{yy}$ some function.  Without scattering,
$\rho_{xx}$ takes a similar form.

We might guess at this point that we can use this resistivity as
$\rho^{\rm CF}_{yy}$ in Eq. \ref{eq:CSrhoyy} to obtain the electrical
conductivity of the even denominator states at $\nu=\frac{1}{2m}$.
Indeed, this is exactly what we did in obtaining Eqs. \ref{eq:sigres}
and \ref{eq:sigres2}.  We will now examine in more detail exactly what
this prescription amounts to.

\subsection{Landau-Silin Chern-Simons Theory }
\label{sub:silin}

Clearly, the above {\it local} energy functional (Eq.
\ref{eq:energyfunc}) can not properly represent a long range
interaction.  More generally, an extension of Fermi-liquid theory was
developed by Silin~\cite{Pines,Nozieres,LandauSilin} that separates out
the Hartree part of the long range interaction and uses the Landau
interaction function $f$ to represent the short range part of the
interaction.  In Landau and Silin's original work~\cite{LandauSilin},
this approach was used to describe electrons with long range Coulomb
interactions.  The prescription they used was to calculated the
conductivity (or polarization $\Pi^v = T^{-1} \sigma T^{-1}$) using the
Boltzmann equation with effective mass $m^*$ and interaction function
$f$.  The long range Coulomb interaction is then added on last via Eq.
\ref{eq:K2sep} (which is equivalent to Thomas-Fermi, or
self-consistent Hartree screening).  The justification for this
prescription is simply that the fermions must respond to the
externally applied force as well as the internally induced force (the
self-consistent Hartree piece).  This approach can be justified in an
explicit perturbation theory~\cite{Pines,Nozieres} for the long
wavelength low frequency limit, and has also been quite successful for
treating stronger interactions, although it has not been rigorously
justified.

In the case of the Chern-Simons Fermi liquid, a similar prescription
can be used~\cite{Simonhalp}.  Here, we should calculate the Composite
fermion conductivity $\sigma^{\rm CF}$ (or polarization $\Pi = T^{-1}
\sigma^{\rm CF} T^{-1}$) using the Boltzmann equation with an effective
mass $m^*$ and interaction function $f$.  The Chern-Simons interaction
is then treated at self-consistent Hartree level to yield an electron
resistivity given by $\rho = \rho^{\rm CF} + \rho^{\rm CS}$ (Eq.
\ref{eq:splitrho}).  The full electromagnetic response can similarly
be calculated by using Eq. \ref{eq:KPi}.  Thus we now have a
full prescription for calculating responses for our Chern-Simons
system through a Landau-Silin Boltzmann framework. (Note that below in
section \ref{sec:Mag} we will see that this prescription neglects
important magnetization terms.  However, for now we will not be
concerned with these effects, and we will see later that our neglect
is justified at least so long as we are concerned with calculating
only the longitudinal responses $K_{00}$ or $\sigma_{xx}$.)

We recall that in section \ref{sub:effmass} above~\cite{HLR,Simonhalp},
there was difficulty in obtaining a response that both satisfied the
sum rules (Kohn's theorem and the $f$-sum rule) and also had low
energy excitations on the right (interaction) energy scale.  We found
that either we used the bare mass $m_{\rm b}$ in our calculation and we
ended up with low energy excitations with energy scale set by $m_{\rm b}$
(instead of $m^*$), or we used an effective mass $m^*$ in our
calculation and we found that the Kohn mode occurred at a frequency
set by $m^*$ instead of a frequency set by $m_{\rm b}$ as is required by
Galilean invariance.  As mentioned above in section \ref{subsub:flcs}
the reason we had this problem is that renormalization of the mass
must imply a nonzero Fermi liquid interaction coefficient $f_1$ (See
Eq. \ref{eq:f1effmass}).  In this Landau-Silin approach it is very
easy to include such an interaction term in our calculation to
restore Galilean invariance (satisfying the sum rules) while keeping
the low energy excitations on the scale of the effective mass.  Thus,
to phenomenologically renormalize the energy scale in this framework
we need only change $f_1$ and $m^*$ together.

In terms of the response function $K$, we write
\begin{eqnarray}
  \label{eq:Pi*}
  K^{-1} &=& [\Pi^*_{f_1 = 0}]^{-1} + {\cal F}_1 + V + C \\ 
  \label{eq:F1def} {\cal F}_1 &=& 
  {\frac{(m^*\!-\!m_{\rm b})}{n_{\rm e} e^2}} \left(
  \begin{array}{cc}  {\frac{\omega^2}{q^2}} & 0 \\ 0 & -1 \end{array}
  \right)_.
\end{eqnarray}
Note that the addition of the ${\cal F}_1$ term is the same as in Eq.
\ref{eq:Pi-2}.  The Landau-Silin result is then achieved by
approximating $\Pi^*_{f_1 = 0} = T^{-1} \sigma T^{-1}$ as the
Boltzmann response of fermions of mass $m^*$ in the mean magnetic
field $\Delta B$.  If we assume all of the other Fermi liquid
coefficients are zero ($f_n = 0$ for $n \ne 1$), then $\Pi^*$
represents the response of free fermions.  The Boltzmann response of
free fermions is calculated explicitly in appendix \ref{app:non} for
the case of $\Delta B = 0$ which corresponds to the case of
$\nu=\frac{1}{2m}$.  Within this framework, it is very easy to include
a phenomenological scattering time as discussed above.  This Boltzmann
approach (neglecting all other Fermi liquid coefficients) was called a
``Modified Semiclassical'' approximation in Ref.
\onlinecite{Simonhalp}.  In section \ref{sub:saw} this type of
approximation is used to make predictions for the results of surface
acoustic wave experiments.

\subsection{Modified RPA (MRPA)}
\label{sub:MRPA}

The Landau-Silin approach discussed above depends on the Boltzmann
equation for calculating responses.  As discussed in appendix
\ref{sub:non}, the Boltzmann equation misses some important features
of the response function, such as the diamagnetic term of the
transverse response.  Furthermore, the Boltzmann equation, being
semiclassical, does not know about the incompressibility of the system
when an integer number of Landau levels are filled.  For this reason,
we would like to define an approximation based on this Landau-Silin
approach that has some of these features properly included (at least
in some approximation). 

In Ref.~\onlinecite{Simonhalp} a Modified RPA (MRPA) was constructed
that attempts to include these effects.  The MRPA is obtained by
setting $\Pi^*$ in Eq. \ref{eq:Pi*} equal to the response $K^{0*}$ of
a system of noninteracting fermions of mass $m^*$ in the mean magnetic
field $\Delta B$.  The response function thus calculated (using $\Pi^*
=K^{0*}$ and Eq.~\ref{eq:Pi*}) will be called
$K^{{\mbox{\tiny{MRPA}}}}$.   Thus, we have
\begin{equation}
\label{eq:MRPABIG}
  [K^{{\mbox{\tiny{MRPA}}}}]^{-1}  = [K_0^*]^{-1} + {\cal F}_1 + V + C  
\end{equation}
Results of such MRPA calculations for the Jain fractions
$\nu=\frac{p}{2mp+1}$ are shown in Refs.~\onlinecite{Simonhalp} and
\onlinecite{SimonballandSong}.  A figure from Ref.
\onlinecite{Simonhalp} is also reproduced in the chapter by Willett in
this volume.

At filling fraction $\nu =\frac{1}{2m}$ there is zero effective field
so the response $K^{0*}$ is diagonal.  We can then use the MRPA
prescription (Eq. \ref{eq:MRPABIG}) to obtain
\begin{equation}
  \label{eq:MRPAresult}
  K_{00} = \frac{1}{\left(\frac{2 \pi \tilde \phi}{q}\right)^2
    \left([K^0_{11}]^{-1} + \frac{m^* - m_{\rm b}}{n} \right)^{-1} +
  ([K_{00}^0]^{-1} - \frac{m^*-m_{\rm b}}{n}\frac{\omega^2}{q^2})  + v(q) }
\end{equation}
When $m_{\rm b} = m^*$ this reduces to the RPA result (Eq.
\ref{eq:K00half}).  We also note that in the limit of $\omega \ll q
\vf$, this reduces to the form of the RPA result (Eq.
\ref{eq:lowenergyRPA}) except that all factors of $m_{\rm b}$ are replaced
by factors of $m^*$.  Thus, as with the RPA result, this response also
has an overdamped mode at $\omega \sim q^3 v(q)$ that will result in
infrared divergences when we try to calculate further corrections in
section \ref{sub:div} below.  In Fig.  \ref{fig:MRPA} the excitation
spectrum for $\nu=\frac{1}{2}$ is shown for such an MRPA calculation
with $m^* = 10 m_{\rm b}$ . The energy scale in that figure is the
renormalized cyclotron frequency, so on that scale the bare cyclotron
mode is off of the top of the graph.

\begin{figure}[htbp]
  \begin{center}
    \leavevmode \epsfxsize=3.5in \epsfbox{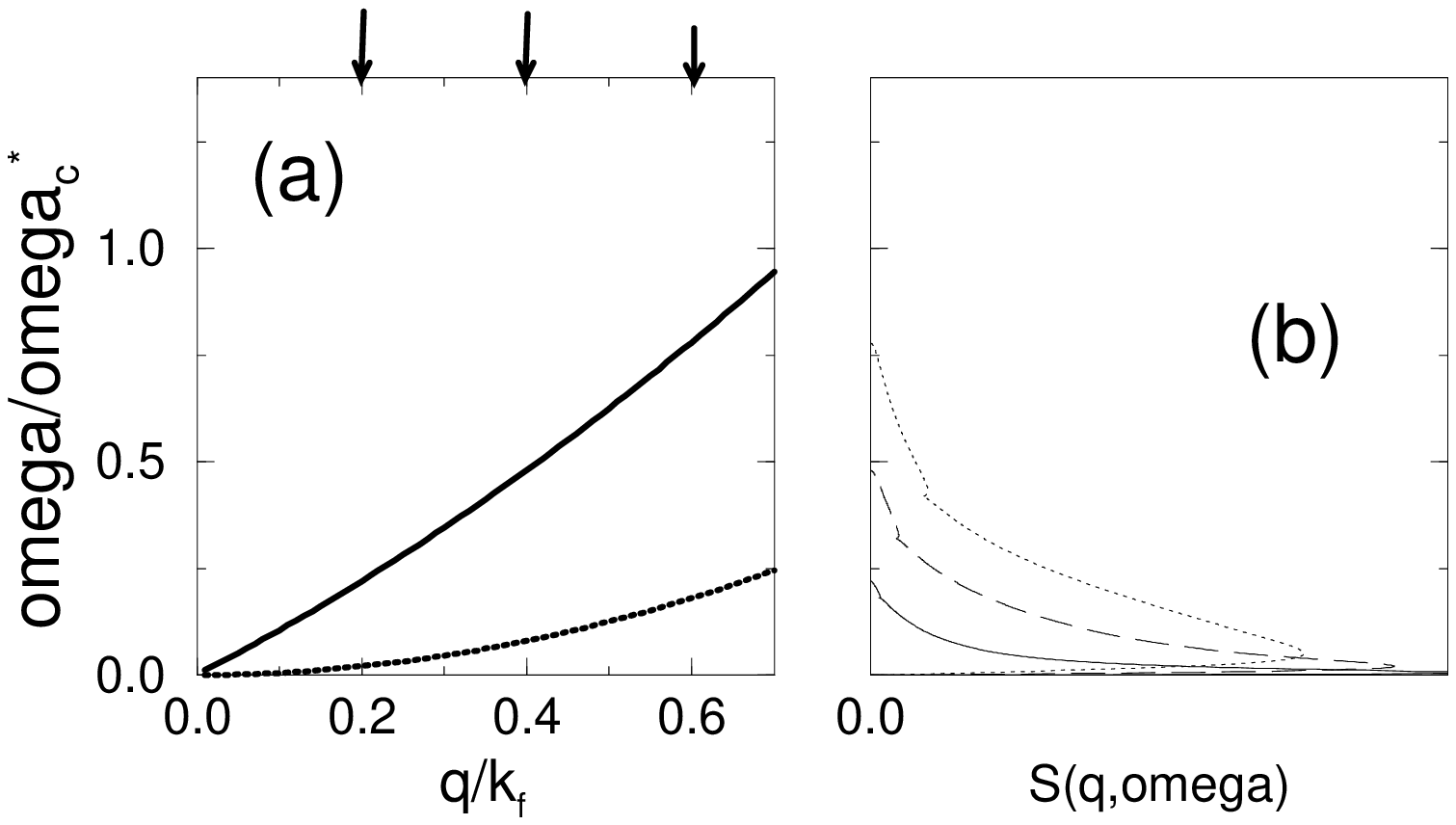}
  \caption{\protect{\small Excitation spectrum of the
      $\nu=\frac{1}{2}$ state in MRPA for $m^*=10m_{\rm b}$}}
  \label{fig:MRPA}
  \begin{minipage}[t]{4.5in}  
    {\small Note that the energy scale is in units of the renormalized
      cyclotron frequency $\omega_{\rm c}^* = \frac{e B}{m^* c}$.  In (a)
      the solid line is the edge of the low energy continuum of
      quasiparticles, the dotted line is the location of the peak in
      the weight (the maximum of the structure factor) of the low
      energy continuum ($\omega \sim q^3 v(q)$).  The cyclotron mode
      is off the top of the graph.  In (b) the structure factor
      $S(q,\omega)$ is shown explicitly for three different
      wavevectors $q/\kf = .2$ (solid), $.4.$ (dashed), and $.6$
      (dotted).  The small cusp in the structure factor is a
      reflection of a similar cusp that occurs in $K^0$ for
      noninteracting fermions.  In comparison to the RPA (See fig.
      \ref{fig:RPA}) the weight is pushed to even lower relative
      frequency.  Here we have used $E_{\rm c}/(\hbar \omega_{\rm c}^*) = 5$ which
      is a reasonable theoretical
      estimate~\cite{MorfanddAmbruminil,Morf}.}
  \end{minipage} 
   \end{center}
\end{figure}
 
Comparisons of results of exact diagonalizations of small systems
projected to the Lowest landau level to results of $K_{00}$ calculated
in the MRPA were quite favorable~\cite{SimonballandSong} for the low
energy excitations at $\nu=\frac{p}{2mp+1}$ for small $p$.  Similar
agreement~\cite{Morf} was found for the low energy modes of small
systems at $\nu=1/2$.

To summarize this section, we have found that by framing the
Chern-Simons theory in a Landau-Silin Fermi liquid description, we
have been able to renormalize the mass of the quasiparticle in a
consistent manner.  Thus, at least phenomenologically, we can have the
low energy excitations on the scale of the interaction strength a
while maintaining the cyclotron mode at the bare cyclotron frequency.
Perhaps more importantly, we have a put the Chern-Simons theory in the
versatile language of the Landau Fermi liquid.  Despite these
successes, and despite the numerical successes in the calculation of
$K_{00}$, we will show in the next section that the (M)RPA does not
properly represent the other elements of the response matrix
($K_{01},K_{10}$, and $K_{11}$) in the limit of $m_{\rm b} \rightarrow 0$.

\mysection{Magnetization and $\mbox{M}^{2}$RPA}
\label{sec:Mag}

We now turn to consider the limit of small band mass $m_{\rm b}$ (or
equivalently large magnetic field $B$) so that the cyclotron frequency
$\frac{e B}{m_{\rm b} c}$ is extremely large.  In particular, in section
\ref{sec:ZFR} we focus on the zero frequency, finite wavevector
electromagnetic response in this limit.  The fact that, in this limit,
the electronic ground state and low energy excitations are constrained
to the lowest Landau level, leads to certain features of the
electronic response which are not properly represented in
approximation schemes such as the mean field or the (M)RPA.  We note
that this problem occurs in the Chern-Simons theory even when
gauge-field fluctuations are not infra-red singular~\footnote{For
  example, if the electron-electron repulsion falls off more slowly
  than $1/r$ there should be no infra-red divergences in the effective
  mass.}  (See section \ref{sec:infrared} below).

We discuss in this section an approach~\cite{SimonStern,SimonNFL} based
on a phenomenological separation of the current into a magnetization
current associated with the cyclotron motion of electrons and
a transport current associated with the guiding center motion.  This
separation, discussed in section \ref{sec:Binding} is achieved by
attaching a magnetization ${\mu_{\mbox{\tiny M}}}$ to each particle.
This magnetization originates from the electrons' orbital motion and
is unrelated to the spin (we have assumed spinless electrons
throughout this paper).  In the limit $m_{\rm b} \rightarrow 0$, the
magnetization ${\mu_{\mbox{\tiny M}}}$ is given by the Bohr
magneton~\footnote{Here it is convenient that the subscript ${\rm b}$ can
  stand for `Bohr' or `bare' to remind us that it is the bare mass
  that enters this expression.}
\begin{equation}
  \mu_{\rm b} = \frac{e \hbar}{2 m_{\rm b} c}
\end{equation}
such that $\mu_{\rm b} B$ represents the noninteracting ground state kinetic
energy $\frac{1}{2} \hbar \omega_{\rm c}$.  The proposed separation
procedure combined with approximations similar to those made in the
MRPA, results in an approximation we call the $\mbox{M}^2$RPA,
discussed in section \ref{sec:M2RPA}, that yields a response functions
that correctly describes the $m_{\rm b} \rightarrow 0$ limit.  In section
\ref{sub:fit} we then revisit Fermi liquid theory and see how we can
fit the $\mbox{M}^2$RPA into the framework of a Landau theory.  We
note that the Lowest Landau level approach of Shankar and
Murthy~\cite{Shankar} explicitly demonstrates this magnetization
attachment.

\subsection{Zero Frequency Response}
\label{sec:ZFR}

In this section we shall examine the form of the zero frequency finite
wavevector response in the high magnetic field (or $m_{\rm b}
\rightarrow 0$) limit.  An acceptable approximation for calculating
the response of the $\nu= \frac{1}{2m}$ state must make correct
predictions in this limit.  We will show below that the usual
Chern-Simons approaches do not do this.

Consider the $\nu=\frac{1}{2m}$ state in the limit $m_{\rm b}
\rightarrow 0$.  In this limit the cyclotron gap $e B/(m_{\rm b} c)$
between Landau levels becomes large so we expect such a system to be
restricted to the lowest Landau level.  If we apply a weak external
static scalar potential at wavevector $q$ to the system, the resulting
state should remain in the lowest Landau level so the induced density
fluctuation should depend only on the interaction strength, and not on
the bare mass $m_{\rm b}$.  Thus, $K_{00}$, the density-density
response, should be independent of the bare mass in this limit (or
more properly, should scale as $(m_{\rm b})^0$ plus ${\cal O}(m_{\rm
  b})$ corrections).  However, the resulting density inhomogeneity
will yield a transverse current called the magnetization current,
given by (here and below the speed of light $c=1$) $ \vec
j_{\mbox{\scriptsize{mag}}} = \hatn z \times \nabla \vec M $ with
$\vec M$ the magnetization density.  For noninteracting particles in
the lowest Landau level, the kinetic energy density is $ E \equiv \vec
M \cdot \vec B = \frac{1}{2} \hbar \omega_{\rm c} n_{\rm e} $ so that
the magnetization per particle is $|\vec M|/n_{\rm e}= \mu_{\rm b}$,
the Bohr magneton.  More generally, when interactions are taken into
account, we let the magnetization per particle be given by a quantity
$\mum$ which must become $\mu_{\rm b}$ in the $m_{\rm b} \rightarrow
0$ limit where the system becomes projected to the Lowest Landau
level.  We can thus write the magnetization current as~\footnote{When
  projected to the lowest Landau level, the projected current and
  density operators satisfy $P {\bf j} P = \mu_{\rm b} ({\bf {\hat z}}
  \times \nabla P n P)$ where $P$ is the projection operator.  In
  other words, for projected states, all of the current is
  magnetization current.}
\begin{equation}
  \label{eq:magn}
  \vec j_{\mbox{\scriptsize{mag}}} = \mum (\hatn z \times \nabla n)
\end{equation}
with $n(\vec r)$ the local electron density.  The physical
interpretation of this magnetization current as follows.  Each
particle in the lowest Landau level can be thought of as a particle in
a cyclotron orbit.  When the density of particles is uniform, the
local currents of all of these orbits cancel and there is no net
current in the system.  However, when there is a density
inhomogeneity, these local currents do not quite cancel and a net
magnetization current results.  Note that this magnetization current
associated with density gradients can be modeled by imagining that a
small magnetization $\mum$ (equivalent to a current loop) is attached
to each quasiparticle.

Using Eq. \ref{eq:magn} we see that in the limit $m_{\rm b} \rightarrow 0$,
when we apply the weak static scalar potential
$A_{0}^{\mbox{\scriptsize{ext}}}(q)$ to the system and we look at the
leading current response we find a magnetization current $\mu_{\rm b}
{\bf{\hat z}} \times i {\bf q} K_{00}
A_{0}^{\mbox{\scriptsize{ext}}}$.  Thus, if $q$ is finite we expect
\begin{equation}
\label{eq:Klim}
\lim_{m_{\rm b} \rightarrow 0} K_{10}/K_{00} = i q \mu_{\rm b} 
\end{equation}
This result is not~\footnote{In the MRPA approach it is clear that this
  ratio cannot be determined by the bare mass since the low energy
  physics is controlled by $m^*$ (the whole purpose of the MRPA was to
  get $m_{\rm b}$ out of the low energy physics).  On the other hand in RPA
  this ratio is indeed on the scale of $m_{\rm b}$ (since that is the only
  scale in the problem), but it can be shown that the coefficient out
  front is incorrect (not to mention the other problems with RPA).}
contained in works based on the Chern-Simons approach previous to that
of Ref.~\onlinecite{SimonStern}.

We can also consider applying a weak external transverse vector
potential $A_1^{\mbox{\scriptsize{ext}}}$ at wavevector $q$ and zero
frequency.  This transverse field generates a magnetic field $\delta B
= i qA_1$ at wavevector $q$.  The variation in the total magnetic
field $B(\vec r) = B_{1/2} + \delta B(\vec r)$ will make the kinetic
energy $\frac{1}{2}\hbar \omega_{\rm c}(\vec r) = \mu_{\rm b} B(\vec r)$
positionally dependent thus attracting electrons to the regions of
minimal magnetic field when $m_{\rm b} \rightarrow 0$.  This attraction is
not modeled in the Chern-Simons fermion picture at the mean field or
(M)RPA level.  

Formally, if the applied variation in magnetic field generates a
density fluctuation $j_0(q)$, we can write the energy cost as $ \delta
E = j_0 (\delta B) \mum + \frac{1}{2} K_{00} j_0^2 $ where $K_{00}$ is
independent of $m_{\rm b}$ as discussed above.  Here, $j_0 (\delta B) \mum$
is just the change in local cyclotron energy which can be thought of
as an effective scalar potential for the fermions.  This term would
occur quite naturally if we were to imagine that a magnetization
$\mum$ were attached to each fermion.  The term $K_{00} j_0^2$ is due
to the Coulomb interactions within the lowest Landau level.  Again
note that $\mum$ must become $\mu_{\rm b}$ in the $m_{\rm b} \rightarrow 0$ limit,
but more generally can include pieces on the interaction scale.
Minimizing the energy $\delta E$ with respect to $j_0$ yields the
density $ j_0 = - (\delta B) \mu_m K_{00} = - i q \mum K_{00} A_1 $
from which we conclude that that the leading term of $K_{01}$ is given
by $i q \mum K_{00}$ (in accordance with the symmetry requirement of
the matrix $K$).

Finally, once we have determined the density fluctuation due to this
local magnetic field fluctuation, we again realize that this density
fluctuation results in a magnetization current, so that we have a
leading piece of $K_{11}$ given by $K_{00} q^2 \mum^2$.

\subsection{Binding Magnetization to Composite Fermions}
\label{sec:Binding}

As suggested by the above discussion, the necessary correction to the
composite fermion picture involves attaching a magnetization $\mum$ to
each composite fermion so that it properly represents a particle in
the lowest Landau level.  Attaching magnetization to each particle can
also be interpreted as attaching a current loop to each particle
associated with the electrons' cyclotron motion. Thus the total
current would include both a piece from the motion of the
particle-currentloop composite and a piece from the current loop
itself. To this end, we define a transport current~\footnote{The
  division into ${\bf j}_{\mbox{\tiny{trans}}}$ and ${\bf
    j}_{\mbox{\tiny{mag}}}$ has some degree of arbitrariness.  Note
  that the definitions in the present paper allow for a nonzero
  transverse component of ${\bf j}_{\mbox{\tiny{trans}}}$ in
  equilibrium for an inhomogeneous interacting electron system.}
\begin{equation}
  \label{eq:jtrans}
  \vec j_{\mbox{\scriptsize{trans}}} = \vec
  j_{\mbox{\scriptsize{total}}} - \vec j_{\mbox{\scriptsize{mag}}}
\end{equation}
which is the current of magnetized gauge transformed fermions, whereas
the magnetization current, as discussed above (see Eq.  \ref{eq:magn})
is the current associated with the attached current loops.

In addition, particles bound to magnetization should experience an
effective potential associated with any local changes in the magnetic
field.  Thus we define the effective scalar potential
\begin{equation}
  \label{eq:Aeff}
  A_0^{\mbox{\scriptsize{eff}}} = A_0 + \mum \delta B.
\end{equation}
This interaction of the bound magnetization with the magnetic field
should be thought of as the effective potential associated with the
local change in the cyclotron energy.

If we keep the conventions that all perturbations are applied with
$\vec q \| \hatn x$, and use the Coulomb gauge again, we can rewrite
Eqns. \ref{eq:jtrans} and \ref{eq:Aeff} as
\begin{eqnarray}
  j_{\mbox{\scriptsize{total}}} &=& M \label{eq:MJ}
  j_{\mbox{\scriptsize{trans}}} \\ A_{\mbox{\scriptsize{eff}}} &=&
  M^\dagger A \label{eq:MA}
\end{eqnarray}
where
\begin{equation}
  \label{eq:Mdef2}
  M = \left[ \begin{array}{cc} 1 & 0 \\ iq \mu_M & 1
\end{array} \right]_.
\end{equation}  
In these equations, all currents are written as two vectors
$(j_0,j_y)$ and vector potentials are written as two vectors
($A_0,A_y$).  The matrix $M$ should be thought of as an operator that
attaches magnetization.  As discussed above, in the limit $m_{\rm b}
\rightarrow 0$, we must have ${\mu_{\mbox{\tiny M}}} \rightarrow
\mu_{\rm b}$ in the matrix $M$, but more generally we can allow corrections
on the interaction scale.  In the rest of this section, however, we will
focus on the $m_{\rm b} \rightarrow 0$ limit and consider ${\mu_{\mbox{\tiny
      M}}} = \mu_{\rm b}$.

\subsection{Magnetized Modified RPA ($\mbox{M}^2\mbox{RPA}$)}
\label{sec:M2RPA}

As discussed above, the (M)RPA approach does not properly model the
magnetization effects discussed in section \ref{sec:ZFR}.  This error
is presumably due to the fact that when we take the mean field
solution as a starting point for a perturbation theory for the
Chern-Simons fermions, we lose the fact that the original electrons
travel in local cyclotron orbits.  In the approach discussed
here~\cite{SimonStern}, we will recover this physics by artificially
attaching magnetization to each particle by hand.  This attachment is
not an exact transformation, but is rather a way of modeling behavior
that is lost when we take the mean field as a starting point.
However, as we will see below, within a Landau-Fermi liquid theory
picture, this attachment seems to give the correct quasiparticles for
the system.

The magnetized particles have the same interactions (Both Chern-Simon
$C$ and Coulomb $V$) as the particles in the traditional Chern-Simons
fermion picture.  However, here, the magnetized fermions now respond
to the effective potential and the motion of these magnetized fermions
yields only the transport current response.  We thus define a matrix
$\tilde K$ to be the {\it transport} current response~\footnote{The
  reader is cautioned that this is not the same as the $\tilde K$
  defined in HLR~\cite{HLR}, which is what we call $\Pi$.}  of the
electrons to the external {\it effective} potential.  In other words,
\begin{equation}
  K = M \tilde K M^\dagger. 
\end{equation}
The `Magnetized Modified RPA' or $\mbox{M}^2$RPA is then defined by
setting $\tilde K$ equal to $K^{{\mbox{\tiny{MRPA}}}}$.  Thus we have
\begin{equation}
  \label{eq:M2def}
  K^{\M2} = M K^{{\mbox{\tiny{MRPA}}}} M^\dagger = M \left(
  [K^{0*}]^{-1} + {\cal F}_1 + C + V \right)^{-1} M^\dagger.
\end{equation}
with ${\cal F}_1$ defined by Eq. \ref{eq:F1def}. It should be noted that
\begin{equation} K_{00}^{\M2} =
  K_{00}^{{\mbox{\tiny{MRPA}}}}
\end{equation}
and therefore the exact diagonalizations~\cite{SimonballandSong,Morf}
that agreed well with calculations of $K_{00}$ in the MRPA agree
equally well with predictions of the $\mbox{M}^2$RPA. However, the
MRPA and $\mbox{M}^2$RPA differ at finite $q$ in their predictions for
the other elements of the matrix $K$.  For example,
\begin{equation}
  \label{eq:K10M2}
  K_{10}^{\M2} = K_{10}^{{\mbox{\tiny{MRPA}}}} + i q \mum
  K_{00}^{{\mbox{\tiny{MRPA}}}} {}_.
\end{equation}
It should be noted however, that all finite $q$ experimental
tests~\cite{WillettReview} of the Chern-Simons theory to date have
measured only $K_{00}$ and therefore do not distinguish between the
MRPA and the $\mbox{M}^2$RPA.  As required, in the limit $m_{\rm b}
\rightarrow 0$, the $\mbox{M}^2$RPA correctly describes the static
response properties described above.  For example, Eq. \ref{eq:K10M2}
clearly satisfies Eq. \ref{eq:Klim}.  

We note that we expect that the $\mbox{M}^2$RPA, in addition to
describing the $\nu=\frac{1}{2m}$ Fermi liquid states, should properly
describe the magnetization effects for the Jain series of quantized
states $\nu=\frac{p}{2mp+1}$ at least for small $p$.

\subsection{Fitting into Fermi Liquid Theory}
\label{sub:fit}

We now turn to discuss how the $\mbox{M}^2$RPA fits into the general
picture of a Fermi liquid theory of the $\nu=\frac{1}{2m}$ state.  In
essence, we will show that $\mbox{M}^2$RPA roughly amounts to adopting
the Fermi liquid picture discussed above in section \ref{sec:FLT} as
describing the dynamics of magnetized composite fermion quasiparticles
rather than unmagnetized ones.  We recall that in conventional Fermi
liquid theory the effective mass of the quasiparticle can be highly
renormalized from the effective mass of the bare particle.  (In the
case of ${}^3$He, the quasiparticle effective mass is approximately
three times the bare mass, such that the quasiparticle is quite
different from the original particle).  In our composite fermion
system, our quasiparticle will not only have a renormalized mass, but
also a renormalized magnetization.

For the Chern-Simons theory, in addition to separating the long ranged
part of the interaction $(C + V)$, for the magnetized fermions, further
separation should be carried out to remove the magnetization effects.
To this end we define a response function $\tilde \Pi$ by  (See
Eq. \ref{eq:KPi} for the definition of $\Pi$) 
\begin{equation}
  \label{eq:tildePi}
  \Pi = M \tilde \Pi M^\dagger. 
\end{equation}
and correspondingly we define a conductivity
\begin{equation}  
  \label{eq:Pirhocf}
\tilde \Pi =
T^{-1} \tilde \sigma^{\rm CF} T^{-1}
\end{equation}

By definition, $\tilde \Pi$ relates the transport current of the {\it
  magnetized} quasiparticles to the {\it effective} total vector
potential, including both external and internally induced
contributions (See Eqns. \ref{eq:MJ}, \ref{eq:MA} and
\ref{eq:Atotal}).  For the Chern-Simons system it is now $\tilde \Pi =
T^{-1} \tilde \sigma^{\rm CF} T^{-1}$ which we claim is given by a
Landau-Boltzmann equation describing the dynamics of quasiparticles
with the finite effective mass $m^*$ interacting via a residual short
ranged interaction $f$.

We now have a prescription for calculating the response $K$ of the
Chern-Simons Fermi liquid given the effective mass $m^*$ and the
interaction function $f(\theta)$.  To reiterate, the prescription is
to solve the Boltzmann equation (Eq. \ref{eq:kin1}) to obtain the
magnetized composite fermion conductivity $\tilde \sigma_{\rm CF}$.  The
response $K$ can then be obtained by using Eqns. \ref{eq:Pirhocf},
\ref{eq:tildePi} and \ref{eq:KPi}.

\subsubsection{Separating Singular Fermi Liquid Coefficients}
\label{sec:sep}

As discussed above, one expects that the effective mass, which
determines the energy scale of the low energy excitations, should be
set by the Coulomb interaction scale.  Similarly, one
expects~\cite{Ady} that the interaction function $f(\theta)$ should be
on the interaction scale ({\em i.e.\/}, proportional to $1/m^*$).  However, as
discussed above, we must satisfy the constraints Eqs.
\ref{eq:f1effmass} and \ref{eq:f0comp} on the values of $f_1$ and
$f_0$ respectively.  From Eq.  \ref{eq:f1effmass} it is clear that
$f_1$ is on the larger scale $1/m_{\rm b}$ rather than the interaction
scale.  However, we now also claim that the constraint Eq.
\ref{eq:f0comp} fixes $f_0$ to be on the scale $1/m_{\rm b}$ also.  This
counterintuitive result is due to the fact that the compressibility
derivative $\frac{d \mu}{d n}$ is taken at fixed $\Delta B$.  One can
understand this~\cite{Ady,Kim} by realizing that the Fermi liquid
theory uses the mean field zero effective field solution for its
ground state.  When a particle is added or subtracted, in order to
maintain a Fermi liquid ({\em i.e.\/}, zero effective field), the external
field must increased by $\tilde \phi$ flux quanta to compensate for
the added Chern-Simons field.  Thus, at fixed $\Delta B=0$, the
magnetic field is linked to the density $n$ via $B = \tilde \phi n
\phi_0$.  In the limit $m_{\rm b} \rightarrow 0$, the interaction energy
between the magnetization ${\bf M} = \mu_{\rm b} n$ and the external field
is given by $ E = {\bf M} \cdot {\bf B} = \pi \tilde \phi \hbar^2
n^2/m_{\rm b}$.  Of course this can also be thought of as the cyclotron
energy.  Differentiating this with respect to $n$ we obtain a
magnetization contribution to the chemical potential
\begin{equation}
  \mu^{\mbox{\scriptsize{mag}}} = \frac{2 \pi \tilde \phi \hbar^2 n}{m_{\rm b}}=
  \hbar \omega_{\rm c}
\end{equation}
such that the magnetization contribution $\tilde f_0$ to the zeroth
Fermi liquid coefficient $f_0$ is given by 
\begin{equation}
  \tilde f_0 = \frac{d \mu{\mbox{\scriptsize{mag}}}}{d n} = \frac{2
    \pi \tilde \phi \hbar^2}{m_{\rm b}}
\end{equation}
which is also the inverse compressibility of free electrons of mass
$m_{\rm b}$ at constant $\Delta B$.  The coefficient $f_0$ is written $f_0 =
\tilde f_0 + \delta f_0$ where $\tilde f_0$ is ${\cal O}(m_{\rm b}^{-1})$
and $\delta f_0$ is on the smaller interaction scale.  As mentioned in
Ref.~\onlinecite{Ady}, in the limit $m_{\rm b} \rightarrow 0$, the requirement
that the low energy spectrum is independent of $m_{\rm b}$ forces the other
interaction coefficients ($f_l$ for $l \ne 0,1$) to be on the
interaction scale.  In addition we note that using the Pauli exclusion
principle a sum rule can be derived~\cite{SimonNFL} for the remaining
Fermi liquid coefficients $f_l$ for $l \ne 0,1$.

Since in the limit of $m_{\rm b} \rightarrow 0$, $\tilde f_0$ and $f_1$ are
on the bare mass scale whereas all other coefficients $f_l$ (as well
as $\delta f_0$) are expected to be on the smaller interaction scale,
we will separate out the contributions of these two coefficients by
writing
\begin{equation}
  \label{eq:corrs}
  \tilde \Pi^{-1} = [\tilde \Pi^*]^{-1} + \tilde {\cal F}_0 + {\cal
    F}_1
\end{equation}
where 
\begin{equation}
  \label{eq:F0def} \tilde {\cal F}_0 = 
  \left(
  \begin{array}{cc}  \tilde f_0 & 0 \\ 0 & 0 \end{array}
  \right)_.
\end{equation}
is analogous to Eq. \ref{eq:Pi-1} and ${\cal F}_1$ is given by Eq.
\ref{eq:F1def}.  As described in section \ref{sub:simple} above, the
function $\tilde \Pi^*$ is to be calculated using a Landau-Boltzmann
equation representing quasiparticles with effective mass
$m^*$ and interaction coefficients $f_l$ except that $f_1$ is
artificially set to zero and the magnetic contribution $\tilde f_0$ is
subtracted off of $f_0$.  Once again, the form of Eq.  \ref{eq:corrs}
looks like the form of Eq.  \ref{eq:KPi} where we have separated two
interaction terms and defined the remaining response $\tilde \Pi^*$ to
be the response of a similar Fermi liquid with those interactions
removed.

The separation of the coefficients $\tilde f_0$ and $\tilde f_1$ are
just the prescriptions given in Eq. \ref{eq:tf0t} and \ref{eq:tf1t}
respectively.  Having made this separation, we expect that the
response $\tilde \Pi^*(q,\omega)$ is independent of $m_{\rm b}$ in the limit
$m_{\rm b} \rightarrow 0$ and is well behaved for all values of $q/m_{\rm b}$.
The transformation Eqns.~\ref{eq:KPi}, \ref{eq:Pi*}, \ref{eq:F1def},
\ref{eq:tildePi}, and \ref{eq:corrs} do not in themselves involve any
approximations, and may be considered simply as a means of defining a
new `irreducible' response function $\tilde \Pi^*(q, \omega)$.

\subsubsection{Relation to $\mbox{M}^2$RPA}
\label{sec:relation}

To relate this Fermi liquid approach to the $\mbox{M}^2$RPA we note
the identity $C + \tilde {\cal F}_0 = M^\dagger{}^{-1} C M^{-1}$ which
holds in the limit $m_{\rm b} \rightarrow 0$.  This identity is a statement
of the fact that if you allow the magnetization to see the
Chern-Simons magnetic field as well as the external magnetic field,
then the $1/m_{\rm b}$ contribution to $f_0$ will vanish since the
magnetization now sees zero magnetic field on average.  We will also
need $V = M^{\dagger} V M$ and ${\cal F}_0 = M^{\dagger} \tilde {\cal
  F}_0 M$ which is just the statement that a density-density
interaction does not care whether or not the particles are magnetized.
Using these identities, we find that $\mbox{M}^2$RPA defined in Eq.
\ref{eq:M2def} is equivalent to approximating $\Pi^*$ by $K^{0*}$, the
response of a free Fermi gas of particles of mass $m^*$, and
calculating the response using Eqns.~\ref{eq:KPi}, \ref{eq:tildePi},
and \ref{eq:corrs}.

We note that in Fermi liquid theory, the Landau-Boltzmann equation
does not correctly describe the Landau diamagnetic contribution to the
transverse static response.  Similarly, we suspect that here the
function $\tilde \Pi^*_{11}$ derived from the Landau-Boltzmann
equation lacks a term of the form $q^2 \chi$ where $\chi$ is some
appropriate Landau susceptibility which we expect to be on the scale
of the interaction strength. As usual, if we fix the ratio $\omega/q$
to be nonzero, and take $q \rightarrow 0$, this diamagnetic term
becomes negligible.  However, when $\tilde \Pi^*$ is approximated as
$K^{0*}$ for the $\mbox{M}^2$RPA, this diamagnetic contribution is
included at least approximately.

To summarize this section, we have found that the magnetization
contributions to the response can be properly obtained by declaring
the Fermi liquid quasiparticle to be a magnetized object.  This
difference in the magnetization of the quasiparticle compared to the
magnetization of the bare particle is similar to the renormalization
of the particle mass that typically occurs in a Fermi liquid theory,
but here is uniquely a result of the Lowest Landau level properties of
the quantum Hall system.

\mysection{Perturbative Approaches and Trouble in the Infrared}

\label{sec:Pert}
 
The MRPA and $\mbox{M}^2$RPA) discussed above are
semi-phenomenological approximations.  In this section, we will
consider a more systematic perturbative approach to the Chern-Simons
problem.  We will see in section \ref{sub:small} that the proper
expansion parameter is essentially $\tilde \phi$, the number of flux
quanta attached to each fermion.  Admittedly, $\tilde \phi$ is not
small in the physical case.  However, considering it to be a small
parameter allows us to organize a perturbation expansion.  In section
\ref{sub:diag} we will write down the diagrammatic rules for such a
perturbative expansion and we will re-derive the RPA as well as
obtaining the RPA screened gauge field propagator and the resulting
fermion self-energy at lowest order.  We will find in section
\ref{sub:div} that the self-energy at $\nu=\frac{1}{2m}$ is singular
due to infrared gauge field fluctuations.  We discuss the effect of
this divergence on the effective mass (which will turn out to diverge)
and on the physical response (which remains well behaved).  Finally,
in section \ref{sub:dFLT} we try to see how the results obtained in
this section are consistent with the Fermi liquid picture discussed
above in the proceeding to sections.  The main realization of this
section is that although the single quasiparticle properties are
singular, smooth excitations of the Fermi surface (which are those
involved in transport at small $q$ and $\omega$) are well behaved so
that we can in some cases work with a finite effective mass.

Throughout this section, for simplicity, we will focus on the
$\nu=\frac{1}{2m}$ Fermi liquid like state such that there is no
residual magnetic field.  We will also use a Lagrangian approach to
analyze the problem.  Aspects of such field theoretical formalisms
were developed by a number of authors in a number of contexts
including the bosonic picture of the fractional quantized Hall
effect~\cite{ZhangReview} and anyon
superconductivity~\cite{Anyons,AnyonSupercon,AnyonicCS}.  This field
theoretical formalism was first used to describe composite fermions by
Lopez and Fradkin who focused on the fractional quantized Hall
states~\cite{Lopez}.  (The interested reader is encouraged to read
their chapter in this book.)  A brief discussion of the Lagrangian
approach is also given by Halperin, Lee, and Read~\cite{HLR} in the
context of the even denominator states.

The action describing the Chern-Simons Hamiltonian (Eq.
\ref{eq:transformedH}) at $\nu=\frac{1}{2m}$ is given by (In this
section we will often set $\hbar = e = c = 1$ for convenience,
returning these factors sporadically, when they are helpful to our
understanding)
\begin{eqnarray} \nonumber
  &  & S  \,\,\,\, =  \,\,\,\,  \int d\vec r\int dt\left\{\psi^*(i\partial_t-a_0)\psi+
 \frac{1}{2m_{\rm b}}|(i\vec\nabla+\frac{e}{c}\vec a - \frac{e}{c} \vec A)\psi|^2 \right.
  \\ &  & +   \left. \frac{1}{\tilde{\phi}\phi_0}
 a_0 (\nabla \times \vec a) + 
 \frac{1}{\tilde \phi^2 \phi_0^2} \int d \vec r'[\nabla\times \vec
 a(\vec r)]  v(\vec r - \vec r')
  [\nabla\times  \vec a(\vec r')]\right\} \,\,\,\,\,\,\,\,\,\,
 \label{eq:actionfull}
\end{eqnarray}
Here, the explicit dependence of the fields on $\vec
r$ and $t$ is  omitted wherever there is no risk of confusion, and 
$\partial_t\equiv \partial/ \partial t$.  The first two terms in the
action are just the action for a free fermion coupled to a gauge
field.  The $a_0 (\nabla \times \vec a)$ term is included to attach
the Chern-Simons flux to the fermions.  This can be seen by
integrating out $a_0$ explicitly (which is trivial since the action is
linear in $a_0$) to yield the usual constraint $ (\tilde \phi \phi_0)
n(\vec r) = (\tilde \phi \phi_0) \psi^* \psi = \nabla \times \vec a $.
In other words, $a_0$ is just a Lagrange multiplier for this
constraint.  The last term in the action is just the Coulomb
interaction between density at position $\vec r$ and density at
position $\vec r'$ rewritten using the constraint. 

\subsection{The Question of a Small Parameter}
\label{sub:small}

The usual way to proceed at this point is to treat the coupling
between the fermions and the gauge field as weak and perform a
perturbative diagrammatic expansion.  But before making any
calculations with this action, we should attempt to identify the
putative small parameter of the theory.  To this end, it is convenient
to re-express the action in terms of dimensionless variables and
follow an argument by Stern and Halperin~\cite{Ady}.  We express length
in units of the magnetic length by defining $\tilde \vec r=\vec r/l_B$,
and time in units of the inverse cyclotron frequency by defining $\tilde
t=\omega_{\rm c}t$. We then rescale the fields accordingly, by defining
$\tilde{\psi}=l_B\psi$, $\tilde{\vec a}= l_B \vec a/\phi_0$,
$\tilde{\vec A} = l_B \vec A/\phi_0$, and
$\tilde{a_0}=a_0/(\hbar\omega_{\rm c})$.  Specializing to the case of the
Coulomb interaction $v(r) = e^2/(\epsilon r)$, the dimensionless
action then becomes,
\begin{eqnarray}
   \nonumber
 \tilde{S}&=&\int d \tilde \vec r \int d \tilde t 
 \left[\tilde{\psi}^*(i\partial_{\tilde t}-
 \tilde a_0)\tilde{\psi}+\frac{1}{2}|(i\tilde{\nabla}+\tilde{\vec a} -
 \tilde {\vec A})
 \tilde{\psi}|^2+
 \right. \\  &+& \frac{1}{\tilde{\phi}} \tilde a_0(\tilde{\nabla}
 \times\tilde{a})  + \left. \frac{E_{\rm c}}{\hbar\omega_{\rm c}}\int d \tilde \vec
   r'
 [\tilde{\vec\nabla}\times \tilde{\vec a}(\tilde \vec r)]\frac{1}
 {|\tilde \vec r-\tilde \vec r'|}
  [\tilde{\vec\nabla}\times \tilde{\vec a}(\vec r')]\right ]
 \label{eq:actionless}
\end{eqnarray}
where $E_{\rm c} = e^2/(\epsilon l_B)$ is the typical Coulomb energy,
and where, again, the dependence of fields on $\tilde \vec r,\tilde t$
was omitted in most places, for brevity.  As clearly seen from Eq.
\ref{eq:actionless}, the problem includes two dimensionless
parameters. The first, $\tilde{\phi}$, is the number of flux quanta
attached to each fermion. The second, $E_{\rm c}/(\hbar\omega_{\rm
  c})$ is the ratio of the typical Coulomb energy to the cyclotron
energy.  In the action (Eq. \ref{eq:actionless}), $\tilde\phi$ is the
fermion--gauge field coupling constant, while $E_{\rm c}/(\hbar
\omega_{\rm c})$ is the parameter controlling the gauge field
fluctuations.  The relevant physical values of $\tilde\phi$ are
$\tilde{\phi}=2,4,6...$, none of which are small. The physical value
of $E_{\rm c}/(\hbar\omega_{\rm c})$ can be assumed to be much smaller
than one corresponding to the high magnetic field (or small $m_{\rm
  b}$) limit.

A perturbation expansion in which $\tilde\phi$ is held fixed at an
even integer, while $E_{\rm c}/(\hbar\omega_{\rm c})$ is turned on is
singular due to the macroscopic degeneracy of the lowest Landau level.
In the action (Eq. \ref{eq:actionless}) this singularity is reflected
by the fact that for even ${\tilde\phi}$, a small value of $E_{\rm
  c}/(\hbar\omega_{\rm c})$ leads to strong gauge field fluctuations,
making the mean field starting point invalid.

Thus, the perturbation expansion starts from $\tilde{\phi}=e^2=0$,
{\em i.e.\/}, from the problem of non--interacting electrons at zero magnetic
field (which we know how to solve). The two parameters should then be
turned on together~\footnote{The actual procedure for turning these on
  is discussed in Ref.~\onlinecite{Ady}.} such that the problem becomes
that of weakly interacting anyons at weak magnetic field.  When
$\tilde\phi=2$ and $e$ gets to its physical value, the problem is that
of interacting electrons at $\nu=1/2$.

While this procedure is probably a good start towards understanding
this Chern-Simons system in the perturbative limit, it cannot solve
the problem of the largeness of the physically relevant value of
$\tilde\phi$.  This problem is most clear in the limit $E_{\rm
  c}/(\hbar\omega_{\rm c})\rightarrow 0$. In that limit, the physics
of the system depends in a non trivial way on $\tilde\phi$. For even
values of $\tilde\phi$, the action (Eq. \ref{eq:actionless}) describes
fermions in a strong magnetic field in a filling factor smaller than
one, whereas for $\tilde \phi$ odd, the action describes bosons in a
strong magnetic field.  For any integer $\tilde \phi$, in the limit
$E_{\rm c}/(\hbar\omega_{\rm c})\rightarrow 0$, the fermion ground
state and low energy excitations are confined to the lowest Landau
level, and are therefore independent of the bare mass $m_{\rm b}$. In
contrast, for non--integer values of $\tilde\phi$ the action (Eq.
\ref{eq:actionless}) describes anyons in a magnetic field. Since
anyon's wavefunctions are non--analytic, they are not confined to the
lowest Landau level and their low energy excitations are not
independent of the bare mass.  A perturbation expansion in
$\tilde\phi$ is likely to miss, at least partially, this aspect.

Many --- if not most --- analytic calculations performed in the field
of composite fermion physics are in some sense perturbations in small
$\tilde \phi$ as described above.  RPA, for example is lowest order
perturbation theory made self consistent.  A very closely related
expansion~\cite{Aim,Kim} is to introduce $N$ species of fermions and
expand in $1/N$ (of course, at the end of the day, we must set $1/N=1$
which is not small, but nonetheless this gives us a slightly different
avenue for analyzing the problem).  Several alternative approaches
have also been attempted and should be briefly mentioned.
Renormalization group calculations~\cite{RG} and
Bosonization~\cite{Marston} have been used to nonperturbatively
describe the low energy properties of the system.  The Eikonal
approximation is another nonperturbative approach that has been
attempted~\cite{Kvesh}.  Finally, we mention that several recent
approaches attempting to re-describe the composite fermion system as a
system of neutral
dipoles~\cite{Shankar,Pasquier,DHLee,Adydipole,Adylongdipole} will be
described in depth in section \ref{sec:wavefunction} below.  Although
many of these approaches differ drasticly with each other, they seem
to be converging on a single picture of the composite fermion Fermi
liquid.  In this section we will focus on the perturbation theory in
small $\tilde \phi$, which is by far the most studied direction.  At
the end of the day, however, we will want to imagine turning up $\tilde
\phi$ to its physical value.

\subsection{Diagrammatics}
\label{sub:diag}

We now return back to the action (Eq. \ref{eq:actionfull}) and
consider a perturbative expansion as discussed above.  Thus we need to
figure out what sort of diagrams we can draw.  To this end, we rewrite
the action in a convenient form
\begin{equation} \label{eq:action123}
   S = \int \!  dt \, \, ({\cal L}_{\rm f} + {\cal L}_{\rm af} + {\cal L}_{\rm a})
\end{equation}
with ${\cal L}_{\rm f}$ the the Lagrangian for free noninteracting fermion,
${\cal L}_{\rm af}$ the coupling between the fermions and the gauge field,
and ${\cal L}_{\rm a}$ the bare gauge field Lagrangian (In Eq.
\ref{eq:actionfull}, the first line is ${\cal L}_{\rm f} + {\cal L}_{\rm af}$
and the second line is ${\cal L}_{\rm a}$).  It is now convenient to rewrite
the bare gauge field Lagrangian as
 \begin{equation}
 \label{eq:effg}
 {\cal L}_{\rm a} =  \int \! d\vec r   \,\,\,  \delta a_{\mu}[({\cal D}^0)^{-1}]_{\mu \nu}
  \delta a_{\nu}
 \end{equation}
 where, $[{\cal D}^0]^{-1}$ is the Fourier transform of
 \begin{equation}
     ({\cal D}^0)^{-1}  = \left( \begin{array}{cc} 0 & \frac{i q}{2 \pi
     \tilde \phi} \\ \frac{- i q}{2 \pi \tilde \phi} & \frac{v(q)
     q^2}{(2 \pi \tilde \phi)^2} \end{array}  \right)
 \end{equation}
 which is known as the bare gauge field propagator.  Here, we
 have defined 
 \begin{equation}
 \delta  a_\mu = a_\mu - A_\mu
 \end{equation}
 for simplicity of notation so that the mean field situation is now
 $\delta a_\mu =0$, and we have used Coulomb gauge $\nabla
 \cdot \delta \vec a = 0$ and $2 \times 2$ matrix notation.  
We note that generally, a gauge field propagator is defined as
\begin{equation}
\label{eq:corr1}
  {\cal D}_{\mu \nu}(\vec r, t) = 
\langle T \delta a_\mu(\vec r, t) \delta a_\nu(0,0) \rangle 
\end{equation}
with $T$ the time ordering operator.  Here, ${\cal D}^0$ is this
correlator for the free gauge field.  We note that the bare gauge
field propagator is nothing but the interaction matrices from
section \ref{sec:RPA} above
 \begin{equation}
   {\cal D}^0 = -(C + V)
 \end{equation}
 which further justifies our earlier derivation of these interaction
 matrices.  Diagrammatically, we will draw ${\cal D}^0$ as a dotted line 
 $$ 
%
%   THIS IS THE FIGURE WITH THE BARE PROPAGATOR
%
{\cal D}^0_{\mu \nu}  =  \mu \mbox{
 \font\thinlinefont=cmr5
 \mbox{\beginpicture
 \setcoordinatesystem units <1.00000cm,1.00000cm>
 \unitlength=1.00000cm
 \linethickness=1pt
 \setplotsymbol ({\makebox(0,0)[l]{\tencirc\symbol{'160}}})
 \setshadesymbol ({\thinlinefont .})
 \setlinear
 %
 % Fig POLYLINE object
 %
 \linethickness=1pt
 \setplotsymbol ({\makebox(0,0)[l]{\tencirc\symbol{'160}}})
 \setdashes < 0.075cm>
 \plot  1.905 25.241  5.080 25.241 /
 \linethickness=0pt
 \putrectangle corners at  1.858 25.288 and  5.127 25.195
 \endpicture}} \nu
 $$
 We also have a bare (free) propagator for the fermions (from ${\cal
   L}_{\rm f}$) which we write in the usual way~\cite{LandauII} as
 \begin{equation}
   G_{\rm f}^0(\vec k,\omega) = \frac{1}{\hbar \omega - \xi_{\vec k} + 
 i 0^+ \mbox{sgn}(\omega)}
 \end{equation}
 with $0^+$  a positive infinitesimal, and 
 \begin{equation}
 \xi_k = \frac{\hbar^2 k^2}{2 m_{\rm b}} - \mu \approx \vfns (k -
 \kf)
 \end{equation}
 with $\mu$ the chemical potential and $\vfns = \pf/m_{\rm b}$ the bare Fermi
 velocity.  (The generalization of $G^0_{\rm f}$ to describe free fermions
 in finite magnetic field should be straightforward~\cite{LandauII}). 
We draw this fermion propagator as a solid line 
 $$
 G_{\rm f}^0  = \mbox{
 \font\thinlinefont=cmr5
 \mbox{\beginpicture
 \setcoordinatesystem units <1.00000cm,1.00000cm>
 \unitlength=1.00000cm
 \linethickness=1pt
 \setplotsymbol ({\makebox(0,0)[l]{\tencirc\symbol{'160}}})
 \setshadesymbol ({\thinlinefont .})
 \setlinear
 %
 % Fig POLYLINE object
 %
 \linethickness=1pt
 \setplotsymbol ({\makebox(0,0)[l]{\tencirc\symbol{'160}}})
 \putrule from  9.525 25.241 to 11.430 25.241
 %
 % arrow head
 %
 \plot 10.922 25.114 11.430 25.241 10.922 25.368 /
 %
 %
 % Fig POLYLINE object
 %
 \linethickness=1pt
 \setplotsymbol ({\makebox(0,0)[l]{\tencirc\symbol{'160}}})
 \putrule from 11.271 25.241 to 12.541 25.241
 \linethickness=0pt
 \putrectangle corners at  9.478 25.288 and 12.588 25.195
 \endpicture}
 }  \,\,\,\,\,\, (\vec k,\omega)
$$

\noindent To write the vertices, we now look to the interaction piece of the
Lagrangian
        \begin{equation}
         \label{eq:af}
         {\cal L}_{\rm af} =  
                \int \! d\vec r \, \left(-   \delta a_0 \, (n -
           \bar n)  - \vec J \cdot \vec \delta \vec a  + 
         \frac{n}{2 m_{\rm b}} |\vec \delta \vec a|^2  
          \right)
        \end{equation}
with $n = \psi^* \psi$ the density, and $\vec J = \frac{1}{2 m_{\rm b}}
[\psi^* (-i \nabla) \psi + \mbox{h.c.} ]$ the paramagnetic part of the
current (the physical current being given here by $\vec J -
\frac{n}{m_{\rm b}} \delta \vec a$).  The first two terms give interaction vertices
which we write as
 \begin{eqnarray}
   v_0(\vec k,\vec k') &=& 1 \\
   v_1(\vec k,\vec k') &=&  \vec J \times \frac{\vec k -
   \vec k'}{|\vec k - \vec k'|} = \frac{1}{m_{\rm b}} \frac{\vec k \times \vec
   k'}{|\vec k - \vec k'|}
 \end{eqnarray}
 which is written diagrammatically as

 \hspace*{30pt}
 \font\thinlinefont=cmr5
 \begingroup\makeatletter\ifx\SetFigFont\undefined%
 \gdef\SetFigFont#1#2#3#4#5{%
   \reset@font\fontsize{#1}{#2pt}%
   \fontfamily{#3}\fontseries{#4}\fontshape{#5}%
   \selectfont}%
 \fi\endgroup%
 \mbox{\beginpicture
 \setcoordinatesystem units <1.00000cm,1.00000cm>
 \unitlength=1.00000cm
 \linethickness=1pt
 \setplotsymbol ({\makebox(0,0)[l]{\tencirc\symbol{'160}}})
 \setshadesymbol ({\thinlinefont .})
 \setlinear
 %
 % Fig POLYLINE object
 %
 \linethickness=1pt
 \setplotsymbol ({\makebox(0,0)[l]{\tencirc\symbol{'160}}})
 \putrule from  2.540 24.130 to  5.080 24.130
 %
 % Fig POLYLINE object
 %
 \linethickness=1pt
 \setplotsymbol ({\makebox(0,0)[l]{\tencirc\symbol{'160}}})
 \setdashes < 0.075cm>
 \plot  3.810 24.130  3.810 25.400 /
 %
 % Fig TEXT object
 %
 \put{\SetFigFont{10}{14.4}{\rmdefault}{\mddefault}{\updefault}$\mu$} [lB] at  3.969 25.2
 %
 % Fig TEXT object
 %
 \put{\SetFigFont{10}{14.4}{\rmdefault}{\mddefault}{\updefault}$v_{\mu}(\vec
   k,\vec k') =$ } [lB] at  -0.5 24.2
 %
 % Fig TEXT object
 %
 \put{\SetFigFont{10}{14.4}{\rmdefault}{\mddefault}{\updefault}$\vec k$} [lB] at  2.064 23.971
 %
 % Fig TEXT object
 %
 \put{\SetFigFont{10}{14.4}{\rmdefault}{\mddefault}{\updefault}$\, \vec k'$} [lB] at  5.080 23.971
 \linethickness=0pt
 \putrectangle corners at  0.159 25.1 and  5.127 23.933
\endpicture}

\noindent Finally, the last term in Eq. \ref{eq:af} 
gives the interaction vertex 

\hspace*{30pt} \font\thinlinefont=cmr5
 \begingroup\makeatletter\ifx\SetFigFont\undefined%
 \gdef\SetFigFont#1#2#3#4#5{%
   \reset@font\fontsize{#1}{#2pt}%
   \fontfamily{#3}\fontseries{#4}\fontshape{#5}%
   \selectfont}%
 \fi\endgroup%
 \mbox{\beginpicture
 \setcoordinatesystem units <1.00000cm,1.00000cm>
 \unitlength=1.00000cm
 \linethickness=1pt
 \setplotsymbol ({\makebox(0,0)[l]{\tencirc\symbol{'160}}})
 \setshadesymbol ({\thinlinefont .})
 \setlinear
 %
 % Fig POLYLINE object
 %
 \linethickness=1pt
 \setplotsymbol ({\makebox(0,0)[l]{\tencirc\symbol{'160}}})
 \putrule from  2.540 24.130 to  5.080 24.130
 %
 % Fig POLYLINE object
 %
 \linethickness=1pt
 \setplotsymbol ({\makebox(0,0)[l]{\tencirc\symbol{'160}}})
 \setdashes < 0.075cm>
 \plot  3.810 24.130  2.440 25.200 /
 %
 % Fig POLYLINE object
 %
 \linethickness=1pt
 \setplotsymbol ({\makebox(0,0)[l]{\tencirc\symbol{'160}}})
 \plot  3.810 24.130  4.980 25.200 /
 %
 % Fig TEXT object
 %
 \put{\SetFigFont{10}{14.4}{\rmdefault}{\mddefault}{\updefault}$\mu$} [lB] at  2.840 24.900
 %
 % Fig TEXT object
 %
 \put{\SetFigFont{10}{14.4}{\rmdefault}{\mddefault}{\updefault}$\nu$} [lB] at  5.080 24.900
 %
 % Fig TEXT object
 %
 \put{\SetFigFont{10}{14.4}{\rmdefault}{\mddefault}{\updefault}$w_{\mu
     \nu} = \frac{1}{m_{\rm b}} \delta_{\mu 1} \delta_{\nu 1} = \,\, $} [lB] at
     -0.700 24.5
 \linethickness=0pt
 \putrectangle corners at  0.000 25.2 and  5.127 24.073
\endpicture}

We now can go ahead and start writing diagrams.  Perhaps the first
thing we should do is to recover the response $K^0$ of noninteracting
fermions coupled to an external gauge field.  Without using the gauge
propagator ({\em i.e.\/}, for noninteracting fermions), we can draw two
diagrams~\footnote{With great apologies, I must warn the reader not to
  confuse $D$, the current-current correlator for ${\cal D}$ the gauge
  field propagator.  Often they are both written as $D$.}  , $D^0$,
the current correlator, and $E$ the diamagnetic term. 
The current-current correlator is given diagrammatically as
 \begin{equation}
   \label{eq:currentcorrelator2}
   D^0_{\mu \nu}(\vec q,\omega) = \int \frac{d \vec k}{(2 \pi)^2} \int
   \frac{d \Omega}{2 \pi} v_\mu(\vec k,\vec q - \vec k) v_\nu(\vec q -
   \vec k, \vec k) 
 G^0_{\rm f}(\vec k, \Omega) G^0_{\rm f}(\vec q - \vec k,  \omega - \Omega)
\end{equation}
\hspace*{20pt}
\font\thinlinefont=cmr5
\begingroup\makeatletter\ifx\SetFigFont\undefined%
\gdef\SetFigFont#1#2#3#4#5{%
  \reset@font\fontsize{#1}{#2pt}%
  \fontfamily{#3}\fontseries{#4}\fontshape{#5}%
  \selectfont}%
\fi\endgroup%
\mbox{\beginpicture
\setcoordinatesystem units <0.50000cm,0.50000cm>
\unitlength=0.50000cm
\linethickness=1pt
\setplotsymbol ({\makebox(0,0)[l]{\tencirc\symbol{'160}}})
\setshadesymbol ({\thinlinefont .})
\setlinear
%
% Fig ELLIPSE
%
\linethickness= 0.500pt
\setplotsymbol ({\thinlinefont .})
\setsolid
\ellipticalarc axes ratio  1.270:0.635  360 degrees 
        from 12.065 25.400 center at 10.795 25.400
%
% Fig POLYLINE object
%
\linethickness= 0.500pt
\setplotsymbol ({\thinlinefont .})
\setdashes < 0.035cm >
\putrule from 11.113 26.035 to 10.636 26.035
%
% arrow head
%
\setsolid
\plot 11.144 26.162 10.636 26.035 11.144 25.908 /
%
%
% Fig POLYLINE object
%
\linethickness= 0.500pt
\setplotsymbol ({\thinlinefont .})
\setdashes < 0.035cm>
\plot  9.525 25.400  9.366 25.400 /
%
% Fig POLYLINE object
%
\linethickness= 0.500pt
\setplotsymbol ({\thinlinefont .})
\plot 12.065 25.400 12.224 25.400 /
\linethickness= 0.500pt
\setplotsymbol ({\thinlinefont .})
\setsolid
%
% Fig CONTROL PT SPLINE
%
% open spline
%
\plot   10.795 24.765 11.033 24.765
         /
\plot 11.033 24.765 11.271 24.765 /
%
% arrow head
%
\plot 10.763 24.638 11.271 24.765 10.763 24.892 /
%
%
% Fig TEXT object
%
\put{\SetFigFont{10}{7.2}{\rmdefault}{\mddefault}{\updefault}= } [lB] at  6.826 25.241
\linethickness=0pt
\putrectangle corners at  6.826 26.082 and 12.270 24.733
\endpicture}

\noindent (Here and below we use 
the usual diagrammatic rules~\cite{LandauII} to
write the diagrams as integrals).  This is really just the Fourier
transform of the correlator
\begin{equation}
  \label{eq:currentcorrelator0}
  D^0_{\mu \nu}(\vec r, t; \vec r', t') = \langle T j_\mu(\vec r,t)
  j_\nu(\vec r',t') \rangle
\end{equation}
with $T$ the time-ordering operator and where here $j_0$ is the density
operator and $j_1$ is the transverse paramagnetic current operator.  

The diamagnetic term, on the other hand, is given by

\begin{minipage}{0.85in}
\font\thinlinefont=cmr5
\begingroup\makeatletter\ifx\SetFigFont\undefined%
\gdef\SetFigFont#1#2#3#4#5{%
  \reset@font\fontsize{#1}{#2pt}%
  \fontfamily{#3}\fontseries{#4}\fontshape{#5}%
  \selectfont}%
\fi\endgroup%
\mbox{\beginpicture
\setcoordinatesystem units <0.50000cm,0.50000cm>
\unitlength=0.50000cm
\linethickness=1pt
\setplotsymbol ({\makebox(0,0)[l]{\tencirc\symbol{'160}}})
\setshadesymbol ({\thinlinefont .})
\setlinear
%
% Fig POLYLINE object
%
\linethickness= 0.500pt
\setplotsymbol ({\thinlinefont .})
\setdashes < 0.035cm>
\putrule from  3.61 23.495 to  4.01 23.495
%
% Fig POLYLINE object
%
%\linethickness= 0.500pt
%\setplotsymbol ({\thinlinefont .})
%\setdashes < 0.035cm>
%\putrule from  3.810 25.400 to  4.128 25.400
%
% arrow head
%
\setsolid
\plot  3.620 25.273  4.128 25.400  3.620 25.527 /
\linethickness= 0.500pt
\setplotsymbol ({\thinlinefont .})
%
% Fig CONTROL PT SPLINE
%
% open spline
%

\plot    3.810 23.495  3.334 23.812
         3.222 23.897
         3.125 23.991
         3.044 24.095
         2.977 24.209
         2.924 24.333
         2.904 24.399
         2.887 24.467
         2.874 24.538
         2.865 24.611
         2.859 24.687
         2.857 24.765
         2.861 24.842
         2.872 24.914
         2.891 24.981
         2.917 25.043
         2.991 25.152
         3.096 25.241
         3.159 25.278
         3.230 25.311
         3.308 25.338
         3.393 25.360
         3.486 25.378
         3.587 25.390
         3.695 25.398
         3.810 25.400
         3.925 25.398
         4.033 25.390
         4.134 25.378
         4.227 25.360
         4.312 25.338
         4.390 25.311
         4.461 25.278
         4.524 25.241
         4.629 25.152
         4.703 25.043
         4.729 24.981
         4.748 24.914
         4.759 24.842
         4.763 24.765
         4.761 24.687
         4.755 24.611
         4.746 24.538
         4.733 24.467
         4.716 24.399
         4.696 24.333
         4.643 24.209
         4.576 24.095
         4.495 23.991
         4.398 23.897
         4.286 23.812
         /
\plot  4.286 23.812  3.810 23.495 /
%
% Fig TEXT object
%
\put{\SetFigFont{10}{7.2}{\rmdefault}{\mddefault}{\updefault}$E_{\mu \nu} =$} [lB] at  0.111 24.2
\linethickness=0pt
\putrectangle corners at  1.111 25.447 and  4.794 23.448
\endpicture}

\end{minipage} 
\begin{minipage}{3.6in}
\begin{equation} = w_{\mu \nu} \int  \frac{d\vec k}{(2 \pi)^2} \int \frac{d\omega}{2
 \pi}  G^0_{\rm f}(\vec k,\omega) = \frac{n_{\rm e}}{m_{\rm b}} \delta_{\mu 1}
 \delta_{\nu 1} \end{equation} 
\end{minipage}

Since these are the only things we can draw without the gauge
propagator, we have the noninteracting response function given by 
\begin{equation}
  \label{eq:K0D0}
    K^0 = D^0 + E
\end{equation}
or

\hspace*{80pt}\font\thinlinefont=cmr5
\begingroup\makeatletter\ifx\SetFigFont\undefined%
\gdef\SetFigFont#1#2#3#4#5{%
  \reset@font\fontsize{#1}{#2pt}%
  \fontfamily{#3}\fontseries{#4}\fontshape{#5}%
  \selectfont}%
\fi\endgroup%
\mbox{\beginpicture
\setcoordinatesystem units <0.50000cm,0.50000cm>
\unitlength=0.50000cm
\linethickness=1pt
\setplotsymbol ({\makebox(0,0)[l]{\tencirc\symbol{'160}}})
\setshadesymbol ({\thinlinefont .})
\setlinear
%
% Fig ELLIPSE
%
\linethickness= 0.500pt
\setplotsymbol ({\thinlinefont .})
\ellipticalarc axes ratio  1.270:0.635  360 degrees 
        from  5.080 24.924 center at  3.810 24.924
%
% Fig ELLIPSE
%
\linethickness= 0.500pt
\setplotsymbol ({\thinlinefont .})
\ellipticalarc axes ratio  1.270:0.635  360 degrees 
        from 10.795 25.082 center at  9.525 25.082
%
% Fig POLYLINE object
%
\linethickness= 0.500pt
\setplotsymbol ({\thinlinefont .})
\setdashes < 0.035cm>
\plot  2.540 24.924  2.381 24.924 /
%
% Fig POLYLINE object
%
\linethickness= 0.500pt
\setplotsymbol ({\thinlinefont .})
\plot  5.080 24.924  5.239 24.924 /
%
% Fig POLYLINE object
%
\linethickness= 0.500pt
\setplotsymbol ({\thinlinefont .})
\setsolid
\putrule from  9.842 25.718 to  9.366 25.718
%
% arrow head
%
\plot  9.874 25.845  9.366 25.718  9.874 25.590 /
%
%
% Fig POLYLINE object
%
\linethickness= 0.500pt
\setplotsymbol ({\thinlinefont .})
\setdashes < 0.035cm>
\plot  8.255 25.082  8.096 25.082 /
%
% Fig POLYLINE object
%
\linethickness= 0.500pt
\setplotsymbol ({\thinlinefont .})
\plot 10.795 25.082 10.954 25.082 /
%
% Fig POLYLINE object
%
\linethickness= 0.500pt
\setplotsymbol ({\thinlinefont .})
\setdashes < 0.035cm > 
\putrule from 14.070 24.289 to 14.415 24.289
\setsolid
%
% Fig POLYLINE object
%
%\linethickness= 0.500pt
%\setplotsymbol ({\thinlinefont .})
%\putrule from 14.287 26.194 to 14.605 26.194
%
% arrow head
%
\plot 14.097 26.067 14.605 26.194 14.097 26.321 /
%
%
% Fig POLYLINE object
%
%\linethickness= 0.500pt
%\setplotsymbol ({\thinlinefont .})
%\setdashes < 0.035cm > 
%\putrule from 14.087 24.289 to 14.487 24.289
%\setsolid
%
% Fig POLYLINE object
%
\linethickness= 0.500pt
\setplotsymbol ({\thinlinefont .})
\putrule from 14.287 26.194 to 14.605 26.194
%
% arrow head
%
\plot 14.097 26.067 14.605 26.194 14.097 26.321 /
\linethickness= 0.500pt
\setplotsymbol ({\thinlinefont .})
%
% Fig CONTROL PT SPLINE
%
% open spline
%
\plot    9.525 24.448  9.763 24.448
         /
\plot  9.763 24.448 10.001 24.448 /
%
% arrow head
%
\plot  9.493 24.320 10.001 24.448  9.493 24.575 /
\linethickness= 0.500pt
\setplotsymbol ({\thinlinefont .})
%
% Fig CONTROL PT SPLINE
%
% open spline
%
\plot   14.287 24.289 13.811 24.606
        13.700 24.691
        13.603 24.785
        13.521 24.889
        13.454 25.003
        13.402 25.127
        13.382 25.193
        13.365 25.261
        13.352 25.332
        13.342 25.405
        13.337 25.481
        13.335 25.559
        13.339 25.636
        13.350 25.708
        13.368 25.775
        13.395 25.837
        13.469 25.946
        13.573 26.035
        13.636 26.072
        13.707 26.104
        13.785 26.132
        13.871 26.154
        13.964 26.171
        14.064 26.184
        14.172 26.191
        14.287 26.194
        14.403 26.191
        14.511 26.184
        14.611 26.171
        14.704 26.154
        14.790 26.132
        14.868 26.104
        14.939 26.072
        15.002 26.035
        15.106 25.946
        15.180 25.837
        15.207 25.775
        15.225 25.708
        15.236 25.636
        15.240 25.559
        15.238 25.481
        15.233 25.405
        15.223 25.332
        15.210 25.261
        15.193 25.193
        15.173 25.127
        15.121 25.003
        15.054 24.889
        14.972 24.785
        14.875 24.691
        14.764 24.606
         /
\plot 14.764 24.606 14.287 24.289 /
\linethickness= 0.500pt
\setplotsymbol ({\thinlinefont .})
%
% Fig CONTROL PT SPLINE
%
% open spline
%
\plot   14.287 24.289 13.811 24.606
        13.700 24.691
        13.603 24.785
        13.521 24.889
        13.454 25.003
        13.402 25.127
        13.382 25.193
        13.365 25.261
        13.352 25.332
        13.342 25.405
        13.337 25.481
        13.335 25.559
        13.339 25.636
        13.350 25.708
        13.368 25.775
        13.395 25.837
        13.469 25.946
        13.573 26.035
        13.636 26.072
        13.707 26.104
        13.785 26.132
        13.871 26.154
        13.964 26.171
        14.064 26.184
        14.172 26.191
        14.287 26.194
        14.403 26.191
        14.511 26.184
        14.611 26.171
        14.704 26.154
        14.790 26.132
        14.868 26.104
        14.939 26.072
        15.002 26.035
        15.106 25.946
        15.180 25.837
        15.207 25.775
        15.225 25.708
        15.236 25.636
        15.240 25.559
        15.238 25.481
        15.233 25.405
        15.223 25.332
        15.210 25.261
        15.193 25.193
        15.173 25.127
        15.121 25.003
        15.054 24.889
        14.972 24.785
        14.875 24.691
        14.764 24.606
         /
\plot 14.764 24.606 14.287 24.289 /
%
% Fig TEXT object
%
\put{\SetFigFont{10}{7.2}{\rmdefault}{\mddefault}{\updefault}$K^0$} [lB] at  3.651 24.765
%
% Fig TEXT object
%
\put{\SetFigFont{10}{7.2}{\rmdefault}{\mddefault}{\updefault}+} [lB] at 12.541 24.924
%
% Fig TEXT object
%
\put{\SetFigFont{10}{7.2}{\rmdefault}{\mddefault}{\updefault} = } [lB] at  6.509 24.924
\linethickness=0pt
\putrectangle corners at  2.335 26.240 and 15.272 24.242
\endpicture}

We can now consider the propagation of the gauge field to construct
the RPA series 
\begin{equation}
  K^{\rm RPA} = K^0 + K^0 {\cal D}^0 K^0 + K^0 {\cal D}^0 K^0 {\cal D}^0 K^0 + ...
\end{equation}
which is precisely the same as the RPA prescription discussed above in
section \ref{sec:RPA}.  We write this diagrammatically as

\font\thinlinefont=cmr5
\begingroup\makeatletter\ifx\SetFigFont\undefined%
\gdef\SetFigFont#1#2#3#4#5{%
  \reset@font\fontsize{#1}{#2pt}%
  \fontfamily{#3}\fontseries{#4}\fontshape{#5}%
  \selectfont}%
\fi\endgroup%
\mbox{\beginpicture
\setcoordinatesystem units <0.50000cm,0.50000cm>
\unitlength=0.50000cm
\linethickness=1pt
\setplotsymbol ({\makebox(0,0)[l]{\tencirc\symbol{'160}}})
\setshadesymbol ({\thinlinefont .})
\setlinear
%
% Fig ELLIPSE
%
\linethickness= 0.500pt
\setplotsymbol ({\thinlinefont .})
\ellipticalarc axes ratio  1.270:0.635  360 degrees 
        from  9.525 24.765 center at  8.255 24.765
%
% Fig POLYLINE object
%
\linethickness= 0.500pt
\setplotsymbol ({\thinlinefont .})
\setdashes < 0.075cm>
\plot  6.985 24.765  6.826 24.765 /
%
% Fig POLYLINE object
%
\linethickness= 0.500pt
\setplotsymbol ({\thinlinefont .})
\plot  9.525 24.765  9.684 24.765 /
%
% Fig TEXT object
%
%
\put{\SetFigFont{10}{7.2}{\rmdefault}{\mddefault}{\updefault}$K^0$} [lB] at  8.096 24.606
%
% Fig ELLIPSE
%
\linethickness= 0.500pt
\setplotsymbol ({\thinlinefont .})
\setsolid
\ellipticalarc axes ratio  1.270:0.635  360 degrees 
        from  3.493 24.924 center at  2.223 24.924
%
% Fig ELLIPSE
%
\linethickness= 0.500pt
\setplotsymbol ({\thinlinefont .})
\ellipticalarc axes ratio  1.270:0.635  360 degrees 
        from 15.558 24.765 center at 14.287 24.765
%
% Fig ELLIPSE
%
\linethickness= 0.500pt
\setplotsymbol ({\thinlinefont .})
\ellipticalarc axes ratio  1.270:0.635  360 degrees 
        from 20.161 24.765 center at 18.891 24.765
%
% Fig ELLIPSE
%
\linethickness= 0.500pt
\setplotsymbol ({\thinlinefont .})
\ellipticalarc axes ratio  1.270:0.635  360 degrees 
        from  9.684 22.066 center at  8.414 22.066
%
% Fig ELLIPSE
%
\linethickness= 0.500pt
\setplotsymbol ({\thinlinefont .})
\ellipticalarc axes ratio  1.270:0.635  360 degrees 
        from 13.811 22.066 center at 12.541 22.066
%
% Fig ELLIPSE
%
\linethickness= 0.500pt
\setplotsymbol ({\thinlinefont .})
\ellipticalarc axes ratio  1.270:0.635  360 degrees 
        from 18.098 22.066 center at 16.828 22.066
%
% Fig POLYLINE object
%
\linethickness= 0.500pt
\setplotsymbol ({\thinlinefont .})
\setdashes < 0.075cm>
\plot  0.953 24.924  0.794 24.924 /
%
% Fig POLYLINE object
%
\linethickness= 0.500pt
\setplotsymbol ({\thinlinefont .})
\plot  3.493 24.924  3.651 24.924 /
%
% Fig POLYLINE object
%
\linethickness= 0.500pt
\setplotsymbol ({\thinlinefont .})
\plot 13.018 24.765 12.859 24.765 /
%
% Fig POLYLINE object
%
\linethickness= 0.500pt
\setplotsymbol ({\thinlinefont .})
\plot 15.558 24.765 15.716 24.765 /
%
% Fig POLYLINE object
%
\linethickness= 0.500pt
\setplotsymbol ({\thinlinefont .})
\plot 17.621 24.765 17.462 24.765 /
%
% Fig POLYLINE object
%
\linethickness= 0.500pt
\setplotsymbol ({\thinlinefont .})
\plot 20.161 24.765 20.320 24.765 /
%
% Fig POLYLINE object
%
\linethickness= 0.500pt
\setplotsymbol ({\thinlinefont .})
\plot 15.558 24.765 17.462 24.765 /
%
% Fig POLYLINE object
%
\linethickness= 0.500pt
\setplotsymbol ({\thinlinefont .})
\plot  7.144 22.066  6.985 22.066 /
%
% Fig POLYLINE object
%
\linethickness= 0.500pt
\setplotsymbol ({\thinlinefont .})
\plot 13.811 22.066 13.970 22.066 /
%
% Fig POLYLINE object
%
\linethickness= 0.500pt
\setplotsymbol ({\thinlinefont .})
\plot 15.558 22.066 15.399 22.066 /
%
% Fig POLYLINE object
%
\linethickness= 0.500pt
\setplotsymbol ({\thinlinefont .})
\plot 18.098 22.066 18.256 22.066 /
%
% Fig POLYLINE object
%
\linethickness= 0.500pt
\setplotsymbol ({\thinlinefont .})
\plot 13.811 22.066 15.399 22.066 /
%
% Fig POLYLINE object
%
\linethickness= 0.500pt
\setplotsymbol ({\thinlinefont .})
\plot  9.684 22.066 11.271 22.066 /
%
% Fig TEXT object
%
\put{\SetFigFont{10}{7.2}{\rmdefault}{\mddefault}{\updefault} = } [lB] at  5.239 24.765
%
% Fig TEXT object
%
\put{\SetFigFont{10}{7.2}{\rmdefault}{\mddefault}{\updefault}$K^{\rm RPA}$} [lB] at  1.394 24.765
%
% Fig TEXT object
%
\put{\SetFigFont{10}{7.2}{\rmdefault}{\mddefault}{\updefault}+} [lB] at 11.271 24.606
\put{\SetFigFont{10}{7.2}{\rmdefault}{\mddefault}{\updefault}$K^0$} [lB] at 14.129 24.606
%
% Fig TEXT object
%
\put{\SetFigFont{10}{7.2}{\rmdefault}{\mddefault}{\updefault}$K^0$} [lB] at 18.733 24.606
%
% Fig TEXT object
%
\put{\SetFigFont{10}{7.2}{\rmdefault}{\mddefault}{\updefault}+} [lB] at 21.907 24.606
%
% Fig TEXT object
%
\put{\SetFigFont{10}{7.2}{\rmdefault}{\mddefault}{\updefault}+} [lB] at  5.239 21.907
%
% Fig TEXT object
%
\put{\SetFigFont{10}{7.2}{\rmdefault}{\mddefault}{\updefault}$+ \ldots$} [lB] at 19.526 21.749
%
% Fig TEXT object

%
\put{\SetFigFont{10}{7.2}{\rmdefault}{\mddefault}{\updefault}$K^0$} [lB] at  8.255 21.907
%
% Fig TEXT object
%
\put{\SetFigFont{10}{7.2}{\rmdefault}{\mddefault}{\updefault}$K^0$} [lB] at 12.383 21.907
%
% Fig TEXT object
%
\put{\SetFigFont{10}{7.2}{\rmdefault}{\mddefault}{\updefault}$K^0$} [lB] at 16.669 21.907
\linethickness=0pt
\putrectangle corners at  0.518 25.588 and 20.907 21.402
\endpicture}

One other thing we might consider is the RPA screened gauge propagator
${\cal D}^{\rm RPA}$ which we draw diagrammatically as a boldface
dotted line.   We have
\begin{eqnarray}  \label{eq:DRPAresult}
  {\cal D}^{\rm RPA} &=& {\cal D}^0 + {\cal D}^0 K^0 {\cal D}^0 + {\cal
    D}^0 K^0 {\cal D}^0 K^0 {\cal D}^0 + \ldots  \\
    &=& {\cal D}^0 + {\cal D}^0 K^{\rm RPA} {\cal D}^0
\end{eqnarray}
\noindent or 

\hspace*{5pt} \font\thinlinefont=cmr5
\begingroup\makeatletter\ifx\SetFigFont\undefined%
\gdef\SetFigFont#1#2#3#4#5{%
  \reset@font\fontsize{#1}{#2pt}%
  \fontfamily{#3}\fontseries{#4}\fontshape{#5}%
  \selectfont}%
\fi\endgroup%
\mbox{\beginpicture
\setcoordinatesystem units <0.50000cm,0.50000cm>
\unitlength=0.50000cm
\linethickness=1pt
\setplotsymbol ({\makebox(0,0)[l]{\tencirc\symbol{'160}}})
\setshadesymbol ({\thinlinefont .})
\setlinear
%
% Fig ELLIPSE
%
\linethickness= 0.500pt
\setplotsymbol ({\thinlinefont .})
\ellipticalarc axes ratio  1.270:0.635  360 degrees 
        from 18.415 24.765 center at 17.145 24.765
%
% Fig POLYLINE object
%
\linethickness= 0.500pt
\setplotsymbol ({\thinlinefont .})
\setdashes < 0.075cm>
\plot  1.905 24.606  1.905 24.606 /
%
% Fig POLYLINE object
%
\linethickness= 0.500pt
\setplotsymbol ({\thinlinefont .})
\plot  7.620 24.765 10.478 24.765 /
\plot 10.478 24.765 10.319 24.765 /
%
% Fig POLYLINE object
%
\linethickness= 0.500pt
\setplotsymbol ({\thinlinefont .})
\plot 13.970 24.765 15.875 24.765 /
%
% Fig POLYLINE object
%
\linethickness= 0.500pt
\setplotsymbol ({\thinlinefont .})
\plot 18.415 24.765 20.320 24.765 /
%
% Fig POLYLINE object
%
\linethickness=4pt
\setplotsymbol ({\makebox(0,0)[l]{\tencirc\symbol{'162}}})
\setdots < 4pt >
\plot  1.905 24.765  5.080 24.765 /
%
% Fig TEXT object
%
\put{\SetFigFont{10}{7.2}{\rmdefault}{\mddefault}{\updefault}$=$} [lB] at  6.509 24.606
%
% Fig TEXT object
%
\put{\SetFigFont{10}{7.2}{\rmdefault}{\mddefault}{\updefault}$+$} [lB] at 11.589 24.606
%
% Fig TEXT object
%
\put{\SetFigFont{10}{7.2}{\rmdefault}{\mddefault}{\updefault}$K^0$} [lB] at 16.986 24.606
%
% Fig TEXT object
%
\put{\SetFigFont{10}{7.2}{\rmdefault}{\mddefault}{\updefault}$+ \ldots$} [lB] at 21.273 24.606
%
% Fig TEXT object
%
\put{\SetFigFont{10}{7.2}{\rmdefault}{\mddefault}{\updefault}${\cal D}^{\rm RPA}$} [lB] at  3.334 25.082
\linethickness=0pt
\putrectangle corners at  1.679 25.430 and 21.273 24.100
\endpicture}

To gain intuition for the meaning of this screened propagator, 
we write an effective Lagrangian for the pieces ${\cal L}_{f} + {\cal
  L}_{\rm af}$ 
\begin{equation}
  \label{eq:effact}
 {\cal L}_{{\rm f}+{\rm af}}^{\rm eff} =  -e \int d\omega \int d\vec q \, \,  \delta
  a_{\mu}(\omega,\vec q) \, K^0_{\mu \nu} \, 
  \delta a_{\nu}(-\omega,-\vec q)
\end{equation}
In field theorist language, we have integrated out the fermions at
1-loop order (the 1 loop being the above diagrams of $D^0$ and $E$).
To see that this effective Lagrangian is correct, all we need to do is
recall that the physical current should be given by $j_\mu = - \delta
S/\delta a_{\mu} = e K^0_{\mu \nu} \delta a_\nu$.

To find the propagator ${\cal D}$ at 1-loop, or RPA level, we simply write an
effective Lagrangian for the full action \ref{eq:action123}
\begin{equation}
{\cal L}_{{\rm f}+{\rm af}+{\rm a}}^{\rm eff} =  \int d\omega \int d\vec q  \, \, \delta
  a_{\mu}(\omega,\vec q) \, ({\cal D}^{-1})_{\mu \nu}
  \, \delta a_{\nu}(-\omega,-\vec q)
\end{equation}
Combining Eq. \ref{eq:effact} and \ref{eq:effg} we immediately obtain the RPA
result (Eq. \ref{eq:DRPAresult}) 
\begin{equation}
   [{\cal D}^{\rm RPA}]^{-1} = [{\cal D}^0]^{-1}  - K^0 
\end{equation}

We now examine some of the features of the gauge field propagator
${\cal D}$.  First of all, there is a direct relation between the exact
value of $K_{00}$ and the exact value of ${\cal D}_{11}$ (which is
formally defined as the correlator Eq. \ref{eq:corr1}).  This should
be obvious since the constraint requires $\phi_0 \tilde \phi n = 2 \pi
\tilde \phi n = \nabla \times \vec a = q a_1$ so (in shorthand
notation)
\begin{equation}
  \label{eq:gaugefieldprop}
  {\cal D}_{11} = \langle \delta a_1 \delta a_1 \rangle  = \left[\frac{2 \pi \tilde \phi}{q}
  \right]^2 \langle \rho \rho \rangle = \left[\frac{2 \pi \tilde \phi}{q}
  \right]^2 K_{00}
\end{equation}
Thus, the low energy long wavelength form of ${\cal D}_{11}$ should
have a pole at $\omega \sim i q^3 v(q)$ as does $K_{00}$ (This pole
does not go away in MRPA although the coefficients may change).

The other important feature of the gauge field propagator is that the
high energy form of of ${\cal D}_{00}$ is given roughly by
\begin{equation} 
\label{eq:DKq}
  {\cal D}_{00} \approx \frac{(2 \pi \tilde \phi)^2}{q^2} K_{11} = \frac{(2 \pi
  \tilde \phi)^2}{q^2} \frac{n_{\rm e}}{m_{\rm b}} \frac{\omega^2}{\omega_{\rm c}^2 - (\omega
  + i 0^+)^2}_.
\end{equation}
This high energy form of $K_{11}$ is guaranteed by Kohn's theorem and
the $f$-sum rule as discussed above in section \ref{subsub:sumrules}. 

The next step beyond RPA in standard perturbation theory is to
calculate self-energy corrections.  The self energy $\Sigma$ is
defined in term of the 
 exact one--particle Green's function $G_{\rm f}(\vec k,\omega)$ by
\begin{equation}
\label{eq:exactG}
  G_{\rm f}(\vec k,\omega) = \frac{1}{\hbar \omega - \xi_{\vec k} -
    \Sigma(k,\omega)} 
\label{gf}
\end{equation}

We will now use the conventional perturbative
expansions~\cite{LandauII} to calculate the self energy $\Sigma$ of the
fermion.  We can write an infinite series of diagrams 
contributing to the self energy.  However, for us, it will be a
sufficient challenge to write down the first nontrivial term.

The simplest self energy we could draw is the following diagram

\hspace*{.4in} \begin{minipage}{.3in}
\font\thinlinefont=cmr5
\begingroup\makeatletter\ifx\SetFigFont\undefined%
\gdef\SetFigFont#1#2#3#4#5{%
  \reset@font\fontsize{#1}{#2pt}%
  \fontfamily{#3}\fontseries{#4}\fontshape{#5}%
  \selectfont}%
\fi\endgroup%
\mbox{\beginpicture
\setcoordinatesystem units <0.50000cm,0.50000cm>
\unitlength=0.50000cm
\linethickness=1pt
\setplotsymbol ({\makebox(0,0)[l]{\tencirc\symbol{'160}}})
\setshadesymbol ({\thinlinefont .})
\setlinear
%
% Fig POLYLINE object
%
\linethickness= 0.500pt
\setplotsymbol ({\thinlinefont .})
\putrule from  3.493 23.495 to  4.128 23.495
%
% Fig POLYLINE object
%
\linethickness=1pt
\setplotsymbol ({\makebox(0,0)[l]{\tencirc\symbol{'160}}})
\setdashes < 0.075cm>
%
% Fig CONTROL PT SPLINE
%
% open spline
%
\plot    3.810 23.495  3.334 23.812
         3.222 23.897
         3.125 23.991
         3.044 24.095
         2.977 24.209
         2.924 24.333
         2.904 24.399
         2.887 24.467
         2.874 24.538
         2.865 24.611
         2.859 24.687
         2.857 24.765
         2.861 24.842
         2.872 24.914
         2.891 24.981
         2.917 25.043
         2.991 25.152
         3.096 25.241
         3.159 25.278
         3.230 25.311
         3.308 25.338
         3.393 25.360
         3.486 25.378
         3.587 25.390
         3.695 25.398
         3.810 25.400
         3.925 25.398
         4.033 25.390
         4.134 25.378
         4.227 25.360
         4.312 25.338
         4.390 25.311
         4.461 25.278
         4.524 25.241
         4.629 25.152
         4.703 25.043
         4.729 24.981
         4.748 24.914
         4.759 24.842
         4.763 24.765
         4.761 24.687
         4.755 24.611
         4.746 24.538
         4.733 24.467
         4.716 24.399
         4.696 24.333
         4.643 24.209
         4.576 24.095
         4.495 23.991
         4.398 23.897
         4.286 23.812
         /
\plot  4.286 23.812  3.810 23.495 /
%
% Fig TEXT object
%
\put{\SetFigFont{10}{7.2}{\rmdefault}{\mddefault}{\updefault}$\Sigma^{(1)}  =$ } [lB] at  0.270 24.198
\linethickness=0pt
\putrectangle corners at  0.270 25.527 and  4.889 23.368
\endpicture}
\end{minipage}
\begin{minipage}{3.6in}
\begin{equation}
 = \int \frac{d\vec k'}{(2 \pi)^2} \int \frac{d
  \Omega}{2 \pi} w_{\mu \nu}{\cal D}_{\mu \nu}(\vec k', \Omega)
\end{equation}
\end{minipage}

\noindent which is a frequency and wavevector independent constant and can
therefore be considered as just a renormalization of the chemical
potential (and is therefore uninteresting).  In the diagram we have
drawn the bare gauge propagator, although we could equally
well have used the RPA screened propagator, and we still would have
found just a trivial constant.

A more nontrivial contribution to the self energy which occurs at the
same order is given by the diagram

\font\thinlinefont=cmr5
\begingroup\makeatletter\ifx\SetFigFont\undefined%
\gdef\SetFigFont#1#2#3#4#5{%
  \reset@font\fontsize{#1}{#2pt}%
  \fontfamily{#3}\fontseries{#4}\fontshape{#5}%
  \selectfont}%
\fi\endgroup%
\mbox{\beginpicture
\setcoordinatesystem units <0.50000cm,0.50000cm>
\unitlength=0.50000cm
\linethickness=1pt
\setplotsymbol ({\makebox(0,0)[l]{\tencirc\symbol{'160}}})
\setshadesymbol ({\thinlinefont .})
\setlinear
%
% Fig CIRCULAR ARC object
%
\linethickness=4pt
\setplotsymbol ({\makebox(0,0)[l]{\tencirc\symbol{'162}}})
\setdots < 4pt >
\circulararc 177.945 degrees from  6.150 22.760 center at  4.128 22.660
%
% Fig POLYLINE object
%
\linethickness= 0.500pt
\setplotsymbol ({\thinlinefont .})
\setsolid
\putrule from  1.705 22.660 to  6.550 22.660
%
% Fig POLYLINE object
%
\linethickness= 0.500pt
\setplotsymbol ({\thinlinefont .})
\setsolid
\plot  4.069 22.460  4.763 22.660  4.069 22.860 /
%
% Fig TEXT object
%
\put{\SetFigFont{10}{7.2}{\rmdefault}{\mddefault}{\updefault}$G_{\rm f}^0$} [lB] at  3.651 21.566
%
% Fig TEXT object
%
\put{\SetFigFont{10}{7.2}{\rmdefault}{\mddefault}{\updefault}$\cal D$} [lB] at  3.969 25.082
%
% Fig TEXT object
%
\put{\SetFigFont{10}{7.2}{\rmdefault}{\mddefault}{\updefault}$\Sigma(\vec
  k,\omega) =$ } [lB] at  -1.111 23.971
\linethickness=0pt
\putrectangle corners at  -1.111 25.311 and  6.397 22.009
\endpicture}
\begin{equation} \label{eq:sel}
   = i \int \frac{d \vec k'}{(2 \pi)^2} \int \frac{d \Omega}{2
  \pi} v_\mu(\vec k,\vec k') v_{\nu}(\vec k',\vec k) {\cal D}_{\mu \nu}(|\vec
  k - \vec k'|, \omega - \Omega) G^0_{\rm f}(\vec k',\Omega)
\end{equation}
We can think of this as the process by which a free fermion of
momentum $\vec k$ emits a gauge fluctuation of momentum $\vec k - \vec
k'$, both the fermion and the gauge field propagate, and then they
interact (recombine) again at some later time.  Note that by symmetry,
the ${\cal D}_{01}$ and ${\cal D}_{10}$ terms vanish from the self
energy and we need only consider ${\cal D}_{00}$ and ${\cal D}_{11}$.

In the very lowest order calculation that we could attempt, we could
use the bare gauge field propagator ${\cal D}^0$ in Eq.  \ref{eq:sel}.
However, we find that the ${\cal D}^0_{11}$ is zero and the only
contribution is from the ${\cal D}^0_{00} = v(q)$ term which results
in the usual self energy of free fermions interacting with themselves
via a Coulomb interaction (We consider this contribution to be
uninteresting).

At RPA level, we use the RPA value of the propagator ${\cal D}$ in Eq.
\ref{eq:sel} (as we have indicated in the diagram).  At this level,
the self energy Eq. \ref{eq:sel} is both ultraviolet and infrared
divergent.

We will begin by looking at the contribution $\delta \Sigma$ from the
${\cal D}_{00}$ term to the self energy Eq. \ref{eq:sel}.  We obtain
the result~\cite{HLR}
\begin{equation}
  \delta \Sigma(\vec k,\omega) \sim  \ln(R q_{\rm max}) (\xi_k - \omega) + \ldots
\end{equation}
with $R$ the size of the system and $q_{\rm max}$ some ultraviolet cutoff.
Since this term vanishes at $\omega = \xi_k$, it does not contribute
to the effective mass of the fermion (see Eq. \ref{eq:mstar} below).
The origin of this ultraviolet divergence is related to the energy
required to instantaneously attach two flux quanta to a fermion.
Turning on these two flux quanta suddenly has the effect of creating
an electric field of strength $\tilde \phi/r$ a distance $r$ away.
This field excites cyclotron oscillations at $\omega_{\rm c}$.  Integrating
$1/r$ gives an energy cost for making this attachment that goes as
\begin{equation}
  E_0 \sim \omega_{\rm c} \tilde \phi \ln(R q_{\rm max})
\end{equation}
This issue is discussed in more detail by HLR~\cite{HLR}.  A version of
the Chern-Simons theory that attaches the flux quanta
adiabatically~\cite{NickReadOld} instead of suddenly, would presumably
be able to stay completely in the lowest Landau level, and would not
create these cyclotron excitations and would hence not have this
singularity.  Although such a lowest Landau level Chern-Simons theory
has not been completely developed, recent work by Shankar and Murthy
seems promising along this direction~\cite{Shankar}.

\subsection{Infrared Divergences}
\label{sub:div}
\label{sec:infrared}

Most of the discussions of divergences in the Chern-Simons theory focus
on the infrared problems, which are quite physical and have very
interesting ramifications.  These divergences come from the low energy
overdamped mode of the transverse gauge field propagator ${\cal
  D}_{11}$.  From here on, we will focus only on this term of the
gauge field propagator since it is the only piece giving us these
divergences (and hence this piece dominates).   

We begin by plugging in the form of ${\cal D}_{11}$ given by Eqs.
\ref{eq:gaugefieldprop} and \ref{eq:lowenergyRPA} into Eq.
\ref{eq:sel} to obtain the self energy~\cite{Ady,HLR} (for the case of
Coulomb interactions)
\begin{eqnarray} 
  \label{eq:self}
  \Sigma(k,\omega) &=& \frac{\tilde \phi^2}{2 \pi} \frac{\epsilon
    \hbar^2 \kf}{m_{\rm b} e^2} \hbar \omega \ln\left( \frac{4 e^2 \kf}{\tilde \phi^2 \epsilon
    \omega} \right) + i \frac{\tilde \phi^2}{4} \frac{ \epsilon
    \hbar^2 \kf}{m_{\rm b} e^2} \hbar \omega \\
  &\sim& \omega \ln (i \omega)
\end{eqnarray}
A detailed description of this calculation is given in Ref. 
\onlinecite{Ady}.  This self energy is mostly frequency dependent,
similar to the self energy arising from electron--phonon interaction
in metals, and in contrast to the self energy resulting from
electron--electron interaction in metals (which is mostly $k$ dependent).

For the case of short-range interactions of the form $ v(r) \sim
r^{-\eta} $ with $1 < \eta \le 2$, it is found that the self energy is
even more singular.  The reason for the increased singularity is that
density fluctuations on long length scales are easy to make when there
are not long range interactions.  These long range (infrared) density
fluctuations couple to the gauge field ({\em i.e.\/}, create local Chern-Simons
fields) which interacts strongly with the fermions.  Performing the
calculation (Eq. \ref{eq:sel}) we obtain
\begin{equation}
  \Sigma \sim (i \omega)^{\frac{2}{\eta + 1}}.
\end{equation}

On the other hand, if one considers longer range interactions ($v(r)
\sim r^{-\eta}$ with $\eta < 1$), the interactions suppress long range
density fluctuations and there there is no singular part of the self
energy (One must then consider all of the pieces of the self energy
since no single divergent piece dominates).  We note also that for
quantized states away from $\nu=\frac{1}{2}$, the
divergences are cut off by the effective cyclotron frequency and the
perturbation theory becomes better defined.

Once we have this self-energy, the excitation modes $\tilde \omega(k)$
of the system are given by the poles of the energy denominator of the
exact Green's function (Eq. \ref{eq:exactG}).  To find the modes, we
set
\begin{equation}
  \label{eq:polehunt}
 \hbar \tilde \omega(k) - \vfns (k - \kf) - \Sigma(\tilde \omega(k),k) = 0
\end{equation}
(where we have approximated $\xi_k \approx \vfns (k - \kf)$).  The
effective mass $m^*$ is then defined through the effective Fermi
velocity $\hbar \kf/ m^*$, which, in turn, is the group velocity
$d \tilde \omega/dk |_{k=\kf}$ of the excitations $\tilde
\omega(\vec k)$.  The effective mass is then given by 
\begin{eqnarray}
\label{eq:mstardef}
  m^*(\omega) &=& \kf\left(\left. {\frac{d \tilde \omega(k)}{d k}}\right|_{k=\kf}
\right)^{-1}  
  = 
m_{\rm b} \frac{1 - \frac{\partial \Sigma}{\partial \hbar \omega}|_{\omega
    = 0}}{1 + \frac{\partial \Sigma}{\partial \xi_k}|_{k=\kf}} \\
&\sim& \frac{(\tilde \phi \hbar)^2}{2 \pi} \frac{\epsilon \kf}{e^2}
  \left| \ln \omega \right|   \label{eq:mstar}
\end{eqnarray}
in the low frequency limit 
for the case of Coulomb interactions,  and 
\begin{equation}
  m^* \sim \omega^{-\frac{\eta -1}{\eta + 1}}
\end{equation}
for in the low frequency limit for the case of short range
interactions $(v(r) \sim r^{\eta}$ with $1 < \eta \le 2)$.  (For
interactions longer ranged than Coulomb there is no divergence of the
effective mass.)  It should be noted that whereas the self energy 
given above is only a first order approximation, it is argued by Stern
and Halperin~\cite{Ady} that this expression (Eq. \ref{eq:mstar}) for
the effective mass is in fact exact in the $\omega \rightarrow 0$
limit, at least for the Coulomb case.

The physical significance of this effective mass is most evident
slightly away from $\nu=\frac{1}{2m}$.  Here, the fractional Hall gaps
for the fractions $\nu = p/(2mp+1)$ are given by~\cite{Ady,HLR}
\begin{equation}
  E_g(\nu) = \frac{\hbar e \Delta B}{m^*(\nu) c}
\end{equation}
where $m^*(\nu)$ is the effective mass (Eq. \ref{eq:mstar}) calculated
self consistently at frequency $\omega = E_g(\nu)/\hbar$.  Thus, for
filling fraction $p/(2mp+1)$ for very large $p$, we should measure
gaps that go as $E_g(p) \sim 1/(p \ln p)$.  For small $p$, on the
other hand, the divergences are cut off so these pieces do not
dominate.  In principle, one could then perform a systematic
perturbation expansion without experiencing any infrared problems.
However, it would remain a problem to eliminate $m_{\rm b}$ from the low
energy physics (See however the approach by Shankar and
Murthy~\cite{Shankar}).

The divergence of the effective mass at $\nu=\frac{1}{2m}$ is also
reflected in the singular behavior of the low energy specific
heat~\cite{HLR,KimCV}.  For a normal Fermi liquid we have $C_v \sim T
m^*$.  For the Chern-Simons Fermi liquid with Coulomb interactions,
the effective mass diverges as $\log \omega$ for excitations on the
scale $\omega$.  Thus, the specific heat is given by $C_v \sim T \log
T$.  For short range interactions~\footnote{Note that for short-range
  interaction, we obtain an incorrect coefficient in front of the
  power law by using a quasiparticle picture due to the fact that the
  individual quasiparticle is not well defined.}, we similarly obtain
$C_v \sim T^{2/(1 + \eta)}$.

Another important quantity we can extract from the self energy is the
quasiparticle-quasiparticle scattering time $\tau_{qp}$.  Usually one
uses the rule that $\tau_{qp}^{-1} = \mbox{Im} \Sigma$.  However, here, we
should be a bit more careful.  More accurately, it is the imaginary
part of the complex excitation frequency $\tilde \omega$ that
determines the lifetime of the excitations via $\tau_{qp}^{-1} = \mbox{Im
  }\tilde \omega$.  Defining $\omega = \mbox{Re }(\tilde \omega(k))$
we find that the imaginary part of $\tilde \omega$ is given by
\begin{eqnarray}  
  \frac{1}{\tau_{qp}} \equiv \mbox{Im } \tilde \omega 
&\sim& \frac{\omega}{\ln \omega}  ~~~~~~ \mbox{Coulomb} \\
&\sim&  \omega  ~~~~~~ \mbox{Short Range} \label{eq:taushort}
\end{eqnarray}
which is a somewhat greater lifetime than the usual expression
$\mbox{Im} \Sigma$ would lead one to believe.  In another
language, the quasi-particle lifetime is lengthened because $\tau_{qp}^{-1}
= Z \mbox{Im } \Sigma$ rather than $\mbox{Im } \Sigma$ where $Z^{-1} =
1 - \frac{d \Sigma}{d \omega}$ is the so-called wavefunction
renormalization.  Thus, $\omega\tau_{qp}\sim\ln{|\omega|}\gg 1$ for
the Coulomb case, and the quasi-particle remains well defined as
we approach the Fermi surface.  However, for short range interactions,
$\omega \tau_{qp} \approx 1$ so that the quasiparticle
is not well defined.

One might be concerned that the singular behavior of the self-energy
would result in a divergent response function.  However, the self
energy, like the exact Green's function itself, is not a gauge
invariant quantity.  When one calculates the gauge invariant
electromagnetic response, one finds that singular vertex corrections
exactly cancel the singular self energy terms to give a finite
nonsingular response~\cite{Kim,Kim2}.  For example, the sum of the
following three divergent diagrams is convergent

\hspace*{40pt} \font\thinlinefont=cmr5
\begingroup\makeatletter\ifx\SetFigFont\undefined%
\gdef\SetFigFont#1#2#3#4#5{%
  \reset@font\fontsize{#1}{#2pt}%
  \fontfamily{#3}\fontseries{#4}\fontshape{#5}%
  \selectfont}%
\fi\endgroup%
\mbox{\beginpicture
\setcoordinatesystem units <0.50000cm,0.50000cm>
\unitlength=0.50000cm
\linethickness=1pt
\setplotsymbol ({\makebox(0,0)[l]{\tencirc\symbol{'160}}})
\setshadesymbol ({\thinlinefont .})
\setlinear
%
% Fig CIRCULAR ARC object
%
\linethickness=4pt
\setplotsymbol ({\makebox(0,0)[l]{\tencirc\symbol{'162}}})
\setdots < 4pt>
\circulararc 180.000 degrees from  8.160 25.781 center at  7.334 25.781
%
% Fig CIRCULAR ARC object
%
\linethickness=4pt
\setplotsymbol ({\makebox(0,0)[l]{\tencirc\symbol{'162}}})
\circulararc 180.000 degrees from  2.985 24.733 center at  3.810 24.733
%
% Fig ELLIPSE
%
\linethickness= 0.500pt
\setplotsymbol ({\thinlinefont .})
\setsolid
\ellipticalarc axes ratio  1.270:0.635  360 degrees 
        from 12.065 25.241 center at 10.795 25.241
%
% Fig ELLIPSE
%
\linethickness= 0.500pt
\setplotsymbol ({\thinlinefont .})
\ellipticalarc axes ratio  1.270:0.635  360 degrees 
        from  8.572 25.241 center at  7.303 25.241
%
% Fig ELLIPSE
%
\linethickness= 0.500pt
\setplotsymbol ({\thinlinefont .})
\ellipticalarc axes ratio  1.270:0.635  360 degrees 
        from  5.080 25.241 center at  3.810 25.241
%
% Fig POLYLINE object
%
\linethickness= 0.500pt
\setplotsymbol ({\thinlinefont .})
\setdashes < 0.1905cm>
\plot  9.525 25.241  9.366 25.241 /
%
% Fig POLYLINE object
%
\linethickness= 0.500pt
\setplotsymbol ({\thinlinefont .})
\plot 12.065 25.241 12.224 25.241 /
%
% Fig POLYLINE object
%
\linethickness=4pt
\setplotsymbol ({\makebox(0,0)[l]{\tencirc\symbol{'162}}})
\setdots < 4pt >
\plot 10.795 25.826 10.795 24.556 /
%
% Fig POLYLINE object
%
\linethickness= 0.500pt
\setplotsymbol ({\thinlinefont .})
\setdashes < 0.1905cm>
\plot  6.064 25.273  5.905 25.273 /
%
% Fig POLYLINE object
%
\linethickness= 0.500pt
\setplotsymbol ({\thinlinefont .})
\plot  8.604 25.273  8.763 25.273 /
%
% Fig POLYLINE object
%
\linethickness= 0.500pt
\setplotsymbol ({\thinlinefont .})
\setsolid
\plot  6.111 25.432  6.587 25.607 /
%
% Fig POLYLINE object
%
\linethickness= 0.500pt
\setplotsymbol ({\thinlinefont .})
\plot  6.350 25.845  6.096 25.463 /
%
% Fig POLYLINE object
%
\linethickness= 0.500pt
\setplotsymbol ({\thinlinefont .})
\putrule from  7.620 25.876 to  7.144 25.876
%
% arrow head
%
\plot  7.652 26.003  7.144 25.876  7.652 25.749 /
%
%
% Fig POLYLINE object
%
\linethickness= 0.500pt
\setplotsymbol ({\thinlinefont .})
\plot 10.270 24.655  9.842 25.004 /
%
% Fig POLYLINE object
%
\linethickness= 0.500pt
\setplotsymbol ({\thinlinefont .})
\plot 10.223 24.638  9.716 24.718 /
%
% Fig POLYLINE object
%
\linethickness= 0.500pt
\setplotsymbol ({\thinlinefont .})
\plot 11.906 24.924 11.271 24.765 /
%
% Fig POLYLINE object
%
\linethickness= 0.500pt
\setplotsymbol ({\thinlinefont .})
\plot 11.906 24.924 11.430 24.511 /
%
% Fig POLYLINE object
%
\linethickness= 0.500pt
\setplotsymbol ({\thinlinefont .})
\plot 11.335 25.798 11.748 25.512 /
%
% Fig POLYLINE object
%
\linethickness= 0.500pt
\setplotsymbol ({\thinlinefont .})
\plot 11.286 25.830 11.811 25.813 /
%
% Fig POLYLINE object
%
\linethickness= 0.500pt
\setplotsymbol ({\thinlinefont .})
\plot  9.842 25.718 10.175 26.003 /
%
% Fig POLYLINE object
%
\linethickness= 0.500pt
\setplotsymbol ({\thinlinefont .})
\plot  9.857 25.654 10.319 25.718 /
%
% Fig POLYLINE object
%
\linethickness= 0.500pt
\setplotsymbol ({\thinlinefont .})
\putrule from  4.128 25.876 to  3.651 25.876
%
% arrow head
%
\plot  4.159 26.003  3.651 25.876  4.159 25.749 /
%
%
% Fig POLYLINE object
%
\linethickness= 0.500pt
\setplotsymbol ({\thinlinefont .})
\setdashes < 0.1905cm>
\plot  2.540 25.241  2.381 25.241 /
%
% Fig POLYLINE object
%
\linethickness= 0.500pt
\setplotsymbol ({\thinlinefont .})
\plot  5.080 25.241  5.239 25.241 /
%
% Fig POLYLINE object
%
\linethickness= 0.500pt
\setplotsymbol ({\thinlinefont .})
\setsolid
\plot  4.985 24.956  4.572 24.892 /
%
% Fig POLYLINE object
%
\linethickness= 0.500pt
\setplotsymbol ({\thinlinefont .})
\plot  4.921 24.924  4.746 24.623 /
%
% Fig POLYLINE object
%
\linethickness= 0.500pt
\setplotsymbol ({\thinlinefont .})
\plot  2.819 24.845  2.445 25.004 /
%
% Fig POLYLINE object
%
\linethickness= 0.500pt
\setplotsymbol ({\thinlinefont .})
\plot  2.809 24.845  2.650 25.195 /
%
% Fig POLYLINE object
%
\linethickness= 0.500pt
\setplotsymbol ({\thinlinefont .})
\plot  8.287 25.607  8.477 25.290 /
%
% Fig POLYLINE object
%
\linethickness= 0.500pt
\setplotsymbol ({\thinlinefont .})
\plot  8.319 25.654  8.714 25.527 /
\linethickness= 0.500pt
\setplotsymbol ({\thinlinefont .})
%
% Fig CONTROL PT SPLINE
%
% open spline
%
\plot    7.144 24.606  7.382 24.606
         /
\plot  7.382 24.606  7.620 24.606 /
%
% arrow head
%
\plot  7.112 24.479  7.620 24.606  7.112 24.733 /
\linethickness= 0.500pt
\setplotsymbol ({\thinlinefont .})
%
% Fig CONTROL PT SPLINE
%
% open spline
%
\plot    3.810 24.606  4.048 24.606
         /
\plot  4.048 24.606  4.286 24.606 /
%
% arrow head
%
\plot  3.778 24.479  4.286 24.606  3.778 24.733 /
%
%
% Fig TEXT object
%
\put{\SetFigFont{6}{7.2}{\rmdefault}{\mddefault}{\updefault}+} [lB] at  5.356 25.082
%
% Fig TEXT object
%
\put{\SetFigFont{6}{7.2}{\rmdefault}{\mddefault}{\updefault}+} [lB] at  8.849 25.082
\linethickness=0pt
\putrectangle corners at  2.335 26.672 and 12.270 23.842
\endpicture}

\noindent This nonsingular behavior of the physical response function
is also obtained in several other approaches (including
nonperturbative approaches) to analyzing the Chern-Simons
problem~\cite{Aim,Marston,Kvesh}.  Furthermore, it has been shown by
Kim et al~\cite{KimB} that one can write down a so-called quantum
Boltzmann equation for the response of the Chern-Simons system that is
nonsingular.  In essence the realization is that although there is a
singularity in the energy required to excite a single quasiparticle,
smooth deformations of the Fermi surface (which are what are required
for physical response functions at long wavelength and low frequency)
are quite well behaved.  In another language, one finds that even for
short range interactions, the transport scattering time $\tau_{tr}$
(which enters the response functions) is given by~\cite{Kim} (compare
Eq.  \ref{eq:taushort}) $ \tau_{tr}^{-1} \sim \omega^{4/(1 + \eta)} $
such that $\omega \tau_{tr} \ll 1$ yielding a nonsingular response.

\subsection{Divergent Fermi Liquid Theory}
\label{sub:dFLT}

Since excitation of a single quasiparticle is extremely singular in
the case of short range interactions, we are prevented from writing
down a conventional Landau Fermi liquid description for such a system
(although a non-conventional Boltzmann-like description can be written
down in the language of the quantum Boltzmann equation~\cite{KimB} ---
even for the case of short range interactions).  For the case of
Coulomb interactions, however, we might hope that the divergences are
not sufficiently singular to prevent us from describing the system in
the conventional Landau Fermi liquid format.  We would hope that the
result of such a calculation would return us to the phenomenological
Landau Fermi liquid approach discussed above in section \ref{sec:FLT}
and \ref{sub:fit}.  For the rest of this section, we focus on the
Coulomb case.

As discussed above in section \ref{sec:FLT}, Landau's Fermi liquid
theory describes the system in terms of the parameters, $m^*$ and
$f(\theta)$, which can be written formally in terms in terms of one--
and two-- particle Green's functions~\cite{Nozieres}.  As discussed
above, the effective mass is singular and must be written as a
function of frequency.  Similarly, perturbative calculation of the
Landau quasiparticle interaction function $f(\theta)$ is singular.
Although one might imagine regularizing this divergence in any one of
a number of different ways, one very natural method was proposed in
Ref.\onlinecite{Ady}, where the interaction function $f(\theta)$ is
written as a function of the energy exchange between the
quasi--particles.  (In conventional, non--singular, Fermi liquid
theory, the Landau interaction involves zero energy exchange, since
both quasi--particles lie on the Fermi surface).  A natural lowest
order calculation then yields the interaction function between
quasiparticles (See Ref.~\onlinecite{Ady} for details)
\begin{equation}
  \label{eq:whatisf}
  f(\theta;\omega) = \frac{2 \pi \kf^2}{m_{\rm b} m^*(\omega)} \left( \cos^2
  \frac{\theta}{2} \right) {\cal D}_{11}(2 \kf \sin \frac{\theta}{2}, \omega)
\end{equation}
For zero frequency exchange, this function takes the singular form
\begin{equation}
  \label{eq:fdelta}
  \lim_{\omega \rightarrow 0} f(\theta;\omega) = \frac{(2
    \pi \hbar )^2}{m_{\rm b}} \delta(\theta)
\label{deltafunc}  
\end{equation}
The effect of such a singular interaction function is very simple, and
is immediately understood when Eq.  \ref{deltafunc} is substituted
into Eq. \ref{eq:eps12} to yield
\begin{equation}
  \delta \epsilon_1(\theta) = \frac{m^*}{m_{\rm b}} \nu(\theta) \gg \nu(\theta)
\end{equation}
Substituting into the Boltzmann transport equation (Eq. \ref{eq:kin1}) in this
approximation then yields
\begin{equation}
  -i\omega \nu(\theta) + \left(iq\vfns \cos(\theta) - \Delta \omega_{\rm c}
  \frac{\partial}{\partial \theta} \right) \nu(\theta) = -e
   \vec E \cdot \hatn n(\theta) +
  \mbox{$I$}(\theta)
\end{equation}
where $\vfns = \pf/m_{\rm b}$ is the {\it bare} Fermi velocity, and $\Delta
\omega_{\rm c} = e \Delta B/m_{\rm b} c$ is the {\it bare} fermion cyclotron
frequency.  The delta function $f(\theta)$ generates then a
transport equation that is the same as that for a system where the
effective mass is {\it re}-renormalized back to the band mass and the
quasi--particles on the Fermi surface are non--interacting!  Thus, if
this approximation (Eq. \ref{deltafunc}) is correct, the long wavelength
low frequency response of the system is that of non--interacting
quasiparticles with an unrenormalized bare mass.  Note that the
magnitude of this delta function yields a value of $f_1$ set by the
bare band mass in agreement with the constraint on $f_1$ given by Eq.
\ref{eq:f1effmass}.   

The expression (Eq. \ref{eq:fdelta}) for the Landau
function$f(\theta;\omega)$ is only an approximation valid to low order
in $\tilde \phi$ as discussed above. We now consider~\cite{Simonun}
what we might conclude about the exact Landau function for the
physical value of $\tilde \phi$ an even integer, in the limit of
$m_{\rm b}\rightarrow 0$.  While $f_0$ and $f_1$ should depend on the bare
mass in the limit $m_{\rm b}\rightarrow 0$ (See section \ref{sec:sep} for
the properties of $f_0$) this should not be the case for all other
Landau parameters, since low $q,\omega$ linear response in that
limit should be constrained to the lowest Landau level, {\em i.e.\/},
independent of the bare mass~\cite{Ady}.  Finally, as discussed above
in section \ref{sub:div}, we expect the diverging effective mass $m^*$
to cancel from linear response functions of the $\nu=\frac{1}{2}$
state.

As demonstrated above, for the diverging $m^*$ to cancel from the
linear response functions, the Landau function should include a
$\delta$--function component.  For the Landau parameters $f_l$ for
$l\ge 2$ to be $m_{\rm b}$ independent, the amplitude of this
$\delta$--function component should be independent of the bare mass,
{\em i.e.\/}, should be determined by the energy scale of electron--electron
interaction $e^2/(l_B \epsilon)$.  Thus by dimensional analysis, we
define a new effective mass
\begin{equation}
  \label{eq:mstarstar}
  \frac{\hbar^2}{m^{**}} =  C(\nu) \frac{e^2}{l_B \epsilon}
\end{equation}
where $C(\nu)$ is an unknown dimensionless number of order 1 which is
only very weakly dependent on $\nu$.  It is this mass $m^{**}$ that
enters the phenomenological Landau-Boltzmann approaches (which we
there call $m^*$) in sections \ref{sec:FLT} and \ref{sec:Mag}.
Indeed, everywhere where a finite effective mass is used for
calculating responses for the $\nu=\frac{1}{2m}$ problem, it is this
finite effective mass $m^{**}$ and not the divergent effective mass
that is being used.  The critical thing to remember is that the
effective mass $m^{**}$ describes smooth deformations of the Fermi
surface, not single quasiparticle excitations, and can therefore be
nonsingular (whereas the true effective mass $m^*(\omega)$
describing single quasiparticle excitations is singular).  The mass
$m^{**}$ yields a re-renormalized Fermi velocity $\vfns^{**} =
\pf/m^{**}$ which determines the edge of the low energy continuum of
quasiparticle excitations.  It is interesting to note that $
\vfns^{**} = C e^2/(\epsilon \hbar)$ is independent of density.  We
note that the effective mass $m^{**}$, since it is non-divergent, is
probably most closely related to the effective mass that determines
the energy gaps of the fractional hall gaps well away from
$\nu=\frac{1}{2m}$ (fractions like $\frac{1}{3}$ and $\frac{2}{5}$).
We note that estimates of this mass have been made by exact
diagonalizations~\cite{MorfanddAmbruminil,Morf,HLR}.

Using this new mass scale, we propose the following conjecture for the
interaction function $f(\theta)$ in the low frequency limit
\begin{equation}
  \label{eq:fform}
  \lim_{\omega \rightarrow 0} f(\theta; \omega) = (2 \pi \hbar)^2
    \left(\frac{1}{m^{**}} - \frac{1}{m^*} \right) \delta(\theta) 
    + f^{**}(\theta)
\label{conjec}
\end{equation}
where $f^{**}(\theta)$ is nonsingular in the low frequency limit.
Here, the $\delta$ function term re-renormalizes the mass from the
divergent $m^*(\omega)$ to the finite value $m^{**}$ which is
appropriate for calculating responses.    The residual term $f^{**}$
gives us the Landau interaction function for these effective
particles of mass $m^{**}$.  Indeed, this is the Landau function we
used in sections \ref{sec:FLT} and \ref{sec:Mag} above (which we just
called $f$ there).  We note that this Landau function must satisfy the
constraints Eq. \ref{eq:f1effmass} and Eq. \ref{eq:f0comp}.  These
require that $m^{**} f^{**}_1/(2 \pi \hbar^2) = (m^{**}/m_{\rm b}) - 1$
and that the leading term of $f_0^{**}$ is given by $\pi \hbar^2
\tilde \phi/m_{\rm b}$.  All other Fourier modes $f_l^{**}$ if
$f^{**}(\theta)$ should be independent of the bare mass. 

To summarize this section, we have found that a perturbative approach
to the Chern-Simons problem at $\nu=\frac{1}{2m}$ forces us to suffer
with infrared divergences of quantities such as the effective mass
$m^*$.  These divergences appear in the fractional Hall gaps and the
specific heat near $\nu=\frac{1}{2}$, but do not appear in the
response.  For this reason, we believe that another {\it finite}
effective mass (which here we call $m^{**}$) can be defined that
should enter into calculations of the response.  This, in some sense,
might justify our earlier use of a finite effective mass (which there
we called $m^{*}$) in calculations in section \ref{sec:FLT} and
\ref{sec:Mag}.  However, more work remains to be done to determine how
well justified this really is.  One alternative possibility that one
might consider is that the response at any given frequency $\omega$
must be calculated with a mass $m^*(\omega)$ appropriate for that
frequency (which may diverge at small $\omega$).  This would make the
edge of the continuum of low energy quasiparticles quite curved.
There is also the possibility that higher Fermi liquid coefficients
conspire such that a great number of collective (yet discrete) zero
sound modes merge into the continuum at low energies~\cite{KimB}.

At fractional Hall states $\frac{p}{2mp+1}$ for small $p$, there is no
divergence of $m^*$ and one could in principle calculate $m^*$
perturbatively.  However, in a straightforward perturbative approach
it is not clear how to get the bare mass to vanish from the answer as
it should in the $m_{\rm b} \rightarrow 0$ limit (See however Ref.
\onlinecite{Shankar}).  Thus, we are still stuck with treating $m^*$
as a phenomenological parameter.  Nonetheless, we would expect it to
obey the scaling form of Eq. \ref{eq:mstarstar}.  Estimates of the
coefficient $C(\nu)$ can be made by exact
diagonalization~\cite{Morf,MorfanddAmbruminil}.

\mysection{Wavefunction Picture of Composite Fermions}
\label{sec:wavefunction}

The success of the Chern-Simons approach should strike us as somewhat
surprising.  We have modeled an electron in a large magnetic field as
a fermion bound to flux and all of a sudden we are getting many of the
right answers at even a crude mean field level --- and we get even
more right answers when we consider corrections to mean field, such as
RPA.  To understand physically why this approach makes sense, it is
useful to think in terms of wavefunctions.  Although the analysis of
wavefunctions is somewhat off of the main line of development of this
paper, it will be useful for developing intuition for the physics.
Those who desire a more thorough discussion of the wavefunction
properties of composite fermions are referred to the works of
Jain~\cite{Jain0,Jain} and Read~\cite{NickRead}.

In the tradition begun by Laughlin~\cite{Laughlin}, we will attempt to
guess trial wavefunctions for the states in which we are interested,
and study the properties of these wavefunctions.  Although such an a
approach is far from systematic, it has greatly deepened our
understanding of the fractional quantum Hall effect.  Indeed, it is
probably not an overstatement to say that Laughlin's
wavefunction~\cite{Laughlin} is the single most important theoretical
step ever made in the theory of fractional quantum Hall effect; and
similarly, it should be recalled that the notion of the composite
fermion was begun by Jain~\cite{Jain0} in the wavefunction language.

The experienced reader is encouraged to skip quickly through section
\ref{sub:wavefunctions} and \ref{sec:laughlin}, which should be review
for anyone well versed in quantum Hall physics.  Those in need of
further introductory details regarding the wavefunction approaches are
referred to Refs.
\onlinecite{Prange,Book2,DasSarmabook,Chakraborty}. In section
\ref{sub:wf2} we discuss the properties of the Rezayi-Read $\nu
=\frac{1}{2}$ wavefunction and  explain how it leads us to a picture
of the composite fermion at $\nu=\frac{1}{2}$ as a neutral dipole.  In
section \ref{sub:smallw} we consider the existence of the cyclotron
scale $R_{\rm c}^*$ in this wavefunction picture and show that it agrees
with the Chern-Simons picture of HLR.   In \ref{sub:jainw} we discuss
the Jain wavefunctions for the fractionally quantized states and in
section \ref{sub:compare} we compare the wavefunction approaches to
the Chern-Simons transforms  discussed above.  In section
\ref{sub:dipole} we perform explicit calculations within this neutral
dipole picture (with no Chern-Simons field in sight) and show that the
results agree with the Chern-Simons approach discussed above.

\subsection{Wavefunctions and Lowest Landau Level Physics}
\label{sub:wavefunctions}

In order to study trial wavefunctions, we work in circular gauge,
$\vec A = \frac{1}{2} \vec B \times \vec r$ and write the position of
the $j^{th}$ particle in its complex representation
as~\cite{GirvinJach} $ z_j = x_j + i y_j $. In this language, we can
write an arbitrary many-electron wavefunction as
\begin{equation}
\label{eq:anal}
  \Psi(z_1,z_2,\ldots,z_N) = 
  f(z_1,z_2,\ldots,z_N) 
\prod_{j=1}^{N}
  e^{-\frac{1}{4} |z_j|^2/l_B^2 }  \,\,\,\,\,\,\,. 
\end{equation} 
The wavefunction $\Psi$ represents electrons restricted to the lowest
Landau level (appropriate for the large $B$ or $m_{\rm b} \rightarrow 0$
limit) if and only if the function $f$ is analytic in all of its
arguments ({\em i.e.\/}, a polynomial containing only nonnegative powers of
the $z$'s and no powers of the $\bar z$'s).  We also point out that
the wavefunction must be antisymmetric under interchange of the
position of any two of the particles in order to be properly
fermionic.

To analyze a lowest Landau level wavefunction we consider fixing the
positions~\cite{Halpacta} of all but one of the electrons (say
electrons $1$ though $N-1$) and move the one remaining test electron
(say electron $N$) adiabatically.  With the positions $z_1, \ldots,
z_{N-1}$ thus fixed, the wavefunction is now an analytic function of
the position $z_N$ of this last electron times an exponential factor.
At certain values of $z_N$ the value of the wavefunction will be zero.
Because of the analyticity condition, as we move $z_N$ in a circle
around any such zero, the wavefunction undergoes a phase rotation of $2
\pi$.  For this reason the zeros of the wavefunction are also referred
to as ``vortices.''  We point out that since the wavefunction is
antisymmetric ({\em i.e.\/}, fermionic), it must go to zero when $z_N$
approaches the position of any other electron.  Thus, there is a
naturally at least one vortex at the position of each electron.

If we now move our test electron adiabatically in a large loop of area
${\cal A}$, the phase accumulated by the electron (the traditional
Aharonov-Bohm phase) should be given by
\begin{equation}
\label{eq:phase1}e^{i \frac{2 \pi}{\phi_0} \oint \vec A \cdot dl} = e^{i 2 \pi
  \Phi/\phi_0}
\end{equation}
where $\Phi = {\cal A}B $ is the flux enclosed.  However, as mentioned
above, the analyticity of the wavefunction also demands that the
accumulated phase is given by $\exp{2 \pi i N_z}$ where $N_z$ is the
number of zeros of the wavefunction enclosed in the loop.  Comparing
these two results, we find that any lowest Landau level wavefunction
must have $B/\phi_0$ zeros per unit area.

To create a density fluctuation in an arbitrary ground state
wavefunction of uniform density, we can introduce a vortex at position $z_0$ by
multiplying the wavefunction by the factor
\begin{equation}
  \label{eq:quasi1}
  \prod_{j=1}^N (z_j - z_0). 
\end{equation}
Clearly, this factor suppresses the amplitude of the wavefunction in
the vicinity of $z_0$ leaving a deficit of electrons (or positive
charge) near $z_0$.  We note that the multiplication of a wavefunction
by this vortex operator increases the degree of the polynomial part
of the wavefunction ($f$), and thus increases the size of the droplet.
This is just another way of saying that the density removed from near
the point $z_0$ is accounted for by an increase in density at the edge
of the system.  In Laughlin's original work~\cite{Laughlin}, a plasma
analogy was used to show that this vortex operator acting on a ground
state wavefunction of the $\nu=\frac{1}{2m+1}$ quantum Hall state
(with $m$ an integer) results in a quasihole of charge $+\nu$.  We
will see next that this is a rather general result.

To evaluate the charge on the vortex, we consider
adiabatically~\footnote{For compressible state there are low energy
  excitations so one cannot have adiabatic motion.  Similarly, the
  argument given below in the next paragraph fails for a compressible
  state that is not a perfect Hall conductor.  Although this issue has
  not been completely sorted out, one possible way to think about such
  compressible states is to consider a limit of a series of
  incompressible states with smaller and smaller gaps.}  moving its
position $z_0$ in a large loop of area ${\cal A}$.  Analogous to the
discussion above, the Aharonov-Bohm phase accumulated by this process
is given by $\exp{(2 \pi i q \Phi/\phi_0)}$ where here $q$ is the
charge on the vortex, and once again, $\Phi = B{\cal A}$ is the flux
inclosed.  On the other hand, from the form of Eq.  \ref{eq:quasi1} it
is clear that the wavefunction picks up a phase of $\exp{(2 \pi i
  N_{\rm e})}$ where $N_{\rm e}$ is the number of electrons enclosed in the loop.
Thus, we conclude that the charge on the vortex is $q = N_{\rm e} \phi_0/B
{\cal A} = \nu$.

As mentioned above, the number of zeros of the wavefunction with
respect to the position of a test electron determines the applied
magnetic field.  In terms of counting these zeros, the introduction of
the vortex (application of the operator Eq. \ref{eq:quasi1} to a
ground state wavefunction) corresponds to the increase of the applied
magnetic field by a single flux quantum~\footnote{This is analogous to
  a real hole being created when a single flux quantum is added to a
  filled Landau level (the $\nu=1$ case).  Here, the increase in field
  by one flux quantum increases the degeneracy of the lowest Landau
  level by one, creating exactly one hole.}.  Indeed, one way to
create such a density fluctuation is to adiabatically insert a single
thin flux quantum through the system at position $z_0$.  As this flux
quantum is turned on, an EMF is created in the direction going around
the flux by Faraday's law.  For a system that is a perfect Hall
conductor, this moves charge away from the point $z_0$
(perpendicular to the EMF).  To calculate the charge moved away from
$z_0$ we write
\begin{equation}
  \frac{1}{c} \frac{d\Phi}{dt} = \oint \vec E \cdot d \vec l = \oint
  \rho \cdot \vec j \cdot d\vec l
\end{equation}
where $\Phi$ is the flux being turned on, $\rho$ is the resistivity
matrix, and the path of integration goes around the point $z_0$.  For
quantized Hall states we can use the fact that the diagonal
resistivity is zero, and this EMF can be integrated with respect to
time to get
\begin{equation} 
  \label{eq:brigade}
  \frac{1}{c} \Delta \Phi = \rho_{xy} \oint d\vec l \cdot
  \int_{t_i}^{t_{\rm f}} dt (\vec j \times \hat z ) \ = \rho_{xy} q
\end{equation} 
where $q$ is the total amount of charge that is moved away from the
point $z_0$ ({\em i.e.\/}, the charge of the created vortex).  Choosing $\Delta
\Phi$ to be a single flux quantum, and recalling that $\rho_{xy} =
h/(e^2 \nu)$, then yields the correct vortex charge of $q=+\nu$.

\subsection{Laughlin's Wavefunction}
\label{sec:laughlin}

Now that we have discussed some of the properties of wavefunctions in
the lowest Landau levels, we turn to writing down trial wavefunctions
for fractional quantized Hall states.  Such trial wavefunction
guesswork might seem at first like trying to find the proverbial
needle (the ground state wavefunction) in the haystack (the huge
Hilbert space of possible wavefunctions).  However, such guesswork has
been extremely successful, and has been justified in retrospect by
comparison to the results of exact diagonalizations of small
systems~\cite{Jain0,Jain,HaldaneChapter,Laughlin,Fano}.

Laughlin's trial wavefunction~\cite{Laughlin} for quantized Hall states
at filling fractions $\nu=\frac{1}{2m+1}$ is given by the expression
\begin{equation}
\label{eq:Laughlin}
\Psi_{\frac{1}{2m+1}}(z_1,z_2,\ldots,z_N) = \prod_{i<j}
 (z_i - z_j)^{(2m+1)} \prod_{i=1}^{N}
  e^{-\frac{1}{4} |z_i|^2/l_B^2 }
\end{equation}
which, for $m$ an integer, is a properly antisymmetric (fermionic)
wavefunction, and has the proper analytic times exponential (lowest
Landau level) form.  This wavefunction has $(N-1)(2m+1)$ zeros with
respect to the position of the $N^{th}$ electron.  Since there must be
$B/\phi_0$ such zeros per unit area we immediately calculate that the
filling fraction for this Laughlin state is indeed $\nu = n_{\rm e} \phi_0/B
= \frac{1}{2m+1}$.  This Laughlin form has been shown to be the exact
ground state wavefunction for a system with a certain type of short
range interaction~\cite{Trugman,HaldaneChapter}.  It has also been
shown to have an extremely high overlap with the exact ground state
wavefunction (calculated numerically) for small systems with Coulomb
interactions~\cite{Laughlin,HaldaneChapter,Fano}.  It should also
be noted that for $m=0$, the Laughlin wavefunction is exactly the
(single Slater determinant) wavefunction for one filled Landau level.

To understand why the Laughlin form is such a good guess for a trial
wavefunction, we once again consider moving one test electron
adiabatically while leaving the position of all the other electrons
fixed.  Here, we see that all of the zeros (vortices) are attached to
the position of the other electrons.  In fact, the Laughlin
wavefunction has a $2m+1^{th}$ order zero as any $z_i$ approaches any
other $z_j$, meaning that the wavefunction amplitude is highly
suppressed as any two electrons approach each other.  The higher the
order of the zero of the wavefunction, the more the electrons stay
away from each other.  Indeed, the reason that this Laughlin
wavefunction has such a good energy is that the zeros are arranged such
that the electrons stay as far away from each other as possible,
thereby lowering the Coulomb energy.  Moreover, for a fixed filling
fraction, there is a fixed number of zeros of the wavefunction, and
the Laughlin form arranges to put {\it all} of these zeros at the
position of the other electrons such that they are efficiently used to
keep electrons away from each other, leaving no zeros ``wasted''.

\subsection{The $\nu=\frac{1}{2}$ Wavefunction}
\label{sub:wf2}

One would naively guess that a good wavefunction for electrons at
filling fraction $\frac{1}{2m}$ would be an analogous Laughlin like
state given by Eq. \ref{eq:Laughlin} only with an exponent $2m$ rather
than $2m+1$.  However, such a wavefunction is symmetric rather than
antisymmetric under particle interchange, and therefore represents a
bosonic rather than a fermionic wavefunction~\footnote{It has indeed
  been shown~\cite{Haldane} that such a wavefunction is extremely good
  for bosons at $\nu=\frac{1}{2}$.}.  What one wants to do to obtain a
good fermionic $\nu=\frac{1}{2m}$ wavefunction is then to alter this
bosonic wavefunction slightly to make it antisymmetric while doing
``minimal damage'' to the good correlation structure of the Laughlin
state~\cite{NickRead}.  To this end one considers multiplying this
bosonic wavefunction by the determinant
\begin{equation}
  \label{eq:Det}
    \mbox{Det}\left[ e^{i \vec k_i \cdot \vec r_j} \right] = \left|
    \begin{array}{ccc} 
e^{i \vec k_1 \cdot \vec r_1} &  e^{i \vec k_1 \cdot \vec r_2} &
\cdots \\ e^{i \vec k_2 \cdot \vec r_1}  & \ddots &  \\
\vdots & & 
\end{array}
\right|_.
\end{equation}  
To have an antisymmetric wavefunction, all of the $k_i$'s must be
different, but to keep the energy minimal~\footnote{For the composite
  fermion wavefunctions, there is an over-completeness which allows us
  to move the center of the Fermi sea to any value of $k$ and still
  have the exact same physical wavefunction. This so called $\bf
  K$-invariance, along with the reason why we choose to keep the
  $k_i$'s minimal to minimize the energy, will be discussed further
  below.}, we will choose the smallest possible values for all the
$k_i$'s so that the determinant (Eq. \ref{eq:Det}) becomes just a
filled Fermi sea, {\em i.e.\/}, the wavefunction of free fermions in zero
magnetic field.  Once this filled Fermi sea is multiplied by the
bosonic Laughlin state the end result is the wavefunction for the
$\nu=\frac{1}{2m}$ state, often called the
Rezayi-Read~\cite{RezayiRead} wavefunction, given by
\begin{equation}
  \label{eq:rezayiread}
 \Psi_{\frac{1}{2m}} =   {\cal P}_{\rm LLL}
  \mbox{Det}\left[ e^{i \vec k_i \cdot \vec r_j} \right]
\prod_{i < j} (z_i - z_j)^{2m} 
\prod_i e^{- \frac{1}{4} |z_i|^2/l_B^2} 
\end{equation}
(where $\vec r_i$ and $z_i$ represent the same coordinate in the
vector and complex notations respectively).  Here we have added the
projection operator ${\cal P}_{\rm LLL}$ out front to re-project this
wavefunction back to the lowest Landau level since the fermionic
determinant has some part in higher Landau levels.  Note that
Jain~\cite{Jain} might prefer to think of this as a fermionic wavefunction
that is ``composite fermionized'' by multiplying by the composite
fermionization factor~\cite{Jain} ({\em i.e.\/}, the bosonic Laughlin
wavefunction), which attaches $2m$ vortices to each electron.  It has
been shown that this Rezayi-Read wavefunction has very good overlap
with the exact ground state for small systems~\cite{RezayiRead,Jain}.

We can compare this procedure of ``minimally damaging'' a bosonic
wavefunction to obtain a fermionic wavefunction with the situation at
zero magnetic field for noninteracting (or weakly interacting)
particles.  In this case the ground state bosonic wavefunction is just
$\Psi(\vec r_1,\ldots,\vec r_n) = 1$.  To get the ground state
fermionic wavefunction, we ``minimally damage'' this good ground state
bosonic state by multiplying by the determinant factor (Eq. \ref{eq:Det})
which makes the wavefunction antisymmetric if all the $k_i$'s are
chosen different, and minimizes the energy if the the $k_i$'s take the
minimum possible value filling a Fermi sea.

It is interesting to consider the effects of the multiplication by the
projected determinant.  Rewriting $\vec r$ in its complex coordinates,
we have $\vec k \cdot \vec r = \frac{1}{2} (k \bar z + \bar k z)$
where the overbar means complex conjugation and $k$ is the complex
coordinate representation for the vector $\vec k$.  Projection to the
lowest Landau level can be achieved by\cite{GirvinJach} replacing
$\bar z$ by $2 l_0^2 \frac{d}{dz}$ where all derivatives are normal
ordered to the left and the derivatives do not act on the exponential
factors $\exp(-\frac{1}{4}|z|^2/l_0^2)$.  The wavefunction can then be
rewritten as
\begin{equation}
 \Psi_{\frac{1}{2m}} =   
  \left[ \prod_i e^ { -\frac{1}{4}
  |z_i|^2/l_B^2} \right]
   \mbox{Det}\left[ e^{i \bar k_j z_k +  i l_B^2 k_j \frac{d}{dz_k} } \right]
\prod_{i < j} (z_i - z_j)^{2m}
\end{equation}
The operator $\exp{(i l_0^2 k \frac{d}{dz})}$ is then a shift operator
for the position $z$ which takes $z \rightarrow z + i l_0^2 k$.  Thus,
the wavefunction can be rewritten as\cite{Haldane,NickRead}
\begin{equation}
  \label{eq:deltaz}
 \Psi_{\frac{1}{2m}} =    {\cal A}
  \left[\prod_i e^{ -\frac{1}{4} |z_i|^2/l_B^2 + i \bar k_i z_i} \right]
\prod_{i < j} ([z_i + i l_B^2 k_i] - [z_j + i l_B^2 k_j])^{2m}
\end{equation}
where ${\cal A}$ is the antisymmetrizing operator that sums over all
possible pairings of the $z_i$'s with the $k_i$'s with odd
permutations added with a minus sign.  It is now clear that the effect
o f the fermionic determinant is to move the position of the zeros of
the wavefunction away from the location of the electrons by a distance
$l_B^2 k$ which is given in terms of the ``momentum'' $k$.  In order
to minimize the Coulomb energy, the distances $l_B^2 k_j$ should be
minimized to keep the zeros of the wavefunction as close to the
electrons as possible.  Furthermore, in order for the wavefunction to
be nonzero after antisymmetrization, all of the $k_j$'s must be
different so the lowest energy configuration should be a filled Fermi
sea.  What is interesting here is that the size of the Fermi sea must
be minimized in order to minimize the potential energy, whereas for
the free electron gas, the Fermi sea is minimized so as to minimize
the kinetic energy.  Surprisingly, boosting the entire Fermi sea via
$\vec k_i \rightarrow \vec k_i + \vec K$ leaves the energy of the
wavefunction (Eq. \ref{eq:deltaz}) completley unchanged! (the net effect
of the transformation is to multiply the wavefunction by $\exp (i \bar
K \sum_j z_j)$ which simply changes the center of mass wavefunction
while keeping the center of mass of the system in the lowest Landau
level thus costing no energy).  This is quite unlike the case of
noninteracting fermions.  This invariance which we will call $\bf
K$-invariance, will be extremely important below.  We will see in section
\ref{sub:dipole} below that preserving this symmetry is essential in
understanding the behaviour of the composite Fermion fermi liquid in
the wavefunction approach.  In terms of a Fermi-Liquid theory, the
statement that it costs no energy to boost the Fermi-sea is equivalent
to saying that there is a Landau-Fermi liquid coefficient $F_1 = m^*
f_1/(2 \pi \hbar^2) = -1$ at least in the long wavelength low
frequency limit (See Eq. \ref{eq:membrane}).  We will return to this
issue in section \ref{sub:dipole}.

The picture that we now have of the composite fermion at
$\nu=\frac{1}{2m}$ is that of an electron ``bound'' to an even number
($2m$) of zeros of the wavefunction, which create its positively charged
correlation-hole.  The charge on the electron is of course $-1$
whereas there is a charge of $+\nu$ for each zero of the wavefunction,
thus resulting in a total charge of $2 m \nu -1 = 0$, {\em i.e.\/}, a neutral
object.  The zeros of the wavefunction are then boosted away a
distance $l_B^2 |k|$ from the precise location of the electron by
the fermionic determinant which makes the wavefunction antisymmetric,
resulting in a dipole moment of the electron-correlation-hole pair.
The Coulomb attraction between the electron and these zeros is some
potential $V(l_B^2 |k|)$.  As usual in a high magnetic field, a
force on a particle results in a drift motion of the particle
perpendicular to the force and proportional to the gradient of the
potential given by~\footnote{Again, the drift velocity is such that the
  total force $\nabla V + \frac{q}{c} \vec v \times \vec B$ vanishes.}
\begin{equation}
  \label{eq:drift}
  v_{\rm drift} =   \frac{c \nabla V \times \vec B}{q |\vec B|^2}
\end{equation}
where $q$ is the charge of the particle in question.  Although the
force on the correlation hole and the force on the electron are
oppositely directed (directed towards each other) their drift
velocities are in the same direction since the charge on the electron
is opposite that of the hole.  Thus, the electron-correlation-hole
pair drift in a direction perpendicular to their separation while
maintaining a fixed separation.  This is very reminiscent of the
advective motion of a vortex-antivortex pair in fluid dynamics.  Since
the velocity of the pair is determined by the gradient of the
potential $dV(l_B^2 |k|)/dk$, we can think of $d^2V(l_B^2
|k|)/dk^2 \sim 1/m^*$ as determining the effective mass of the
pair~\cite{NickRead}.  Note that this effective mass is determined
entirely by the interaction strength between the electron and
correlation-hole.

\subsection{Excitations and Small $\Delta B$}
\label{sub:smallw}

This trial ground state wavefunction can very naturally be extended to
describe excited states.  One simply chooses to replace the Fermi sea
in Eq. \ref{eq:rezayiread} by an excited Fermi sea, by promoting one
of the $k$-vectors from a value below the Fermi level to a value above
the Fermi level.  It has been shown numerically that these trial
excited states have very good overlaps with the exact excited states
of the system~\cite{Jain0,Jain,Haldane}.  

We can also consider what happens to such an excitation when the
magnetic field is not quite quite at $\nu=\frac{1}{2m}$.  Here, the
total charge on the electron-correlation-hole bound state is given by
$e^* = -1 + 2m \nu$ which is no longer zero owing to the fact that
$\nu$ is no longer quite $\frac{1}{2m}$.  We might then expect such a
charged object in a magnetic field to undergo cyclotron motion.
Indeed this will turn out to be the case~\cite{NickRead}.  We once
again consider the structure of the bound state of the electron and
the correlation hole of $2m$ zeros of the wavefunction.  Since the
charges on these two objects are no longer precisely opposite, the two
drift velocities (Eq. \ref{eq:drift}) are slightly unequal with the
electron velocity differing from the correlation-hole velocity by a
factor of $q_{\rm hole}/q_{\rm electron} = (2m\nu)/(-1)$.  Because of this
inequality in velocities, the pair will turn slowly as they move
eventually completing a circle with the faster particle moving around
the outside and the slower particle moving on the
inside~\footnote{Imagine a car whose left wheels move faster than its
  right.}.  The ratio of radii of motion of these two particles
($R_{\rm electron}/R_{\rm hole}$) is given by this same ratio of velocities
(or ratio of charges), $2m\nu$.  The difference in radii, on the
other hand, is given by the distance between the electron and its
correlation-hole.  Since we are interested in excitations near the
Fermi surface, we can set $k \approx \kf$, and the distance between
the electron and its screening vortices is given by $R_{\rm electron} -
R_{\rm hole} = 2 \l_B^2 \kf = l_B/\sqrt{m}$.  With this information we can
solve (with some algebra) to find that the radius of the cyclotron
motion is (for $\nu$ near $\frac{1}{2m}$)
\begin{equation}
 R_{\rm electron} \approx R_{\rm hole}  \approx  \frac{\kf \hbar c}
{B e^*} = \frac{\kf \hbar c}{e B (1 - 2m\nu)} =  \frac{\kf \hbar c}{e \Delta B}
\end{equation}
The first form of this expression is what one might have expected in
the first place --- an object (the bound electron-correlation-hole
pair or composite fermion) of charge $e^* = e (1 - 2 m \nu)$ should have
a cyclotron radius given by $\kf \hbar c/ (e^* B)$.  On the other hand,
this result can also be equivalently rewritten in the form of a
particle of charge $e$ in a reduced magnetic field $\Delta B$ so that
it agrees with the Chern-Simons result in Eq.  \ref{eq:Rc}.

\subsection{Jain's Wavefunctions}
\label{sub:jainw}

We would like to write down wavefunctions analogous to Eq.
\ref{eq:rezayiread} for quantized Hall states that form near
$\nu=\frac{1}{2m}$.  To describe a wavefunction at an arbitrary
filling fraction $\nu$, we once again begin with the bosonic Laughlin
state at filling fraction $\frac{1}{2m}$ whose correlations are
expected to be good, and we would like to ``minimally damage'' this
wavefunction to make it an appropriate wavefunction for the new
filling fraction.  Here, we not only need to make the wavefunction
antisymmetric, but we must also add enough zeros to the wavefunction
to change the filling fraction from $\frac{1}{2m}$ to the target
filling fraction $\nu$.  To this end, we consider the generalization
of Eq. \ref{eq:rezayiread} constructed by replacing the fermionic
determinant of plane wave states representing fermions in zero
effective field ($\nu=\frac{1}{2m}$) with a wavefunction $\Phi_p$ for
fermions in a finite effective field~\footnote{As discussed above, we
  could equivalently think of this as free fermions of charge $1 - 2 m
  \nu$ in the full magnetic field $B$.  The effective filling factor
  $p$ will remain the same.}, Thus we obtain a wavefunction of the
form
\begin{equation}
  \label{eq:ization2}
 \Psi_{\frac{p}{2mp+1}} =   {\cal P}_{\rm LLL}
  \Phi_{p}(\vec r_1,\vec r_2,\ldots,\vec r_N)
\prod_{i < j} (z_i - z_j)^{2m}
\prod_i e^{- \frac{1}{4} |z_i|^2/l_{B_{1/2m}}^2} 
\end{equation}
where the wavefunction $\Phi_p$ is the wavefunction for free fermions
at filling fraction $p$.  Certainly, the resulting wavefunction is
antisymmetric since the bosonic Laughlin state is symmetric and
$\Phi_p$ is antisymmetric.  Furthermore, it can be shown~\cite{Jain}
that the multiplication of the bosonic Laughlin state by the
wavefunction ${\cal P}_{\rm LLL} \Phi_p$ adds just enough zeros of the
wavefunction to change the filling fraction from $\nu=\frac{1}{2m}$ to
$\nu=\frac{p}{2mp+1}$.  This should be easy to believe since we know
that wavefunctions should have $B/\phi_0$ zeros per unit area.  Thus
multiplying a wavefunction corresponding to a field $B_1$ with a
wavefunction corresponding to a field $B_2$ results in a wavefunction
corresponding to a field $B_1 + B_2$.  Here, we have the bosonic
Laughlin state corresponding to a field $B_{1/2m} = \frac{n_{\rm e}
  \phi_0}{2m}$ and the composite fermion wavefunction $\Phi_p$
corresponding to a field $\Delta B = n_{\rm e} \phi_0 p$ which then add to
yield a filling fraction $\nu = \frac{n_{\rm e} \phi_0}{B + \Delta B} =
\frac{p}{2mp+1}$.  (The issue of projection makes the argument
slightly more complicated than this simplified version).

Note also that the magnetic length in the exponential is taken to be
the magnetic length at filling fraction $\nu=\frac{1}{2m}$.  There
will in general be another exponential factor within the wavefunction
$\Phi_p$ with magnetic length $l_{\Delta B}$.  These two will combine
together to yield an exponential with the proper magnetic length for a
system in field $B = B_{1/{2m}} + \Delta B$ (since $l_B^{-2} =
l_{B_{1/(2m)}}^{-2} +l_{\Delta B}^{-2}$).

The prescription outlined here for constructing wavefunctions at
filling fraction $\nu =\frac{p}{2mp+1}$ is of course precisely that
first proposed by Jain~\cite{Jain}.  It should be noted that, for the
sake of pedagogy, we have arrived at the Jain wavefunctions only after
a discussion of the $\nu=\frac{1}{2m}$ state; whereas historically,
Jain's wavefunction was introduced well before there was even
interest in the even denominator states.  It should further be noted
that Rezayi-Read wavefunction is clearly just the $p \rightarrow
\infty$ limit of the Jain wavefunction.

For $p$ an integer, the wavefunction $\Phi_p$ can be taken as a
trivial fermionic determinant of $p$ filled Landau levels.  As an
example, we consider the $p = 1$ case, corresponding to
$\nu=\frac{1}{2m+1}$.  Here we have
\begin{eqnarray}
\label{eq:recoverlaughlin}
\Psi_{\frac{1}{2m+1}}(z_1,z_2,\ldots,z_N) &=& 
\left[ \prod_{i < j} (z_i - z_j)  
  \prod_i e^{-\frac{1}{4} |z_i|^2/l_{\Delta B}^2 } \right]\times \nonumber \\
 & &  \left[ \prod_{i < j} (z_i
  - z_j)^{2m}  
  \prod_i e^{-\frac{1}{4} |z_i|^2/l_{B_{1/2m}}^2 } \right] 
\end{eqnarray}
The first factor is just the wavefunction for a single filled Landau
level $\Phi_{p=1}$.  The second factor (with the power $2m$) is the
composite-fermionization factor that attaches $2m$ zeros of the
wavefunction to each electron.  Thus, the $\nu=\frac{1}{2m+1}$ state
is written as a composite fermionized $\nu=1$ state.  As discussed
above, the two exponential factors will combine to give the proper
exponential factor for the full magnetic field, and we will be left
with simply the Laughlin wavefunction (Eq. \ref{eq:Laughlin}).  We
note that the composite fermion quasiparticle --- the electron plus
its vortex screening cloud of $2m$ extra zeros --- has a net charge of
$e^* = -1 + 2m \nu$ which is the correct quasiparticle charge,
$-\frac{1}{2m+1}$ for the quasiparticle of the Laughlin state.

In this case of $p=1$, where we recover the Laughlin wavefunction, the
projection is not needed here since the wavefunction $\Phi_{p=1}$ is
already in the lowest Landau level.  For $p$ not equal to $1$, we will
need to project the resulting wavefunction back down to the lowest
Landau level, and the final form of the wavefunction will (after
projection) be quite complex (See for example, Ref.
\onlinecite{Kamilla}).  Nonetheless, it has been shown numerically
that such a prescription results in a trial wavefunction with
extremely high overlap with the exact ground state for small
systems~\cite{Jain0,Jain}.

The composite fermionization described by Eq. \ref{eq:ization2} or Eq.
\ref{eq:rezayiread}, can be thought of in several equivalent ways.  On
the one hand, we can think of the composite-fermionization as
adiabatically inserting $2m$ flux quanta at the position of each
electron.  Equivalently we can think of this factor as the binding of
$2m$ positively charged zeros (vortices) to each electron.  The good
energetics of these composite fermion states can be thought of in
terms of the large binding energy~\cite{MorfanddAmbruminil} of these
electron-correlation-hole objects (which are, of course, the composite
fermions themselves).

\subsection{Wavefunctions vs. Chern-Simons Theory}
\label{sub:compare}

The idea of the composite fermionization factor being equivalent to
adiabatically inserting $2m$ flux quanta at the position of each
electron is certainly quite reminiscent of the Chern-Simons
transformation described in section \ref{sub:CS} above.  To make the
relation more clear we rewrite the Chern-Simons transformation
(Eq. \ref{eq:noanal1}) in analytic coordinates as
\begin{equation}
  \label{eq:noanal2}
  \Psi_{\rm e}(\vec r_1,\vec r_2,\ldots,\vec r_N) = 
 \left[ \prod_{i<j}  \left( \frac{z_i - z_j}{|z_i - z_j|} \right)^{2m}
 \right] \Phi(\vec r_1,\vec r_2,\ldots,\vec r_N). 
\end{equation}
If we consider the $\nu=\frac{1}{2m}$ state, at the mean field level,
the composite fermion wavefunction $\Phi$ is just a filled Fermi sea
(Eq. \ref{eq:Det}).  We then see that the Chern-Simons transformed
wavefunction $\Psi_{\rm e}$ looks very similar to the Rezayi-Read trial
wavefunction (Eq. \ref{eq:rezayiread}).  As with the Rezayi-Read
wavefunction, an extra phase factor of $e^{2m\pi i}$ is included when
one electron wraps around another.  However, in the Chern-Simons
transformation, there is a non-analytic factor $|z_i - z_j|$ in the
denominator which indicates that the wavefunction (Eq.
\ref{eq:noanal2}) is not properly in the lowest Landau level.
However, we must remember that choosing $\Phi$ to be the filled Fermi
sea is just the mean field result of the Chern-Simons theory.  When
the leading fluctuations around mean field are accounted for, it can
be shown that the non-analytic factor in the denominator is canceled,
at least when one is only concerned with the long distance
physics~\cite{Lopez3,Shankar,Zhang}.  However, even after this
cancellation, the factor $\Phi$ may retain pieces outside of the
lowest Landau level.  Obtaining a fully projected Fermi sea within a
Chern-Simons approach remains an open problem.

Another interesting difference between the wavefunction approach and
the Chern-Simons approach is that in the wavefunction approach, the
elementary excitation, the bound electron-correlation-hole has charge
$(1-2 m \nu) e$ with a dipole moment whereas in the Chern-Simons
approach the transformed fermion has charge $e$ with no obvious dipole
moment.  However, as discussed above, many of the important physically
relevant parameters like the cyclotron radius and the effective
filling fraction away from $\nu=\frac{1}{2m}$ turn out the same in
either approach.  

Finally, we note that in the wavefunction approach, there is no
Chern-Simons gauge field.  Thus, one might wonder if the unusual
infrared problems of the gauge-field propagator might be absent in the
dipole-fermion picture of the composite fermion.  We will see below
that these infra-red problems re-appear in a slightly different guise.

\subsection{Response of Neutral Dipole Composite Fermions}
\label{sub:dipole}

Several recent approaches have attempted to bring these two approaches
into more of an agreement.  D.-H. Lee~\cite{DHLee} has constructed a
Chern-Simons theory where particles and holes bind together to form
neutral fermions in a manner more reminiscent of the above
wavefunction arguments. Work by Shankar and Murthy~\cite{Shankar}
separates out the inter-Landau-level excitations and results in a
field theory that is somehow ``projected'' to the lowest Landau level.
Finally, Haldane and Pasquier~\cite{Pasquier} have used a lowest
Landau level algebraic approach.  These three approaches  all seem to
be converging on a coherent picture of a Fermi sea of
neutral dipoles at $\nu = \frac{1}{2}$. The neutral-dipole
Hamiltonians obtained by all of these groups appear to be equivalent
(at least as long as one is only concerned with the long distance
physics).  In this section, we will show (at least roughly) how these
approaches lead to the same results as that of the Chern-Simons
approach.

One natural question to ask is how can we have any DC current if all
of the particles in the system are neutral?  As we discussed in
section \ref{sub:IQHE} above, by Galilean invariance, any clean system
must have a DC Hall conductivity $\sigma_{xy} =\nu \frac{e^2}{h}$, so
we know that current must be carried somehow.  This strange
contradiction is a result of having restricted our attention to the
lowest Landau level~\cite{Sondhi}, which by itself is not Galilean
invariant.  Application of an electric field to the system mixes
Landau levels (no matter how large $\omega_{\rm c}$ is) and it is precisely
these higher Landau level pieces that carry the current~\footnote{It
  may seem strange that we can mix Landau levels even for large
  $\omega_{\rm c}$.  Indeed, as we take $m_{\rm b} \rightarrow 0$ such that
  $\omega_{\rm c} \rightarrow \infty$, the amount of mixing gets small.
  However, the current operator $\vec j = \vec p/m_{\rm b}$ has a factor of
  $m_{\rm b}^{-1}$ which compensates so that the current due to this mixing
  remains finite.}.  Thus, in order to obtain the proper Hall
conductivity, we must include these Landau level excitations, which
complicates matters (see Ref.~\onlinecite{Shankar} for a discussion of this
physics).  Here, we will instead focus on the response $K_{00}(\vec
q,\omega)$ at frequencies well below the cyclotron frequency which can
be obtained without consideration of these higher energy excitations.

We will now give a slightly hand-waving derivation of the electronic
response in the neutral dipole picture.  Note that we will call these
fermionic objects ``dipole fermions'' even at filling fractions away
from $\nu=\frac{1}{2m}$ when they have a charge as well as a dipole
moment.

\subsubsection{Dipole Fermion Variables}

Inspired by the above discussion, we write the electron density
($\rho$) in the form~\cite{Shankar}
\begin{equation}
  \label{eq:densitylong}
  \rho = e^* \rho^{\rm d} - i l_B^2  \nabla \times \vec g
\end{equation}
where 
$\rho^{\rm d}(\vec r) = \sum_i \delta(\vec r - \vec r_i)$ 
is the dipole fermion density and
\begin{equation}
\vec g(\vec r) = \sum_i \left[ \vec p_i + \frac{e^*}{c} \vec A(\vec r_i)
\right]  \delta(\vec r - \vec r_i)
\end{equation}
is the dipole fermion momentum density (We use the notation $\vec g$
rather than writing this as a current $\vec j$ because we do not want
the factor of mass to enter here).  Here, $\vec r_i$ and $\vec p_i$
are the position and momentum of the $i^{th}$ dipole fermion (which
are assumed to obey canonical commutation relations).  The first term
in Eq.  \ref{eq:densitylong} is just the density of these fermions
times their net charge $e^* = -1 + 2 m \nu$ which, of course, vanishes
at filling fraction $\frac{1}{2m}$ where the fermions become neutral
dipoles. The second term is the dipolar term that gives each fermion a
dipole moment perpendicular to its momentum as suggested by the above
discussion.  In that term $e^* \vec A = e \Delta \vec A$ is the vector
potential associated with the field difference from the
$\nu=\frac{1}{2m}$ denominator state.

We note that this form of the electron density has the
appealing feature that, in the long wavelength limit, it satisfies the
proper commutation relations for a density in the lowest Landau
level~\cite{Shankar} given by~\cite{GirvinJach}
\begin{equation}
  [\rho(\vec q),\rho(\vec q')] = i l_B^2 (\vec q \times \vec q')
  \rho(\vec q + \vec q') 
\end{equation}
where we have expanded for small $q$ here.  Since this representation
of $\rho$ satisfies the proper commutation, it is completely
acceptable to work in terms of these dipole fermion variables (at
least for small $q$).  However, the dipole variables have 2 degrees of
freedom per particle whereas the lowest Landau level has only one.
There must therefore be constraints on these degrees of freedom.
These constraints have been written down but it is not at all clear
how to properly handle them in a systematic
way~\cite{Shankar,Pasquier,DHLee} which makes performing a completely
controlled calculation difficult.  The simplest thing we can do is to
neglect the constraint entirely which we will do here. 

For simplicity, we will now focus on filling fraction $\frac{1}{2m}$
where the density becomes
\begin{equation}
  \label{eq:density}
  \rho(\vec q)  =  -i
  \frac{\nu}{2 \pi n}  q g_1(\vec q) = -i l_B^2 \sum_i (\vec q \times
  \vec p_i) e^{-i \vec q \cdot \vec r_i}
\end{equation}
with $g_1 = \hatn q \times \vec g$ the transverse part of the momentum
current of the dipole fermions.  Here, the dipole fermions see no
effective magnetic field (since they are neutral).  For a discussion
of $\nu \ne \frac{1}{2m}$ see Refs.~\onlinecite{Shankar} and
\onlinecite{Adylongdipole}.  We note here that we have $\rho \sim
\nabla \times \vec g$ which appears quite similar to the Chern-Simons
case $\rho \sim \nabla \times \vec a$ with $\vec a$ the Chern-Simons
vector potential.  These two actually act in quite similar ways as we
will see below.

The form of Eq. \ref{eq:density} has been explicitly derived by
several groups~\cite{Shankar,Pasquier,DHLee}.  Here, however, we have
neglected the contribution of high energy ``magnetoplasmon''
oscillations (inter Landau-level excitations).  As mentioned above,
several properties of the system, such as the Hall conductivity,
cannot be correctly obtained without considering the contributions of
these high energy excitations~\cite{Shankar,Sondhi}.  However, for
calculating the response $K_{00}$, we can stay completely in the
lowest Landau level.

\subsubsection{Response Functions}

We now wish to calculate the density-density response function
$K_{00}$ for electrons.  From linear response theory, we know that
$K_{00}(\vec q, \omega)$ is given by the Fourier transform of the time
ordered correlator  (See Eqs. \ref{eq:currentcorrelator0} and
\ref{eq:K0D0} with  $E_{00} = 0$)
\begin{equation} 
\label{eq:Kdef22}
K_{00}(\vec r, t) = \langle T \rho(\vec r,t) \rho(0,0) \rangle
\end{equation}

Using Eqs. \ref{eq:density} and \ref{eq:Kdef22}, we
relate this to a correlator for the dipole fermion current
\begin{equation}
    K_{00}(\vec q,\omega) =  
\left( \frac{\nu q}{2 \pi n}  \right)^2 D^{\rm d}_{11}(\vec
    q,\omega)
  \label{eq:KD}
\end{equation}
where we have defined $D^{\rm d}_{11}(\vec q,\omega)$ to be the Fourier
transform of the time ordered transverse momentum correlator of the
dipole fermions
\begin{equation}
  \label{eq:densitycorrelator}
D^{\rm d}_{11}(\vec r, t) = \langle T g_1(\vec r,t) g_1(0,0) \rangle
\end{equation}
(Again, we have the notation that $g_1$ is the transverse part of
$\vec g$.  Also note that the usually the current correlator [See Eq.
\ref{eq:currentcorrelator0}] is defined with respect to the current
$\vec j$ rather than the momentum current $\vec g$ which differ by
factors of mass).  We note that here, this dipole correlator is
precisely equivalent to the transverse gauge field propagator ${\cal
  D}_{11}$ we found in Eq.  \ref{eq:gaugefieldprop} (compare Eq.
\ref{eq:KD}).  Thus, our choice of notation of using $D^{\rm d}$ for the
dipole current correlator, and ${\cal D}$ as the gauge field
correlator is conveniently chosen to be suggestive that they play the
same role.

Analogous to the usual separation of the response $K_{00}$, into a
polarization and a Hartree long range interaction, we now define the
correlator $\tilde D^{\rm d}_{11}$, which is the part of $D^{\rm d}_{11}$ which is
irreducible with respect to the long range Coulomb interaction between
the dipoles.  Being that the dipole-dipole interaction is
written as
\begin{eqnarray} \label{eq:dint}
  H_{\rm int} &=&  \sum_q v(q) \rho^e(\vec q) \rho^e(-\vec q) \\
  &=& \sum_q v(q) 
\left( \frac{q \nu }{2 \pi
    n} \right)^2 g_1(\vec q) g_1(-\vec q)
\end{eqnarray}
we define the irreducible correlator via
\begin{equation} \label{eq:tildeD}
    [D^{\rm d}_{11}]^{-1}  
= [\tilde D^{\rm d}_{11}]^{-1} - v(q) \left( \frac{q \nu}{2 \pi
    n} \right)^2
\end{equation}

More generally, these correlators should be treated as matrices
$D^{\rm d}_{\mu \nu}$.  However, at filling fractions
$\nu=\frac{1}{2m}$ these matrices are all diagonal and we can consider
one component at a time~\footnote{So long as we continue to ignore all
  constraints.} (see Ref.~\onlinecite{Adylongdipole} for $\nu \ne
\frac{1}{2m}$).  We note that this Hartree separation of the dipole
response can be compared to Eqs.  \ref{eq:K2sep} and \ref{eq:KD} to
show that the electronic polarization is given by
\begin{equation}
  \label{eq:polD}
  \Pi_{00}^v(\vec q,\omega) 
  =    \left( \frac{\nu q}{2 \pi n}  \right)^2 \tilde D^{\rm d}_{11}
(\vec q,\omega). 
\end{equation}
In other words, one can equivalently treat the Coulomb interaction in
terms of its action on the original electrons or in terms of its
action on the dipole fermions.  It is interesting to note that
although the Coulomb interaction looks very weak when written in terms
of the dipoles, it is equally important independent of which variables
it is written in terms of. 

We note that the importance of the irreducible correlator $\tilde D^{\rm d}$
is that at the RPA level (as in a Landau-Silin theory) we will be able
to use a fully local Hamiltonian to calculate this quantity while
neglecting the long range part of the interaction.

\subsubsection{$\bf K$-invariance}

One might at this point be tempted to naively approximate the dipole
fermions as noninteracting fermions~\footnote{Such an approximation has
  been made in various early versions of 
  Refs.~\onlinecite{Shankar,Pasquier,DHLee}.  See 
  Refs.~\onlinecite{Adydipole} and~\onlinecite{Adylongdipole} for a detailed
  discussion of how such an approximation fails.}.  In the long
wavelength, low frequency limit, the correlator $D^0_{11}$ for
noninteracting fermions is a constant (See Eq. \ref{eq:D0} below).
Approximating either $D_{11}$ (or more properly $\tilde D_{11}$) as a
constant ($D^0_{11}$) would lead us to naively believe (Using Eq.
\ref{eq:KD}) that the electron response is given by $K_{00} \sim q^2$
which differs drasticly from the Chern-Simons RPA results obtained in
section \ref{sec:RPA} and \ref{sec:FLT} above.  The apparent reason
for this ``incompressibility'' is that the external fields couple very
weakly to the dipole particles~\footnote{A quantized Hall state has
  $K_{00} \sim q^4$ whereas a compressible state has $K_{00} \sim
  1/v(q)$ whenever $v(q)$ diverges for small $q$ and $K_{00} \sim$
  constant otherwise.}  This apparent contradiction with the result of
Chern-Simons theory has been the source of much confusion in the
community.

The error in this naive approximation is that, as discussed in section
\ref{sub:wf2} above, the dipoles represented by the Rezayi-Read
$\nu=\frac{1}{2}$ wavefunction form a very unusual Fermi liquid that
is invariant under the transformation $\vec p_i \rightarrow \vec p_i +
\vec K$ which corresponds to a boost of the Fermi sea.  In Ref.
\onlinecite{Adylongdipole} it is shown that within the formalism of Shankar
and Murthy, this so-called $\bf K$-invariance is actually a gauge
symmetry.  Since it costs no energy to boost the system uniformly, we
will find that the current response $D^{\rm d}_{11}$ diverges in the long
wavelength limit canceling powers of $q$ in Eq. \ref{eq:KD} (or Eq.
\ref{eq:polD}) so that the system remains compressible ($\Pi_{00}
\sim$ constant or $K_{00} \sim 1/v(q)$) in agreement with the
predictions of the Chern-Simons theory.

Approximating the dipole particles as noninteracting fermions clearly
violates this $\bf K$-invariance since it certainly costs kinetic
energy to boost an entire Fermi sea of noninteracting fermions.  We
again turn to Fermi liquid theory for a proper description of this
deviant Fermi liquid.  In the Landau-Silin sense, we have already
separated out the long range part of the interaction in Eq.
\ref{eq:tildeD} and we should now describe the response $\tilde D^{\rm d}$
(analogous to $\Pi$) in terms of a Landau Fermi liquid theory with
only local interactions (this is all discussed in depth in section
\ref{sec:FLT} above).  Thus, in a Landau Fermi liquid description of
these dipole fermions, we should think of the dipoles as having some
mass $m^*$ and some Landau interaction parameters $F_n$ (See section
\ref{sec:FLT} above).  In order to guarantee that it costs no energy
to boost the whole Fermi sea all we need to do is to force $F_1 = -1$
(See Eq. \ref{eq:membrane}) at least in the long wavelength
limit~\footnote{Examining Eq. \ref{eq:F1part2} we see that for fixed
  effective mass $m^*$, the value $F_1 = -1$ implies that $m_{\rm b}
  \rightarrow \infty$.  This is clearly not the true bare mass of
  the problem (which we have taken to zero!).  This ``effective''
  infinite bare mass makes sense because it should cost zero energy to
  boost the Fermi sea by momentum $\vec K$ yielding an energy cost
  $\vec K^2/(2 m_{\rm b})$.  We note that this divergence of the
  ``effective'' bare mass is the reason we are working with the
  momentum density $\vec g$ rather than the current $\vec j = \vec
  g/m_{\rm b}$.}.  

To see how the value of the Landau parameter $F_1 = -1$ preserves this
$\bf K$-symmetry, we note that the $F_1$ interaction is a
current-current interaction (See Eqs. \ref{eq:membrane},
\ref{eq:currentfourx0} and \ref{eq:currentfoury0} above).  A local
Hamiltonian representing fermions with mass $m^*$ interacting via an
$F_1$ interaction is written as
\begin{equation}
  \label{eq:dHlocal}
   H_{\rm local} = \sum_i \frac{\vec p_i^2}{2m^*} + \frac{F_1}{2
   n m^*} \int d\vec r |\vec g(\vec r)|^2
\end{equation}
where $\vec g$ is the momentum density (Note that this differs
slightly from the discussion above in section \ref{sec:FLT} because
factors of the mass are left out of $\vec g$ whereas
conventionally we put these factors in and call the current $\vec
j$).  For the value of $F_1 = -1$, this local Hamiltonian is
equivalent to the form proposed by Haldane~\cite{Haldane}, 
\begin{equation}
    H_{\rm local}^{(F_1 = -1)} = \frac{1}{2m^*} \sum_i^N \left( \vec p_i - \frac{1}{N} 
\sum_j^N \vec p_j \right)^2
\end{equation}
which is clearly $\bf K$-invariant.  The physical meaning of this
expression is that the momentum of each fermion is measured with
respect to the average momentum.

\subsubsection{Dipole RPA}

We would now like to apply the RPA to this effective local interaction
to obtain a physical response function.  With the experience gained in
section \ref{sec:RPA} and \ref{sec:FLT} above we know quite well how
to treat the momentum-momentum interaction in the local dipole
Hamiltonian (Eq. \ref{eq:dHlocal}).  We separate out the interaction
term and define an irreducible response $D^{{\rm irr}}_{11}$ given by
\begin{equation} 
  \label{eq:DD}
  [\tilde D^{\rm d}_{11}]^{-1} = [\tilde D^{{\rm irr}}_{11}]^{-1} + \frac{F_1}{n m^*}
\end{equation}
Note that this separation of $F_1$ is equivalent to
Eq. \ref{eq:tf1t}.   

The RPA approximation is now just the statement that $\tilde D^{{\rm
    irr}}_{11}$ should be approximated as the momentum correlator of
noninteracting fermions.  We now recall that the transverse momentum
correlator $D^0_{11}$ is closely related to the a response function
$K^0_{11}$ of free fermions (See Eq. \ref{eq:K0D0}).  Note that here
$D^0$ is a correlator of momentum current rather than charge current
so there are two extra factors of mass compared to Eq. \ref{eq:K0D0}.
As discussed in section \ref{sub:diag} above, for noninteracting
fermions of mass $m^*$ we have the relation (accounting for the extra
mass factors)
\begin{equation}
\label{eq:K0*}
 K^{0*}_{11}(\vec q,\omega) = \frac{1}{(m^*)^2} D^{0*}_{11}(\vec q,\omega) - \frac{n}{m^*}_.
\end{equation}
The ${}^*$ again indicates the mass is renormalized.  Using the low
frequency form of $K^0$ given in Appendix \ref{app:non}, yields the
momentum correlator for noninteracting fermions of mass $m^*$,
\begin{equation}
\label{eq:D0}
  D^{0*}_{11} (\vec q,\omega) = n m^*  + \frac{q^2 m^*}{24 \pi} + i \frac{ 2 n
  \omega (m^*)^2}{\kf q} + \ldots
\end{equation}

We now make the RPA and approximate the irreducible correlator $\tilde
D^{{\rm irr}}$ as $D^{0*}$.  It is critical to note that in Eq.
\ref{eq:DD} the leading term in $D^{0*}$ cancels the interaction term
for $F_1 = -1$.  This precise cancellation leads to $\tilde D^{\rm
  d}_{11} \sim q^{-2}$ in the low frequency limit which is singular
and cancels the powers of $q$ in Eq.  \ref{eq:polD} resulting in a
polarization $\Pi^v_{00}$ which is constant in the low frequency long
wavelength limit representing a compressible system as expected.
Carrying out this calculation explicitly using Eqs. \ref{eq:D0},
\ref{eq:DD}, \ref{eq:tildeD} and \ref{eq:KD} yields the low frequency,
long wavelength response
\begin{equation}
  K_{00} =   
  \frac{q^2}{ \left( \frac{2 \pi}{\nu} \right)^2 \left[ \frac{q^2}{24 \pi m^*} 
 - i\frac{\omega \kf
      }{2 \pi q}
  \right]  + q^2 v(q) }
\end{equation}
At higher frequencies (but still well below the cyclotron energy) we
can write our result in terms of the response $K^{0*}$
of noninteracting fermions of mass $m^*$  (using Eq. \ref{eq:K0*})
yielding 
\begin{equation}
  K_{00}  = \frac{1}{\left(\frac{2 \pi}{q \nu}\right)^2 \left[\left( 
  [K^{0*}_{11}]^{-1} + \frac{m^*}{n} \right)^{-1}  \right] +  v(q)} 
\end{equation}
We can compare this result to the result of the Modified RPA
calculation of Eq. \ref{eq:MRPAresult} in the limit of $m_{\rm b}
\rightarrow 0$ ({\em i.e.\/}, if we make the cyclotron energy very
large so we can focus on lowest Landau level behavior).  We see that
these results are quite similar although one finite $q$ correction
term related to $K^0_{00}$ seems to be present in the MRPA calculation
(or in HLR) which is not present in this dipole approach.  It is
believed that this correction term can also be obtained in this dipole
picture~\cite{Adylongdipole} to obtain complete agreement with the
MRPA by imposing the constraint that $\rho^{\rm d} = (q m^*/\omega)
g_x = \rho^e$, which then imposes a relation (via Eq.
\ref{eq:density}) between the transverse and longitudinal parts of
$\vec g$.  This allows the transverse current of $\vec g$ to couple to
the compression mode of the dipoles, hence producing the appropriate
factor of $K_{00}^0$.  However, the details of this are not yet fully
established (See Ref. \onlinecite{Adylongdipole}).

We note however that the correction term vanishes in the limit of
$q,\omega \rightarrow 0$, and whether or not we include it, we always
obtain a form of low frequency, long wavelength, electron response
given by
\begin{equation}
  K_{00} \sim \frac{1}{C_1 \frac{i \omega}{q^3} + [C_2 + v(q)] }
\end{equation}
that agrees with the form (Eq. \ref{eq:lowenergyRPA}) found by HLR
(with $C_1$ and $C_2$ dimensionful constants).  This form guarantees
the existence of the overdamped relaxation characteristic of the HLR
theory of the $\nu=\frac{1}{2}$ state.  This overdamped mode also
guarantees that we will encounter infra-red divergences when we try to
calculate perturbative corrections to quantities such as the fermion
effective mass. To see how this happens we note that the interaction
term $F_1$ (as well as the dipole interaction $q^2 v(q)$) provides our
theory with an interaction vertex with the transverse current.  When
we calculate perturbative corrections, we should use $D^{\rm d}_{11}$
as a transverse current propagator.  We can then calculate the self
energy of the dipole fermion by writing the diagram shown in Eq.
\ref{eq:sel} only using $D^{\rm d}_{11}$ as a propagator rather than
the Chern-Simons gauge field propagator ${\cal D}_{11}$.  {\it But
  these two propagators are precisely the same} (as required by Eq.
\ref{eq:KD} and \ref{eq:gaugefieldprop}).  Thus, we obtain the exact
same divergent diagrams as in the Chern-Simons approach.

In summary, we have found that the picture of the dipole Fermi liquid
seems to be equivalent to that of the Chern-Simons Fermi liquid.  The
picture we obtain from the two models are quite different, but at the
end of the day the physical predictions turn out to be the same.

\mysection{Selected Experiments}
\label{sec:exp}

One of the most exciting features of composite fermion physics is that
it is a rich world for experimentalists as well as theorists.
Literally hundreds of experiments have been performed in the composite
fermion regime.    For a more a thorough discussion of all of the
experiments, see Ref.~\onlinecite{WillettReview}, as well as the
chapters by Willett and Smet in this book.

\subsection{Surface Acoustic Waves}
\label{sub:saw}

The Surface Acoustic Wave experiments of Willett et al~\cite{Willett0}
were the first clear indication of the existence of a compressible
state at $\nu=\frac{1}{2}$.  The theory of Halperin, Lee, and Read was
then developed in close conjunction with the development of these
experiments~\cite{Willett1}.  Since these experiments have been
instrumental in the development of our understanding of
$\nu=\frac{1}{2}$ physics, we will give a brief description of how
this experiment measures the response $K_{00}$.

In the experiment, one applies a surface acoustic wave (SAW) to the
surface of the sample which couples piezoelectrically to the 2DEG
which is only slightly below the surface.  The rough idea is that the
SAW can excite modes in the 2DEG, thereby losing energy to the
electron.  By measuring the attenuation and velocity shift, one can
extract information about the response of the 2DEG at the frequency
and wavevector of the SAW.

To see more explicitly how this happens~\cite{SimonSAW}, we
consider a SAW of wavevector $q$, frequency $\omega = v_s q$ with
$v_s$ the wave velocity, and amplitude $I$.  The electric potential
applied to the 2DEG due to the piezoelectric coupling is $A_0^{\rm ext} =
\zeta(q d) I$ where $d$ is the depth of the 2DEG beneath the surface,
and $\zeta$ is some function related to the piezoelectric coupling,
which is usually assumed to be roughly constant for $qd < 4$ (See,
however, Ref.~\onlinecite{SimonSAW}).  The induced density
modulation is then $j_0 = K_{00} A_0^{\rm ext} = \zeta K_{00} I$ where, of
course, all quantities are at $q$ and $\omega = v_s q$.  The induced
density interacting with the externally applied field results in an
interaction energy density per unit area given by $\delta U =
\frac{1}{2} j_0 A_0^{\rm ext} = \frac{1}{2} \zeta I^2 K_{00}$.  The
mechanical energy density per unit area of the surface, on the other
hand, is given by $U = \gamma q I^2$ with $\gamma$ some materials dependent
constant (note that a wave has energy density $q^2 I^2$, then we
integrate to a depth of roughly one wavelength away from the surface
to get an areal energy density proportional to $q$).  Thus, the
fractional energy shift is given by
\begin{equation} 
\frac{\delta U}{U} = \frac{\zeta(q d)}{2 \gamma q} K_{00}(q,\omega = v_s q).
\end{equation}
This fractional energy shift can be thought of as the fractional
shift in frequency of the SAW.  Thus, the real part is the fractional
velocity shift, and the imaginary part will be the attenuation.  We
typically measure the velocity shift with respect to the shift that
occurs for a system with infinite conductivity.  For such an
infinitely conductive system, (See Eqs.  \ref{eq:K2sep} and
\ref{eq:sigPiV}) we have $K_{00} = 1/v(q)$.  Thus, with respect to
that system as a reference,
\begin{equation}
  \label{eq:deltavoverv0}
  \frac{\delta v_s}{v_s}  + i \frac{\kappa}{q} =  
\frac{\alpha^2}{2} \left[v(q) K_{00}(q, \omega=v_s q) - 1 \right].
\end{equation}
where $\kappa$ is the attenuation (In practice, the attenuation is
difficult to measure accurately).  Here we have made the conventional
definition that $\alpha^2/2 = \zeta(q d) /(2 \gamma q v(q))$.  The
coupling constant $\alpha$, which is a materials dependent
function,~\footnote{For GaAs samples with Coulomb interactions
  unscreened by a gate, $\alpha \approx 3.2 \times 10^{-6}$.  However,
  this coupling depends somewhat~\cite{SimonSAW} on the depth of the
  2DEG beneath the surface of the sample.}  is usually taken to be
roughly constant for $ qd < 4$.

We now use Eq.  \ref{eq:K2sep}, as well as Eq. \ref{eq:sigPiV} to
write
\begin{equation}
    K_{00} = 
    \frac{\epsilon q/(2 \pi)}{ 1 - i
    \sigma_{\rm m}/\sigma_{xx} }
\end{equation}
with 
\begin{equation}
  \label{eq:sigmam}
  \sigma_{\rm m} = \frac{v_s}{v(q) q}
\end{equation}
For the Coulomb form of the interaction, $\sigma_{\rm m} = v_s
\epsilon/(2 \pi)$ (Note, however, $\epsilon$ is weakly dependent on
$qd$ since the 2DEG is near a dielectric interface\cite{SimonSAW}).
We can now insert this expression into Eq. \ref{eq:deltavoverv0} to
obtain the often quoted result~\cite{SimonSAW,OLDSAW}
\begin{equation}
\label{eq:finalsaw}
  \frac{\delta v_s}{v_s}  + i \frac{\kappa}{q} = \frac{\alpha^2/2}{1 +
  i \sigma_{xx}/\sigma_{\rm m}}.
\end{equation}
where $\sigma_{xx}$ is evaluated at $q$ and $\omega  = v_s q$. 

In experiments by Willett~\cite{Willett0,Willett1}, it is found that
the measured surface wave velocity shift at long wavelength do indeed
reflect the DC conductivity $\sigma_{xx}$ of the 2DEG via this
formula. However, in order to obtain a good match between the DC
conductivity and the velocity shift, $\sigma_{\rm m}$ is treated as a
fitting parameter that may be as large as 4 times that predicted by
Eq. \ref{eq:sigmam}.  The reason for this discrepancy is unknown.  One
possible explanation is that long scale inhomogeneities in the system
make the macroscopic DC conductivity not accurately reflect the
microscopic conductivity of the system on the scale of the wavelength
of the SAW~\cite{SimonReslaw}.  Another possibility is that parallel
conduction in the donor layer of the samples screens the Coulomb
interaction, thus altering $v(q)$ and hence $\sigma_{\rm m}$ via Eq.
\ref{eq:sigmam}.  Until these details are sorted out, we should then
probably treat $\sigma_{\rm m}$ as a fitting parameter.

Since we have prescriptions for calculating response functions for the
composite fermion system, we should be able to convert this into a
prediction for the surface wave velocity shift.  We use Eq.
\ref{eq:CSrhoyy} to write $\sigma_{xx}$ of the electrons in terms of
$\rho_{yy}^{\rm CF}$ of the transformed fermions.  Using a
Landau-Boltzmann approach as discussed in section \ref{sec:FLT} above,
for $\nu=\frac{1}{2m}$, Eq.  \ref{eq:rhoyyex} gives us an expression
for $\rho_{yy}$ where we have included an $F_1$ coefficient and
current conserving scattering~\footnote{As mentioned above, it
  actually does not matter whether or not we worry about making sure
  the the scattering conserves current.}  For $q \gg 1/l$ (with $l =
\vf \tau$ a scattering length) we obtain $ \sigma_{xx}(q,\omega = q
v_s) = C q $ with $C$ a constant which is mostly real for $v_s \ll
\vf$.  This is agreement with our earlier result Eq. \ref{eq:sigres}.
This results in a prediction of
\begin{equation}
  \frac{\delta v_s}{v_s} \approx \frac{\alpha^2/2}{1 + \tilde C^2 q^2} 
\end{equation}
where $\tilde C$ must be treated as unknown due to our uncertainty in
the appropriate value of $\sigma_{\rm m}$.  Such a form of the velocity
shift is indeed in agreement with experiment~\cite{Willett1} for $q >
1/l$.  It should be noted, however, that the current experimental
data could also be in agreement with $\delta v_s/v_s \sim q$ for
larger $q$.  At lower $q$ where disorder is more important,
$\sigma_{xx}$ is seen to roll over be a constant (See Eq.
\ref{eq:sigres2}).
 
\begin{figure}[htbp]
  \begin{center}
    \leavevmode
    \epsfxsize=4in
     \epsfbox{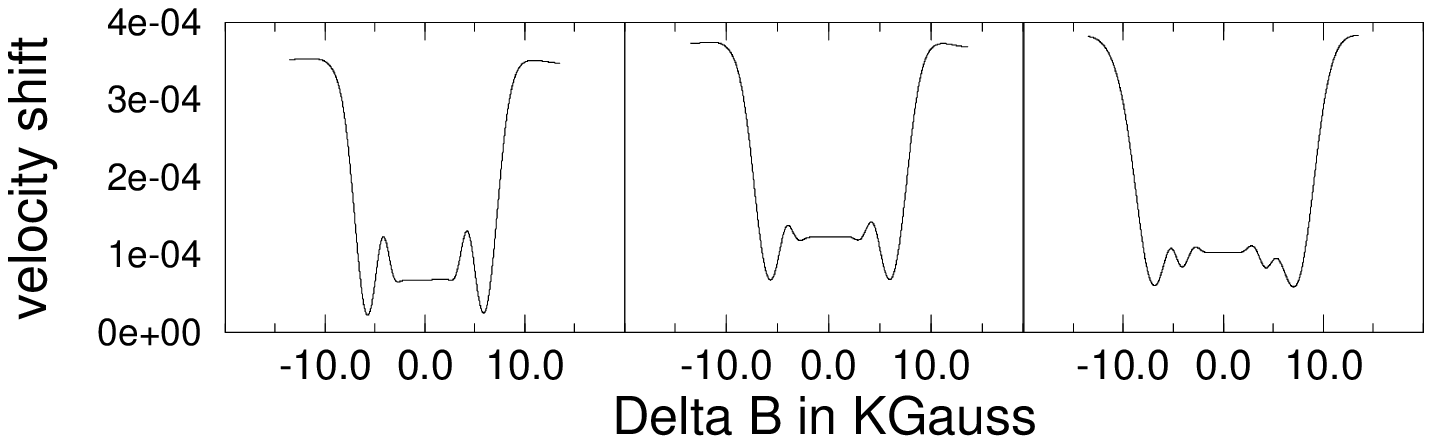}
  \caption{\protect{Predicted SAW velocity shifts for
    various values of the effective mass.}}
  \label{fig:saw}
  \begin{minipage}[t]{4.3in}  
    {\small From left to right $m^* = 5m_{\rm b}$, $10 m_{\rm b}$, and $25 m_{\rm b}$,
      and Parameters are $n=1.64 \times 10^{11}$, frequency $=10.7$
      GHz, scattering length $l = 1 \mu$, density inhomogeneity = $1
      \%$, and $\sigma_{\rm m} = 6.8 \times 10^{-7} \Omega^{-1}$.  These are
      chosen to agree with the experimental work of 
      Ref.~\onlinecite{Willett3}.  One should think of the $y$-axis as
      being somewhat re-scalable due to our uncertainty in the proper
      value of $\sigma_{\rm m}$.  The important agreement with theory is the
      location of the minimum (indicating commensuration of $2R_{\rm c}^*$
      and the SAW wavelength), which is relatively independent of
      $m^*$ up to at least $m^* \approx 15 m_{\rm b}$.}
  \end{minipage} 
   \end{center}
\end{figure}

A more impressive agreement of theory with
experiment~\cite{Willett2,Willett3} occurs when one looks at finite
$\Delta B$.  Resonances seen at finite $\Delta B$ roughly
indicate the commensuration of the SAW wavelength with the composite
fermion cyclotron diameter $2 R_{\rm c}^*$.  Using a Landau-Silin Boltzmann
approach (called the ``Modified Semiclassical'' Approach above in
section \ref{sub:silin}) with a current conserving scattering term as
discussed in section \ref{sec:FLT} above, one can obtain predictions
for the SAW velocity shift near $\nu=\frac{1}{2}$ that agree quite
well with experiments~\cite{Simon2un,WolfleScat}.  There are several
free parameters that must be put into the theory by hand here,
including the effective mass $m^*$, the scattering time $\tau$, the
constant $\sigma_{\rm m}$, as well as the fractional density inhomogeneity
of the sample (which has the effect of smearing the magnetic field
axis since the value of the filling fraction differs slightly from
place to place).  However, making a reasonable guess for $\tau$ and
the fractional density fluctuations (which can be roughly measured by
other experiments) and using a theoretical prediction for $\sigma_{\rm m}$,
we can make predictions for the SAW velocity shifts at various values
of the effective mass which are shown in Fig. \ref{fig:saw}.  Note,
however, that uncertainties in $\sigma_{\rm m}$ will mainly have the effect
of changing the scale of the vertical axis.  (These figures use
parameters appropriate to compare with the experimental data from Ref.
\onlinecite{Willett3}). We see that up $m^* \approx 15 m_{\rm b}$ the
predicted curves are relatively insensitive to the exact value of
$m^*$.  Above about $20 m_{\rm b}$, the location of the resonance and the
shape of the curve changes dramatically as the SAW frequency actually
moves above the lowest excitation mode of the system.  Thus, we can
put an upper bound on the value of the effective mass.

\subsection{Coulomb Drag}

In Coulomb drag experiments~\cite{Eisenstein}, two 2DEGs are positioned
close to each other and interact only via Coulomb forces.  A current
$I$ driven in once layer induces a voltage $V$ in the other layer, and
thus one can define a drag resistivity $\rho_d = V/I$.  To lowest
order in the interaction between the layers the drag can be written
as~\cite{Macdonald,Sterndrag}
\begin{equation}
  \label{eq:drag1}
  \rho_d  =   \frac{\hbar^2}{4 \pi T n_1 n_2 e^2} \int \frac{d\vec q}{(2
  \pi)^2} q^2 |U(q,\omega)|^2 \int_0^\infty d\omega 
\frac{[\Pi^v_{00}]_1(\vec q,\omega)
  [\Pi^v_{00}]_2(\vec q,\omega)}{\sinh^2(\hbar \omega/(2 T))}
\end{equation}
where $T$ is the temperature, $n_i$ is the density in layer $i$,
$[\Pi_{00}^v]_i$ is the density-density component of the polarization
in layer $i$ and $U$ is the screened interlayer interaction~\footnote{As
  discussed in reference~\onlinecite{Sterndrag} one must be careful to
  properly screen both the inter and intra-layer interactions.}.  We
note that the form of Eq. \ref{eq:drag1} requires that $\rho_D$ must
vanish as $T \rightarrow 0$.

Since we have simple prescriptions (RPA or other approximations) for
calculating $\Pi^v$, we can go ahead and use this formula to calculate
predictions for Coulomb drag experiments.  At filling fraction $\nu_1
= \nu_2 =\frac{1}{2}$, as we discovered in section \ref{sec:RPA}
above, $\Pi^v$ is expected~\footnote{Note that here we have a short
  range Coulomb interaction in each layer due to the screening of the
  other layer.}  to have a pole at $\omega \sim i q^3$. 

Plugging the form of $\Pi^v$ into the above expression, we find that the
$q$ integral is dominated by the region near the pole $q_0 \sim
\omega^{1/3}$.  The $\omega$ integral is then dominated by $\omega
\sim T$ so we obtain the result~\cite{Sterndrag,Otherdrag}
\begin{equation}
\label{eq:dragres}
\rho_d \sim T^{4/3}
\end{equation}
for $\nu_1 = \nu_2 = 1/2$.  We compare this to the result for drag
between two electron gases in zero field which scales as $\rho_d \sim
T^2$.  It should be noted that the unusual power law form for the
$\nu=\frac{1}{2}$ drag is a direct reflection of the overdamped mode.
The coefficient of the scaling law (Eq. \ref{eq:dragres}) , as well as
sub-leading corrections have also been calculated in reference
\onlinecite{Sterndrag} within the Chern-Simons theory.

Experiments by Eisenstein et al~\cite{Eisenstein} have measured
$\rho_d$ for the $\nu_1 = \nu_2 =\frac{1}{2}$ state.  For temperatures
above $100 \mbox{ mK}$, the $T^{4/3}$ prediction seems to agree fairly
well with experiment.  Furthermore, the magnitude (the coefficient of
$T^{4/3}$) agrees reasonably well with the prediction of Ref.
\onlinecite{Sterndrag}.  Agreement away from $\nu = \frac{1}{2}$
remains uncertain at the present.

At lower temperature ($T < 100{\mbox mK}$), results of these experiments are
extremely puzzling.  Most notably, $\rho_d$ does not always go to zero
as $T$ goes to zero, and thus cannot be described by Eq.
\ref{eq:drag1}.  One possibility is that the interlayer interaction
(although it may be weak) must be treated beyond second order -- and
perhaps may require nonperturbative treatment.  This may also suggest
the formation of a new type of state at low temperature with
interlayer coherence~\cite{Sterndrag3}.

\subsection{Activation Energy}
\label{subsub:expgaps}

One particularly significant set of experiments were measurements of
activation gaps~\cite{Gaps,Du2} in the Jain series of fractions $\nu =
\frac{p}{2p+1}$.  In these experiments, the longitudinal
resistivity $\rho_{xx}$ is measured as a function of temperature in
the middle of a plateau and is fit to an exponential form $\rho_{xx}
\sim \exp(-E_g/(2 T))$ over some range of temperatures.  It should be
noted that at high temperatures, the transport is not expected to be
activated, and at sufficiently low temperatures, a variable-range
hopping form is expected to be more appropriate~\cite{VariableRange}.
There is thus always some uncertainty in the value of the fitted gap.
It is found that near $\nu=\frac{1}{2}$, the gaps are given roughly by
\begin{equation}
  \label{eq:shif}
  E_g =  \alpha \Delta B  - \Gamma
\end{equation}
with some dimensioned constant $\alpha$ (which may have a weak $B$
dependence).  The prediction of composite fermion theory, would
give~\footnote{As discussed in section \ref{sub:div}, $m^*$ formally
  diverges as $\log \Delta B$ as one approaches $\nu=\frac{1}{2}$.  It
  is suspected, however, that this weak divergence would only be
  important very close to $\nu=\frac{1}{2}$ and is probably not
  observable in any experimental system. We also note that following
  the form of Eq. \ref{eq:mstarstar}, we expect that $m^*$ will might
  scale as $1/\sqrt{B}$ which would become more obvious farther away
  from $\nu=\frac{1}{2}$. }
\begin{equation}
E_g =\hbar \Delta \omega_{\rm c}^*  = \hbar e \Delta B/ (m^* c)
\end{equation}
Thus, the experimental measurement of $\alpha$ is thought to be a
measurement of the effective mass.  We note that although an offset
$\Gamma$ is also found for case of electron cyclotron gaps near zero
magnetic field~\cite{Gaps,Du2}, the current understanding of the effects of
disorder in the Chern-Simons system is quite crude, and there is no
theoretical derivation of the form of Eq.  \ref{eq:shif}.

\subsection{Shubnikov-deHaas Oscillations}

Other sets of experiments\cite{Leadly,Du2} have measured the temperature
dependences of the resistance oscillations around $\nu=\frac{1}{2}$.
Analogous to the situation around zero field, one expects to see
oscillations in the conductivity corresponding to the filling of
successive Landau levels where the disorder is strong enough to
prevent a quantum Hall state from fully forming.  There is, however, a
great deal of difficulty in analyzing this data.

For the case of electrons near zero magnetic field, one can typically
fits the experimental result to the Ando (Landau-Koshelev)
formula~\cite{Ando}
\begin{equation}
\Delta \rho_{xx} = \frac{X}{\sinh{X}} \exp \left(\frac{-\pi}{\omega_{\rm c}^*
    \tau}\right) \cos[2 \pi (\nu - \frac{1}{2} )]
\end{equation}
where $X = 2 \pi^2 k_b T/ (\hbar \omega_{\rm c}^*)$ and $\omega_{\rm c}^* = e
B/(m^* c)$.  This form has been derived for the case of electrons near
zero magnetic field.  By fitting experimental data to this form, one
can extract both an effective mass $m^*$ and a scattering time $\tau$.

Many groups have assumed that this Ando form should also hold for the
oscillations near $\nu = \frac{1}{2}$, where $B$ is replaced by
$\Delta B$ and $\nu$ replaced by $\nu_{\rm CF} = p = \nu/(2 \nu - 1)$.
It is found that this two parameter fit can match the data over a
reasonable range of temperatures if $m^*$ is allowed to vary as a
function of filling fraction.  However, it is not at all clear that
this form is appropriate.  Semiclassical calculations~\cite{Mirlin1}
suggests that the argument of the exponential $\pi/(\tau \Delta
\omega_{\rm c}^*)$ should actually be replaced by $[\pi/(\tau \Delta
\omega_{\rm c}^*)]^4$.  This unusual form is a result of the dephasing
caused by the Chern-Simons gauge field.  The experimental data can
indeed be fit to this form also.  However within this semiclassical
picture, when one tries to estimate the scattering time (based on the
supposed strength of the disorder), one obtains a result that differs
from experiment by a very large factor.

Perhaps the most serious problem with this interpretation of the data
is that the effect of long scale disorder is extremely important in
high~\footnote{In sufficiently low magnetic fields $\rho_{xy}$ is
  typically small enough not to influence $\rho_{xx}$ greatly, and
  these long scale inhomogeneities are much less important.}  magnetic
fields~\cite{SimonReslaw,MirlinPolyakov} essentially mixing the effects
of local $\rho_{xx}$ and $\rho_{xy}$.  Thus, a macroscopic measurement
of the resistivity may not probe the local longitudinal resistivity
at all.

\subsection{Geometric Experiments}
\label{sub:Rc}

As discussed at length above, one of the most surprising features of
the composite fermion theory is the emergence of the very large length
scale corresponding to the composite fermion cyclotron radius $R_{\rm c}^* =
\hbar \kf c/(e \Delta B)$ near filling fraction $\nu=\frac{1}{2}$.
Several experiments (in addition to the SAW experiments) have
attempted to observe this length scale directly by looking for the
commensuration of the cyclotron orbit with an externally defined
static structure.  These experiments should be considered to be
measurements of $\kf$.  Several of these experiments include anti-dot
experiments~\cite{Rc*} and ballistic electron
focusing~\cite{Goldman,Smet,Frost}.  These experiments do indeed
observe the existence of the scale $R_{\rm c}^*$ in agreement with theory.
In all of these geometric resonance experiments it should be noted,
that although the theory of ballistic fermion motion has been
reasonably well developed for electrons near zero
field~\cite{WeissTheory}, the theoretical analysis~\cite{Geisel,Smet}
near $\nu=\frac{1}{2}$ has been performed at the mean field level only
in a semiclassical approximation.  However, the agreement with these
crude calculations seem reasonable.

\mysection{Last Words}

\label{sec:theend}

Throughout this review we have maintained the theme of describing the
even denominator quantum Hall states as unusual Fermi liquids.
Starting with the mean field solution, we treated the Chern-Simons
interaction at RPA level in section \ref{sec:RPA} and found that
although many of the results we found were correct (DC Hall
conductivity, existence of overdamped mode), the energy scales were
incorrect.  Attempts to systematically correct this using perturbation
theory in section \ref{sec:Pert} were plagued with divergences for
$\nu=\frac{1}{2m}$.  However, a detailed study of these divergences
(and where they cancel) encourages us to believe that a Landau Fermi
liquid approach may be a reasonably justified approximation.  We
construct such a Landau approach in section \ref{sec:FLT} and
\ref{sec:Mag} and arrive at a picture of a quasiparticle with highly
renormalized (but finite) effective mass, and magnetization equal to
one Bohr magneton in the limit of large cyclotron energy.  In section
\ref{sec:wavefunction} we turn to the wavefunction picture of
$\nu=\frac{1}{2m}$ and develop the picture of the neutral dipole
composite fermion.  In the end, we discover that this new picture is
quite equivalent to the Chern-Simons approach.  Finally, in section
\ref{sec:exp} we have looked briefly at some of the experiments.

Although this review has covered a great deal of ground, it has left
even more ground uncovered.  Certainly there are large topics of
composite fermion physics that we have not even touched on.  Some of
the more glaring omissions (such as the physics of composite fermion
edge states~\cite{Edges} and the very large number of experiments that
have not already been mentioned here~\cite{WillettReview}) will be
addressed by other chapters of this book.  However, even counting
these works, we have still not discussed many of the more interesting
topics in composite-fermion physics. 

One particularly interesting theoretical topic that has been neglected
is the calculation of one electron Green's function and the physics of
tunneling into composite fermion systems~\cite{Tunneling}.  Also, we
have not said a single word about the spin of the composite
fermion~\cite{Spin}.  Nor have we said much about finite temperatures.
Yet another direction that seems particularly exciting at the moment
is the question of whether the composite fermion Fermi sea is unstable
to forming a paired~\footnote{One such possible state is the so-called
  Pfaffian~\cite{Pfaffian} which is unusual in that it has exotic
  nonabelian statistics.} (BCS-like) state~\cite{Pfaffian,Haldane}.

There are certainly many other works that that we could examine at
this point, and we would never come to the end of this chapter.  Even
more, we could fill several books mulling  over the many outstanding
questions in the field.   Needless to say, composite fermions
will keep many of us busy for years to come.

\section*{Acknowledgments} 

Most of what I know about this subject is a result of my
collaborations with Bertrand Halperin and Ady Stern.  It is a pleasure
to acknowledge this fruitful collaboration.  In preparing this review,
I have also benefited from discussions with many people including
(but not limited to) R. Shankar, G. Murthy, Y. B. Kim, R.  L.
Willett, S.  Kivelson, J. K. Jain, R. K. Kamilla, and N. d'Ambrumenil.

This work was supported by Lucent Technology, NSF Grants No.
DMR-94-16910 and DMR-95-23361, by the ISI foundation, and by the
ESPRIT 8050 Small Structures program.

%%%%%%%%%%%%%%%%%%%%%%%%%%%%%%%%%%%%%%%%%%%%%%%%%%%%%%%%%%%%%%%%%%%%
%%%%%%%%%%%%%%%%%%  APPENDICES %%%%%%%%%%%%%%%%%%%%%%%%%%%%%%%%%%%%%
%%%%%%%%%%%%%%%%%%%%%%%%%%%%%%%%%%%%%%%%%%%%%%%%%%%%%%%%%%%%%%%%%%%%

\appendix
\setcounter{section}{0}
\renewcommand{\theequation}{\Alph{section}.\arabic{equation}}
\setcounter{equation}{0}

\section{Noninteracting Response Functions in Zero Field}
\label{app:non}
\setcounter{equation}{0}
\label{sub:non}

For noninteracting fermions of mass $m^*$ in zero magnetic field, the
current-current correlator (Eq. \ref{eq:currentcorrelator0} or Eq.
\ref{eq:currentcorrelator2}) can be calculated to yield expressions
given in Eqs. 2.18-2.19 of Ref.~\onlinecite{HLR}.  These integrals can
be evaluated exactly to yield the result
\begin{eqnarray}
  K_{00}(q,\omega) &=& \frac{m^*}{2 \pi Q^2} \left(A_- B_- - A_+ B_+
  \right)  \\  
  K_{11}(q,\omega) &=&  \frac{1}{6 \pi
  Q^2 m^*} \left( -Q^2 +   A_+ B_+ C_+ - A_- B_- C_-\right)
\end{eqnarray}
where $Q = q/\kf$, $\Omega = (\omega + i0^+)/(q \vf), \,\,\,\,\, A_\pm =
\Omega \pm Q^2/2, \,\,\,\, B_{\pm} = 1 - \sqrt{1 - Q^2/A_{\pm}^2}$, and
$C_{\pm} = 1 - A_\pm^2/Q^2$.   

It will be useful to write down the long wavelength low frequency limit
of these results.  For $q \ll \kf$ and $\omega \ll q \vf$ the
expressions reduce to 
\begin{eqnarray}
\label{eq:non1}    K^0_{00} &=& \frac{m^*}{2 \pi} ( 1 +
\frac{i \omega}{q \vf} + \ldots)  \\
\label{eq:non2}    K^0_{11} &=& \frac{-q^2}{24 \pi m^*} + \frac{i
  \omega}{q} \frac{2 n_{\rm e}}{\kf} + \ldots  
\end{eqnarray}
(And clearly, in zero field, the off-diagonal components should be
zero).

It is sometimes convenient to use an approximate Boltzmann approach to
calculate the noninteracting response function.  Such an approach is
discussed in detail in section \ref{sec:FLT} in the context of Fermi
liquid theory.  Since such a calculation is particularly simple, we
will demonstrate it here. 

The transport equation in zero magnetic field with all Fermi liquid
coefficients set to zero (so $\delta \epsilon_1=0$) and the scattering
term taken to zero (or $\tau$ taken to $\infty$) is given by $ \left(
  -i \omega + i q \vf \cos \theta \right) \nu(\theta) = -e \vec E
\cdot \hatn n(\theta)$ which is just Eq. \ref{eq:kin1} rewritten.
The function $\nu$ is then given by
\begin{equation}
  \nu(\theta) = \frac{ \frac{-e}{iq \vf} \left( E_x \cos
    \theta + E_y \sin \theta \right)}{-\Omega + \cos \theta},
\end{equation}
with $\Omega = \frac{\omega}{ q \vf }$.  Integrating this expression
in Eq. \ref{eq:fldef} to obtain $\nusub{{\pm 1}}$ and hence the current
via Eqs. \ref{eq:currentfourx0} and \ref{eq:currentfoury0}, yields the
conductivity at finite frequency and wavevector
\begin{eqnarray}
  \label{eq:appres1}
   \sigma^0_{xx} &=& \frac{i k_{\mbox{\tiny{F}}} e^2}{q 2 \pi
    \hbar} \Omega \left[ \frac{1}{\sqrt{1 - \frac{1}{\Omega^2}}} - 1  \right]
  \\
  \label{eq:appres2} \sigma^0_{yy} &=& \frac{i k_{\mbox{\tiny{F}}} e^2}{q 2 \pi
    \hbar} \Omega \left[ 1 - \sqrt{1 - \frac{1}{\Omega^2}} \right].
\end{eqnarray}
and of course the off diagonal conductivity is zero in zero magnetic
field.  Note that $\sigma$ can be written as $q^{-1}$ times a function
only of $\Omega$.  Expressions for Boltzmann conductivity in finite
magnetic field~\cite{Harrison} are given explicitly in Refs.
\onlinecite{Simonhalp} and~\onlinecite{HLR} for example.

To approximately include disorder to the Boltzmann approach one simply
analytically continues $\omega \rightarrow \omega + i/\tau$.  However
as discussed in section \ref{subsub:scat} this is a non current
conserving scattering model.  For most purposes, however, the
difference between results of different scattering models is small. 

Using $K^0 = T \sigma^0 T$ (analogous to Eq. \ref{eq:K2sep} and
\ref{eq:sigPiV} where $v=0$) we can obtain an approximate expression
for the response of electrons in zero field.  We note that at finite
wavevector but very low frequency, the Boltzmann approach and the
exact expression do not agree.  This is a reflection of the fact that
the Boltzmann equation does not model the static compressibility of
the system correctly, nor does it properly obtain the diamagnetic
term~\footnote{This inaccuracy of the Boltzmann equation was the
  motivation for redefining the conductivity in Ref.~\onlinecite{HLR}
  such that the Boltzmann equation correctly obtains the low frequency
  limit.}.  At slightly higher frequencies however, the Boltzmann
expression is quite accurate.

\section{RPA in yet another language}
\label{app:RPAs}
\setcounter{equation}{0}

An RPA treatment of the Chern-Simons interaction was used in earlier
works on anyon superconductivity~\cite{Anyons,AnyonSupercon}.  These
works derive the RPA equations explicitly from a perturbative approach
and obtain a prescription that, although seemingly different, is
precisely equivalent to the HLR form of the RPA.  The different
appearance stems from an apparently different treatment of diamagnetic
terms.  In this appendix we will show how the HLR RPA can be derived
from the earlier form of RPA given in Refs.  \onlinecite{Anyons} and
\onlinecite{AnyonSupercon} by a simple regrouping of the perturbation
series.

In Refs.~\onlinecite{Anyons} and \onlinecite{AnyonSupercon}, the
diamagnetic term is separated out of the noninteracting response
function as (See Eq.  \ref{eq:K0D0}) $ K_0 = D_0 + E $.  In this
approach, an RPA prescription for the full current-current correlator
is given as
\begin{eqnarray}
  D^{-1} &=& D_0^{-1} - W \label{eq:DW} \\ W^{-1} &=& (C + V)^{-1} - E
\end{eqnarray}
where $V$ is the Coulomb interaction and $C$ is the Chern-Simons
interaction~\footnote{In Refs.~\onlinecite{Anyons,AnyonSupercon}
  the Coulomb interactions is set to zero.  Also in these references
  $W$ is called $V$.}.  From this full correlation function, the full
response is given by
\begin{eqnarray}
  K &=& \Lambda + E \label{eq:KLam1}\\ \Lambda &=& (1 + EW) D ( 1 +
  WE)  \label{eq:KLam2}
\end{eqnarray}
Noting that Eq. \ref{eq:DW} is equivalent to $D = D_0 W D + D_0$ can
rewrite Eqs. \ref{eq:KLam1} and \ref{eq:KLam2} as
\begin{eqnarray}
  K &=& (D_0 W D + D_0 + EWD)( 1 +  WE)   + E  \nonumber
  \label{eq:longK1}\\  &=& (D_0 + E) (W DW + W) (W^{-1} + E) 
\end{eqnarray}
where we have used the fact that $EWE=0$. Using $K_0 = D_0 + E$ as
well as the fact that Eq. \ref{eq:DW} implies that $(W^{-1} -
D_0)^{-1} = WDW + W$ we obtain
\begin{eqnarray}
 K  &= & K_0 (W^{-1} - D_0)^{-1} (W^{-1} + E)  \nonumber \\
 &=& K_0 ([C+V]^{-1} - E - D_0) (C+V)^{-1} \nonumber
  \\ &=& K_0 ( [C+V]^{-1} - K_0) [C+V]^{1} \nonumber \\ &=& K_0 ( 1 -
 [C+V] K_0)^{-1} \nonumber \\
 &=&  ( [K^0]^{-1} - C - V )^{-1} 
\end{eqnarray}
which is precisely the HLR prescription given by Eq. \ref{eq:BigRPA}
(except for $-$ signs which are a result of having used a different
convention for $e$).

%%%%%%%%%%%%%%%%%% REFERENCES %%%%%%%%%%%%%%%%%%%%%%%%%%%%%%%%%%%%%%%%%%%

\end{document}